\documentclass[%
 aip,
 amsmath,amssymb,
 reprint,%
]{revtex4-1}
\usepackage{graphicx}
\usepackage{subfigure}
\usepackage{dcolumn}
\usepackage{bm}
\usepackage{color}
\usepackage[T1]{fontenc}
\usepackage{mathptmx}
\usepackage{multirow}
\usepackage{array}
\usepackage{float}
\usepackage{url}
\usepackage{CJKutf8}
\usepackage{titlesec}
\titleclass{\subsubsubsection}{straight}[\subsection]
\newcounter{subsubsubsection}[subsubsection]
\renewcommand\thesubsubsubsection{\thesubsubsection.\alph{subsubsubsection}}
\titleformat{\subsubsubsection}
 {\normalfont\normalsize\bfseries}{\thesubsubsubsection}{1em}{}
\titlespacing*{\subsubsubsection}
{0pt}{3.25ex plus 1ex minus .2ex}{1.5ex plus .2ex}

\begin{document}
\preprint{AIP/123-QED}

\title{Discrete Boltzmann modeling of high-speed compressible flows with various depths of non-equilibrium}

\author{Dejia Zhang \begin{CJK*}{UTF8}{gbsn} (张德佳) \end{CJK*}}
 \affiliation{State Key Laboratory for GeoMechanics and Deep Underground Engineering, China University of Mining and Technology, Beijing 100083, P.R.China}
 \affiliation{Laboratory of Computational Physics, Institute of Applied Physics and Computational Mathematics, P. O. Box 8009-26, Beijing 100088, P.R.China}

\author{Aiguo Xu \begin{CJK*}{UTF8}{gbsn} (许爱国) \end{CJK*}}
 \thanks{Corresponding author: Xu\_Aiguo@iapcm.ac.cn}
\affiliation{Laboratory of Computational Physics, Institute of Applied Physics and Computational Mathematics, P. O. Box 8009-26, Beijing 100088, P.R.China}
\affiliation{HEDPS, Center for Applied Physics and Technology, and College of Engineering, Peking University, Beijing 100871, China}
\affiliation{State Key Laboratory of Explosion Science and Technology, Beijing Institute of Technology, Beijing 100081, China}

\author{Yudong Zhang \begin{CJK*}{UTF8}{gbsn} (张玉东) \end{CJK*}}
\affiliation{School of Mechanics and Safety Engineering, Zhengzhou University, Zhengzhou 450001, P.R.China}

\author{Yanbiao Gan \begin{CJK*}{UTF8}{gbsn} (甘延标) \end{CJK*}}
\affiliation{Hebei Key Laboratory of Trans-Media Aerial Underwater Vehicle, School of Liberal Arts and Sciences, North China Institute of Aerospace Engineering, Langfang 065000, China}

\author{Yingjun Li \begin{CJK*}{UTF8}{gbsn} (李英骏) \end{CJK*}}
\thanks{Corresponding author: lyj@aphy.iphy.ac.cn}
\affiliation{State Key Laboratory for GeoMechanics and Deep Underground Engineering, China University of Mining and Technology, Beijing 100083, P.R.China}%

\date{\today}

\begin{abstract}

The non-equilibrium high-speed compressible flows present wealthy applications in engineering and science.
With the deepening of Thermodynamic Non-Equilibrium (TNE), higher-order non-conserved kinetic moments of the distribution function are needed to capture the main feature of the flow state and evolution process.
Based on the ellipsoidal statistical Bhatnagar-Gross-Krook model,
Discrete Boltzmann Models (DBMs) that consider various orders of TNE effects are developed to study flows in various depths of TNE.
In numerical tests, DBMs including the first up to the sixth order TNE effects are demonstrated.
Specifically, at first, the model's capability to capture large flow structures with zeroth-order TNE effects in two types of one-dimensional Riemann problems is demonstrated.
And, the ability to capture large flow structures with first-order TNE effects is also shown in the Couette flow.
Then, a shock wave structure given by Direct simulation Monte Carlo is used to verify the model's capability to capture fine structures at the level of the mean free path of gas molecules.
Further, we focus on the TNE degree of two colliding fluids mainly decided by two parameters, the relaxation time $\tau$ and relative speeds $\Delta \mathbf{u}$ of two fluids.
Consequently, three numerical tests for flows with various depths of TNE are constructed.
Due to any definition of the TNE strength is dependent on the perspective of investigation, we propose to use a $N$- component vector $\mathbf{S}_{TNE}$ to describe the TNE system from $N$ perspectives.
As specific applications, we
use a three-component vector  $\mathbf{S}_{TNE} = (\tau, \Delta \mathbf{u}, \bm{\Delta_{2}^{*}})$ to roughly characterize three cases for numerical tests in this work.
Then, we check the system TNE behavior from the perspective of
the $xx$ component of the TNE quantity, viscous stress $\bm{\Delta_{2}^{*}}$.
It is found that, for the first two cases, at least up to the second-order TNE effects, i.e., the second-order terms in Knudsen number in the CE expansion,
should be included in the model construction; while for the third case, at least up to the third-order TNE effects should be included.
Similar to $\bm{\Delta_{2}^{*}}$, three numerical tests for flows in various depths of $\bm{\Delta_{3,1}^{*}}$ are constructed.
It is found that from the perspective of $\Delta_{3,1,x}^{*}$, for case 1 and case 3, at least up to the second-order TNE effects should be required; while for case 2, the first-order TNE effects are enough.
These findings demonstrate that
 the inadequacy of focusing only on the few kinetic moments appearing in Navier-Stokes increases with the degree of discreteness and deviation from thermodynamic equilibrium.
Finally, a two-dimensional free jet is simulated to indicate that,
to obtain satisfying hydrodynamic quantities, the DBM should include at least up to the third-order TNE effects.
This study is meaningful for the understanding of the TNE behavior of complex fluid systems and the choice of an appropriate fluid model to handle desired TNE effects.
\end{abstract}

\maketitle

\section{\label{sec:level1} Introduction}

The high-speed compressible flow which contains complex hydrodynamic and thermodynamic non-equilibrium (HNE and TNE
\footnote[1]{ Generally, the non-equilibrium described by hydrodynamic equations is called hydrodynamic non-equilibrium (HNE), and the non-equilibrium described by kinetic theory due to deviation from thermodynamic equilibrium is called thermodynamic non-equilibrium (TNE).
Clearly, the HNE is only one part of TNE.
}
) effects are universal in nature and engineering\cite{2002Boltzmann,Xu2018-Chap2,1960Rarefied,2016ChenPOG,2021XuACTA,2021XuCJCP,2021XuACTAA,Ding2017PRL,Luo2019JFM2,Luo2019JFM2,Ding2018CSB,
Qiu2020POF,Bao2022POF}. Navier-Stokes (NS) equations, based on the continuum hypothesis,  have long been applied to large-scale flows and slow behaviors.\cite{2016Fluid-mechanics} The continuum hypothesis implies that the mean free path of molecules $\lambda$ is negligibly small compared to the characteristic length $L$,  i.e., the Knudsen (Kn) number
\footnote[2]{Knudsen number can be defined as the ratio of the mean free path of molecules $\lambda$ to the characteristic length $L$ i.e., $Kn=\lambda/L$, where
$\lambda=c_{s}\tau$ with the relaxation time $\tau$ and the local speed of sound
$c_{s}$. The characteristic length $L$ depends on macroscopic quantity gradients. That is to say $L=\psi/\mid \nabla \psi \mid$ where $\psi$ represents the macroscopic quantities such as density $\rho$, temperature $T$, velocity $\mathbf{u}$, and pressure $p$. In non-equilibrium flows, the Kn number can also be defined as the ratio of relaxation time $\tau$ to the characteristic time $t_0$. Kn number is one of the common parameters to describe the non-equilibrium degrees of fluid systems from its own perspective. Generally, the larger the Kn number is, the deeper the TNE degree of the system is. However, due to the complexity of TNE behaviors of the system, the Kn number is inadequate in describing the TNE degrees of the system in some cases.}
is negligibly small.
However, in some fluid systems where the average Kn number or local Kn number is not always very small, which challenges the continuum hypothesis.
For example,  in the Inertial Confined Fusion (ICF) system \cite{Manuel2021MRE,Yao2020MRE,Cai2021MRE},  various time-spatial scales are coexisting. Among them, the relatively large mean free path of molecules (relaxation time) leading to high Kn number results in discrete and TNE effects.\cite{2021Shan-kinetic-effects,2020Cai-kinetic-effects} In the aerospace field, the low-density characteristic of gas molecules at high altitudes gives rise to a high Kn number and causes significant TNE or rarefied gas effects\cite{Tsien2012Superaerodynamics}.
Meanwhile, the spacecraft may pass through various flow regimes with different Kn numbers, including continuum regime, slip regime, transition regime, and free molecule flow regime, which creates the necessity for the cross-regime adaptive model.
Moreover, in some mesoscale applications, \cite{Arkilic1997Microelectromechanical} such as Micro-Electro-Mechanical System (MEMS),\cite{1998MEMS,Nie2002JSP} reservoir exploitation in the tight fissure, and heat transfer characteristics researches in micro-nano chips, the large Kn number effects result in at least two kinds of unusual behaviors, (i) large specific surface area and consequently strong near wall viscous effect\cite{2002LimPRE,Lockerby2005POF,2019ZhangGER}
 and (ii) the significant Knudsen layer effect,\cite{2022Zhang-AIPAdv-Slip} which may dominate the overall behavior of fluid systems.

Fundamentally, as shown by Table \ref{table1}, the flow can be divided into different regimes according to the value of Kn number, including the inviscid flow, continuum flow, slip flow, transition flow, and the free-molecular flow\cite{1960Rarefied}.
From Chapman-Enskog (CE) multiscale analysis\cite{1990Chapman}, through retaining various orders of Kn number (that means considering different orders of TNE effects), the Boltzmann equation can reduce to the corresponding macroscopic fluid equations which can be used for flows in the corresponding flow regimes.
For the case where the Kn number approaches 0, the Boltzmann equation reduces to the Euler equations, where there is no viscosity and heat conduction.
From the kinetics point of view, the Euler equations describe the case where the system is always at the thermodynamic equilibrium state, more strictly speaking, the system is always at the thermodynamic quasi-equilibrium state.
With increasing the Kn number, when only the first-order terms in the CE expansion need to be considered, the evolution of three conservative kinetic moments (density, momentum, energy) gives the Navier-Stokes equations.
When the second-order terms need to be considered, the evolution of three conservative kinetic moments gives the Burnett equations.
When the third and higher-order terms need to be considered, the corresponding hydrodynamic equations are generally referred to super-Burnett equations.
\emph{But it should be pointed out that, the Boltzmann equation is more than the corresponding hydrodynamic equations.}
When flows refer to the free-molecular regime, in addition to Direct simulation Monte Carlo (DSMC), the collisionless Boltzmann equation can also be adopted.

\begin{center}
\begin{table*}
\begin{tabular}{|p{1.5cm}<{\centering} |c| p{4cm}<{\centering}| p{5.5cm}<{\centering}| p{3cm}<{\centering}|}
\hline
Kn number& flow regime & Fluid model &CE expansion&order of Kn\\
\hline
Kn $\rightarrow$ 0 & inviscid flow & \multirow{2}{*}\centering{Euler Eqs.} & $f=Kn^0 f^{eq}$&$O(Kn^0)$ \\
\hline
0 $\sim$ 0.001& continuum regime & \multirow{2}{*}\centering{Navier-Stokes Eqs.}&$f=Kn^0 f^{eq}+Kn^1 f^{(1)}$&$O(Kn^1)$\\
\hline
0.001$\sim$0.1 & slip regime & \multirow{2}{*}\centering{NS Eqs. with slip boundary}&$f=Kn^0 f^{eq}+Kn^1 f^{(1)}$&$O(Kn^1)$\\
\hline
0.1$\sim$10 & transition regime & \multirow{2}{*}\centering{Burnett and super-Burnett Eqs. with slip boundary, DSMC}&$f=Kn^0 f^{eq}+Kn^1 f^{(1)}+Kn^2 f^{(2)}(+\cdots)$ &$O(Kn^2)(O(Kn^n))$ \\
\hline
Kn $>$10 & free-molecular regime & \multirow{2}{*}\centering{DSMC,
Collisionless Boltzmann Eqs.}&CE expression is invalid& \\
\hline
\end{tabular}
\caption{The Kn number, flow regimes, fluid models, expressions of $f$ in CE expansion, and the order of Kn number that should be retained in CE expansion, where $n$ represents the order of Kn number.
$f$ and $f^{eq}$($=\frac{n}{2\pi RT}(\frac{1}{2\pi IRT})^{1/2}\exp [-\frac{(\mathbf{v}-\mathbf{u})^2}{2RT}-\frac{\eta^2}{2IRT}]$ are the distribution function and Maxwellian(equilibrium) distribution function, respectively.
For example, when the Kn number of the flow is 0.1$\sim$1, the flow is in the transition regime.
In that case, by retaining to order $O(Kn^2)$ of Kn number (that means considering up to the second-order TNE effects), the Boltzmann equation can reduce to the Burnett equation which can be used to model the flows in the transition regime.
}
\label{table1}
\end{table*}
\end{center}

The traditional hydrodynamic method relies only on the evolution of three conserved moments to capture the main characteristics of a fluid system.
When the system is in a thermodynamic equilibrium state, three conserved moments are adequate to determine the distribution function $f$($f=f^{eq}$, where $f^{eq}$ is the equilibrium distribution function) and all its kinetic moments.
Namely, the whole system behaviors can be characterized by traditional macroscopic quantities (density, velocity, pressure, and temperature).
When the system deviates slightly from the thermodynamic equilibrium state, only relying on three conserved moments can approximate the main characteristics of $f$($f \approx f^{eq}$) and roughly determine the system behaviors.
However, with the deepening of TNE degree, it is entirely inadequate to rely only on the above few macroscopic quantities in order to capture the main characteristics of the system reasonably.
Consequently, to characterize the main feature of the flow state and evolution process properly, we have to rely on partial higher-order non-conserved moments, not only the low-order conserved moments.
The required order of kinetic moment increases with the deeper TNE degree.

Generally, there are three kinds of physical modeling methods (or models) for flows with various depths of TNE, i.e., microscopic, mesoscopic, and macroscopic modeling methods.
As a common macroscopic modeling method for transition flows, the Burnett equations, obtained from some kinetic methods such as Chapman-Enskog analysis, Grad's 13 equations method, etc.,  can be used to characterize transition flows to some extent.\cite{1990Chapman,2005Macroscopic,1936BurnettPLMS,Grad1949CPAM,2003Struchtrup}
When dealing with flows with deeper depths of TNE, super-Burnett equations (or higher-order super-Burnett equations) that involve extremely complex expressions are needed.
However, besides the complexity of theoretical derivation, the derived highly nonlinear Burnett stress and heat flux terms contain higher than second-order derivatives. The latter raises enormous challenges in numerical stability and is demanding on computation cost.\cite{Agarwal2001POF}
In addition, the boundary conditions for Burnett equations are still open problems.
The above factors all hinder the application of Burnett equations in high Ma number flow, direct simulation of large-scale flow, cross-regime problems, etc.
More importantly, as mentioned above, some higher-order kinetic moments which
are extremely valuable to understanding TNE behaviors are not included in the traditional macroscopic modeling method.
The microscopic modeling and simulation methods, such as the well-known Molecular Dynamics (MD) simulation,\cite{Sun2020POF,Ding2021Single,Xie2022POF} are capable of capturing much more behaviors for flows, but are restricted to small spatio-temporal scales due to the huge computing costs.
The mesoscopic method, generally related to kinetic theory in non-equilibrium statistical physics, can be roughly classified into two categories, the numerical method for solving Partial Differential Equation(s) (PDE) and the construction method of the physical model.
Currently, the former includes the direct solution of Boltzmann equations,\cite{1994Molecular,Wagner1992JSP} moment method,\cite{2005Macroscopic,1936BurnettPLMS,Grad1949CPAM} gas-kinetic scheme (unified gas kinetic scheme, discrete unified gas kinetic scheme and unified gas kinetic wave-particle), \cite{Xu1994JCP,Liu2019ACTA,Xu2010JCP,Guo2013PRE} Lattice Boltzmann Method (LBM),\cite{Nie2002JSP,2002LimPRE,Zhang2005PRE,Fei2019POF,Wang2021POF,Huang2022POF,
Wen2020PRE,Gu2022POF} etc.
The frequently used mesoscopic method for transition flow, DSMC, which was firstly proposed by \citet{1994Molecular}, has been promoted by many other researchers for its significant breakthrough in research on the supersonic flow of rarefied gas and heat transfer characteristics in microscale flows, etc.\cite{1994Molecular,Oran1998DIRECT,FAN2001JCP}
However, it is restricted to too much more time consumption and memory demand in the continuum-transition regime because its not `` low enough'' gas densities.
Also, the huge signal-noise ratio in low-speed flows has hampered its application in microscale flows.

The recently proposed Discrete Boltzmann Method (DBM)
\footnote[3]{ The DBM can also be interpreted as the Discrete Boltzmann Model or the Discrete Boltzmann Modeling method according to the context.}
is an effective modeling method mainly for such a ``mesoscale'' dilemma case that the macroscopic models are no longer reasonable or their physical functions are insufficient, and at the same time, the MD simulation can not access due to the too large spatio-temporal scale.\cite{Xu2018-Chap2,Xu2012-FoP-review,2015Xu-PRE,Xu2018-RGD31,2021XuCJCP,2018Gan-pre,Zhang2017Discrete}
As a theoretical modeling method, the primary strategy of DBM is as follows:
Decompose the complex problem into parts.
According to the research requirement, choose a perspective to study one set of kinetic properties.
Therefore, it is required that the kinetic moments describing this set of kinetic properties keep their values unchanged in the process of model simplification.
The research perspective and modeling accuracy should be adjusted according to the actual demand.\cite{2021XuACTA,2021XuCJCP,2021XuACTAA}
Based on the CE analysis,\cite{1990Chapman} via considering different orders of TNE effects (as shown by Table \ref{table1}), DBM can model for flows with various depths of non-equilibrium.
Different from the Kinetic Macroscopic Modeling (KMM) method, the DBM method is a kind of Kinetic Direct Modeling (KDM) method.
The KMM is to obtain the macroscopic model, which has the same physical functions as the DBM, from the kinetic theory. The macroscopic model is described by a set of Generalized Hydrodynamic Equations (GHEs). The GHEs are composed of evolution equations of not only the conservative moments but also the most relevant non-conservative moments.
Firstly, the difficulty of KMM increases sharply when higher-order TNE effects need to be considered. In fact, when considering only up to the third-order TNE effects, the process of deriving GHEs has become extremely difficult, let alone the higher-order cases.
Secondly, even if the GHEs can be finally derived, the GHEs involve stronger nonlinearity and higher-order spatial partial derivatives, and the term number increases sharply as the TNE degree/level rises, which raises the huge challenge for practical numerical simulation. Therefore,
as the TNE level rises, the KMM approach quickly becomes unviable.
As the TNE level rises, the complexity of the DBM approach increases, too, but at a much slower speed. So, it is expected that the DBM can go farther.
Because it does not need to obtain the complex GHEs.
The CE expansion is only used to quickly determine which kinetic moments should keep values in the model simplification process.
It should also mention that the CE expansion is often used to, but not the only option to determine the kinetic moments for keeping values in model simplification.
DBM approach applies also to the case where some other methods, such as the MD, indicate which kinetic moments should keep values in model simplification.

The purpose of DBM is to provide a feasible modeling method beyond the traditional macroscopic modeling for capturing the main features of systems as the non-continuity and TNE degree increase.
In 2012, \citet{Xu2012-FoP-review} pointed out that, under the framework of LBM and under the conditions that do not use non-physical Boltzmann equation and kinetic moments, the non-conservative moments of ($f-f^{eq}$) can be used to describe how and how much the system deviates from the thermodynamic equilibrium, and to check corresponding effects due to deviating from the thermodynamic equilibrium.
This was the starting point for the DBM approach.
In 2015, \citet{2015Xu-PRE} proposed to open phase space using the non-conservative moments of ($f-f^{eq}$) and describe the extent of TNE using the distance between a state point to the origin in the phase space or its sub-space.
In 2018, \citet{Xu2018-RGD31} further developed the non-conservative moment phase space description methodology.
They proposed to use the distance $D$ between two state points to roughly describe the difference between the two states deviating from their thermodynamic equilibriums, and the reciprocal of distance, $1/D$, is defined as a similarity of deviating from thermodynamic equilibrium.
The mean distance during a time interval, $\overline D$, is used to roughly describe the difference between the two corresponding kinetic processes, and the reciprocal of $\overline D$, $1/ \overline D$ is defined as a process similarity.
In 2021, \citet{2021XuCJCP} extended the phase space description methodology to any system characteristics. A set of (independent) characteristic quantities is used to open phase space, and this space and its sub-spaces are used to describe the system properties.
A point in the phase space corresponds to a set of characteristic behaviors of the system.
Distance concepts in the phase space or its sub-spaces are used to describe the difference and similarity of behaviors.
It should be noted that what DBM presents include two parts: i) a series of physical constraints on the model used by the physical problem, and ii) a series of schemes for checking the TNE and picking out as more as possible helpful information from the simulation data. Being different from the LBM extensively studied in the literature,\cite{2002Boltzmann,Guo2013lattice,Huang2015multiphase,Shi2021JFM,
Nie2002JSP,Qian1992EPL,Bhadauria2021POF,Sofonea2018PRE,Tian2011JCP,
Sun2011CMA,Chai2013PRE,Liang2021POF,Yu2021POF,Swift1995PRL,
Osborn1995PRL,Wagner1998PRL} and being similar to the KMM, the specific discretization scheme is not a part of the DBM. The discretization scheme itself is an open research topic.

Physically, the extent of TNE can not be fully described by a single parameter because any definition of TNE strength depends on the perspectives of investigation.
In the DBM phase space description method, in addition to the traditional description by gradients of macroscopic quantities (density, temperature, flow velocity, pressure, etc.), we can also adopt the relaxation time $\tau$, Kn, and the distance concepts in the phase space, to define the TNE strength from their corresponding perspectives.
The descriptions from various TNE perspectives are highly related to each other, but they differ in some ways.
Together, they constitute a more complete characterization of the non-equilibrium state.
Consequently, to obtain an accurate and complete description of the TNE strength of a non-equilibrium system, we should look at the system from $N$ angles and characterize it by a vector composed of $N$ components.
From the point of $\Delta_{2,xx}^{*}$, the $xx$ component of viscous stress $\bm{\Delta_{2}^{*}}$, \citet{2018Gan-pre} performed the multiscale simulations over a wide range of Kn number and characterized the non-equilibrium flows with two additional criteria, i.e., the relative TNE strength and TNE discrepancy instead of the Kn number itself.

Currently, the DBM has been applied in a variety of complex fluid systems such as combustion and detonation,\cite{2016Lin-CNF,2018Mesoscopic,Ji2022JCP,Shan2022JMES,Su2022CTP} fluid instability,\cite{Lai2016PRE,2017Lin-DDBM-RT,Chen2018POF,2019Kinetic,Chen2020POF,Ye2020Entropy,Lin2021PRE,Zhang2021POF,Chen2022FOP,Chen2022PRE} multiphase flow,\cite{gan2011PRE,Gan2015Soft,Zhang2019Matter,Zhang2020FOP} plasma system,\cite{Liu2022JMES} and other non-equilibrium flows\cite{Lin2018Binary,Lin2017SR}.
Based on considering up to the first-order TNE effects, these works provide a new perspective for the investigation of TNE behaviors of complex systems that cannot be obtained by the NS model.
Further, through considering higher-order TNE effects in modeling construction, several DBMs are capable of describing flows with a high Kn number.
\cite{Zhang2017Discrete,2018Gan-pre,2019Zhang-Shakhov,2022Zhang-AIPAdv-Slip,Gan2022JFM}
In 2018, \citet{2018Gan-pre} investigated high-speed compressible flows ranging from continuum to transition regime through a tran-scale DBM in which the second-order TNE effects are considered.
To improve the multi-scale predictive capability of DBMs to describe the thermo-hydrodynamic non-equilibrium intensity, \citet{Gan2022JFM} incorporated more higher-order independent kinetic moments in modeling construction.
The model in \citet{Gan2022JFM} is beyond the third-order super-Burnett level.
However, it is commonly recognized that the TNE behaviors of complex fluid systems are valuable but challenged.
For investigating the complex TNE behaviors of high-speed compressible flows, especially the TNE strength of systems, DBMs that considers various orders of TNE effects are developed.
Among these, up from the first to the sixth order TNE effects are demonstrated.
Meanwhile, it has long been realized that the Bhatnagar-Gross-Krook (BGK) collision operator\cite{BGK1954} in the simplified Boltzmann equation brings a problem that the Prandtl (Pr) number is fixed to unity, which causes the viscosity and heat conductivity to change simultaneously when the relaxation time is adjusted.
To remove this binding under the framework of single-relaxation-time, in this work, the model construction is based on the Ellipsoidal Statistical Bhatnagar-Gross-Krook (ES-BGK) model.\cite{1966ES,Zhang2017Discrete,Zhang2020POF}

The modeling method is presented in Section \ref{Model construction}.
Then, Section \ref{Numerical simulations} shows some numerical tests and results.
Section \ref{Conclusions} concludes the current paper.
Additional information, including tedious derivation, is given in the appendix.

\section{Model construction for DBMs that considers various orders of TNE effects }\label{Model construction}

\begin{figure*}[htbp]
\center\includegraphics*
[width=0.7\textwidth]{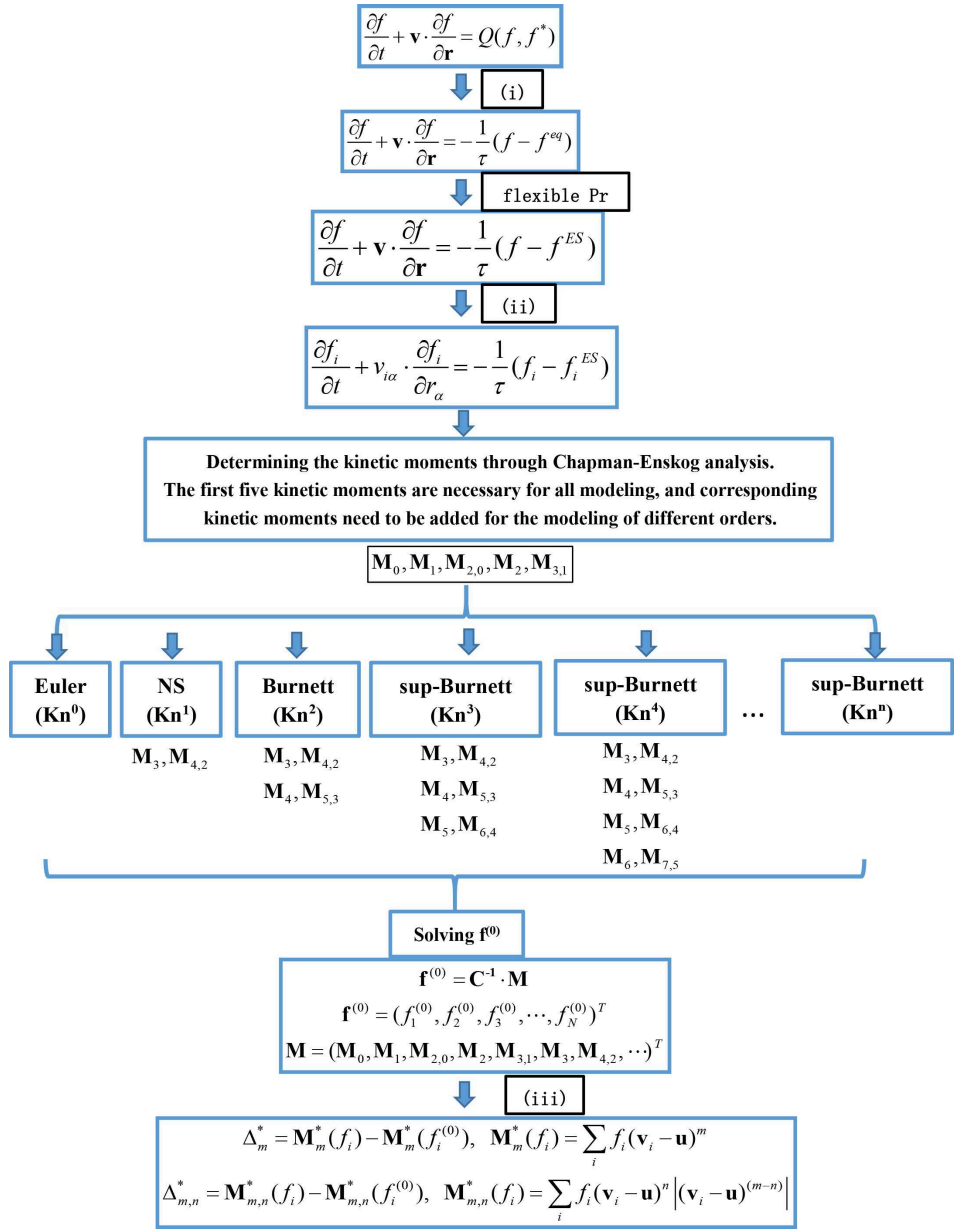}
\caption{ Flow Chart of DBMs with flexible Pr number or specific heat ratio considering various orders of TNE effects.
From the original Boltzmann to a DBM, three fundamental steps are needed.
Through the CE multiscale analysis, the required kinetic moments can be quickly determined.
For example, when considering the zeroth-order TNE effects, only five kinetic moments($\mathbf{M}_0$, $\mathbf{M}_1$, $\mathbf{M}_{2,0}$, $\mathbf{M}_2$, $\mathbf{M}_{3,1}$) are needed.
When considering up to the first-order TNE effects, seven kinetic moments($\mathbf{M}_0$, $\mathbf{M}_1$, $\mathbf{M}_{2,0}$, $\mathbf{M}_2$, $\mathbf{M}_{3,1}$, $\mathbf{M}_3$, $\mathbf{M}_{4,2}$) are needed.
When considering up to the second-order TNE effects, at least the zeroth-order to (5,3)th order kinetic moments(i.e., $\mathbf{M}_0$, $\mathbf{M}_1$, $\mathbf{M}_{2,0}$, $\mathbf{M}_2$, $\mathbf{M}_{3,1}$, $\mathbf{M}_3$, $\mathbf{M}_{4,2}$, $\mathbf{M}_4$, $\mathbf{M}_{5,3}$) are necessary, according to CE multiscale expansion.
Similarly, when developing a DBM in which the third-order (fourth-, fifth-, and sixth-order) TNE effects are considered, two more moments, i.e., $\mathbf{M}_5$ ($\mathbf{M}_6$, $\mathbf{M}_7$ and $\mathbf{M}_8$) and $\mathbf{M}_{6,4}$ ($\mathbf{M}_{7,5}$, $\mathbf{M}_{8,6}$, and $\mathbf{M}_{9,7}$), should be retained, respectively.
} \label{fig001}
\end{figure*}

Based on the ES-BGK single-relaxation model, DBMs that consider various orders of TNE effects with a flexible Prandtl number and specific heat ratio are presented.
For the bulk flow being far from boundary, from the original Boltzmann to a DBM, fas shown by the Flow Chart (Fig.\ref{fig001}), three fundamental steps are needed:
(i) Simplification and modification of the Boltzmann equation; (ii) Discretization of the particle velocity space; and (iii) Checking the TNE state and extracting TNE information.
The first two steps are for making the model simple enough but with sufficient physical function.
The third step is to present schemes for extracting helpful TNE information as more as possible.

\subsection{  Simplification and modification of the Boltzmann equation }

As a kind of mesoscopic method that naturally connects the macroscopic method and microscopic method, the original Boltzmann equation is in principle able to characterize the full spectrum of flow regimes.
However, the complex collision term which contains the high dimensional distribution functions before and after particles collisions, is complicated to solve directly, or its direct solution requires huge computing consumption.
For convenience, \citet{BGK1954} firstly proposed the well-known original BGK collision operator by introducing a local equilibrium distribution function $f^{eq}$ into the collision operator and writing it in a linearized form, i.e., $-\frac{1}{\tau}(f-f^{eq})$.
The starting point for them to obtain the original BGK operator is: on the constrain of single-relaxation time, only keeping the values of the first three low-order conserved moments and following the $H$-theorem. 
Therefore, the original BGK model describes a situation where the molecules' density and collision frequency are high enough, and the system is always in the quasi-equilibrium state.
Namely, the original BGK model characterizes a situation where the Euler equations do, in which the Kn number of the system is much less than 1 and $f \approx f^{eq}$.
However, in the vast majority of cases, the intermolecular correlations within the system are far from being as weak and simple as Boltzmann's equations require.
The large gradient or fast changing behavior of any physical quantity such as density, flow velocity, and temperature drives the system to deviate from the ``quasi-equilibrium" condition required by the original BGK-like.
\footnote[4]{ The BGK-like model refers to the model of Boltzmann equation which is similar in form to the BGK model. }
Therefore, in the strict sense, most of the kinetic behaviors of non-equilibrium flows cannot be described only by the pure kinetic theory based on the original BGK-like models. 
The actually used BGK-like models in the field can be regarded as a modified version incorporating the mean-field theory description. 
The mean field theory has two main responsibilities: (i) supplementing the description of intermolecular interaction potential effect omitted by the Boltzmann equation, and (ii) effectively extending the application scope of the BGK-like model to be suitable for a higher degree of non-equilibrium.
In fact, the BGK-like models used in the studies on non-equilibrium flow can be dynamically regarded as a modified Boltzmann equation.\cite{Xu20220304-DBM-note}

Different in physical function, there are many choices for BGK-like model such as the BGK model,\cite{Li2022CTP} ES-BGK model,\cite{1966ES,Zhang2017Discrete} Shakhov model,\cite{Shakhov1968FD,2019Zhang-Shakhov} Rykov model,\cite{2010LarinaCMMP} and Liu model,\cite{1990LiuPOF} etc.
To remove the bounding that the viscosity and heat conductivity change simultaneously when $\tau$ is adjusted, the ES-BGK model is adopted in the model construction.
Under the above considerations, the simplified Boltzmann equation, i.e., the ES-BGK-Boltzmann equation can be written as follow:
\begin{equation}
\frac{\partial f}{\partial t}+\mathbf{v}\cdot\frac{\partial f}{\partial \mathbf{r}}=-\frac{1}{\tau}(f-f^{ES})
\end{equation}
where $f^{ES}$ is
\begin{equation}
\begin{aligned}
f^{ES}&=n(\frac{m}{2\pi})^{\frac{D}{2}}\frac{1}{\sqrt{\left|\lambda_{\alpha\beta}\right|}}(\frac{m}{2\pi IkT})^\frac{1}{2}\\
& \times \exp[-\frac{m}{2}\lambda_{\alpha \beta}^{-1}(v_{\alpha}-u_{\alpha})(v_{\beta}-u_{\beta})-\frac{m\eta^{2}}{2IkT}]
\label{Eq.fES}
\end{aligned}
\end{equation}
with $m$, $n$, $\mathbf{u}$, and $T$ represent particle mass, particle number density, flow velocity vector, and temperature, respectively. $I$ is the extra degrees and $\eta$ is a free parameter that describes the energy of molecular rotation and vibration. $k$ is the Boltzmann constant and $D$ is the spatial dimension.
The modified term is $\lambda_{\alpha\beta}=kT\delta_{\alpha\beta}+\frac{b}{n}\Delta^{*}_{2,\alpha\beta}$ where $b$ is a flexible parameter related to Pr number, i.e., $\Pr=1/(1-b)$.
$\Delta^{*}_{2,\alpha\beta}$ represents viscous stress.
$\alpha$($\beta$) is the spatial coordinate.
In the ES-BGK model, the viscosity coefficient is $\mu=\Pr \tau P$ and the heat conductivity is $\kappa=c_p \tau P$, where $c_p$ is the specific heat at constant pressure.
Therefore, through adjusting $b$, the $\Pr$ number and $\mu$ are changed on the condition that $\kappa$ is unchanged.

\subsection{ Discretization of the particle velocity space }

The continuous-form Boltzmann equation, which describes the situation where the particle can move in any direction with a value of velocity ranging from $-\infty$ to $+\infty$, is difficult to simulate.
Different from conventional spatiotemporal discretization, DBM discretizes the particle velocity space.
By replacing the velocity space with a limited number of particle velocities, the continuous-form kinetic moment can be converted into the summation form for calculation.
The discrete form of the Boltzmann equation is
\begin{equation}
\frac{\partial f_{i}}{\partial t}+v_{i\alpha}\cdot\frac{\partial f_{i}}{\partial r_{\alpha}}=-\frac{1}{\tau}(f_{i}-f^{ES}_{i})
\label{Eq.Discrete-Boltzmann}
\end{equation}
where $i$ is the kinds of discrete velocities and $i=1$, $2$, $\cdots$, $N$. $N$ represents the total number of discrete velocities.
Therefore, $f_i$ does not represent the probability of velocity $\mathbf{v}_i$.
And, it is not the specific values of $f_i$ that are used when analyzing system behaviors, but the kinetic moments of $f$.
It requires that the reserved kinetic moments should keep their values unchanged after discretizing the velocity space, i.e., $\int f \Psi '(\mathbf{v}) d\mathbf{v}=\sum_{i} f_i \Psi '(\mathbf{v}_i)$,
where $\Psi ' = [1,\mathbf{v},\mathbf{vv},\mathbf{v\cdot v},\mathbf{vvv},\mathbf{vv\cdot v},\cdots]^{T}$ represent the reserved kinetic moments.
According to the CE analysis, the calculation of the kinetic moment of $f$ can be transformed into the calculation of the kinetic moment of $f^{eq}$.
Therefore, the constrain that should be obeyed in the discretization process is $\int f^{eq} \Psi ''(\mathbf{v}) d\mathbf{v}=\sum_{i} f^{eq}_i \Psi ''(\mathbf{v}_i)$.

Different from the standard LBM, the DBM distinguishes the physical modeling process and the selection process of discrete formats.
The standard LBM inherits a concise physical image of ``propagation+collision'' in a given way of ``virtual particles'' in the lattice gas method.
This simple image is helpful for its efficiency in the computational simulation of LBM.
However, this image imposes an additional ``burden'' on its interpretation using kinetic theory.
DBM is a kind of physical model construction method which gives the physical constraints required by the study of physical problems.
There is no restriction on the specific discrete scheme for DBM.
After obtaining a DBM, just like other models such as NS, it is necessary to choose an appropriate discrete scheme for simulation.

Mathematically, through solving the inverse matrix, the values of $f^{ES}_{i}$ can be confirmed.
Specifically, we write those kinetic moments (as shown in Appendixes \ref{sec:AppendixesB}) into a matrix form, i.e.,
\begin{equation}
\mathbf{C}\cdot\mathbf{f}^{ES}=\mathbf{\hat{f}}^{ES}
\tt{,}
\end{equation}
where $\mathbf{f}^{ES}$ and $\mathbf{\hat{f}}^{ES}$ represent vectors of dimension $N_{m} \times 1$ in velocity space and moment space, respectively.
$N_{m}$ is the number of kinetic moments.
$\mathbf{C}$ is the transformation matrix from moment space to velocity space, and its elements are determined by the DVM which we choose.
The discrete form of $\mathbf{f}^{ES}$ can be obtained as follow.
\begin{equation}
\mathbf{f}^{ES}=\mathbf{C}^{-1}\cdot \mathbf{\hat{f}}^{ES}
\tt{,}
\end{equation}
where $\mathbf{C}^{-1}$ is the inverse matrix of $\mathbf{C}$ obtained from Mathematica.

The elements of $\mathbf{\hat{f}}^{ES}$ depend on the specific depth of TNE.
For example, as shown by the Flow Chart \ref{fig001}, when constructing a DBM in which only the zeroth-order TNE effects are considered, five moments ($\mathbf{M}_0$, $\mathbf{M}_1$, $\mathbf{M}_{2,0}$, $\mathbf{M}_2$, $\mathbf{M}_{3,1}$, correspond to nine components, i.e., $N_m=9$) are enough.
When considering up to the second-order TNE effects, at least the zeroth-order to (5,3)th order kinetic moments(i.e., $\mathbf{M}_0$, $\mathbf{M}_1$, $\mathbf{M}_{2,0}$, $\mathbf{M}_2$, $\mathbf{M}_{3,1}$, $\mathbf{M}_3$, $\mathbf{M}_{4,2}$, $\mathbf{M}_4$, $\mathbf{M}_{5,3}$, correspond to nine components, i.e., $N_m=25$) are necessary, where ``5,3'' means that the fifth-order tensor is contracted to a third-order tensor.
Similarly, when developing a DBM in which the third-order (fourth-, fifth-, and sixth-order) TNE effects are considered, two more moments, i.e., $\mathbf{M}_5$ ($\mathbf{M}_6$, $\mathbf{M}_7$, and $\mathbf{M}_8$) and $\mathbf{M}_{6,4}$ ($\mathbf{M}_{7,5}$, $\mathbf{M}_{8,6}$, and $\mathbf{M}_{9,7}$), should be retained, respectively.
The kinetic moments are obtained by integrating $\mathbf{v}$ and $\eta$ with the continuous-form $f^{ES}$ (Eq. (\ref{Eq.fES})) through some softwares such as Mathematica.
The specific form of these formulas can be seen in Appendixes \ref{sec:AppendixesB}.

To determine the specific values of $\mathbf{f}^{ES}$, we also need to choose discrete velocity models (DVMs).
The construction of DVM depends on the number of reserved kinetic moments, the numerical stability, and computational efficiency.
To construct DBMs considering various orders of TNE effects, corresponding DVMs are selected, as shown in Table \ref{table-DVM}.
For improving computational efficiency, the total number of the discrete velocities $N$ is chosen to equal the number of moments $N_m$.
For example, to construct a DBM considering up to the second-order TNE effects with extra freedom of degree, at least 25 kinetic moments need to be considered, and a DVM with a total of 25 discrete velocities is chosen.
For highlighted, we call the DBM which considers up to the $s$-th order TNE effect the ``$s$-th model'', e.g., the 2-nd DBM represents a model that up to the second-order TNE effects are included in the modeling.

\begin{figure}[htbp]
\center\includegraphics*
[width=0.4\textwidth]{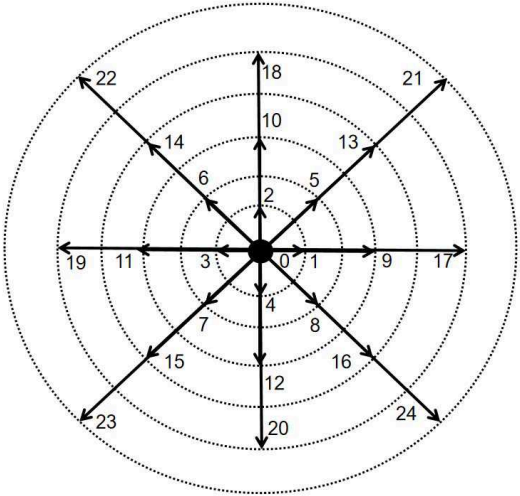}
\caption{Sketch of D2V25 model used in the present paper. The numbers in the figure represent the index $i$ in Eq. (\ref{Eq.Discrete-Boltzmann}).} \label{fig002}
\end{figure}

\begin{center}
\begin{table}
\begin{tabular}{| m{2.7cm}<{\centering} | m{2.7cm}<{\centering} |m{2.7cm}<{\centering}| }
\hline
Name & DVM(D2VN) & order of Kn bumber   \\
\hline
1-st order DBM & D2V16 & $O(Kn^1)$   \\
\hline
2-nd order DBM & D2V25  & $O(Kn^2)$  \\
\hline
3-rd order  DBM & D2V36  & $O(Kn^3)$  \\
\hline
4-th order  DBM & D2V49  & $O(Kn^4)$   \\
\hline
5-th order  DBM & D2V64  & $O(Kn^5)$   \\
\hline
6-th order  DBM & D2V81 & $O(Kn^6)$   \\
\hline
\end{tabular}
\caption{
The selected DVMs for various DBMs.
For improving computational efficiency, the total number of the discrete velocities $N$ is chosen to equal the number of moments $N_m$.
The third column represents the order of Kn number that needs to be considered when constructing an $s$-order DBM.
For example, when constructing a 2-nd order  DBM (i.e., retaining to order $O(Kn^2)$ of Kn number), a DVM with 25 discrete velocities is needed.
}
\label{table-DVM}
\end{table}
\end{center}
Sketches of the DVMs are as follow:
 \[\mathbf{v}_{i}=
\left\{
\begin{array}{lll}
(0,0), &M&=0, \quad i =0  \tt{,} \\
Mc[\textup{cos}\frac{(i-j)\pi}{2},\textup{sin}\frac{(i-j)\pi}{2}],  &M&=odd, i=4M $-$3  \sim  4M \tt{,} \\
Mc[\textup{cos}\frac{(2i-a)\pi}{4},\text{sin}\frac{(2i-a)\pi}{4}], &M&=even,  i=4M $-$3  \sim  4M \tt{.}  \\
\end{array} \label{Eq:DDBM-DVM1}
\right.
\]
 where $M$ represents the number of turns of the DVMs and $j$=4$M$-3, $a$=4$M$-7. $c$ is the discrete velocity. It should be noticed that when $N$ is even, there is no zeroth velocity (\emph{i}=0) in DVMs.
For understanding, we show the sketch of D2V25 which can be seen in Fig. \ref{fig002} (here $M$=0,1,2,3,4,5,6, $N=25$). The model of D2V36 is $M$=1,2,$\cdots$,9 and $N=36$. Other DVMs can be obtained similarly.

The specific values of  D2V25  are given in the following equations:
\[\mathbf{v}_{i}=(v_{ix},v_{iy})=
\left\{
\begin{array}{lll}
(0,0), &i& =0\tt{,} \\
c[\textup{cos}\frac{(i-1)\pi}{2},\textup{sin}\frac{(i-1)\pi}{2}],     &i& = 1-4 \tt{,} \\
2c[\textup{cos}\frac{(2i-1)\pi}{4},\textup{sin}\frac{(2i-1)\pi}{4}],  &i& = 5-8 \tt{,} \\
3c[\textup{cos}\frac{(i-9)\pi}{2},\textup{sin}\frac{(i-9)\pi}{2}],    &i& = 9-12 \tt{,} \\
4c[\textup{cos}\frac{(2i-9)\pi}{4},\textup{sin}\frac{(2i-9)\pi}{4}],  &i& = 13-16 \tt{,} \\
5c[\textup{cos}\frac{(i-17)\pi}{2},\textup{sin}\frac{(i-17)\pi}{2}],     &i& = 17-20 \tt{,} \\
6c[\textup{cos}\frac{(2i-17)\pi}{4},\textup{sin}\frac{(2i-17)\pi}{4}],  &i& = 21-24 \tt{.} \\
\end{array} \label{Eq:DDBM-DVM1}
\right.
\]
and the $\eta$ is flexible.
In this work, the sketch of $\eta$ in D2V25 is $\eta_{i}=\eta_{0}$ for $i=1-4$, $\eta_{i}=2\eta_{0}$ for $i=5-8$, and $\eta_{i}=0$ for $i=0$ and $i=9-24$.

\subsection{ Checking the TNE state and extracting TNE information. }

The most important process in constructing a DBM is providing a method for describing the TNE state and extracting TNE information.
In the traditional macroscopic fluid model, the commonly used parameters for TNE strength are Kn number, viscosity, heat conduction, and the gradients of macroscopic quantity, etc.
They all characterize the TNE strength of systems from their own perspectives.
However, they are all highly condensed, averaged, and coarse-grained description methods.
Some specific information can not be investigated directly through them, such as the internal energy in various degrees of freedom, viscous stress, heat flux, or higher-order kinetic moments.
Based on non-equilibrium statistical physics, DBM provides a more detailed description of TNE behaviors of complex fluid systems through the evolution of non-conserved kinetic moments of $(f-f^{eq})$.
Through defining various characteristic quantities which can describe the TNE state from different perspectives, the fundamental information of a specific non-equilibrium state and the non-equilibrium effects of flow can be extracted.
The fundamental characteristic quantities are written as follows:

\begin{equation}
\bm{\Delta}^{*}_{m}=\sum_{i}(f_{i}-f^{eq}_{i}) \underbrace { \mathbf{v}^{*}_{i}\mathbf{v}^{*}_{i} \cdots \mathbf{v}^{*}_{i} }_m
\tt{,} \label{Eq:DDBM-NF28}
\end{equation}
\begin{equation}
\bm{\Delta}^{*}_{m,n}=\frac{1}{2}\sum_{i}(f_{i}-f^{eq}_{i})(\mathbf{v}^{*}_{i}\cdot\mathbf{v}^{*}_{i}+\eta_{i}^{2})^{(m-n)/2} \underbrace { \mathbf{v}^{*}_{i} \cdots \mathbf{v}^{*}_{i} }_n
\tt{,}
\end{equation}
Here, $\mathbf{v}^{*}_{i}=\mathbf{v}_{i}-\mathbf{u}$ represents the central velocity, where $\mathbf{u}$ represents the macro flow velocity.
Mathematically, $\bm{\Delta^{*}_{m}}$ is $m$-order tensor and the subscript $m$ represents the number of $\mathbf{v}^{*}_{i}$.
$\bm{\Delta}^{*}_{m,n}$ means the $m$-order tensor contract to the $n$-order tensor with $n$ the number of $\mathbf{v}^{*}_{i}$.
For example, the TNE quantities that can be extracted in a 2-nd order DBM are as follows:
\begin{equation}
\bm{\Delta}^{*}_{2}=\sum_{i}(f_{i}-f^{eq}_{i})\mathbf{v}^{*}_{i}\mathbf{v}^{*}_{i}
\tt{,} \label{Eq:DDBM-NF28}
\end{equation}
\begin{equation}
\bm{\Delta}^{*}_{3,1}=\frac{1}{2}\sum_{i}(f_{i}-f^{eq}_{i})(\mathbf{v}^{*}_{i}\cdot\mathbf{v}^{*}_{i}+\eta_{i}^{2})\mathbf{v}^{*}_{i} \tt{,}
\end{equation}
\begin{equation}
\bm{\Delta}^{*}_{3}=\sum_{i}(f_{i}-f^{eq}_{i})\mathbf{v}^{*}_{i}\mathbf{v}^{*}_{i}\mathbf{v}^{*}_{i} \tt{,}
\end{equation}
\begin{equation}
\bm{\Delta}^{*}_{4,2}=\frac{1}{2}\sum_{i}(f_{i}-f^{eq}_{i})(\mathbf{v}^{*}_{i}\cdot\mathbf{v}^{*}_{i}+\eta_{i}^{2})\mathbf{v}^{*}_{i}\mathbf{v}^{*}_{i} \tt{,}
\end{equation}
\begin{equation}
\bm{\Delta}^{*}_{4}=\sum_{i}(f_{i}-f^{eq}_{i})\mathbf{v}^{*}_{i}\mathbf{v}^{*}_{i}\mathbf{v}^{*}_{i}\mathbf{v}^{*}_{i} \tt{,}
\end{equation}
\begin{equation}
\bm{\Delta}^{*}_{5,3}=\frac{1}{2}\sum_{i}(f_{i}-f^{eq}_{i})(\mathbf{v}^{*}_{i}\cdot\mathbf{v}^{*}_{i}+\eta_{i}^{2})\mathbf{v}^{*}_{i}\mathbf{v}^{*}_{i}\mathbf{v}^{*}_{i} \tt{.}
\end{equation}
Physically, $\bm{\Delta^{*}_{2}}=\Delta^{*}_{2,\alpha\beta}\mathbf{e}_{\alpha}\mathbf{e}_{\beta}$ represents viscous stress tensor, and $\bm{\Delta}^{*}_{3,1}=\Delta^{*}_{3,1}\mathbf{e}_{\alpha}$ indicates heat flux tensor, with $\mathbf{e}_{\alpha}$ the unit vector in the $\alpha$ direction.
The last four higher-order non-equilibrium quantities contain more condensed information.
$\bm{\Delta}^{*}_{3}=\Delta^{*}_{3\alpha\beta\gamma}\mathbf{e}_{\alpha}\mathbf{e}_{\beta}\mathbf{e}_{\gamma}$ and $\bm{\Delta}^{*}_{4,2}=\Delta^{*}_{4,2\alpha\beta}\mathbf{e}_{\alpha}\mathbf{e}_{\beta}$ represent the flux of viscous stress ($\bm{\Delta^{*}_{2}}$) in $\mathbf{e}_{\gamma}$ direction and the flux of heat flux ($\bm{\Delta}^{*}_{3,1}$) in $\mathbf{e}_{\beta}$ direction, respectively. From this perspective, $\bm{\Delta}^{*}_{4}=\Delta^{*}_{4\alpha\beta\gamma\chi}\mathbf{e}_{\alpha}\mathbf{e}_{\beta}\mathbf{e}_{\gamma}\mathbf{e}_{\chi}$ ($\bm{\Delta}^{*}_{5,3}=\Delta^{*}_{5,3\alpha\beta\gamma}\mathbf{e}_{\alpha}\mathbf{e}_{\beta}\mathbf{e}_{\gamma}$) indicates the flux of $\bm{\Delta}^{*}_{3}$ ($\bm{\Delta}^{*}_{4,2}$) in $\mathbf{e}_{\chi}$ ($\mathbf{e}_{\gamma}$) direction.
The TNE quantities of various orders of DBMs can also be extracted similarly.
When the sixth order non-equilibrium effects are considered, the non-equilibrium quantities $\bm{\Delta^{*}_{2}}$ to $\bm{\Delta^{*}_{8}}$, and $\bm{\Delta}^{*}_{3,1}$ to $\bm{\Delta}^{*}_{9,7}$ can be extracted.

Moreover, all the independent components of TNE characteristic quantities ($\bm{\Delta}_{m}^{*}$, $\bm{\Delta}_{m,n}^{*}$ etc.) constitute a high-dimensional phase space, in which the origin represents thermodynamic equilibrium state, and a specific point in phase space indicates a specific TNE state.
In the phase space, the distance $D$ between two state points is used to roughly describe the difference between the two states deviating from their thermodynamic equilibriums, and the reciprocal of distance, $1/D$, can define as the similarity of the two states deviating from thermodynamic equilibrium.
The mean distance during a time interval, $\overline D$, is used to roughly describe the difference between the two corresponding kinetic processes, and the reciprocal of $\overline D$, $1/ \overline D$ is defined as a process similarity.
Other coarse-grained quantities of TNE strength can also be defined according to the specific requirement.\cite{2021XuCJCP,2021XuACTA,2021XuACTAA}

It is clear that the definition of any non-equilibrium strength depends on the perspective of the investigation.
Complex systems need to be investigated from multiple perspectives.
If we look at the system from $N$ angles, there are $N$ kinds of non-equilibrium strengths.
Therefore, if the $N$ non-equilibrium strengths are taken as components to introduce a non-equilibrium strength vector,  $\mathbf{S}_{TNE}$,  it should be more accurate and specific to use this vector to describe the non-equilibrium strength of the system.

It should be pointed out that the mechanism of TNE in near-wall flow may be greatly different from that in the bulk flow. Consequently, the DBM for near-wall flow needs one more step, construction of kinetic boundary conditions. An example is referred to \citet{2022Zhang-AIPAdv-Slip}.

\section{ Numerical simulations and results }\label{Numerical simulations}

\begin{figure}[htbp]
\center\includegraphics*
[width=0.4\textwidth]{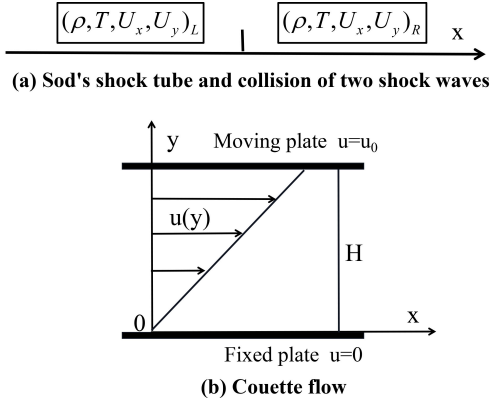}
\caption{ (a) Configurations of two kinds of Riemann problems. (b) Configuration of Couette flow. }
\label{fig003}
\end{figure}

In this section, four types of numerical validations of DBMs are performed.
In a word, we focus on flow scales from large to fine, physical quantities ranging from macroscopic to mesoscopic, and spatial dimensions from one-dimensional to two-dimensional.
(i) To show the model's capability to capture large flow structures of macroscopic quantities with zero-order TNE effects, comparisons between the simulation results and the analytical solutions of the one-dimensional Riemann problems (Sod's shock tube, collision of two strong shock waves) are performed by a 2-nd  order DBM.
And, the effect of $\Pr$ number on large flow structure with first-order TNE effects in Couette flow is investigated.
Configurations of Riemann problems and Couette flow are shown in Fig. \ref{fig003}, and the major initial quantities can be seen in Table \ref{table2}, where the subscript ``L'' (R) is the left (right) side of the flow field.
(ii) Then, a comparison between DBM simulation and DSMC result, which shows the capability of DBMs to capture fine flow structure of density profile at the level of the mean free path of gas molecules, is presented.
Specifically, a right-propagating shock wave with Ma=1.45 is simulated by a 2-nd order DBM, and the shock wave structure is compared with results from DSMC.
(iii) Further, the fine structures of TNE quantities in a simulation of head-on collisions between two compressible fluids are captured.
Through adjusting the relaxation time $\tau$ and relative speeds of two colliding fluids, three cases for flows in various strengths of $\bm{\Delta_{2}^{*}}$ are simulated by DBMs that consider various orders of TNE effects, and simulation results are compared with analytical solutions.
Similarly, through adjusting the relaxation time $\tau$ and relative pressure of two colliding fluids, three cases in various strengths of $\bm{\Delta_{3,1}^{*}}$ are also constructed.
The performances of different DBMs for describing various depths of TNE effects are shown.
(iv) Finally, the two-dimensional large flow structures of macroscopic quantities are captured.
The two-dimensional free jet is simulated by four DBMs that consider various orders of TNE effects.

Considering computational efficiency, numerical stability, and calculation accuracy, the first-order forward difference scheme and the second-order nonoscillatory nonfree dissipative (NND) scheme\cite{zhang1991nnd} are adopted to calculate the temporal and spatial derivatives in Eq. (\ref{Eq.Discrete-Boltzmann}), respectively.

\begin{center}
\begin{table*}[htbp]
\begin{tabular}{p{6cm}<{\centering}cp{8cm}<{\centering}}
\hline
Items& Pr number &Initial condition\\
\hline
\hline
1.Sod's shock tube& \multirow{2}{*}\centering{Pr=0.5,0.8,1.0,2.0,5.0}{}&
$\left\{
\begin{array}{l}
(\rho,T,U_{x},U_{y})_{L}=(1.0,1.0,0.0,0.0)\\
(\rho,T,U_{x},U_{y})_{R}=(0.125,0.8,0.0,0.0)
\end{array}
\right.$\\
\hline
2.The collision of two strong shock waves& \multirow{2}{*}\centering{Pr=0.8,1.0,2.0}{}&
$\left\{
\begin{array}{l}
(\rho,T,U_{x},U_{y})_{L}=(5.99924,76.8254,19.5975,0.0) \\
(\rho,T,U_{x},U_{y})_{R}=(5.99242,7.69222,-6.19633,0.0)
\end{array}
\right.$\\
\hline
3.Couette flow& \multirow{2}{*}\centering{Pr=0.8,1.0,2.0}{}&
$\left\{
\begin{array}{l}
(\rho,T,U_{x},U_{y})=(1.0,1.0,0.0,0.0) \\
u_{upper}=0.8,u_{bottom}=0.0,T_{upper}=T_{bottom}=1.0.
\end{array}
\right.$\\
\hline
\hline
\end{tabular}
\caption{The main initial conditions of flow field of Sod's shock tube, collision of two strong shock wave, and Couette flow.}
\label{table2}
\end{table*}
\end{center}

\subsection{ Description of large flow structure: Riemann problem and Couette flow }

\subsubsection{Sod's shock tube}\label{Sod' shock tube}
\begin{figure}[htbp]
\centering
\subfigure{
\begin{minipage}{4.2cm}
	\includegraphics[width=4.5cm]{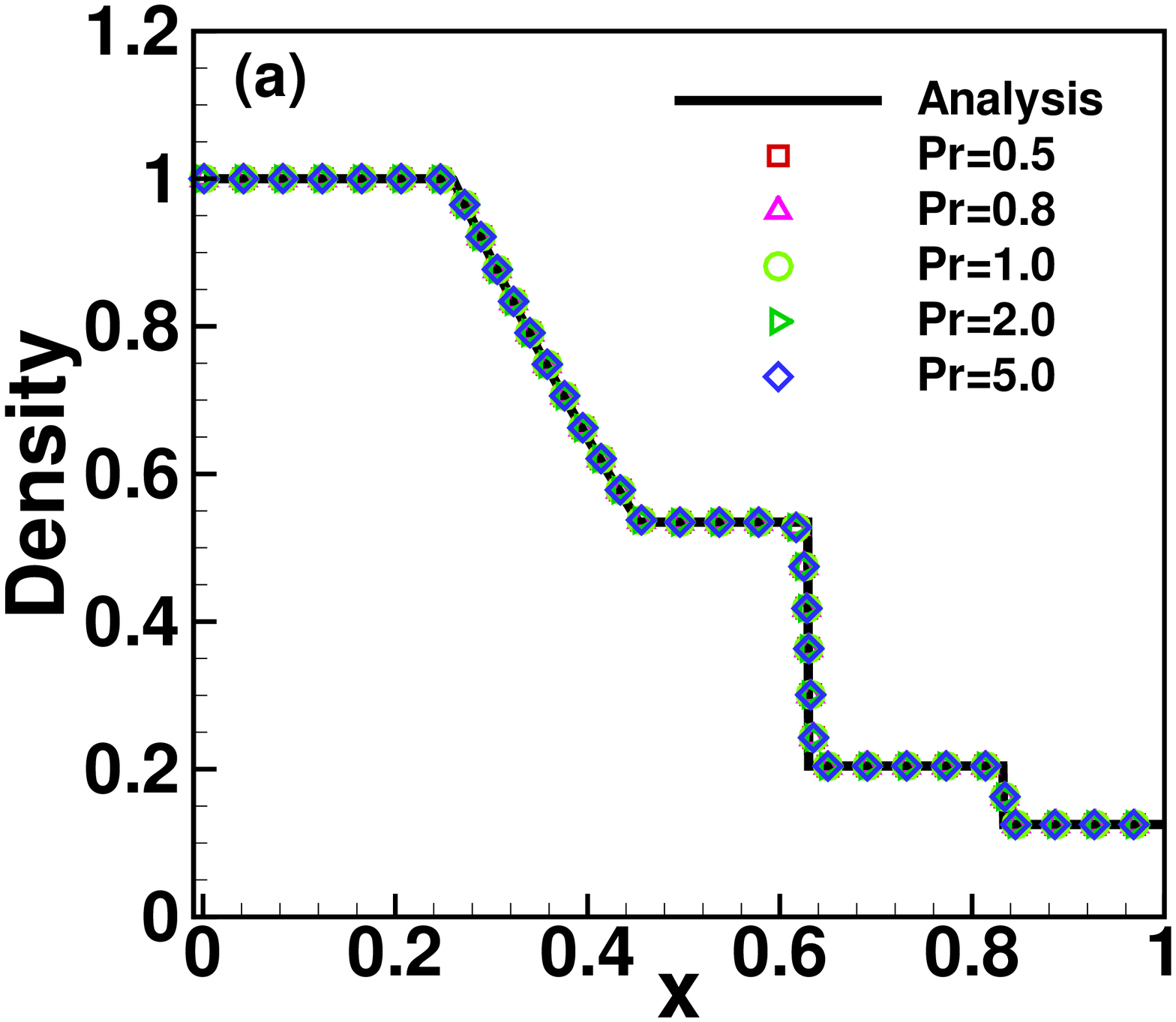}
\end{minipage}
\begin{minipage}{4.2cm}
	\includegraphics[width=4.5cm]{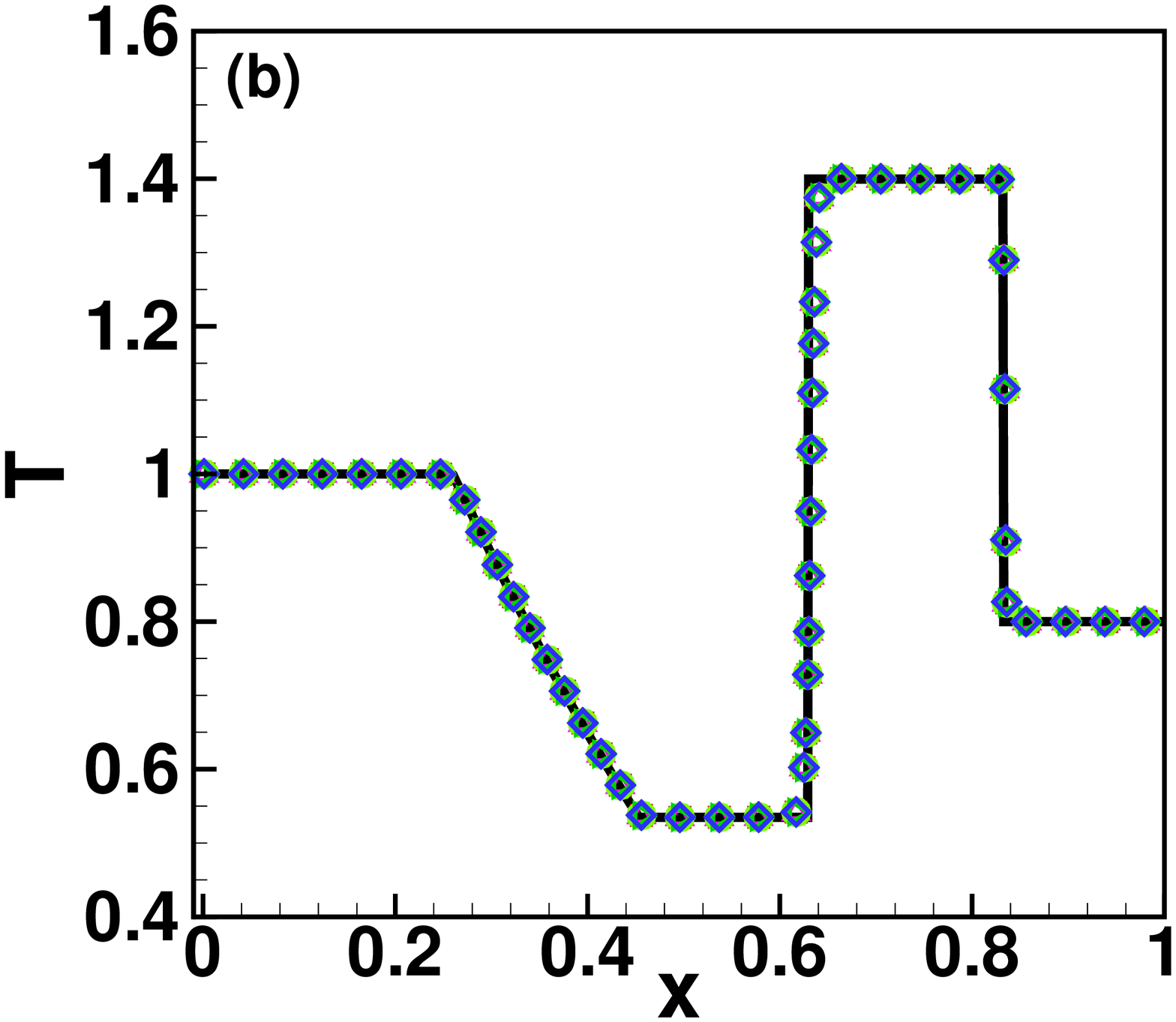}
\end{minipage}
}
\subfigure{
\begin{minipage}{4.2cm}
	\includegraphics[width=4.5cm]{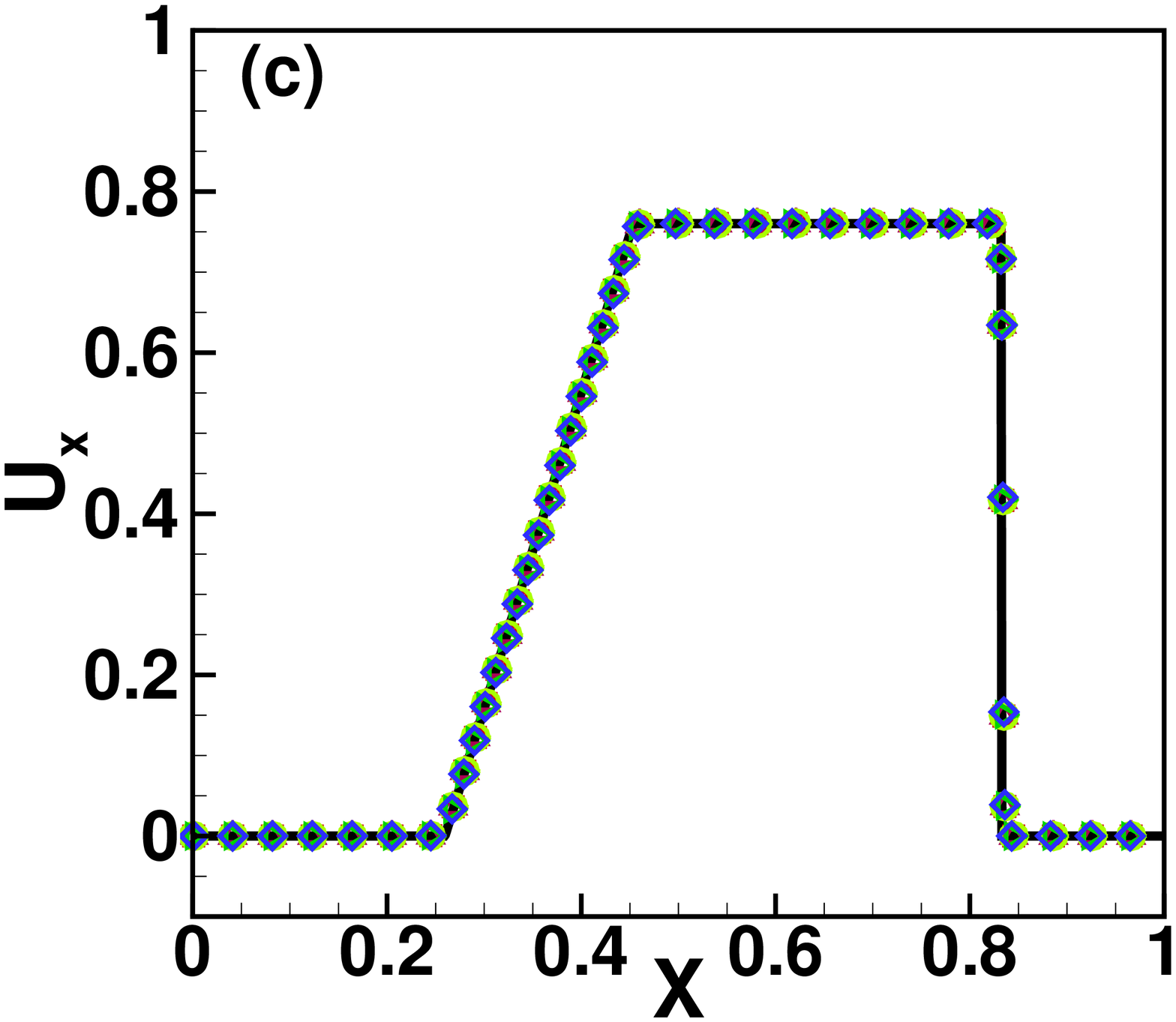}
\end{minipage}
\begin{minipage}{4.2cm}
	\includegraphics[width=4.5cm]{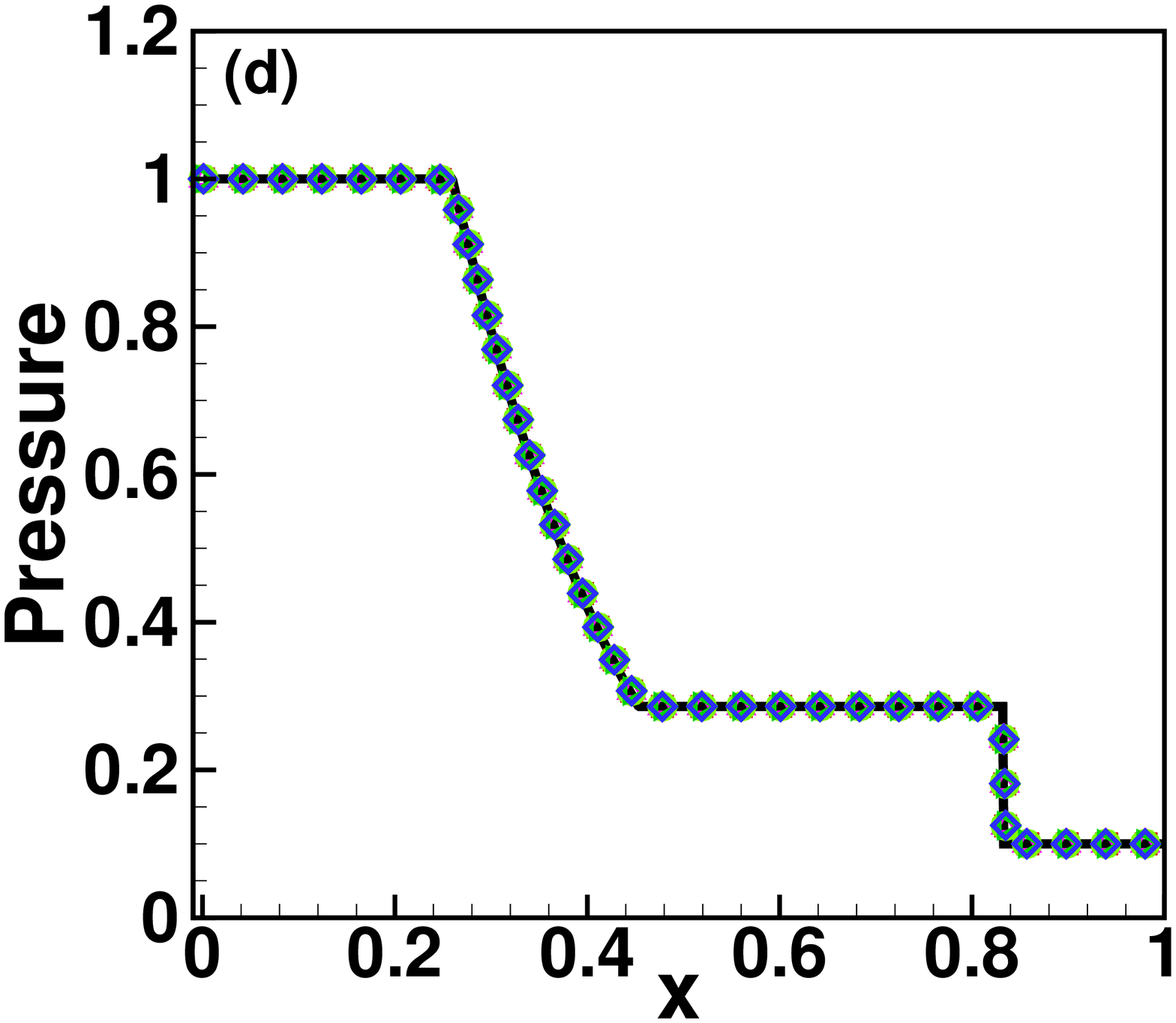}
\end{minipage}
}
\caption{
The large flow structure of macroscopic quantities of the Sod's shock tube, at $t=0.17$. (a) Density profile, (b) temperature profile, (c) $U_{x}$ profile, and (d) pressure profile.
The lines indicate Riemann solutions, and the simulation results are
 denoted by circles, squares, and triangles, corresponding to $\Pr$ = 0.5, 0.8, 1.0, 2.0, and 5.0, respectively.}
\label{fig004}
\end{figure}

The initial conditions of other quantities in Sod's shock tube are $c=0.8$, $m=1$, $\tau=4\times 10^{-6}$, $I=0(\gamma=2.0)$, $\Delta t=2\times10^{-6}$, $\Delta x=\Delta y=10^{-3}$, and $\eta=0$. The grid size is $N_{x}\times N_{y}=1000\times 1$.
In the $x$ direction, the zero gradient boundary condition is adopted in this simulation.
Shown in Fig. \ref{fig004} are the comparisons between the simulation results (the lines) and Riemann analytical solutions (the symbols) of density (a), temperature (b), velocity (c), and pressure (d) profiles at $t=0.17$, with $\Pr$ = 0.5, 0.8, 1.0, 2.0, and 5.0, respectively.
Clearly, the left-propagating rarefaction wave, contact discontinuity, and right-propagating shock wave are all captured accurately by DBM.
Then, because the large structure of macroscopic quantities depends on Euler equations which do not involve the effect of viscosity and heat flux.
Therefore, results from various $\Pr$ numbers (corresponding to various viscosity coefficients) exhibit almost consistent profiles.

\subsubsection{The collision of two strong shock waves}
\begin{figure}[htbp]
\centering
\subfigure{
\begin{minipage}{4.2cm}
	\includegraphics[width=4.5cm]{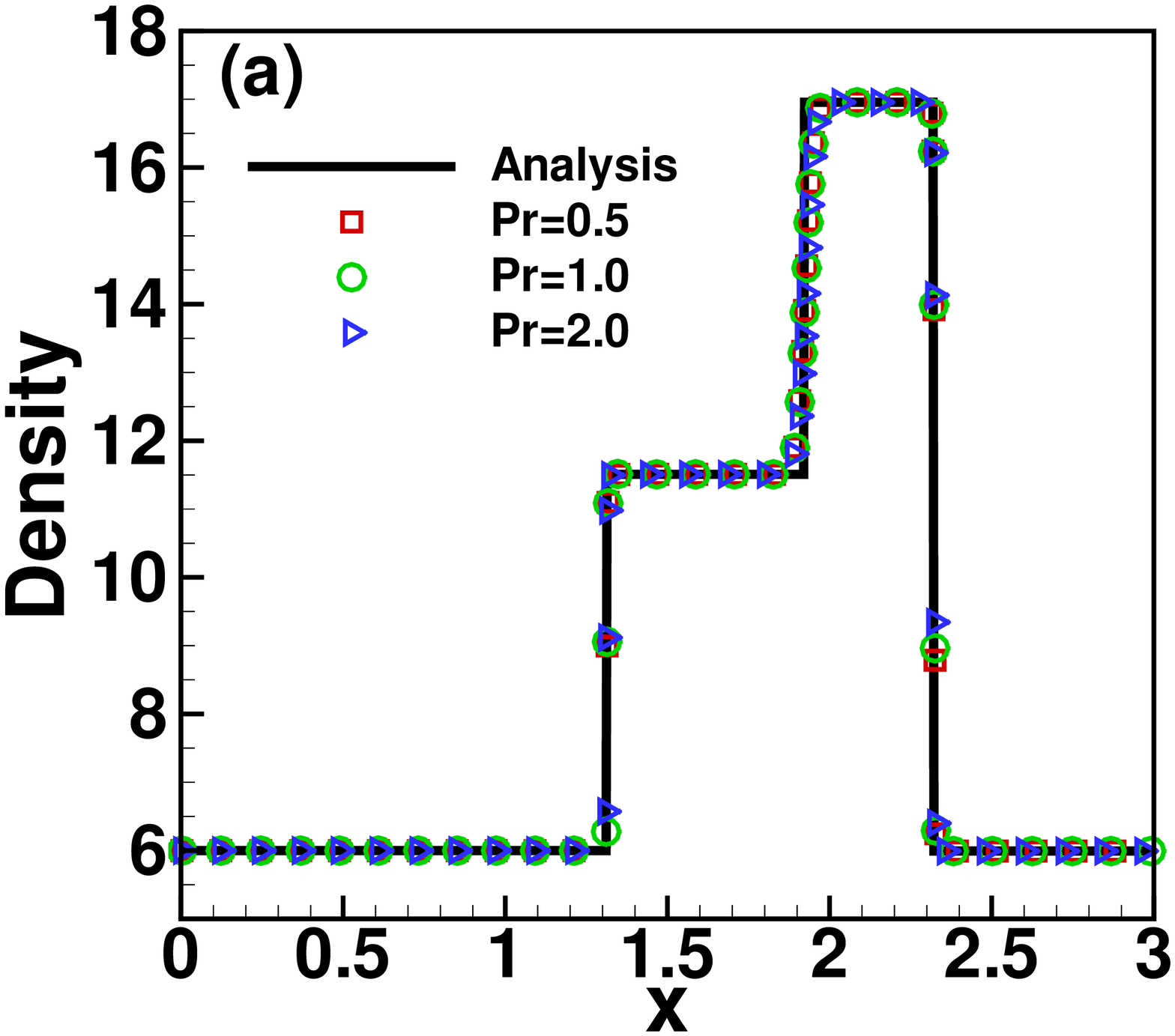}
\end{minipage}
\begin{minipage}{4.2cm}
	\includegraphics[width=4.5cm]{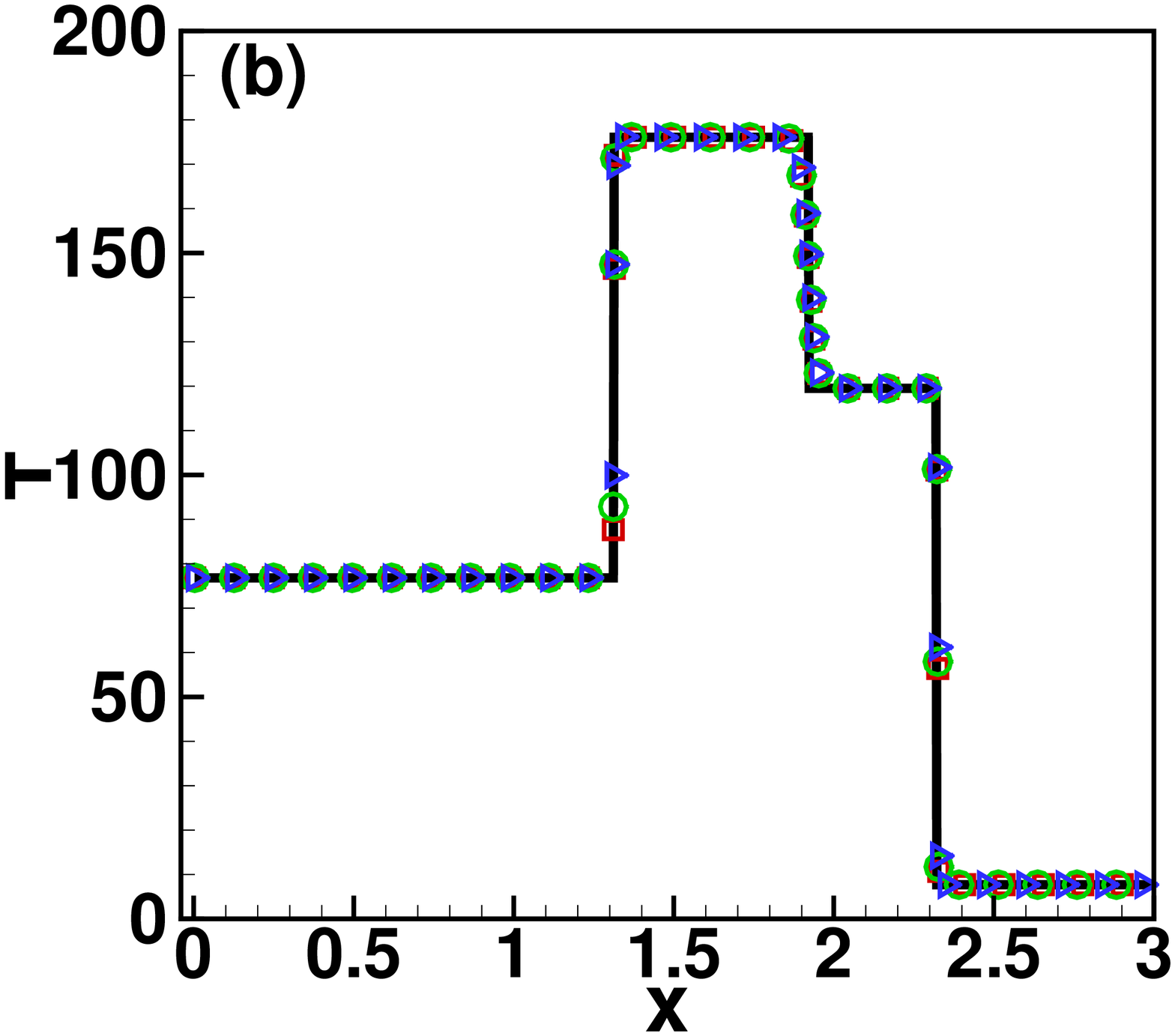}
\end{minipage}
}
\subfigure{
\begin{minipage}{4.2cm}
	\includegraphics[width=4.5cm]{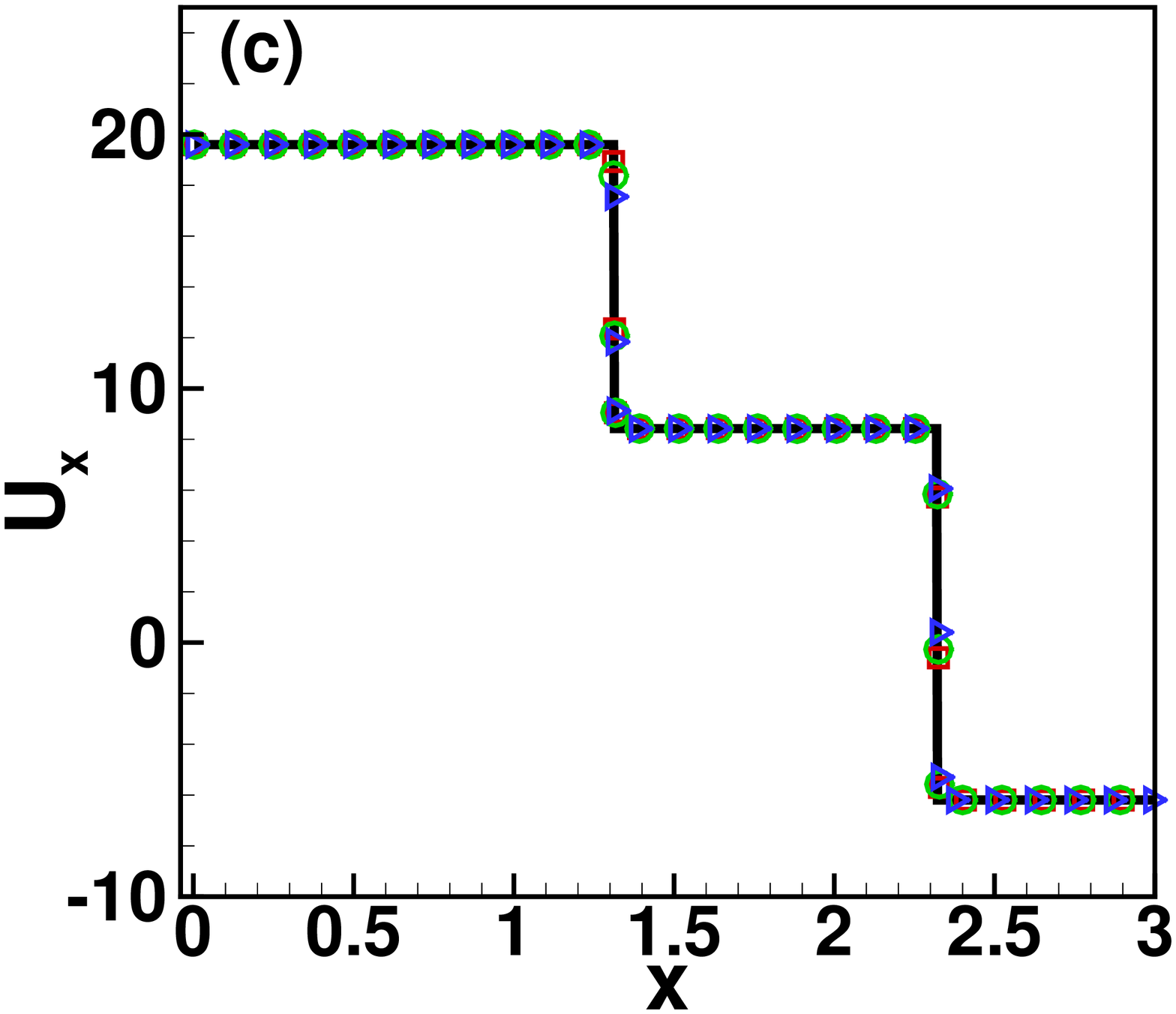}
\end{minipage}
\begin{minipage}{4.2cm}
	\includegraphics[width=4.5cm]{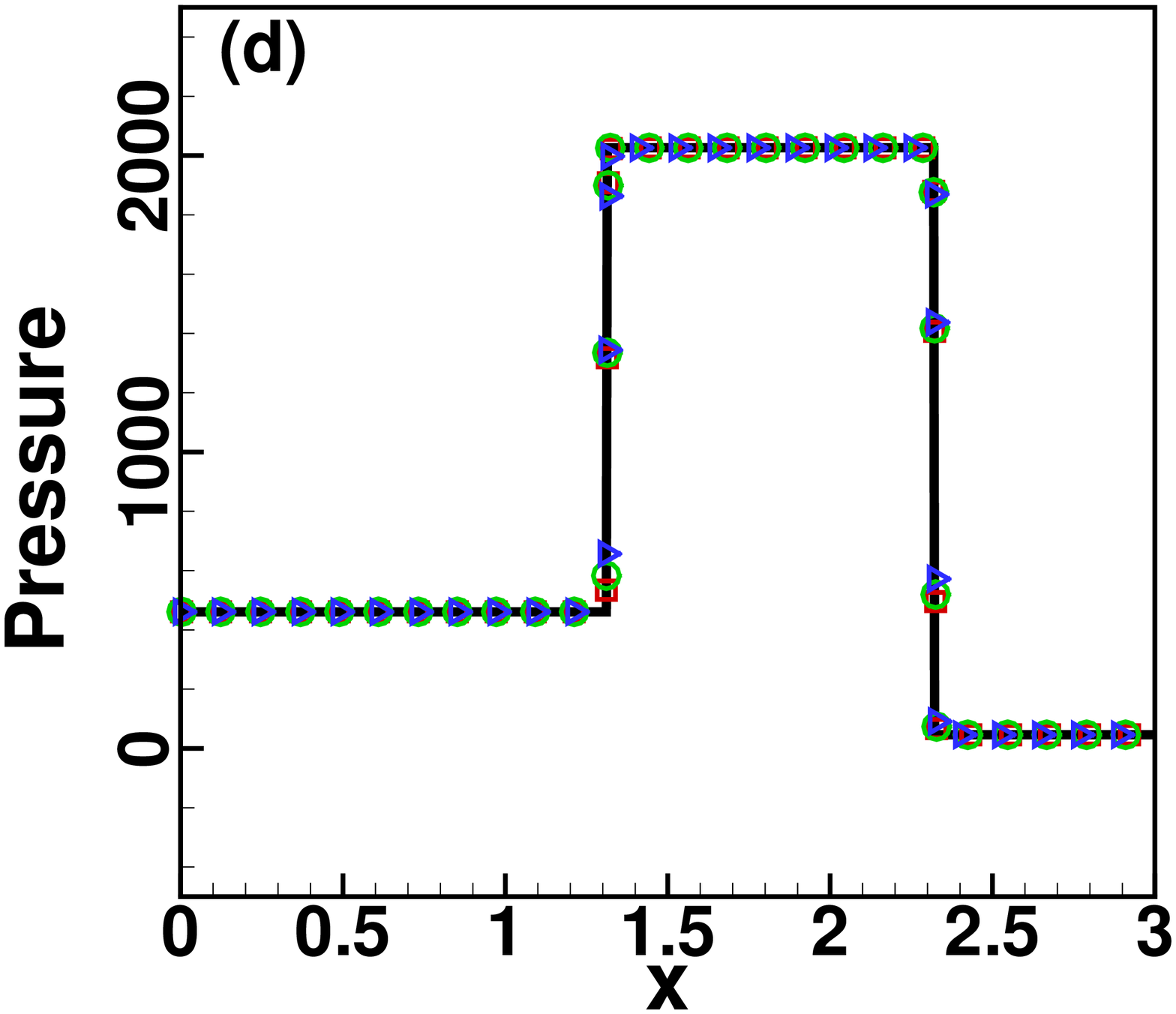}
\end{minipage}
}
\caption{
The large flow structure of macroscopic quantities of collision of two strong shock waves, at $t=0.05$. (a) Density profile, (b) temperature profile, (c) $U_{x}$ profile, and (d) pressure profile.
The lines indicate Riemann solutions, and the simulation results are denoted by circles, squares, and triangles, corresponding to $\Pr$ = 0.5, 1.0, and 2.0, respectively.}
\label{fig005}
\end{figure}

To further verify the robustness of the model in capturing strong shock with a high Ma number and the precision for compressible flows, we consider the collision of two strong shock waves.
The initial conditions of other quantities are $c=7.2$, $m=1$, $\tau=2\times 10^{-5}$, $I=0  (\gamma=2.0)$, $\Delta t=2\times10^{-6}$, $\Delta x=\Delta y=3\times 10^{-3}$,$\eta=0$. The grid size is $N_{x}\times N_{y}=1000\times 1$.
The zero gradient boundary condition is adopted in this simulation.
Shown in Fig. \ref{fig005} are the comparisons between the simulation results (the lines) and Riemann analytical solutions (the symbols) of density (a), temperature (b), velocity (c), and pressure (d) profiles at $t=0.05$, with $\Pr$ = 0.5, 1.0, and 2.0, respectively.
It is clear that a left-propagating shock wave and a right-propagating shock wave are both captured accurately by DBM, which indicates the proposed DBM is applicable to compressible flows with strong shock wave interaction.

\subsubsection{Couette flow}\label{Sjogreen's problem}

\begin{figure}[htbp]
\centering
\subfigure[]{
\begin{minipage}{6cm}
\centering
	\includegraphics[width=7cm]{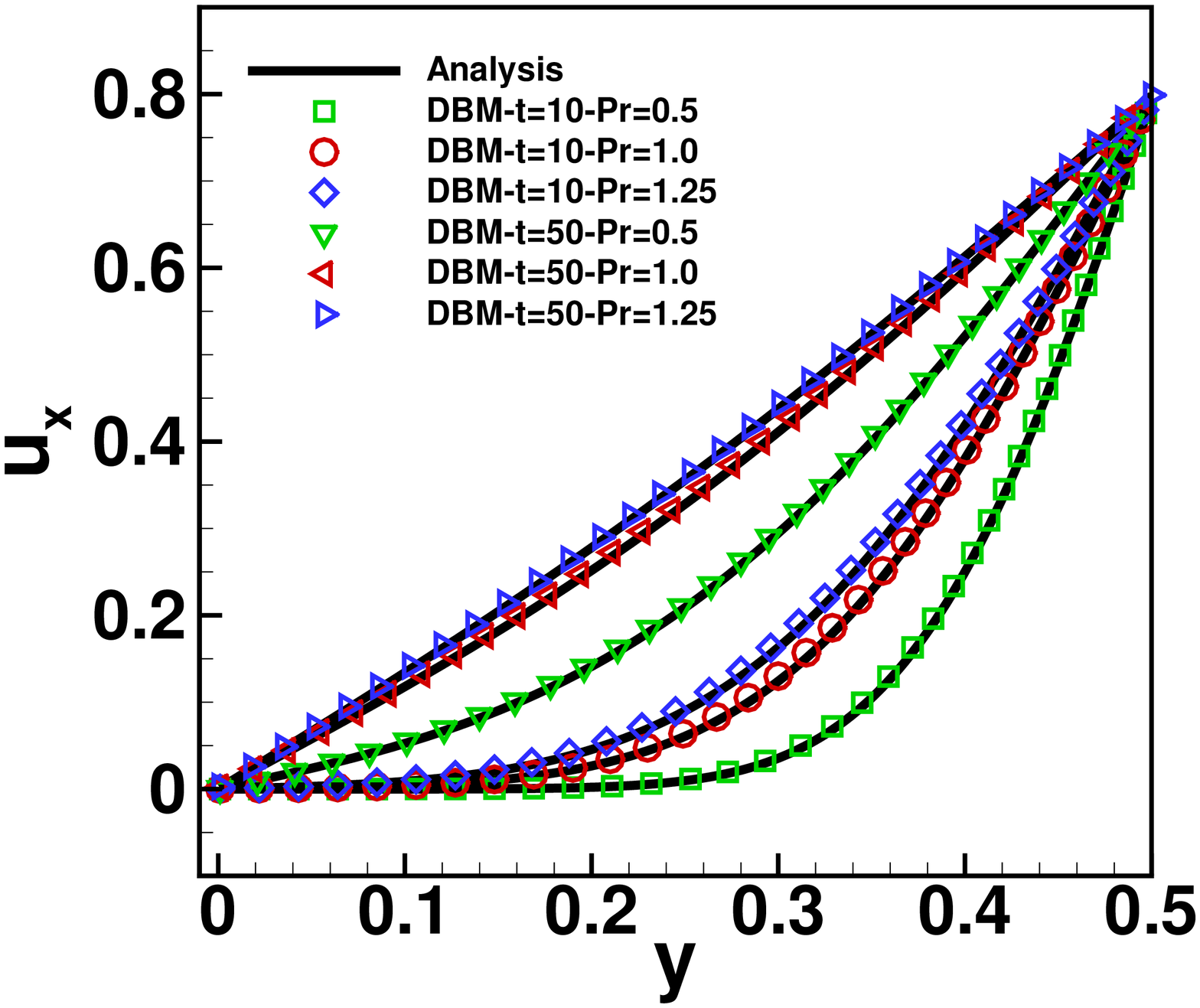}
	\label{fig006a}
\end{minipage}
}
\subfigure[]{
\begin{minipage}{6cm}
\centering
	\includegraphics[width=7cm]{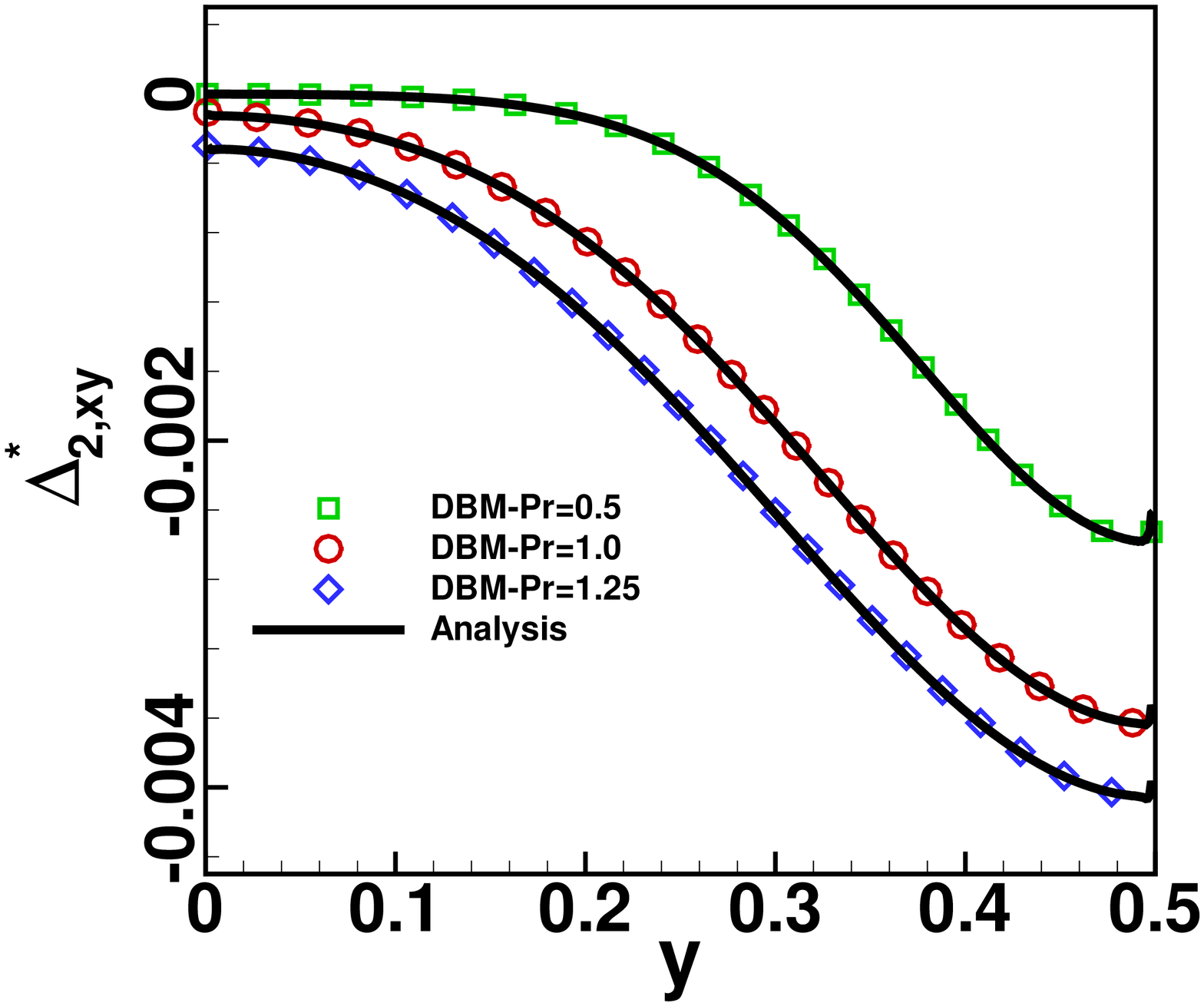}
	\label{fig006b}
\end{minipage}
}
\caption{ (a) Comparison of velocity $u_x$ along the $y$ direction between DBM results and analysis solution at two different time ($t=10, 50$), with three various $\Pr$ numbers ($\Pr=0.5,1.0,1.25$).
The green (red, blue) symbols represent DBM results with $\Pr=0.5$ ($\Pr=1.0$, $\Pr=1.25$), at time $t=10$ and $t=50$, respectively.
The black lines indicate the corresponding analytical solutions.
(b) Comparison of $\Delta_{2,xy}^{*}$ (the $xy$ component of viscous stress) between DBM results and analytical solutions with three different $\Pr$ numbers ($\Pr=0.5,1.0,1.25$), at time $t=15$. }
\label{fig006}
\end{figure}

The Couette flow is a classical physical problem that verifies the effect of viscosity on momentum transport between two layers of fluid.\cite{Zhang2018CTP,2022Zhang-AIPAdv-Slip}
In the Couette flow, the two infinite plates are filled with viscous fluid.
When the upper plate moves along the $x$ direction at a fixed velocity, the upper fluid will drive the lower fluid under the effect of viscosity.
The profile of $u_x$ along the $y$ direction follows the below analytical solution:
\begin{equation}
u_x(y)=\frac{y}{H}u_{0}+\frac{2}{\pi}\sum^{\infty}_{j=1}[\frac{(-1)^j}{j}\exp(-j^2 \pi ^2 \frac{\mu t}{\rho H^2})\sin(\frac{j\pi y}{H})] \tt{.}
\end{equation}
Initial parameters are: $c=1.0$, $\tau=1\times 10^{-3}$, $I=0(\gamma=2.0)$, $\Delta t=1\times10^{-4}$, $\Delta x=\Delta y=1\times 10^{-3}$, and $\eta=2$.
The grid size is $N_{x}\times N_{y}=1\times 500$.
In this simulation, the non-equilibrium extrapolation boundary is adopted in the $y$ direction.
Figure \ref{fig006a} shows the agreement of $u_x$ profile along the $y$ direction between DBM results and the analysis solution at two different times ($t=10, 50$).
In order to investigate the effects of $\Pr$ number on shear between two layers of fluid, three working conditions with different $\Pr$ numbers are given in the figure.
The green (red, blue) symbols represent DBM results with $\Pr=0.5$ ($\Pr=1.0$, $\Pr=1.25$), at time $t=10$ and $t=50$, respectively.
The black lines indicate the corresponding analytical solutions.
It can be observed that the larger the $\Pr$ number, the stronger the shear effect, resulting in a faster evolution of velocity profile.
The shear strength between three cases can also be seen in Fig. \ref{fig006b}, in which the larger $\Pr$ number, the stronger the strength of $\Delta_{2,xy}^{*}$.

\subsection{ Description of fine flow structure: comparison between DBM and DSMC of a shock wave structure }
\begin{figure}[htbp]
\center\includegraphics*
[width=0.4\textwidth]{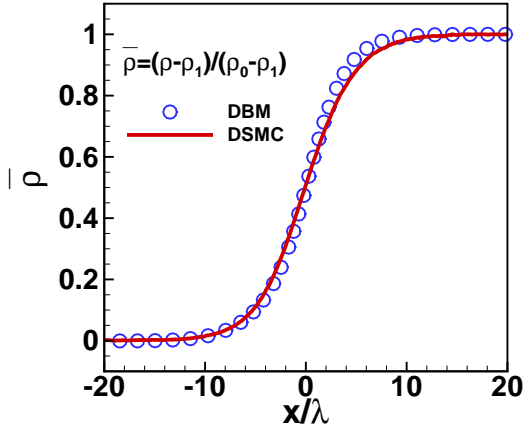}
\caption{ Comparison of DBM simulation and DSMC simulation of a shock structure.
The red lines represent the results of the normalized density profile from the DSMC code.
The blue circles indicate results from a 2-nd order DBM.
}
\label{fig007}
\end{figure}

The problem of flow characteristics at discontinuous interfaces of a shock wave has always been regarded as a typical example to verify the reliability and accuracy of the models.\cite{1994Molecular,Torrilhon04regularized13-moment,Li2007Gas}
In the following part, a right-propagating shock wave with Ma=1.45 is simulated by a 2-nd order DBM, and the comparisons of a shock wave structure between DBM and DSMC are shown.
The dimensionless conditions of macroscopic quantities in initial time are as follows:
\[
\left\{
\begin{array}{l}
(\rho,u_x,T)^{1}_{x}=(1.64871,0.736742,1.44324) \tt{,} \\
(\rho,u_x,T)^{0}_{x}=(1.0,0.0,1.0) \tt{.}
\end{array}
\right.
\]
where the index ``0'' (``1'') indicates wavefront (wave rear).
The dimensionless process from the real quantities to dimensionless quantities is shown in Appendixes \ref{sec:AppendixesC}.
Other parameters are: $c=0.8$, $\eta=5$, $I=1$, $b=0$, $\tau=1.017$, $\Delta x = \Delta y = 2.5\times 10^{-1}$, $\Delta t = 1\times 10^{-3}$, $N_x \times N_y=2000 \times 1$.
Figure \ref{fig007} shows the normalized density profile of a shock wave structure between DBM simulation and DSMC results.
The red lines are the results of the density profile from the DSMC code.
The blue circles indicate results from a 2-nd order DBM.
Agreement on the shape of the shock wave structure can be found between DBM simulation and DSMC results, indicating the model's capability to capture fine structures at the level of the mean free path of gas molecules.

\subsection{ Performance of the DBMs for describing various depths of TNE effects }

\subsubsection{ Viscous stress }

In this section, head-on collisions between two compressible fluids are simulated.
Physically, the TNE behaviors of a system are driven by many factors such as relaxation time $\tau$, density gradient, temperature gradient, velocity gradient, pressure gradient, etc.
According to the analytical expression of viscous stress (as shown by Eq. (\ref{Eq.analysis-1})), the most powerful factors that control the strengths and structures of $\Delta_{2,\alpha\beta}^{*}$ are $\tau$ and velocity gradient.
Generally, the greater the value of $\tau$ and velocity gradient, the greater the Kn number, and the farther the system deviates from equilibrium states.
When the Kn number is small enough, lower-order macroscopic models such as the Euler equations or NS equations are valid.
However, in some cases, although the Kn number is small enough, the lower-order models are no longer effective, and the higher-order model should be adopted.
From the perspective of complex system analysis, the reason is that it is incomplete to describe the non-equilibrium strength of systems from only one angle.
The TNE indicators ($\tau$, Kn, Ma, gradients of macroscopic quantity, etc.) all describe TNE behaviors from their own perspective.
These TNE indicators are highly related to each other, but they differ in some ways.
Together, they constitute a more complete description of the non-equilibrium state.
Based on the above consideration, through adjusting the relaxation time $\tau$ and relative speeds $u_{0}$ of two colliding fluids, flows across a wide range of Kn number and $\Delta_{2,xx}^{*}$ strength are constructed.
And below we use the three-component vector $\mathbf{S}_{TNE}=(\tau, \Delta \mathbf{u}, \bm{\Delta_{2}^{*}})$ to roughly describe the strength of non-equilibrium.
The initial conditions and the resulting $\Delta_{2,xx}^{*}$ strength for the three cases are shown in Table \ref{table3}.
The initial configurations are as follows:
\begin{equation}
\rho(x,y)=\frac{\rho_L+\rho_R}{2}-\frac{\rho_L-\rho_R}{2}\text{tanh}(\frac{x-N_{x}\Delta_{x}/2}{L_{\rho}})\tt{,}
\end{equation}
\begin{equation}
u_{x}(x,y)=-u_{0}\text{tanh}(\frac{x-N_{x}\Delta_{x}/2}{L_{u}})\tt{,}
\end{equation}
\begin{equation}
u_{y}(x,y)=0\tt{,}
\end{equation}
\begin{equation}
p(x,y)=p_L=p_R\tt{.}
\end{equation}
where $u_0$ is the collision velocity. $L_{\rho}$ and $L_{u}$ are the widths of transition layers of density and velocity, respectively. $\rho_L$ ($\rho_R$) and $p_L$ ($p_R$) represent the density and pressure away from the interface of the left (right) fluid. The computational length of this one-dimensional simulation is 0.4, divided into 8000 uniform meshes.
The initial conditions of other quantities are $p_L=p_{R}=2$, $\eta=0$, $I=0$, $b=0$, $\Delta x = 5\times 10^{-5}$, $\Delta t = 1\times 10^{-6}$, $L_u=L_{\rho}=160$, $N_x \times N_y=8000 \times 1$.

\begin{center}
\begin{table*}
\begin{tabular}{ | m{2.5cm}<{\centering} | m{2.5cm}<{\centering}| m{2.5cm}<{\centering}| m{2.5cm}<{\centering} | m{2.5cm}<{\centering}|m{2.5cm}<{\centering}|  }
\hline
 & density  & pressure & velocity & $\tau$ & $\Delta_{2,xx}^{*}$ strength \\
\hline
case1 & \multirow{3}{2cm}{$\rho_L=2 \rho_R=2$} & \multirow{3}{2cm}{$p_L=p_R=2$} & $u_0=0.0$ & $1\times 10^{-4}$ & weaker \\
case2 &  & & $u_0=0.5$ & $1\times 10^{-4}$ & moderate \\
case3 &  & & $u_0=0.5$ & $1\times 10^{-3}$ & stronger \\
\hline
\end{tabular}
\caption{Initial conditions and the resulting $\Delta_{2,xx}^{*}$ strength of collisions of two fluids.}
\label{table3}
\end{table*}
\end{center}

\begin{figure*}[htbp]
\centering
\subfigure[ ]{
\begin{minipage}{5.5cm}
\centering
	\includegraphics[width=5.5cm]{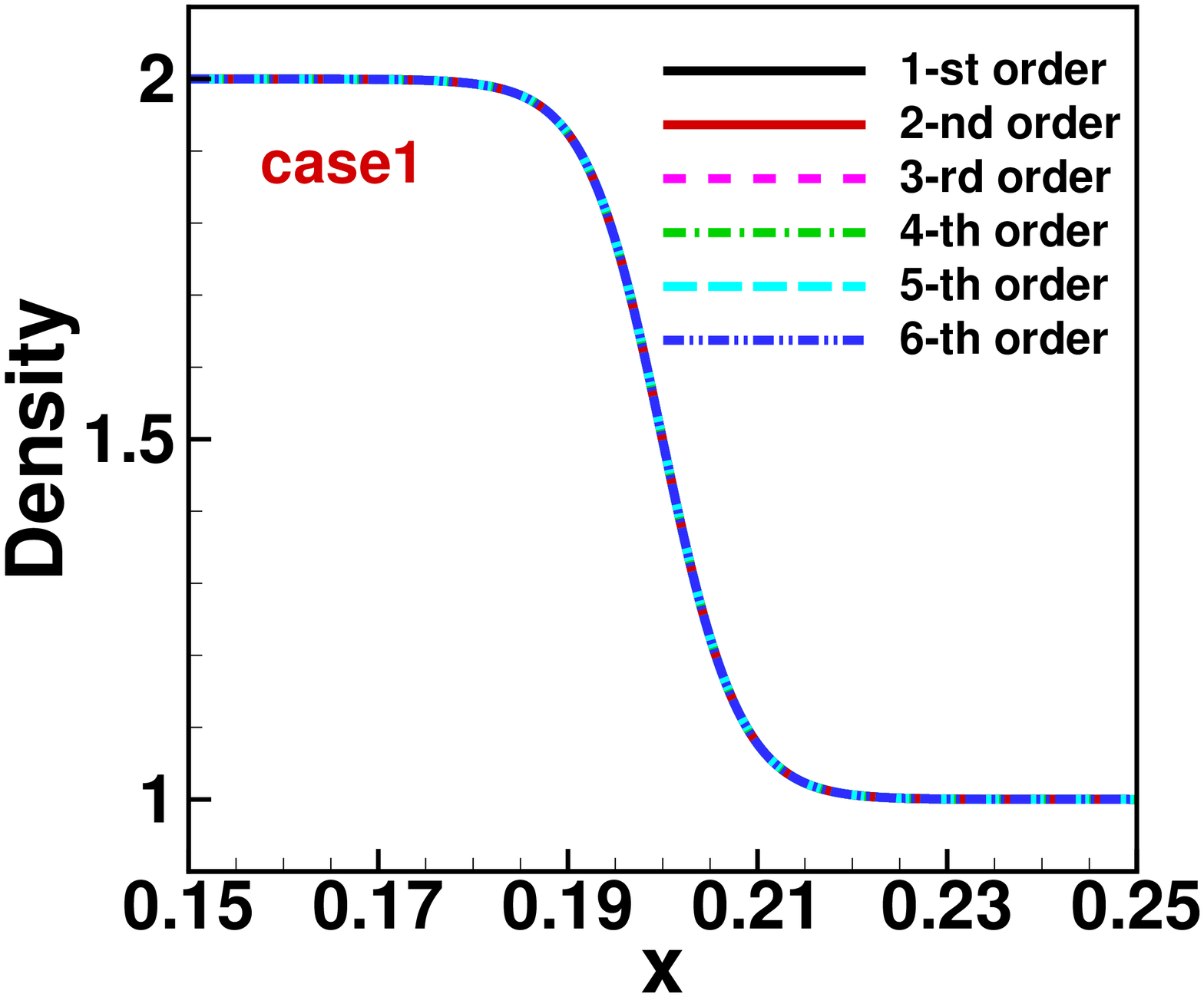}
	\label{fig008a}
\end{minipage}
}
\centering
\subfigure[]{
\begin{minipage}{5.5cm}
\centering
	\includegraphics[width=5.5cm]{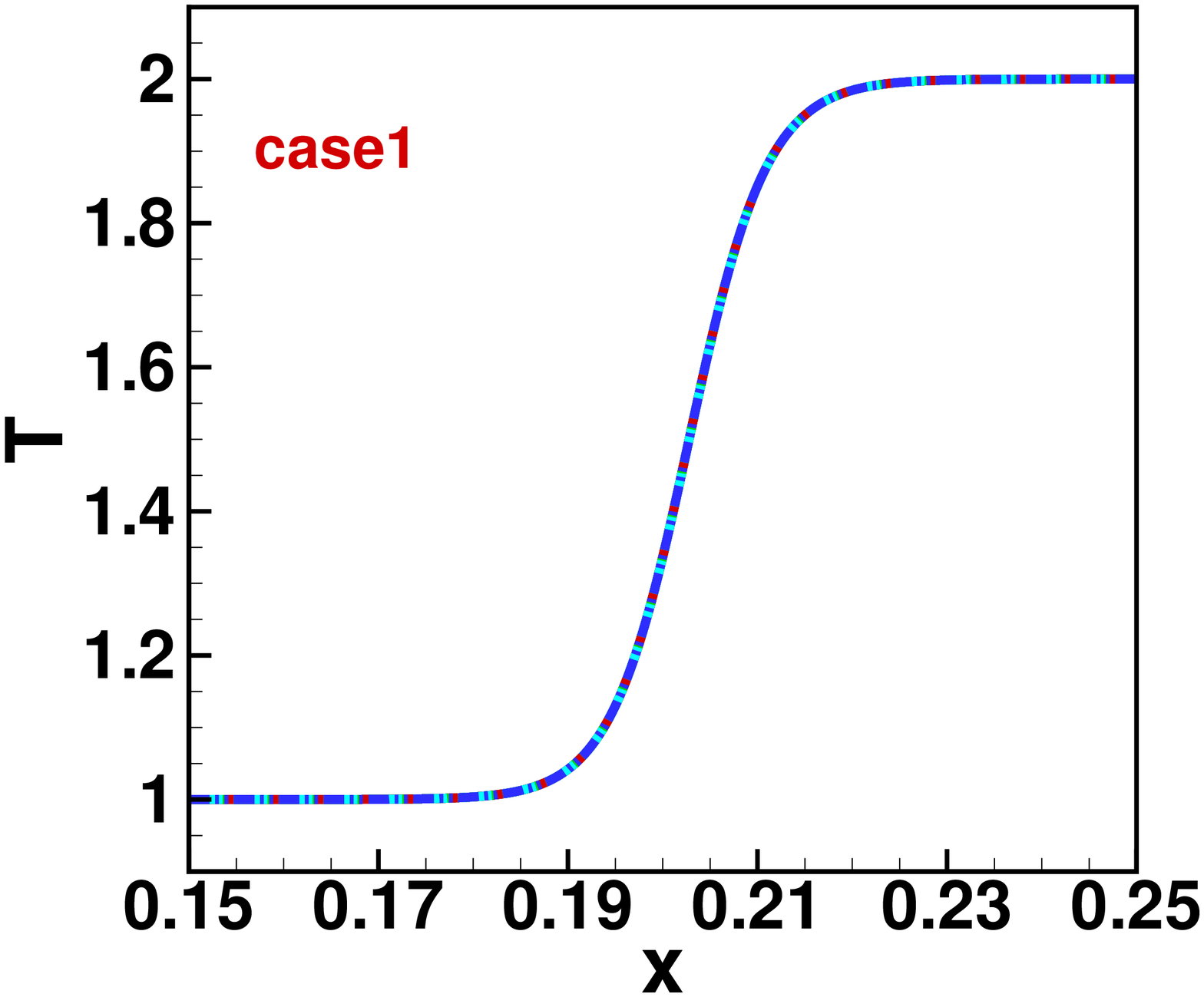}
	\label{fig008b}
\end{minipage}
}
\subfigure[]{
\begin{minipage}{5.5cm}
\centering
	\includegraphics[width=5.5cm]{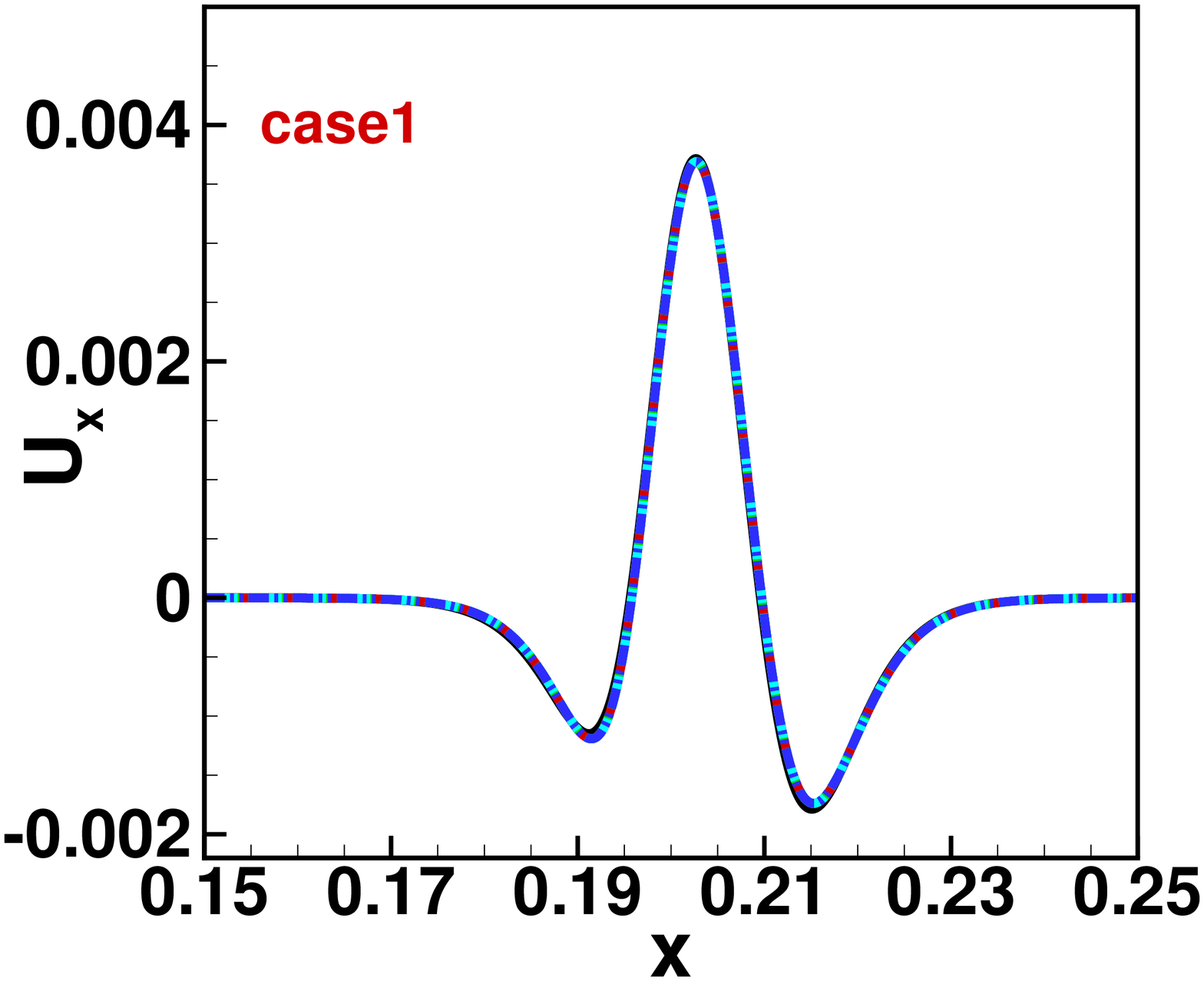}
	\label{fig008c}
\end{minipage}
}
\subfigure[]{
\begin{minipage}{5.5cm}
\centering
	\includegraphics[width=5.5cm]{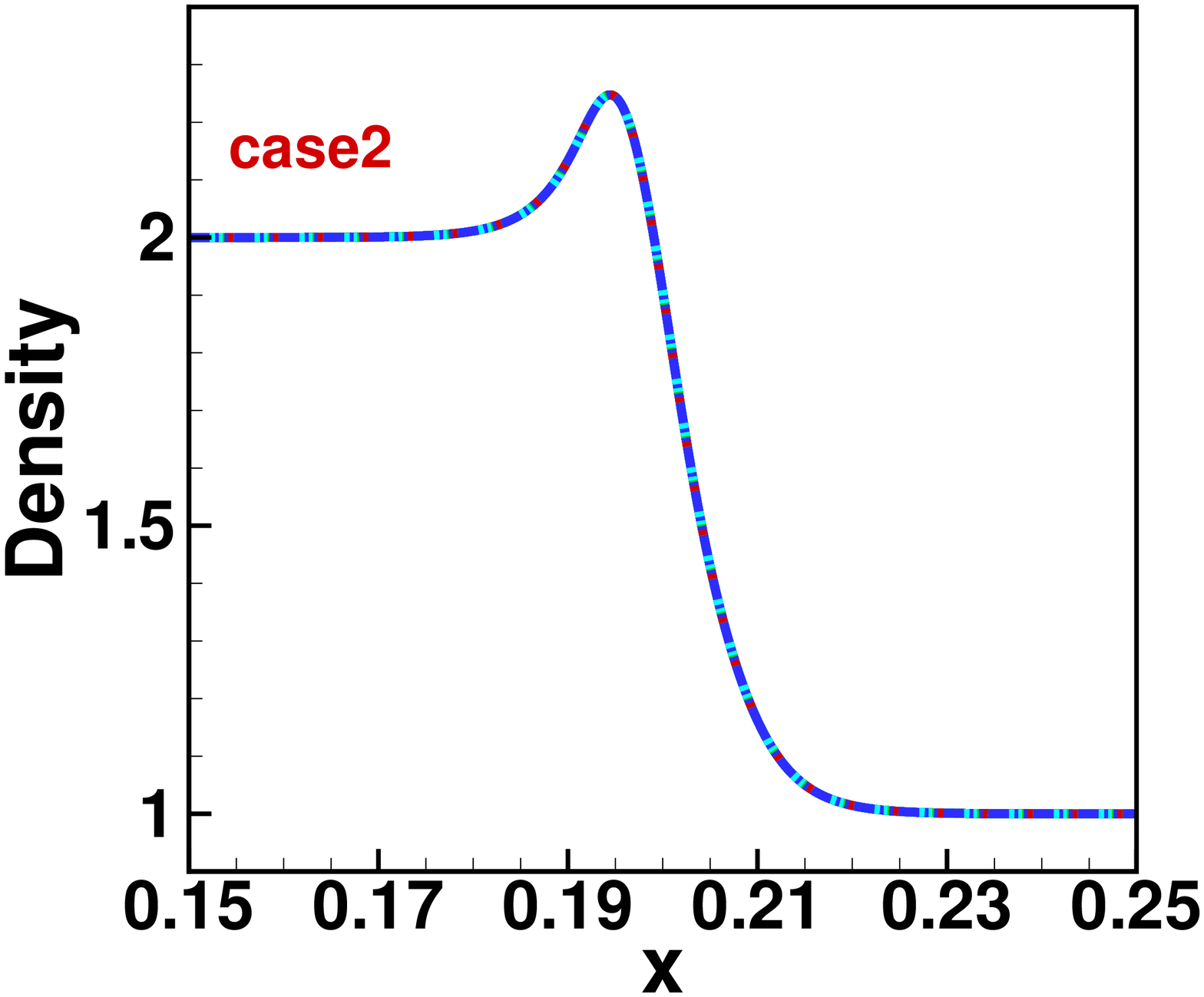}
	\label{fig008d}
\end{minipage}
}
\subfigure[]{
\begin{minipage}{5.5cm}
\centering
	\includegraphics[width=5.5cm]{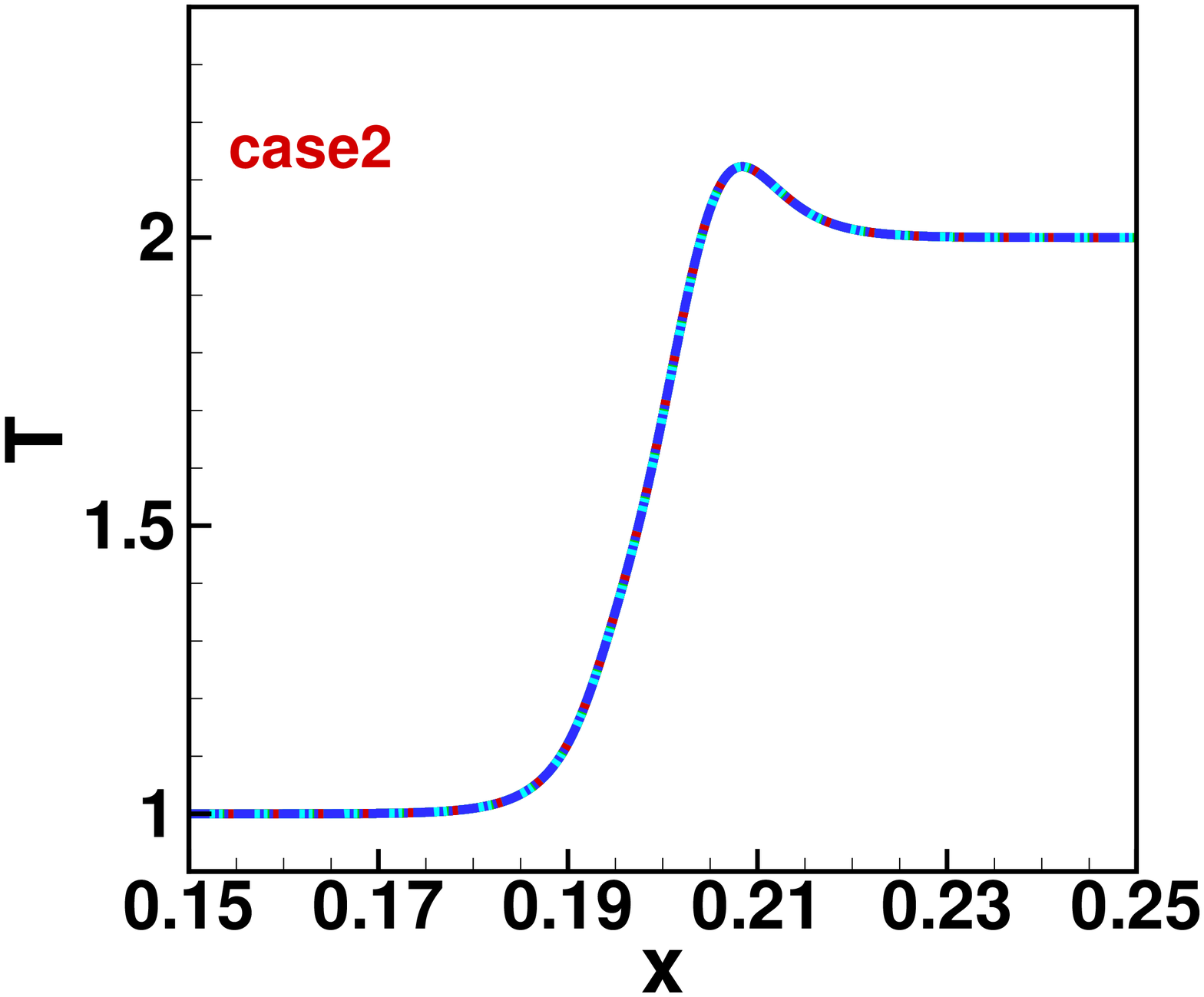}
	\label{fig008e}
\end{minipage}
}
\subfigure[]{
\begin{minipage}{5.5cm}
\centering
	\includegraphics[width=5.5cm]{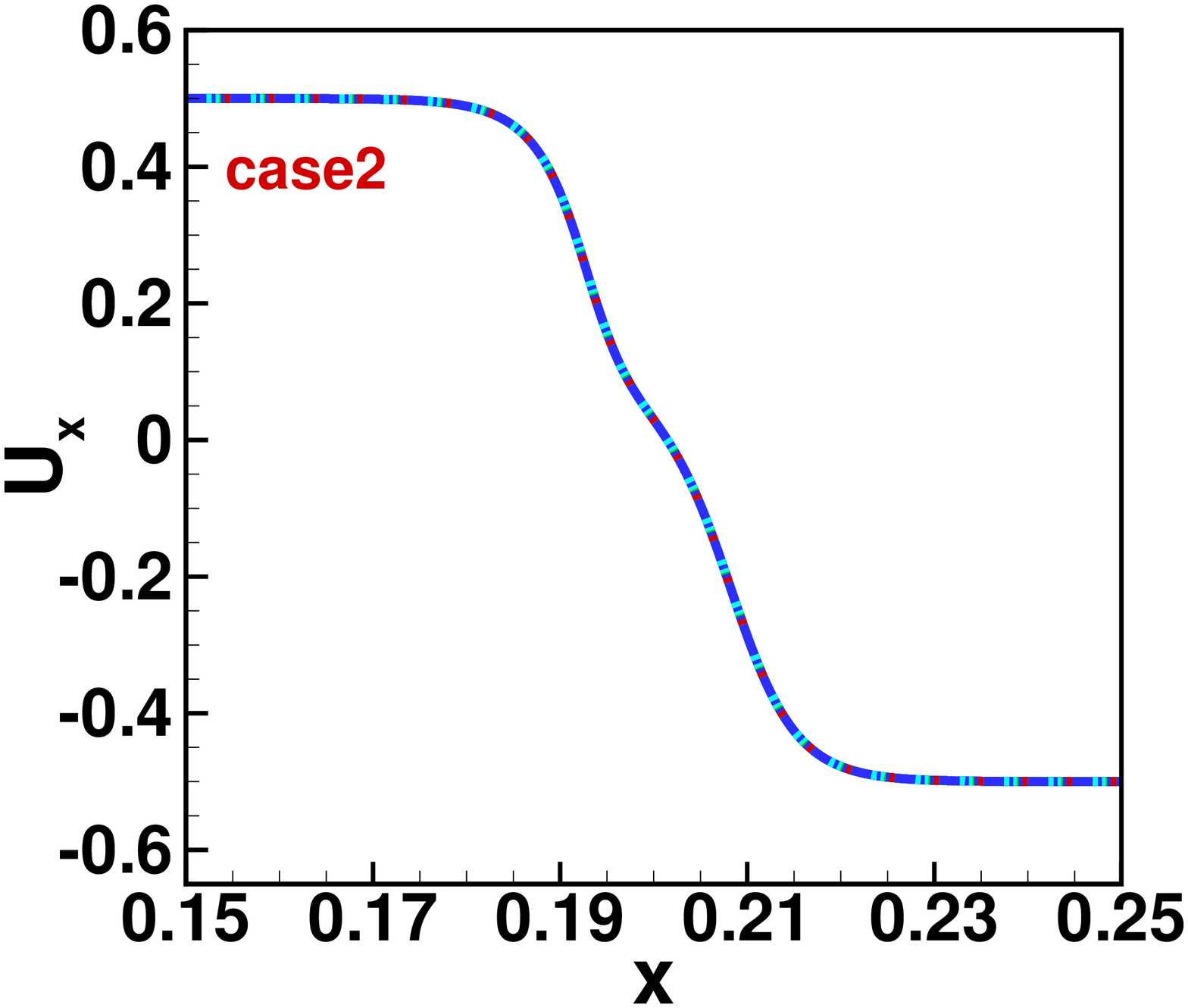}
	\label{fig008f}
\end{minipage}
}
\subfigure[]{
\begin{minipage}{5.5cm}
    \centering
	\includegraphics[width=5.5cm]{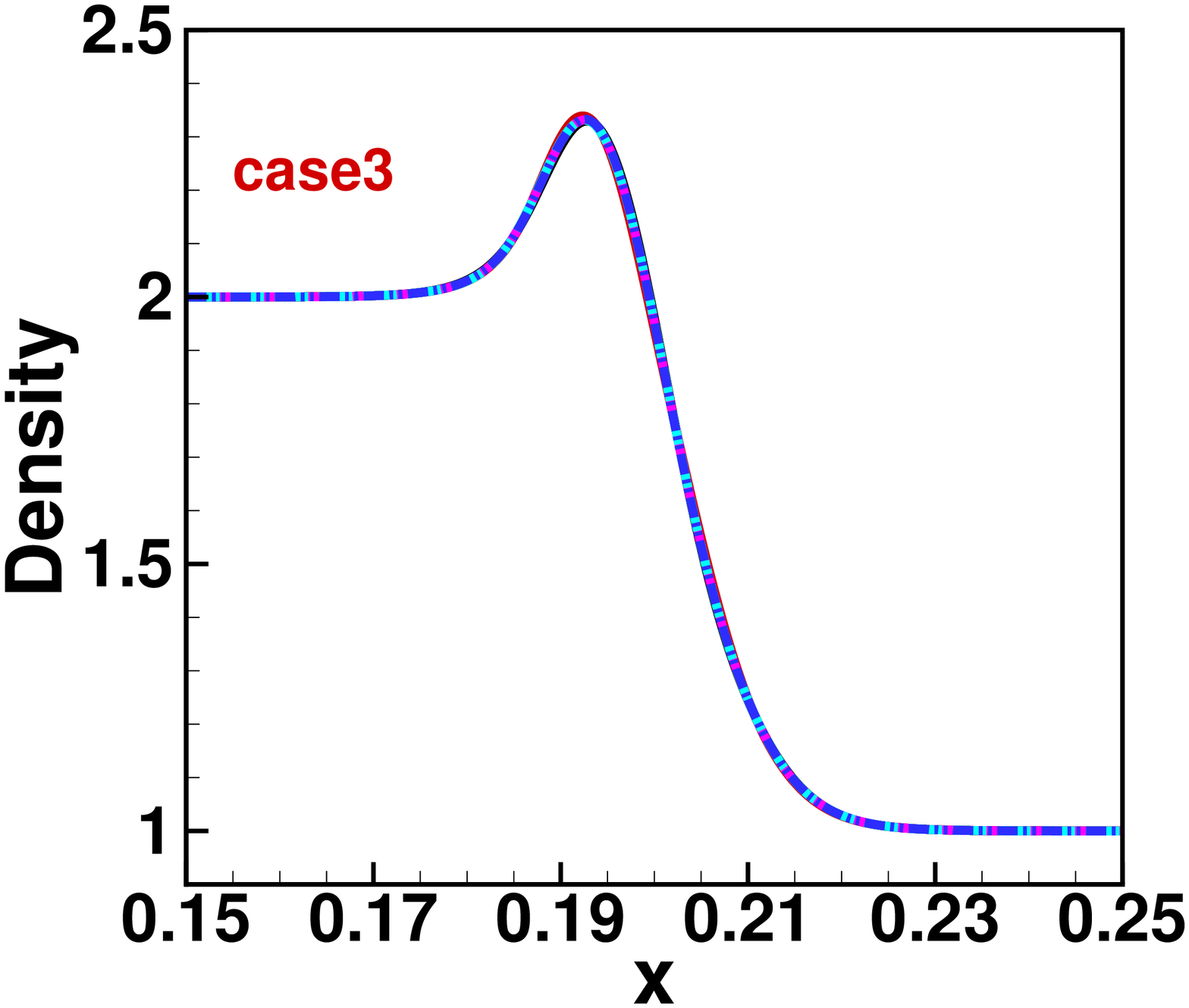}
	\label{fig008g}
\end{minipage}
}
\subfigure[]{
\begin{minipage}{5.5cm}
    \centering
	\includegraphics[width=5.5cm]{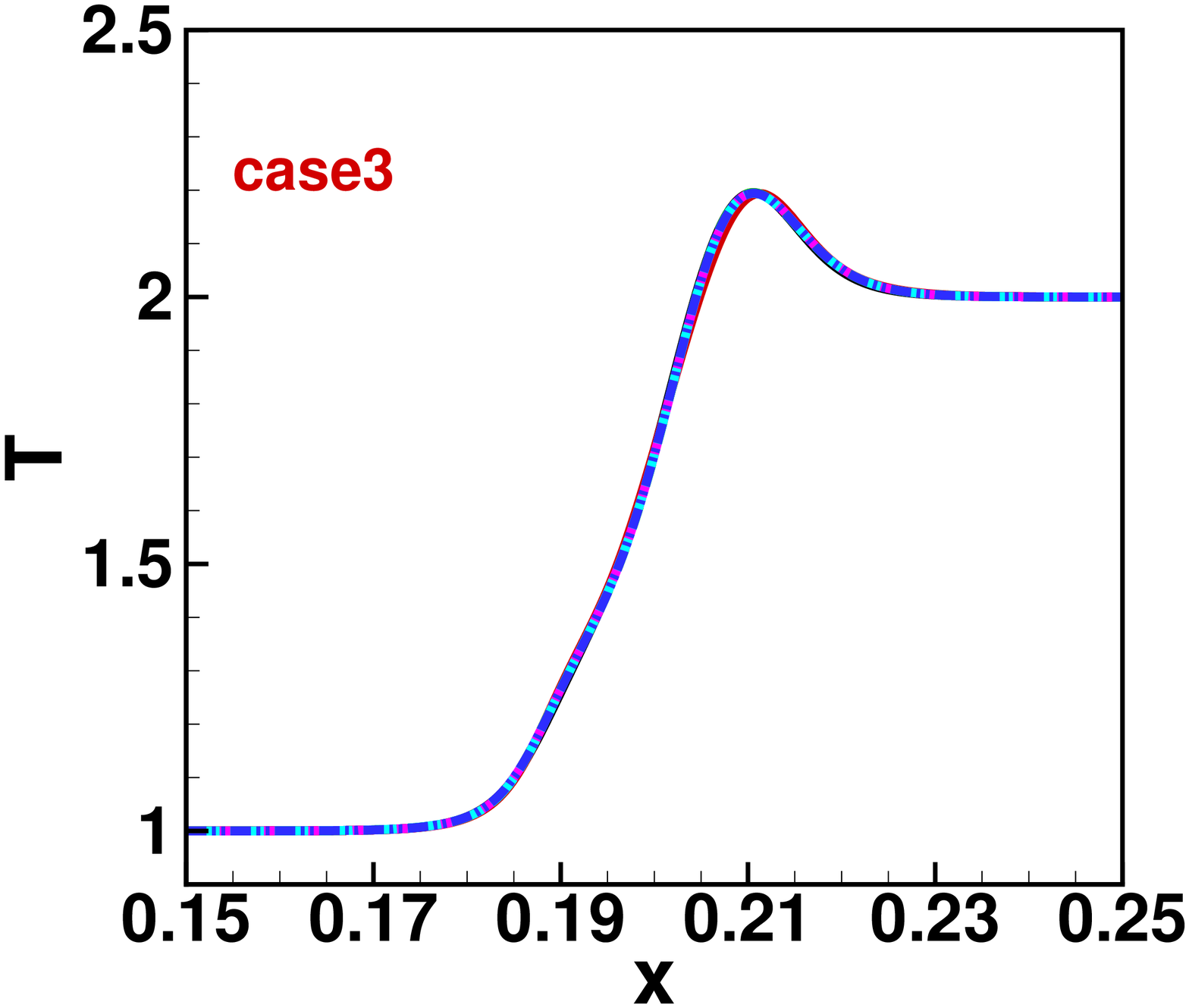}
	\label{fig008h}
\end{minipage}
}
\subfigure[]{
\begin{minipage}{5.5cm}
\centering
	\includegraphics[width=5.5cm]{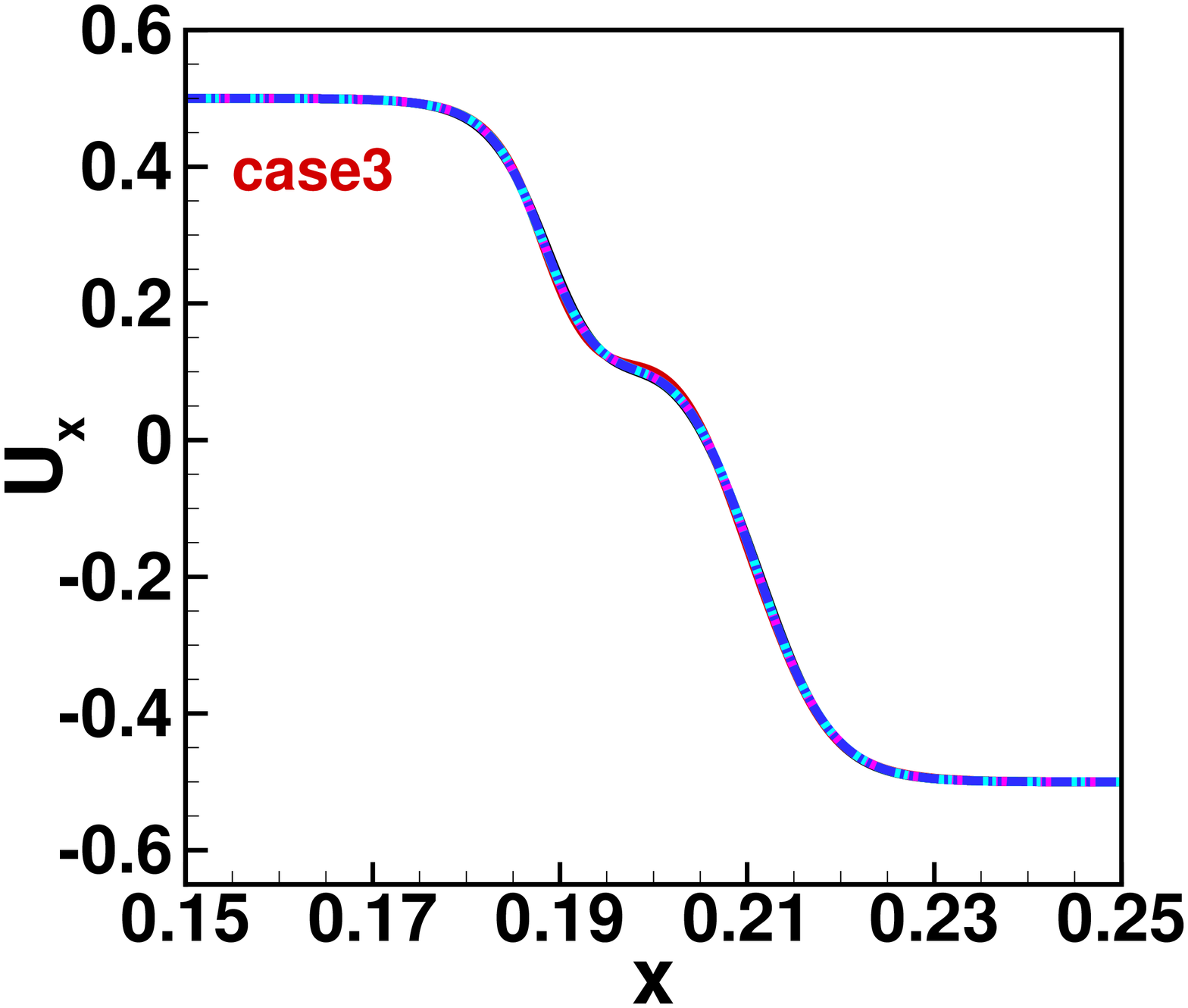}
	\label{fig008i}
\end{minipage}
}
\caption{ Profiles of macroscopic quantities around the interface with DBMs considering various orders of TNE effects. 
The lines with black (red, pink, green, light blue, and blue) color represent the results from 1-st order DBM (2-nd, 3-rd, 4-th, 5-th, and 6-th order DBM), respectively.
(a) Density profiles for case 1. (b) Temperature profiles for case 1. (c) Velocity profiles for case 1.
(d) Density profiles for case 2. (e) Temperature profiles for case 2. (f) Velocity profiles for case 2.
(g) Density profiles for case 3. (h) Temperature profiles for case 3. (i) Velocity profiles for case 3.  }
\label{fig008}
\end{figure*}

Figure \ref{fig008} shows the profiles of macroscopic quantities around the interface obtained from various DBMs at different times ($t=0.005$ for case1 and case 2,  $t=0.007$ for case 3).
The first, second, and third rows correspond to case 1, case 2, and case 3, respectively.
It can be seen that results from various DBMs are consistent at the same time.
Namely, when focusing only on the traditional macroscopic quantities, lower-order models are enough for the three cases.
However, in the following discussion, it can be found that although the profiles of macroscopic quantities obtained by various DBMs are consistent, profiles of some TNE quantities (such as the $\Delta_{2,xx}^{*}$) may show significant differences.
To characterize these TNE quantities properly, the higher-order DBMs should be adopted.

\begin{figure}[htbp]
\centering
\subfigure[]{
\begin{minipage}{8cm}
\centering
	\includegraphics[width=7cm]{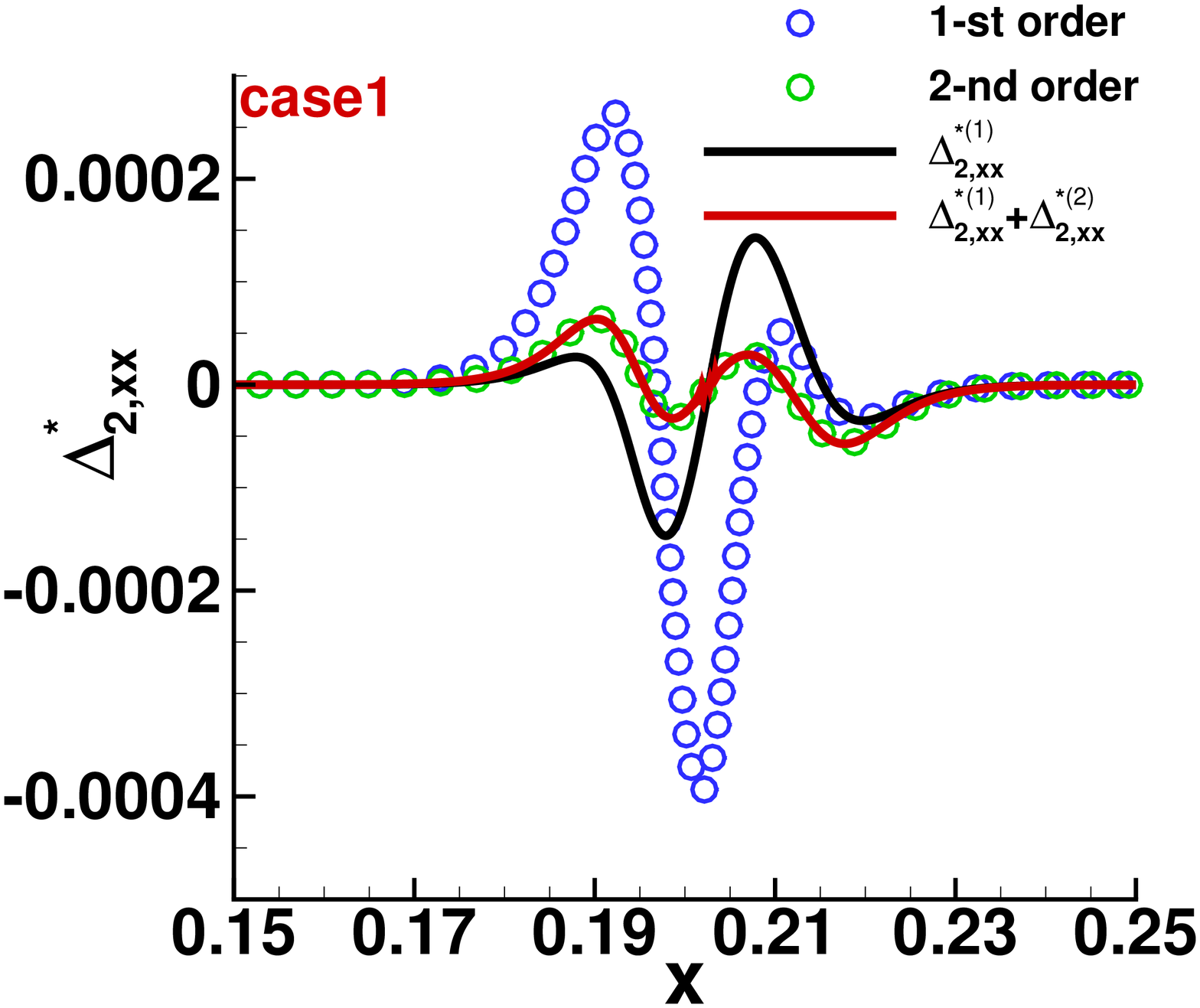}
	\label{fig009a}
\end{minipage}
}
\subfigure[]{
\begin{minipage}{8cm}
\centering
	\includegraphics[width=7cm]{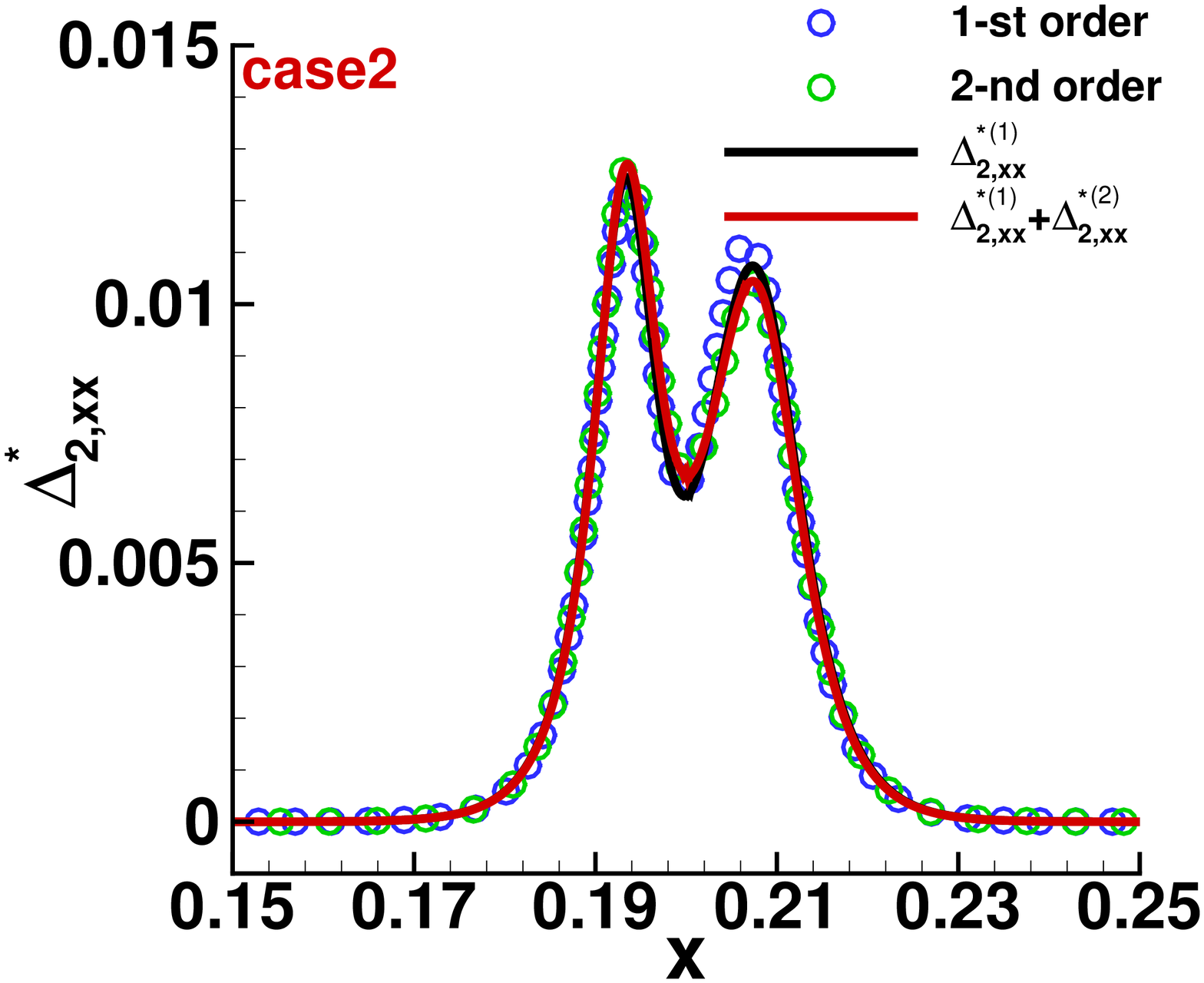}
	\label{fig009b}
\end{minipage}
}
\subfigure[]{
\begin{minipage}{8cm}
\centering
	\includegraphics[width=7cm]{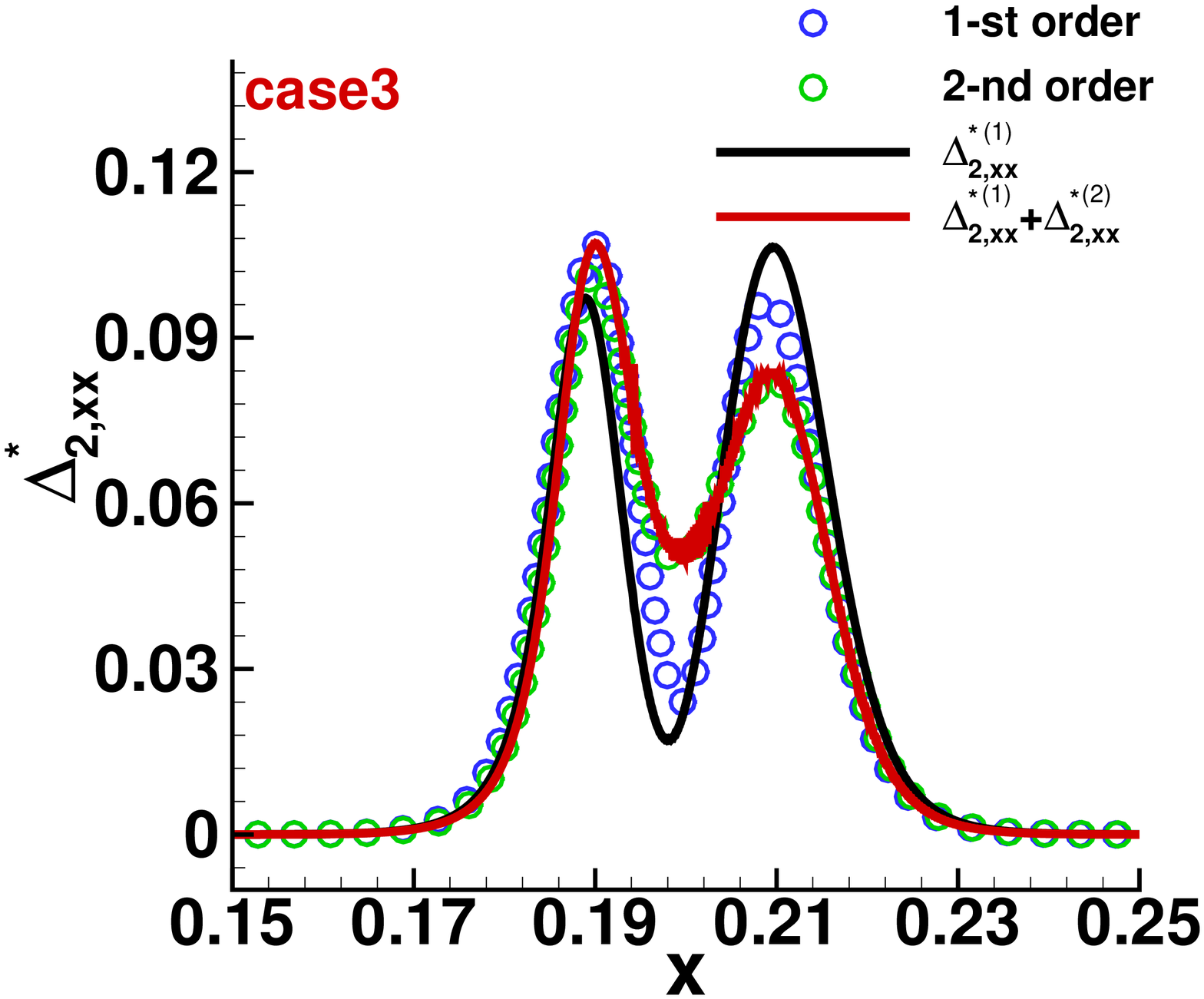}
	\label{fig009c}
\end{minipage}
}
\caption{Comparison of $\Delta_{2,xx}^{*}$ between DBM simulation results (blue circles from 1-st order DBM, green circles from 2-nd order DBM) and two kinds of analytical solutions (the black lines are at first-order accuracy and the red lines are at second-order accuracy).
(a) Results for case 1,
(b) results for case 2,
and (c) results for case 3.
}
\label{fig009}
\end{figure}

Figure \ref{fig009} shows the simulation results of  $\Delta_{2,xx}^{*}$ at different times ($t=0.005$ for case1 and case 2,  $t=0.007$ for case 3) where two DBMs (the 1-st order and 2-nd order DBMs) are used.
The blue circles represent results from 1-st order DBM and green circles from 2-nd order DBM.
The first, second, and third rows correspond to the three cases, respectively.
For comparisons, analytical solutions at first-order accuracy (black lines) and at second-order accuracy (red lines) calculated from Eqs. (\ref{Eq.analysis-1}) and (\ref{Eq.analysis-2}) are plotted, respectively.
From case 1 to case 3, what we can see is that the $\Delta_{2,xx}^{*}$ strengths increase gradually because the TNE driving force ($\tau$ or gradient of velocity) increases.
Meanwhile, the TNE effects are pronounced around the contact interface where the amplitudes of quantity gradient ($\bm{\nabla} \rho$, $\bm{\nabla} u$, $\bm{\nabla} p$, and $\bm{\nabla} T$) reach their local maxima.
And the TNE effects are negligible in the region far from the interface.

Moreover, differences between results obtained from various DBMs and analytical solutions indicate the performance of various DBMs in describing flows with various depths of non-equilibrium.
Figure \ref{fig009a} shows the comparison for case 1 between simulation results of two various DBMs and two kinds of analytical solutions with different accuracy.
As shown by Fig. \ref{fig009a}, the results of analytical solutions between first-order accuracy and second-order accuracy show a significant difference.
Theoretically, as shown by Eq. (\ref{Eq.analysis-1}), the first-order term of viscous stress (i.e. $\Delta_{2,xx}^{*(1)}$) is weak due to the small velocity gradient and $\tau$.
However, because of the existence of density and temperature gradients, the second-order term of viscous stress (i.e., $\Delta_{2,xx}^{*(2)}$) is pronounced.
In that case, $\Delta_{2,xx}^{*(2)}$ cannot be ignored compared to $\Delta_{2,xx}^{*(1)}$, i.e., the relative TNE strength ($\Delta_{2,xx}^{*(2)}/\Delta_{2,xx}^{*(1)}$) is considerably large.
Consequently, the profiles of the two analytical solutions deviate from each other.
Meanwhile, the 1-st order DBM considers only the first-order of TNE effects (it retains only the first-order term of viscous stress, i.e., $\Delta_{2,xx}^{*(1)}$).
Therefore, its results can not match the analytical solution which is at second-order accuracy.
However, when a 2-nd order DBM that considers up to the second-order of TNE effects (i.e., $\Delta_{2,xx}^{*(2)}$ is retained) is adopted, agreements between DBM results and second-order analytical solutions can be seen.
Therefore, a 1-st order DBM is not suitable for the cases where $\Delta_{2,xx}^{*(2)}$ is not negligible, whereas a 2-nd order DBM is suitable.

Figure \ref{fig009b} shows the comparison for case 2, in which the velocity gradient is larger than that in case 1, and the resulting $\Delta_{2,xx}^{*}$ strength is dozens of times than that of case 1.
Due to the larger velocity gradient, $\Delta_{2,xx}^{*(2)}$ can be negligible compared to $\Delta_{2,xx}^{*(1)}$, i.e., the relative TNE strength is considerably small.
Consequently, analytical solutions between first-order accuracy and second-order accuracy are almost consistent.
In that case, the 1-st order DBM and 2-nd order DBM all present satisfactory simulation results.

For further investigation of TNE effects, we take the relaxation time $\tau$ ten times larger than that in case 2.
Consequently, the $\Delta_{2,xx}^{*}$ strength is about ten times larger.
As can be seen from Fig. \ref{fig009c}, the first-order analytical solution shows great differences with the second-order one, demonstrating that $\Delta_{2,xx}^{*(2)}$ can no longer be ignored compared to $\Delta_{2,xx}^{*(1)}$.
Naturally, the 1-st order DBM can not provide satisfactory simulation results.
Agreements between simulation results from a 2-nd order DBM and the second-order analytical solution can be seen.
Therefore, with increasing TNE strength, the 1-st model gradually loses its capability to describe the TNE quantities properly.

\begin{figure}[htbp]
\centering
\subfigure[]{
\begin{minipage}{8cm}
\centering
	\includegraphics[width=7cm]{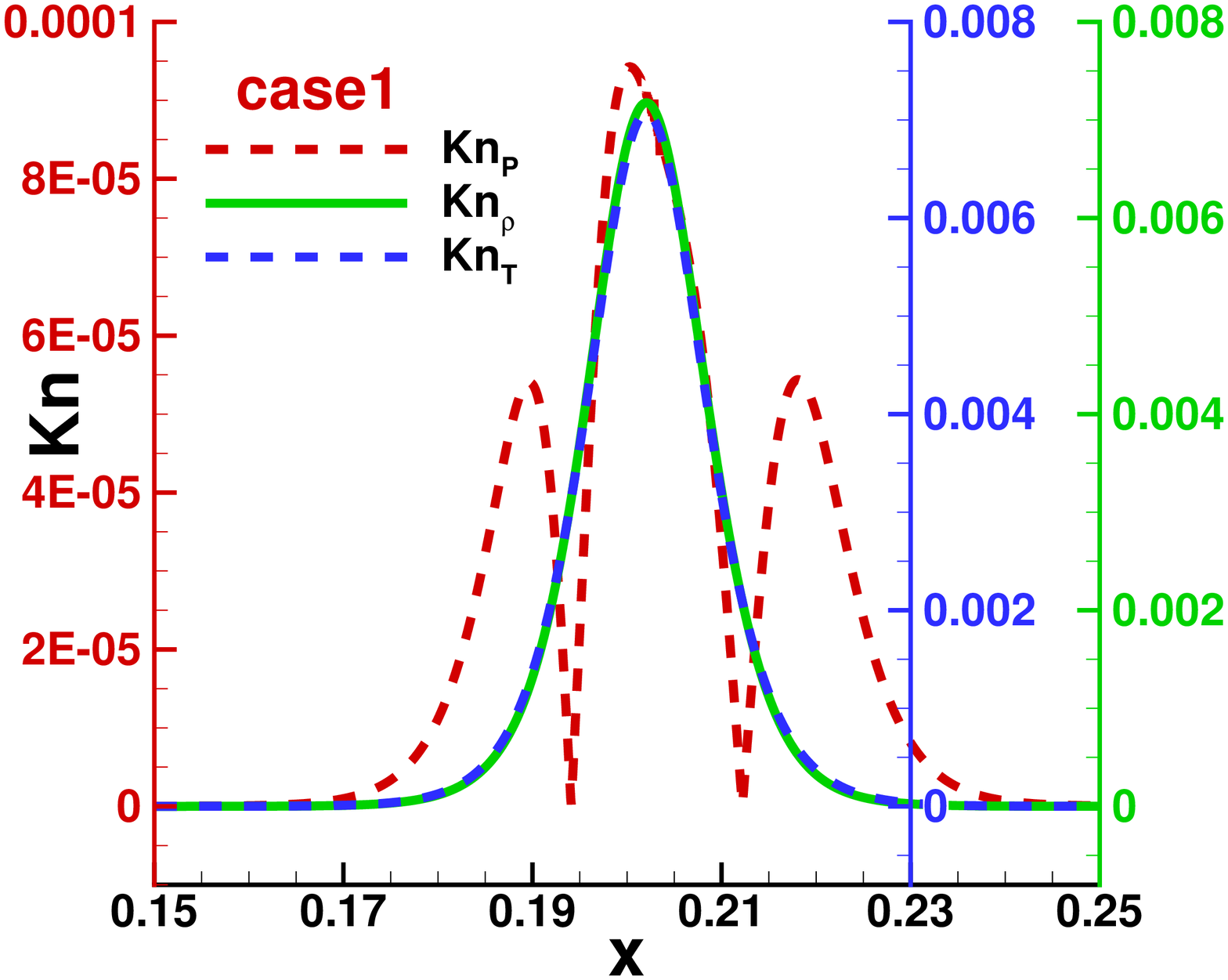}
	\label{fig010a}
\end{minipage}
}
\subfigure[]{
\begin{minipage}{8cm}
\centering
	\includegraphics[width=7cm]{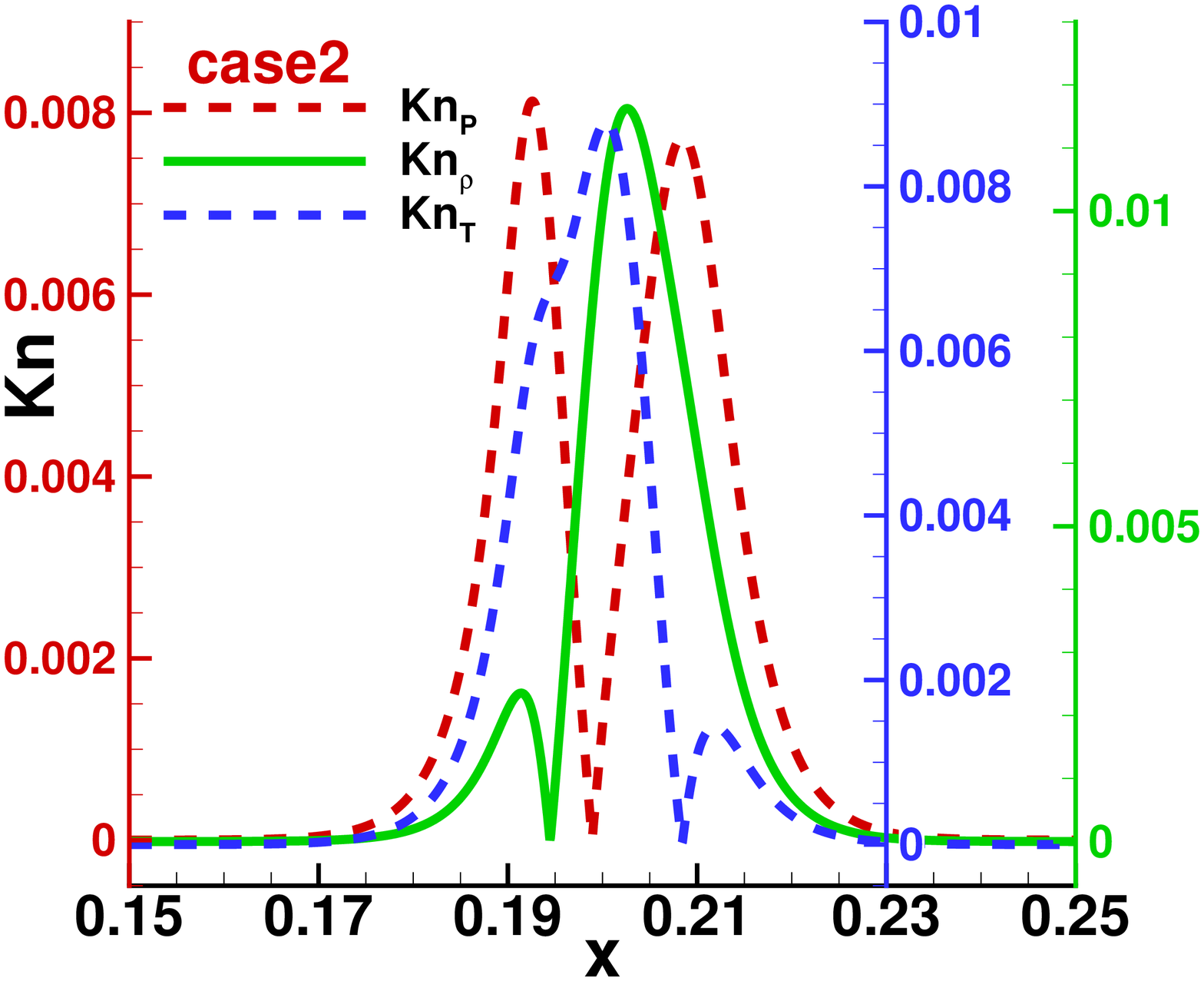}
	\label{fig010b}
\end{minipage}
}
\subfigure[]{
\begin{minipage}{8cm}
\centering
	\includegraphics[width=7cm]{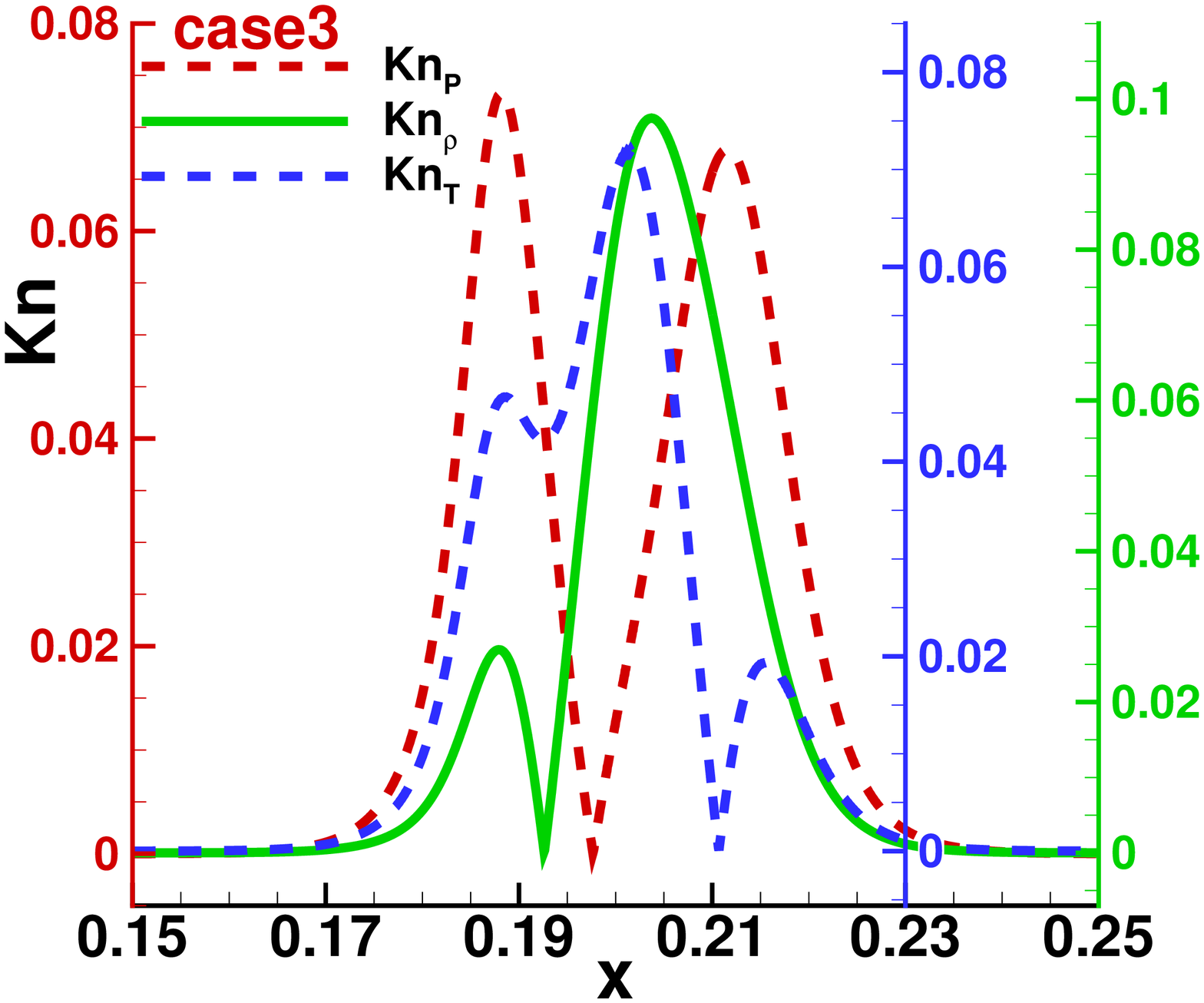}
	\label{fig010c}
\end{minipage}
}
\caption{Local Kn numbers calculated from pressure (red lines), density (green lines), and temperature (blue lines), respectively.
(a) Results from case 1,
(b) results from case 2,
and (c) results from case 3.  }
\label{fig010}
\end{figure}

Interestingly, the 1-st order DBM shows satisfactory results in case 2, where the Kn number is larger.
In contrast, it shows unsatisfactory results in case 1, where the Kn number is smaller.
The reason is that $\Delta_{2,xx}^{*(2)}$ is not negligible compared to $\Delta_{2,xx}^{*(1)}$.
In that case, the relative TNE strength and TNE discrepancy should also be adopted to further characterize the TNE strength, instead of the Kn number itself.\cite{2018Gan-pre}
Physically, it is difficult to describe the TNE strength and choose the suitable fluid model for simulation from only one perspective, e.g., the Kn number.
The local Kn numbers around the interface for three cases are presented in Figs. \ref{fig010a}-\ref{fig010c} to show further the differences in Kn numbers obtained from various perspectives.
The local Kn numbers are calculated from equation $Kn=\lambda / L=c_{s} \tau / (\phi / \bm{\nabla}  \phi)$, where $c_{s}$, $L$, and $\phi$ are the local speed of sound, characteristic length, and characteristic quantity, respectively.
From case 1 to case 3 (as shown in Fig. \ref{fig010a} to \ref{fig010c}), similar to $\Delta_{2,xx}^{*}$, values of the local Kn number increase with the gradients of macroscopic quantities.
Moreover, in Fig. \ref{fig010a}, Kn numbers calculated from various characteristic quantities present significant distinctions, e.g., the maximum between the red line and blue line differ dozens of times (across the inviscid flow and slip flow).
Their shapes are also significantly different.

\begin{figure}[htbp]
\centering
\subfigure[]{
\begin{minipage}{6cm}
\centering
	\includegraphics[width=7cm]{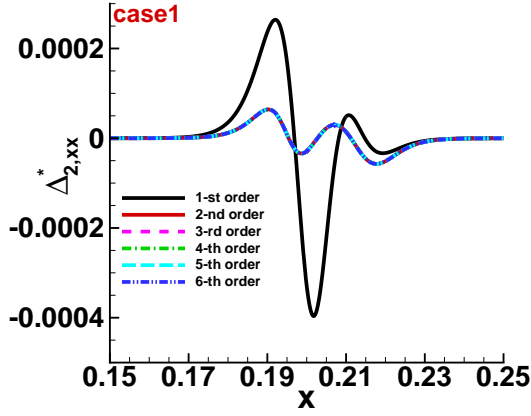}
	\label{fig011a}
\end{minipage}
}
\subfigure[]{
\begin{minipage}{6cm}
\centering
	\includegraphics[width=7cm]{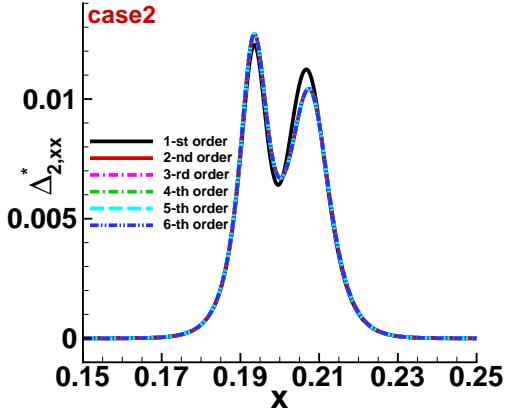}
	\label{fig011b}
\end{minipage}
}
\subfigure[]{
\begin{minipage}{6cm}
\centering
	\includegraphics[width=7cm]{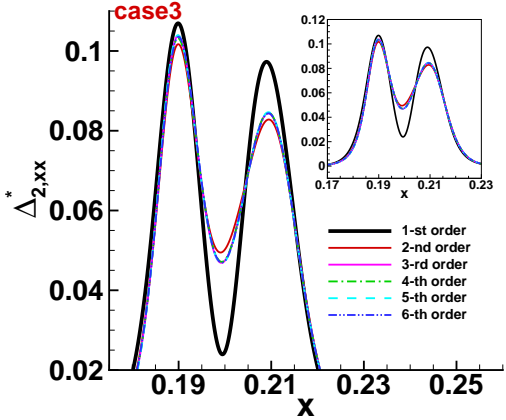}
	\label{fig011c}
\end{minipage}
}
\caption{Simulation results of $\Delta_{2,xx}^{*}$ from various DBMs: (a) from case 1, (b) from case 2, and (c) from case 3.
The lines with black (red, pink, green, light blue, and blue) color represent the results from 1-st order DBM (2-nd, 3-rd, 4-th, 5-th, and 6-th order DBM), respectively.}
\label{fig011}
\end{figure}

However, what we can see is that the analytical solutions from Eqs. (\ref{Eq.analysis-1}) and (\ref{Eq.analysis-2}) are not the real solutions because of their dependence on macroscopic quantities obtained from simulation.
Hence, additional measures are needed to verify the reliability and authenticity of simulation results.
In the following part, the above three cases are simulated by six various DBMs.
Figure \ref{fig011} shows the simulation results obtained from various DBMs, in which Figs. \ref{fig011a}, \ref{fig011b}, \ref{fig011c} are from case 1, case 2, and case 3, respectively.
The lines with black (red, pink, green, light blue, and blue) color represent the results from 1-st order DBM(2-nd, 3-rd, 4-th, 5-th, and 6-th order DBM), respectively.
For case 1, as shown in Fig. \ref{fig011a}, except for the result from the 1-st order DBM, results of other DBMs converge together, indicating that to simulate accurately the case 1, at least the second-order TNE effects should be considered.
Further, as shown in Fig. \ref{fig011b}, although results from the 1-st order DBM show agreement with the analytical solution (as shown in Fig. \ref{fig009b}), discernible difference around the peak is found between the black line and higher-order ones because of the large gradients of macroscopic quantities.
Therefore, similar to case 1, at least the second-order TNE effects should be considered in case 2.
Different understanding can be obtained from case 3, as shown in Fig. \ref{fig011c}, in which the $\Delta_{2,xx}^{*}$ strength is about ten times larger than that in case 2.
Simulation results of the 1-st order DBM (black line) show significant differences with results from higher-order DBMs.
At the same time, because of the large gradients of macroscopic quantities, discernible differences would appear between the result of 2-nd order DBM (red line) and results from higher-order DBMs.
Consequently, although agreements between a 2-nd order DBM and analytical solution are shown in Fig. \ref{fig009c}, higher-order TNE effects, at least up to the third-order, should be considered to obtain more accurate results for case 3.
Another important conclusion can be obtained by comparing Fig. \ref{fig008} and Fig. \ref{fig011}.
What we can see is the profiles of macroscopic quantities obtained from various DBMs are consistent, whereas the TNE quantity ($\Delta_{2,xx}^{*}$) from various DBMs shows apparent distinctions.
Physically, with the deepening of TNE degrees, it is inadequate to characterize the whole system's behaviors only by conserved moments.
We also have to rely on partial higher-order non-conserved moments to capture the main feature of the flow state and evolution process.
The required order of kinetic moments increases with the deeper TNE degree.

For easier understanding, the TNE strengths of three cases obtained from various views are summarized in Table \ref{table5}.
Then, fluid models that should be chosen when focusing on different physical quantities are also listed.

\begin{table*}
\centering
\begin{tabular}{|m{3cm}<{\centering}|m{3cm}<{\centering}|m{3cm}<{\centering}|m{3cm}<{\centering}|m{3cm}<{\centering}|}
\hline
                  & View & Case 1 & Case 2 & Case 3  \\
\hline
\multirow{5}{*}{TNE strength} & $\tau$ and $\nabla$$\phi$ & smaller & moderate & larger  \\
\cline{2-5}
                  & $\tt{Kn}_{\rho}$  & smaller  & moderate & larger    \\
\cline{2-5}
                  & $\Delta_{2,xx}^{*}$ & weaker & moderate & stronger    \\
\cline{2-5}
                  & $\Delta_{2,xx}^{*(2)}$/$\Delta_{2,xx}^{*(1)}$ & stronger & weaker  & moderate    \\
\cline{2-5}
                  &$\Delta_{2,xx}^{*(3)}$/$\Delta_{2,xx}^{*(2)}$  & $\approx 0$ & $\approx 0$ & $\neq 0$    \\
\hline
\multirow{2}{*}{The proper models} & macroscopic quantities  & 1-st order model & 1-st order model & 1-st order model    \\
\cline{2-5}
                  & $\Delta_{2,xx}^{*}$ & 2-nd order model & 2-nd order model & 3-rd order model   \\
\hline
\end{tabular}
\caption{ In the case of analysing the viscous stress: the TNE strength of three cases obtained from various views, and the proper models that should be adopted in the corresponding views.
}
\label{table5}
\end{table*}

\subsubsection{ Heat flux}

The performance of various DBMs to describe higher-order heat flux is also verified similarly.
In the following part, we use the three-component vector $\mathbf{S}_{TNE}=(\tau, \Delta T, \bm{\Delta_{3,1}^{*}})$ to roughly describe the strength of non-equilibrium.
The initial configurations are as follows:
\begin{equation}
\rho(x,y)=\frac{\rho_L+\rho_R}{2}-\frac{\rho_L-\rho_R}{2}\text{tanh}(\frac{x-N_{x}\Delta_{x}/2}{L_{\rho}})\tt{,}
\end{equation}
\begin{equation}
p(x,y)=\frac{p_L+p_R}{2}-\frac{p_L-p_R}{2}\text{tanh}(\frac{x-N_{x}\Delta_{x}/2}{L_p})\tt{,}
\end{equation}
\begin{equation}
u_{x}(x,y)=-u_{0}\text{tanh}(\frac{x-N_{x}\Delta_{x}/2}{L_{u}})\tt{,}
\end{equation}
\begin{equation}
u_{y}(x,y)=0\tt{.}
\end{equation}

The initial conditions of other quantities are $p_L=p_{R}=2$, $\eta=0$, $I=0$, $b=0$, $\Delta x = \Delta y = 5\times 10^{-5}$, $\Delta t = 1\times 10^{-6}$, $L_u=L_{\rho}=L_{p}=160$, $N_x \times N_y=8000 \times 1$.
Parameters, the resulting $\Delta_{3,1,x}^{*}$ strength and the values of Kn number in the three cases are listed by Table \ref{table4}.

\begin{center}
\begin{table*}
\begin{tabular}{ | m{2.5cm}<{\centering} | m{2.5cm}<{\centering}| m{2.5cm}<{\centering}| m{2.5cm}<{\centering} | m{2.5cm}<{\centering}|m{2.5cm}<{\centering}| }
\hline
 & temperature  & pressure & velocity & $\tau$ & $\Delta_{3,1,x}^{*}$ strength \\
\hline
case1 & $T_L=T_R=1$ & $2 p_L=p_R=2$ & $u_0=0.5$ & $2\times 10^{-4}$ & weaker \\
case2 & $T_L=2 T_R=1.2$ & $ p_L=p_R=1.2$ & $u_0=0.5$ & $8\times 10^{-5}$ & moderate \\
case3 & $T_L=2 T_R=1.2$ & $ p_L=p_R=1.2$ & $u_0=0.5$ & $8\times 10^{-4}$ & stronger \\
\hline
\end{tabular}
\caption{Initial parameters and the resulting $\Delta_{3,1,x}^{*}$ strength of collisions of two fluids.}
\label{table4}
\end{table*}
\end{center}

\begin{table*}
\centering
\begin{tabular}{|m{3cm}<{\centering}|m{3cm}<{\centering}|m{3cm}<{\centering}|m{3cm}<{\centering}|m{3cm}<{\centering}|}
\hline
                  & View & Case 1 & Case 2 & Case 3  \\
\hline
\multirow{4}{*}{TNE strength} & $\tt{Kn}_{\rho}$  & moderate  & smaller & larger    \\
\cline{2-5}
                  & $\Delta_{3,1,x}^{*}$ & weaker & moderate & stronger    \\
\cline{2-5}
                  & $\Delta_{3,1,x}^{*(2)}$/$\Delta_{3,1,x}^{*(1)}$ & stronger & weaker  & moderate    \\
\cline{2-5}
                  &$\Delta_{3,1,x}^{*(3)}$/$\Delta_{3,1,x}^{*(2)}$  & $\approx 0$ & $\approx 0$ & $\approx 0$    \\
\hline
\multirow{2}{*}{The proper models} & macroscopic quantities  & 1-st order model & 1-st order model & 1-st order model    \\
\cline{2-5}
                  & $\Delta_{3,1,x}^{*}$ & 2-nd order model & 1-st order model & 2-nd order model    \\
\hline
\end{tabular}
\caption{ In the case of analysing the heat flux: the TNE strength of three cases obtained from various views, and the proper models that should be adopted in the corresponding views.
}
\label{table6}
\end{table*}

\begin{figure}[htbp]
\centering
\subfigure[]{
\begin{minipage}{8cm}
\centering
	\includegraphics[width=7cm]{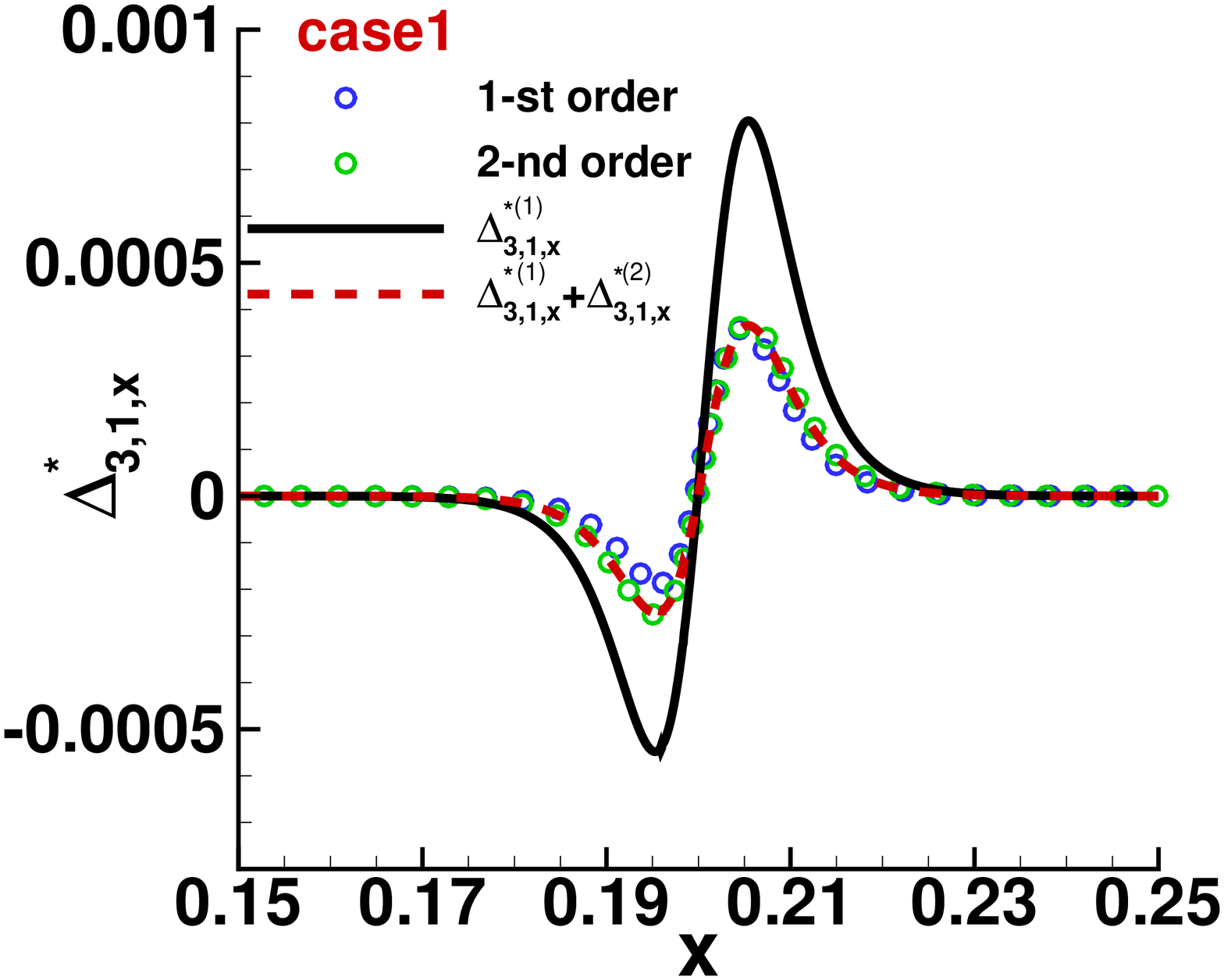}
	\label{fig0012a}
\end{minipage}
}
\subfigure[]{
\begin{minipage}{8cm}
\centering
	\includegraphics[width=7cm]{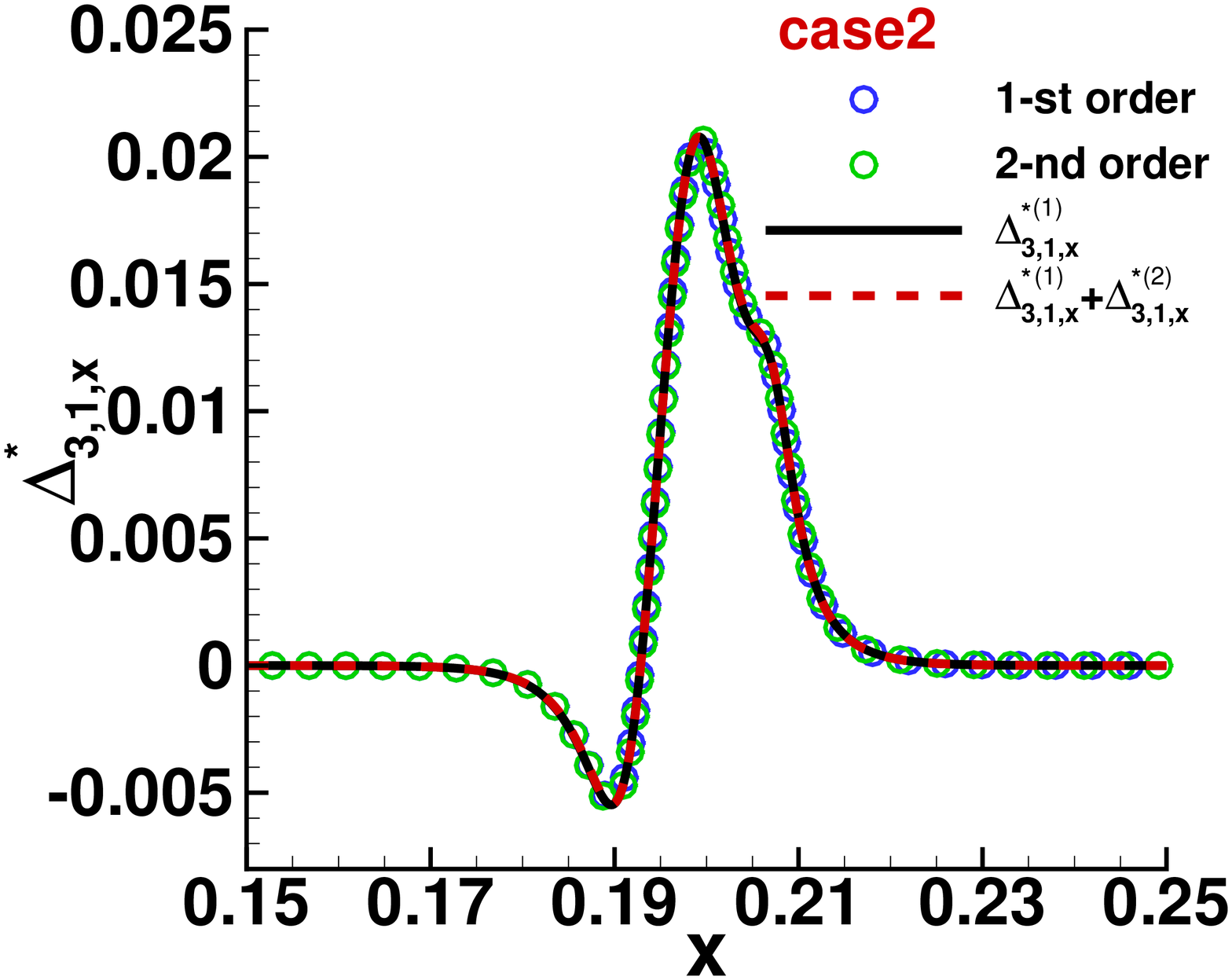}
	\label{fig0012b}
\end{minipage}
}
\subfigure[]{
\begin{minipage}{8cm}
\centering
	\includegraphics[width=7cm]{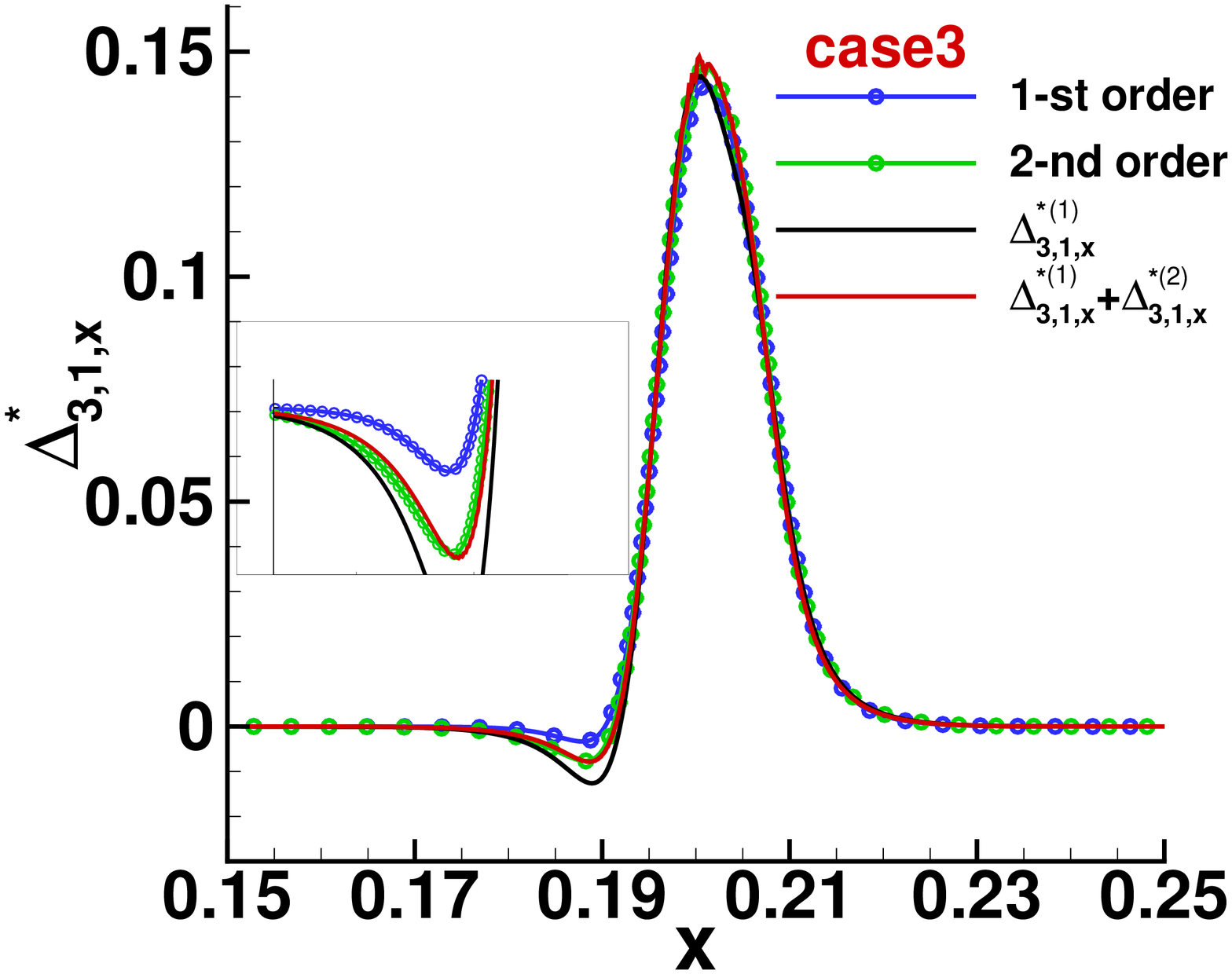}
	\label{fig0012c}
\end{minipage}
}
\caption{Comparison of $\Delta_{3,1,x}^{*}$ between DBM simulation results (blue circles from 1-st order DBM, green circles from 2-nd order DBM), analytical solutions at first-order accuracy (black lines),  and analytical solutions at second-order accuracy (red lines).
(a) Results for case 1,
(b) results for case 2,
and (c) results for case 3.
Two enlarged views show discernible differences at the bottom.
}
\label{fig0012}
\end{figure}

\begin{figure}[htbp]
\centering
\subfigure[]{
\begin{minipage}{8cm}
\centering
	\includegraphics[width=7cm]{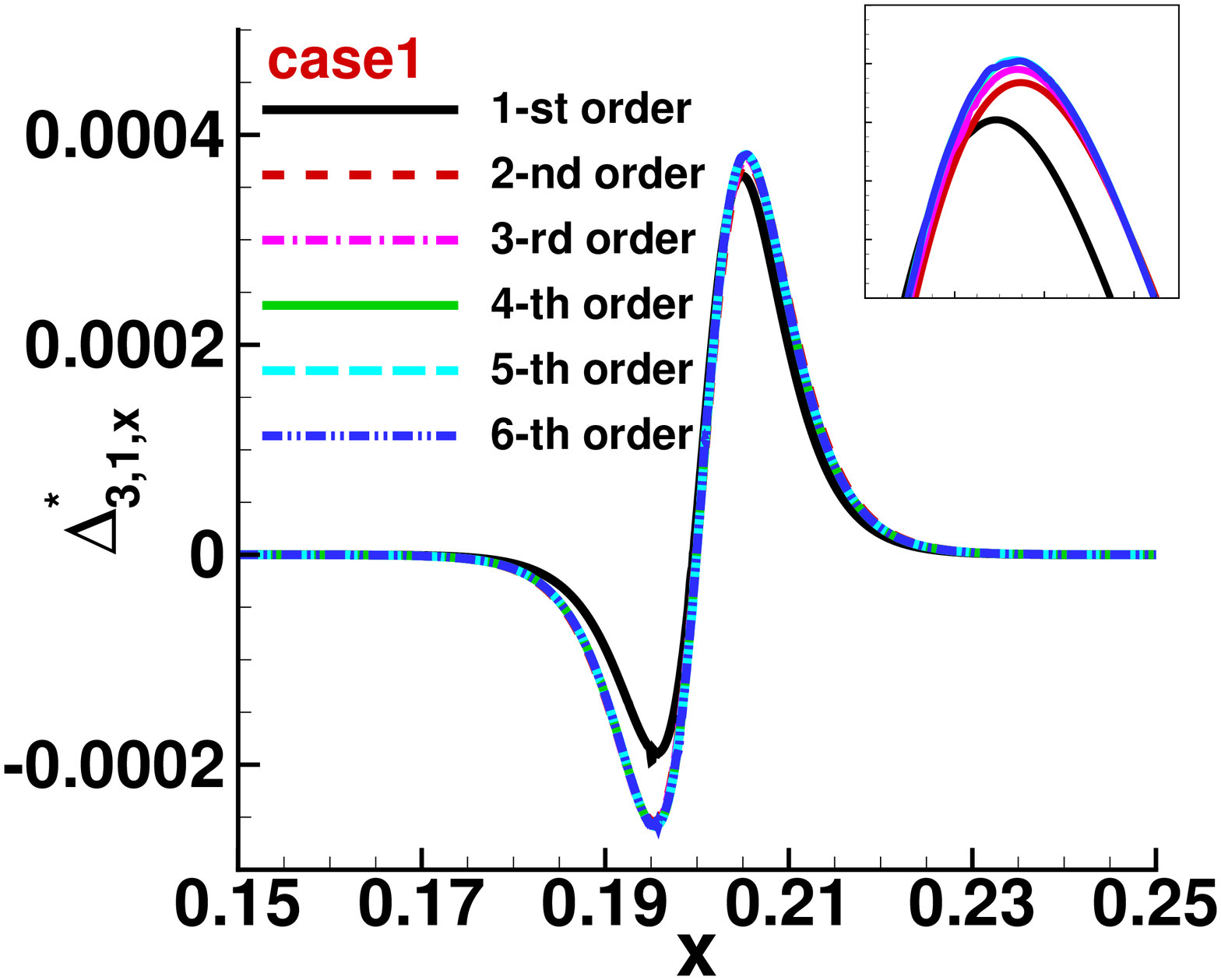}
	\label{fig0013a}
\end{minipage}
}
\subfigure[]{
\begin{minipage}{8cm}
\centering
	\includegraphics[width=7cm]{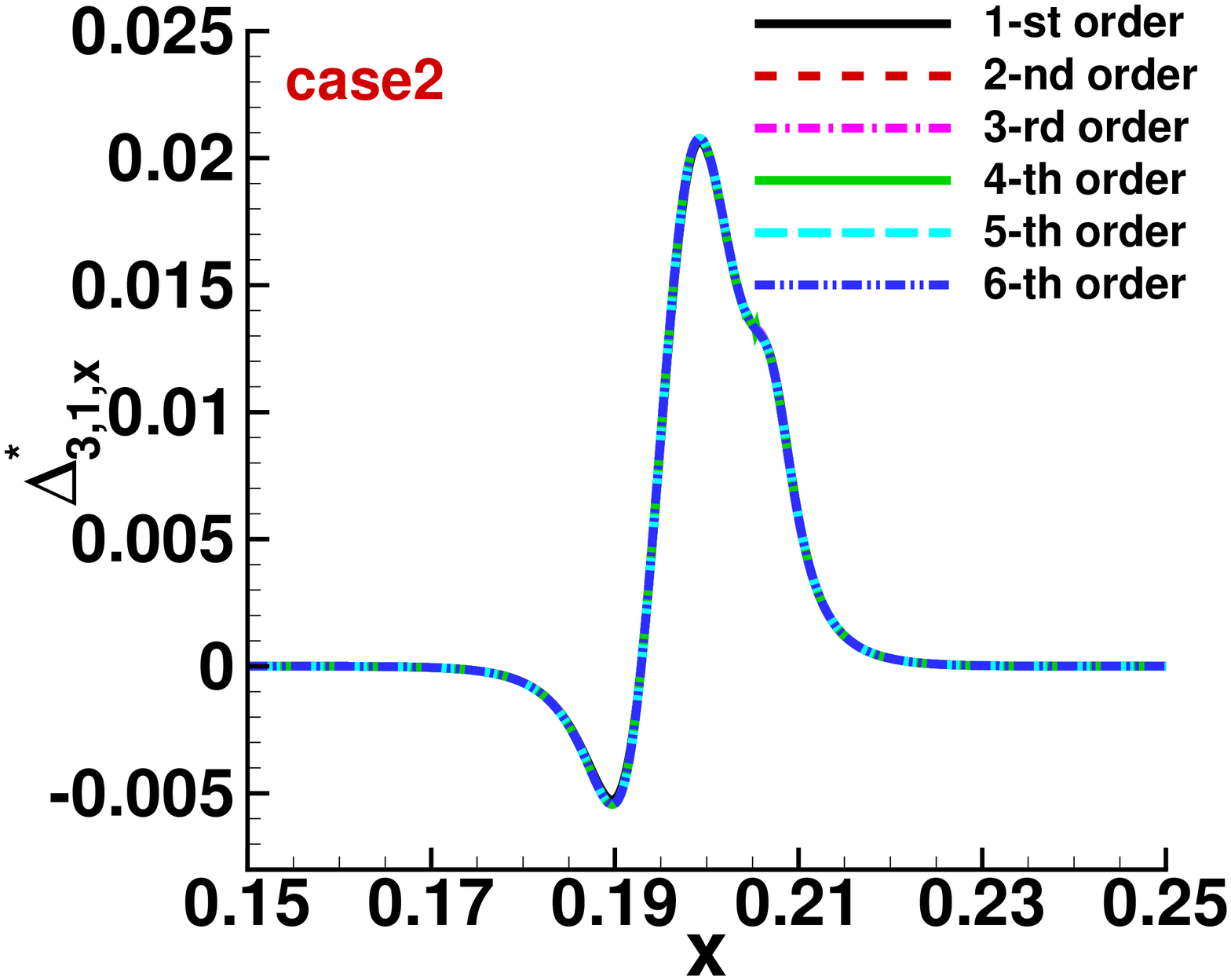}
	\label{fig0013b}
\end{minipage}
}
\subfigure[]{
\begin{minipage}{8cm}
\centering
	\includegraphics[width=7cm]{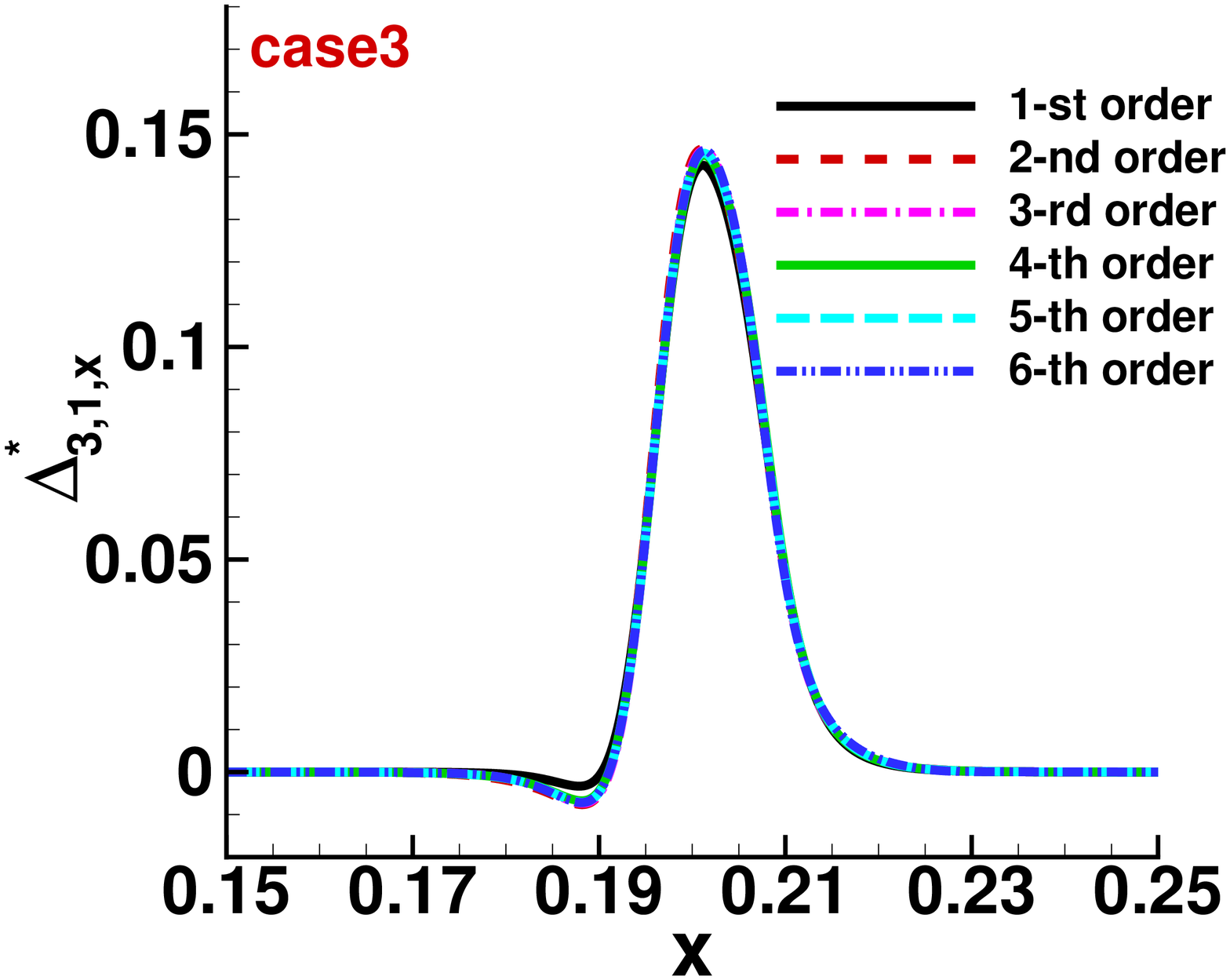}
	\label{fig0013c}
\end{minipage}
}
\caption{Simulation results of $\Delta_{3,1,x}^{*}$ from various DBMs: (a) from case 1, (b) from case 2, and (c) from case 3.
The lines with black (red, pink, green, light blue, and blue) color represent the results from 1-st order DBM (2-nd, 3-rd, 4-th, 5-th, and 6-th order DBM), respectively.}
\label{fig0013}
\end{figure}

Figure \ref{fig0012} shows the comparisons of $\Delta_{3,1,x}^{*}$ strength between DBM simulation results and analytical solutions.
The blue (green) circles represent results from 1-st order DBM (2-nd order DBM), and the black (red) lines indicate analytical solutions at first-order (second-order) accuracy calculated from Eqs.  (\ref{Eq.analysis2-1}) and (\ref{Eq.analysis2-2}).
The first (second, third) row corresponds to results from case 1 (case 2, case 3) at time $t=0.00018$ ($t=0.007$ and $t=0.0028$).
The enlarged view shows discernible differences at the bottom.
Similar to $\Delta_{2,xx}^{*}$, from case 1 to case 3, what we can see is that the strengths of $\Delta_{3,1,x}^{*}$ also increase because the TNE driving force increases.
With TNE degree increase, the 1-st order DBM gradually fails to describe the situation where $\Delta_{3,1,x}^{*(2)}$ cannot be ignored, whereas the 2-nd order DBM is acceptable.
Figure \ref{fig0013} shows simulation results from six various DBMs.
It can be seen in Figs. \ref{fig0013a} and \ref{fig0013c}, results from 1-st order DBM are obviously different from those of higher-order DBMs.
As shown in the enlarged view Fig. \ref{fig0013a}, results from higher-order DBMs also show slight differences near the peak region where the gradients of macroscopic quantity are significant.
Ignoring the slight differences, at least a 2-nd order DBM should be adopted for case 1 and case 3, whereas for case 2, a 1-st order DBM is enough.
Similarly to viscous stress, the TNE strengths of three cases obtained from various views are summarized in Table \ref{table6}.
The selected fluid models when focusing on different views are also listed.

\subsection{Fluid jet}
The fluid jet is encountered in many fields such as water conservancy, hydropower engineering, aerospace, and energy machinery.
It refers to a situation where fluids with a certain initial velocity are ejected from various forms of orifices or nozzles and mixed with the surrounding fluid (the same fluid or different)\cite{Eggers2008Physics}.
The most studied case is the free jet, in which the fluid spouts from the nozzle and enters an infinite space where there are fluids with the same characteristics.
The traditional simulations of fluid jets are always based on the hypothesis of equilibrium or near-equilibrium.
However, the narrow entrances lead to large gradients of macroscopic quantities and large local Kn numbers.
Consequently, the accurate simulations of the fluid jet have become a challenge.
In this paper, accurate simulations for free jets are conducted using four single-fluid DBMs: the 1-st order DBM, 2-nd order DBM, 3-rd order DBM, and 4-th order DBM.
The initial field of a free jet is shown by Fig. \ref{fig0014}, which is composed of a rectangle flow field with length $0.114$ and height $0.08$, and a rectangle entrance with length $0.048$ and height $0.006$ on the left side of the flow field.
The numbers in Fig. \ref{fig0014} represent the type of boundary conditions adopted in this simulation, i.e., the index ``1'' is the outflow boundary, ``2'' the inflow boundary, and ``3'' the solid wall boundary.
Considering computational efficiency and accuracy, the continuous flow field is discretized into uniform meshes with $N_x \times N_y=570 \times 400$, and the entrance $N_x \times N_y=30 \times 240$. The initial conditions of macroscopic quantities in the free jet are:

\begin{figure}[tbp]
\center\includegraphics*
[width=0.4\textwidth]{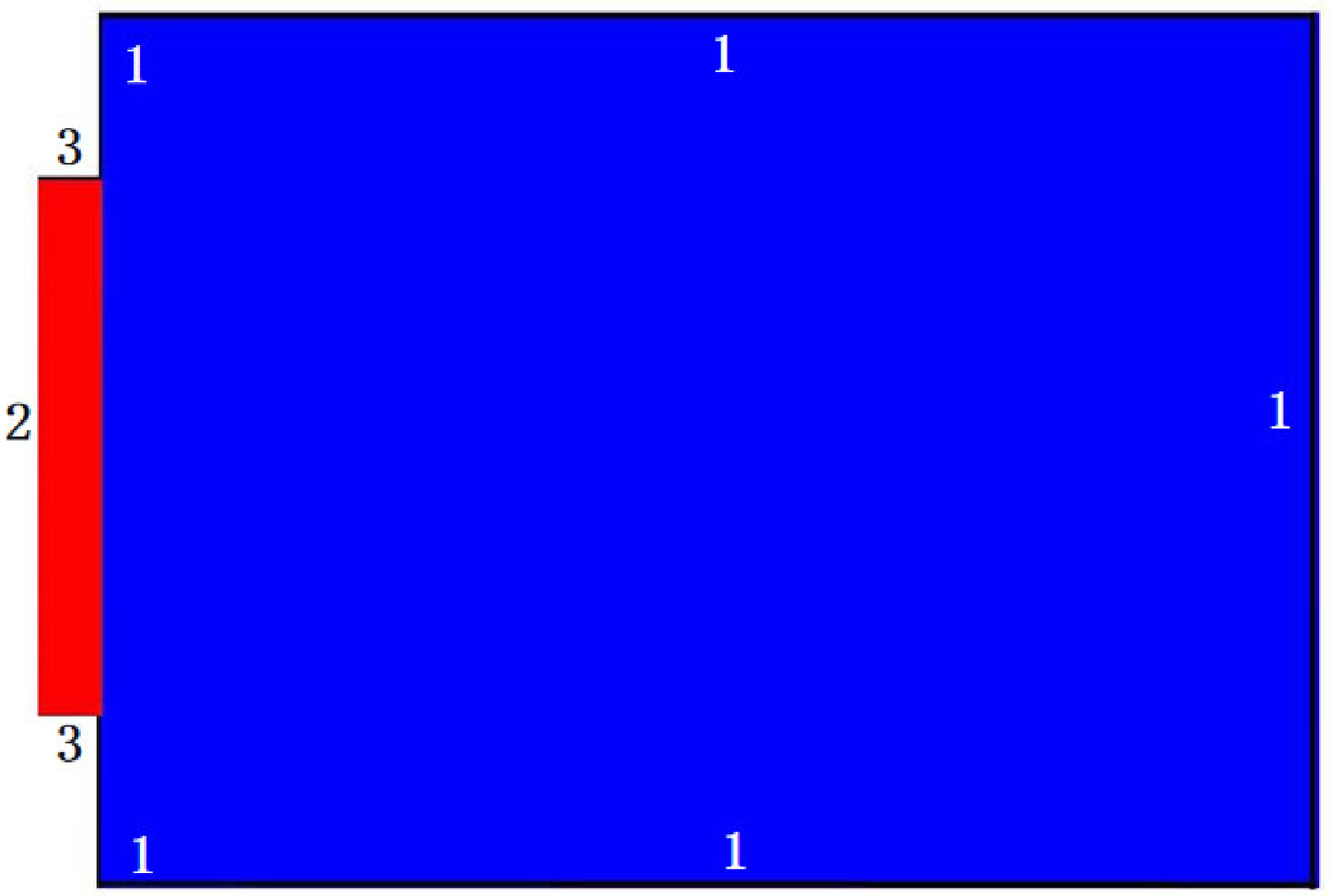}
\caption{The initial configuration of the free jet: the red part and blue part represent a rectangle entrance and a rectangle flow field, respectively. Indexes represent the types of boundary conditions, i.e., ``1'' is the outflow boundary, ``2'' the inflow boundary, and ``3'' is the solid wall boundary.} \label{fig0014}
\end{figure}

\[
\left\{
\begin{array}{l}
(\rho,u_x,u_y,p)^{e}_{x,y}=(1.28,0.3774,0.0,1.32096) \tt{,} \\
(\rho,u_x,u_y,p)^{f}_{x,y}=(0.1358,0.0,0.0,1.0) \tt{.}
\end{array}
\right.
\]
where ``e'' (``f'') means entrance (flow field). Other parameters used in the paper are: $m=1$, $\tau=2\times 10^{-5}$, $I=0$, $\eta=0$, $b=0$, $\Delta t=4\times 10^{-6}$, and $\Delta x=\Delta y=2\times 10^{-4}$.
Figure \ref{fig0015} shows density contours at three different times ($t=0$, 0.02, and 0.1, respectively).
It can be observed that after the heavy fluid ejects into the lighter fluid, a pair of vortexes is generated because of the Kelvin-Helmholtz instability.\cite{Gan2019FOP}
\begin{figure}[htbp]
\centering
\subfigure[]{
\begin{minipage}{8cm}
\centering
	\includegraphics[width=8cm]{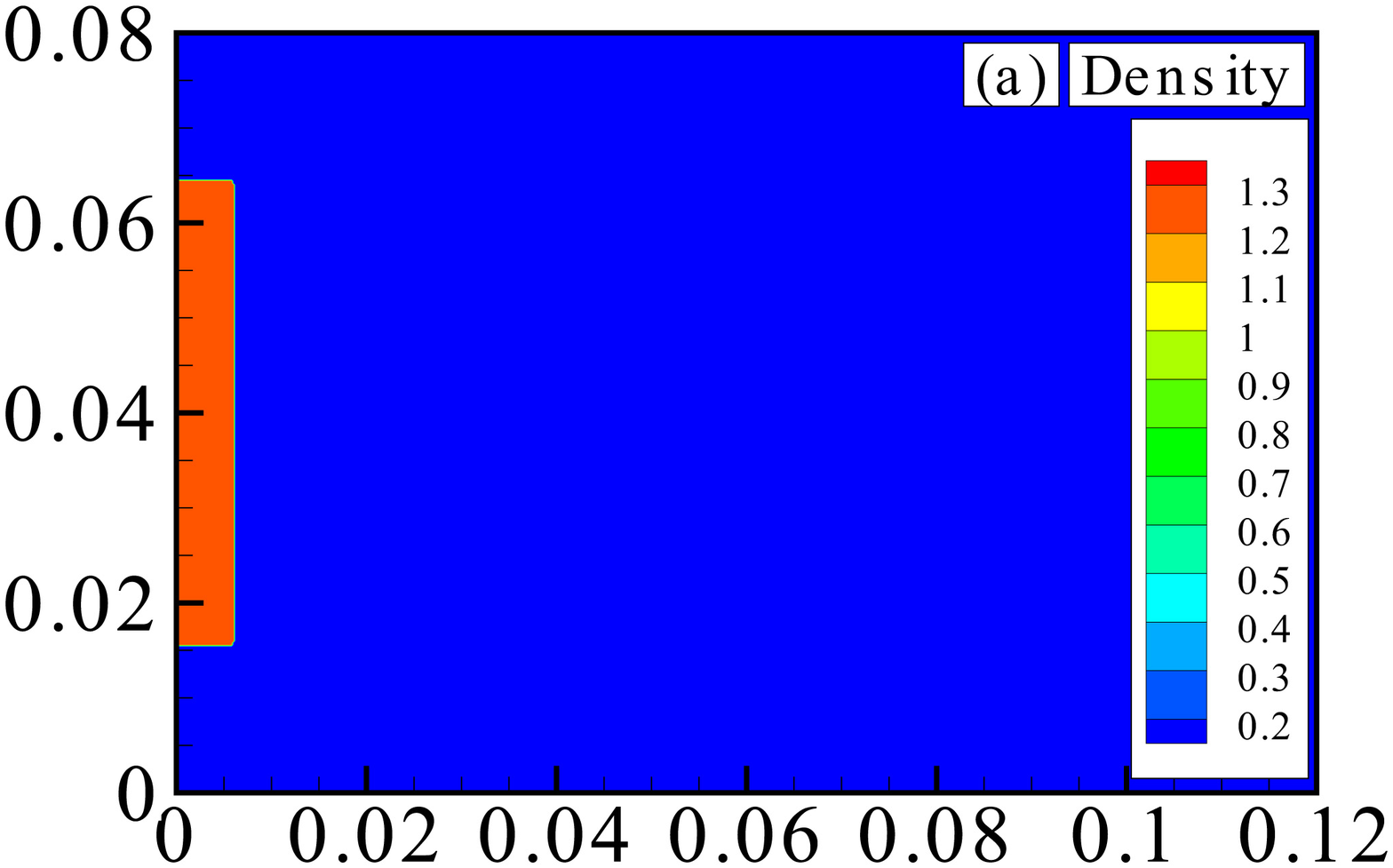}
	\label{fig0015a}
\end{minipage}
}
\subfigure[]{
\begin{minipage}{8cm}
\centering
	\includegraphics[width=8cm]{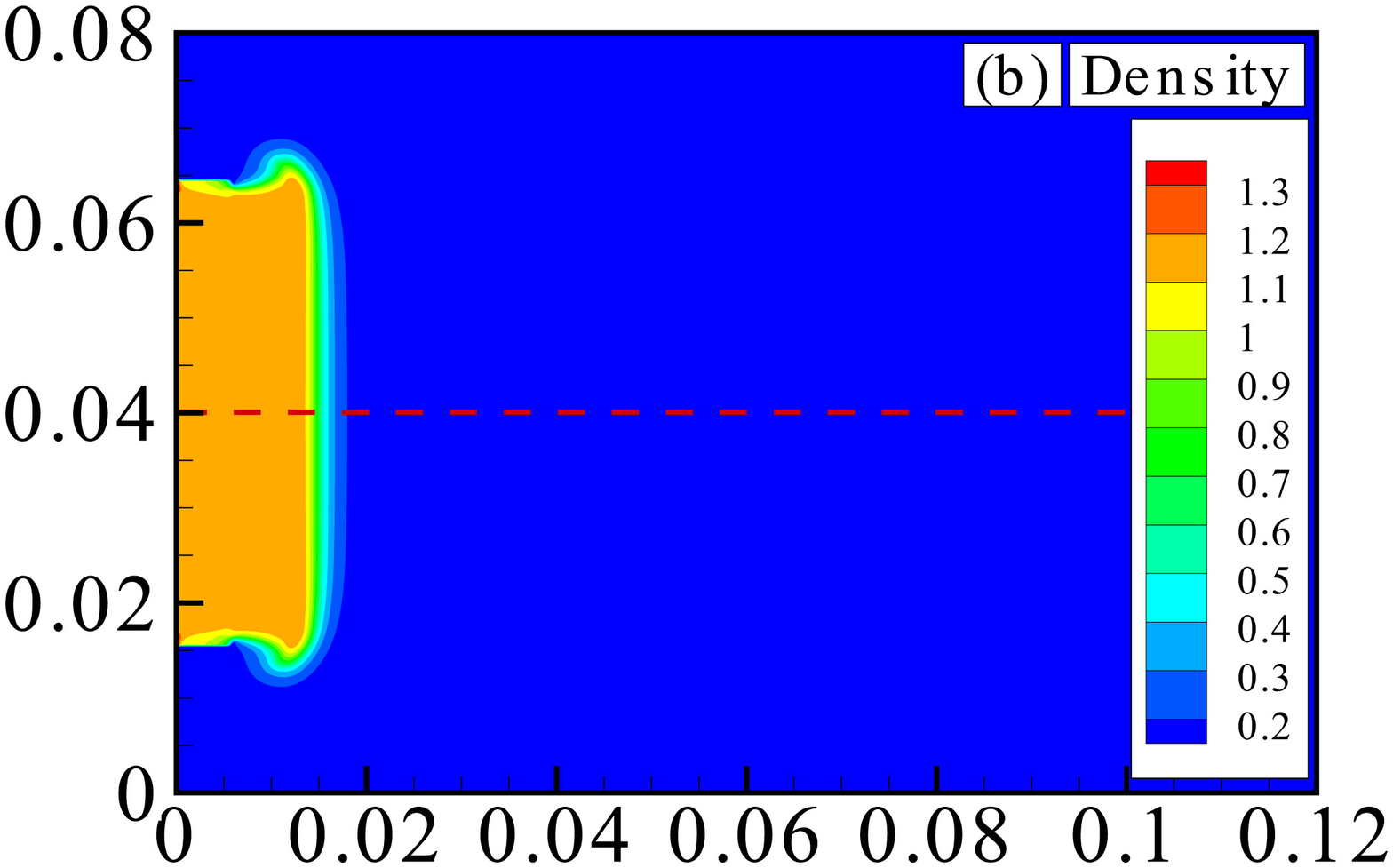}
	\label{fig0015b}
\end{minipage}
}
\subfigure[]{
\begin{minipage}{8cm}
\centering
	\includegraphics[width=8cm]{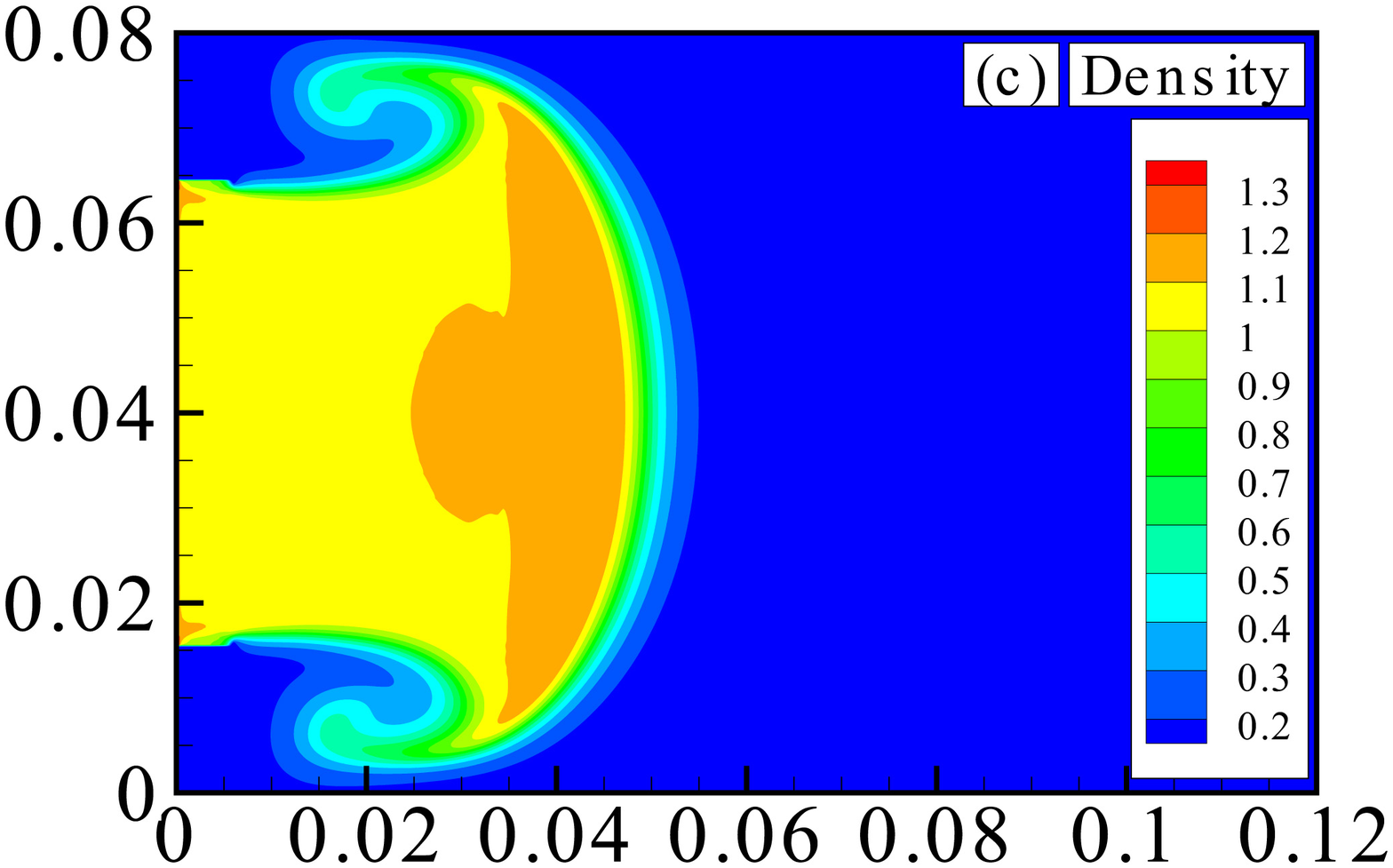}
	\label{fig0015c}
\end{minipage}
}
\caption{Density contours at different times: (a) $t=0$, (b) $t=0.02$, and (c) $t=0.1$.
The color from blue to red indicates the increase in density.} \label{fig0015}
\end{figure}

\begin{figure*}[htbp]
\centering
\subfigure[]{
\begin{minipage}{7cm}
\centering
	\includegraphics[width=7cm]{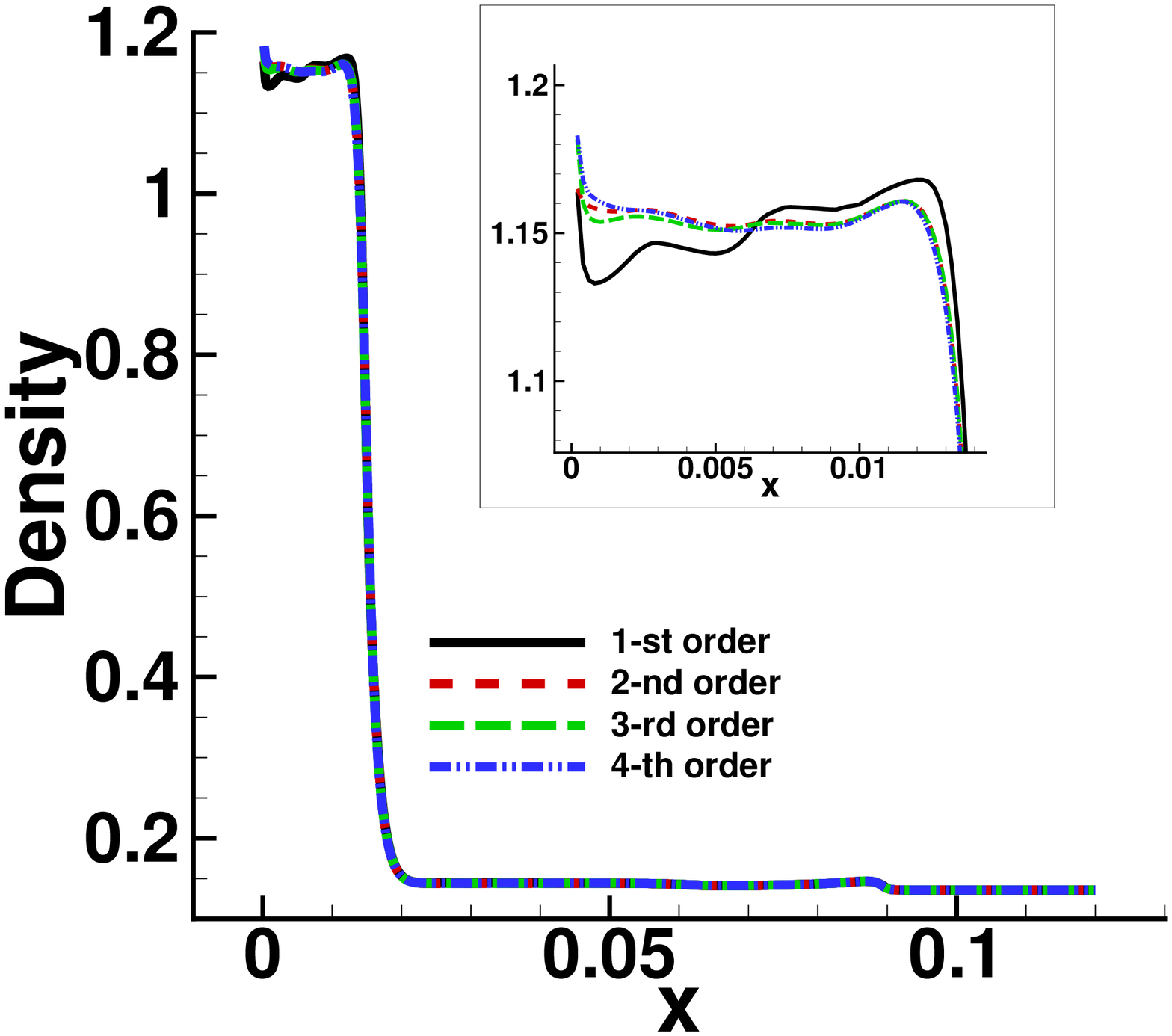}
	\label{fig0016a}
\end{minipage}
}
\subfigure{
\begin{minipage}{7cm}
\centering
	\includegraphics[width=7cm]{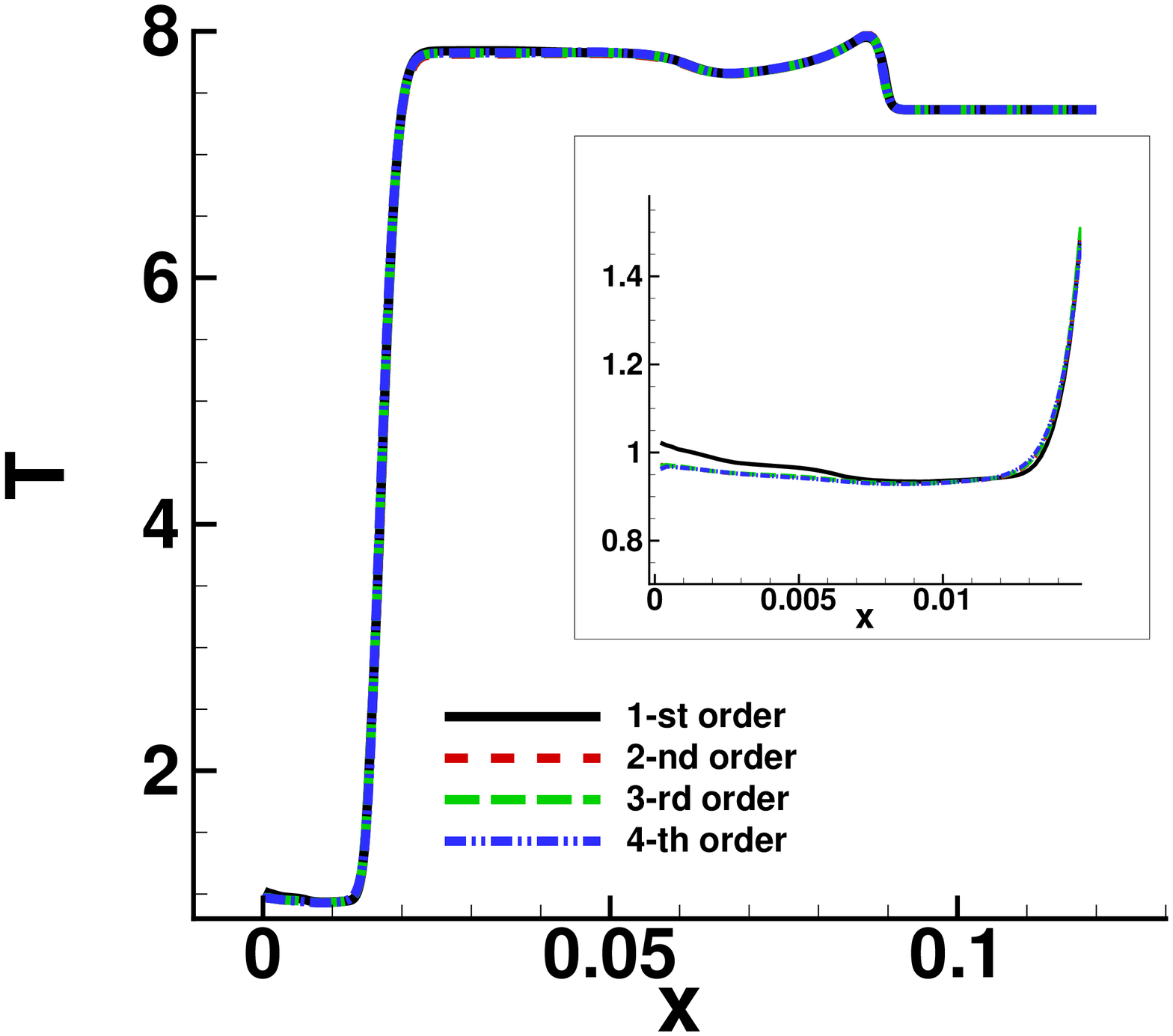}
	\label{fig0016b}
\end{minipage}
}
\subfigure[]{
\begin{minipage}{7cm}
\centering
	\includegraphics[width=7cm]{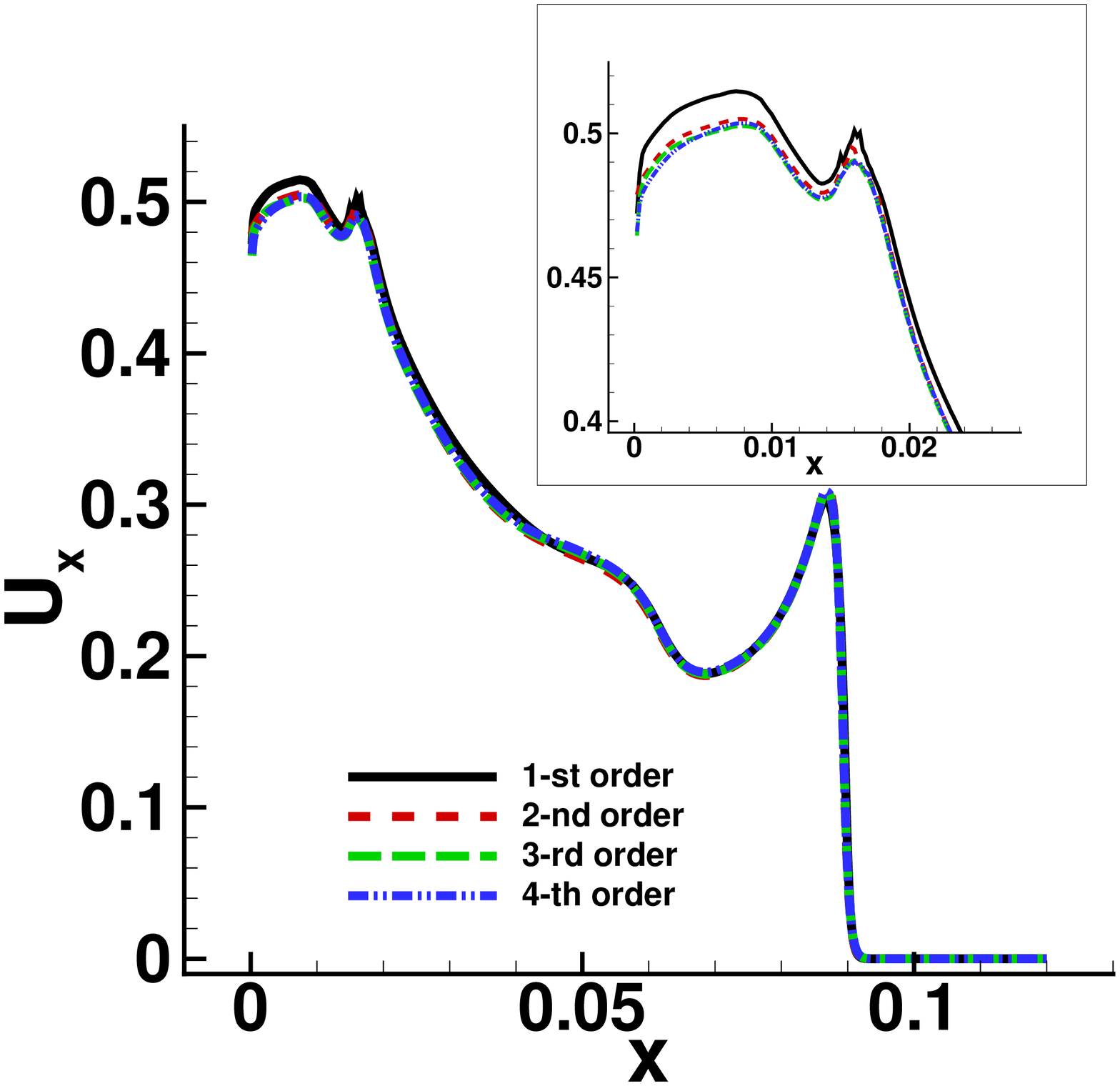}
	\label{fig0016c}
\end{minipage}
}
\subfigure[]{
\begin{minipage}{7cm}
\centering
	\includegraphics[width=7cm]{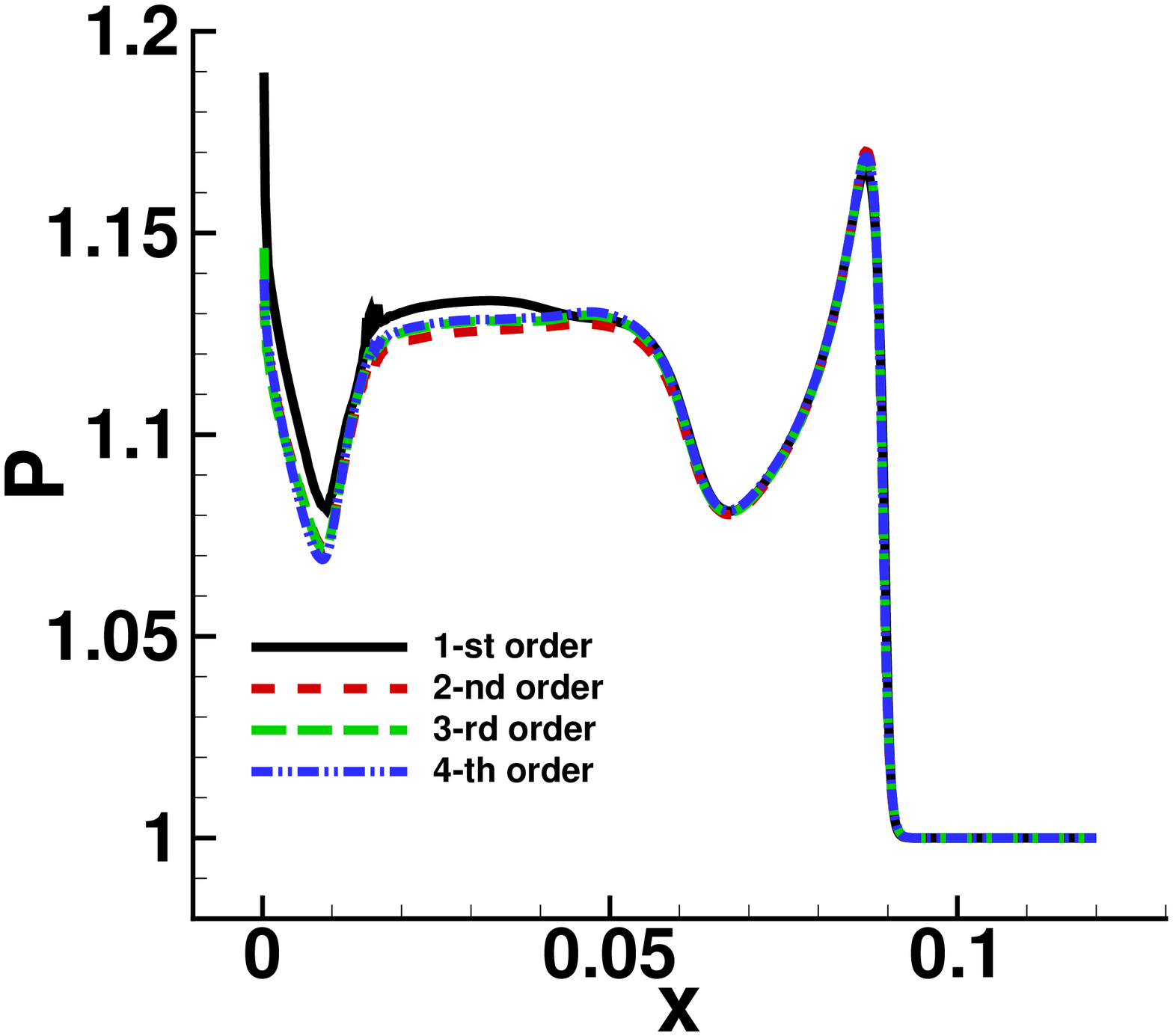}
	\label{fig0016d}
\end{minipage}
}
\caption{Profiles of quantities at $t=0.02$ along the $x$ direction, at $y=L_y/2$ (the red line in Fig. \ref{fig0015}(b)). 
(a) Density profiles, (b) temperature profiles, (c) velocity profiles, and (d) pressure profiles.
Results of the black line (red, green, and blue) are from 1-st order (2-nd, 3-rd, and 4-th order) DBM. Enlarged views in the figures show the slight difference between various lines.}
\label{fig0016}
\end{figure*}

For comparison, four DBMs are used to simulate this problem.
Figure \ref{fig0016} shows the profiles of various macroscopic quantities at time $t=0.02$, along the $x$ direction at $y=L_y/2$ (the red line in Fig. \ref{fig0015b}).
Figures \ref{fig0016a}-\ref{fig0016d} represent profiles of density, temperature, velocity, and pressure, respectively.
The black (red, green, and blue) lines represent results from the 1-st order DBM (2-nd, 3-rd, and 4-th order DBM, respectively).
Enlarged views in the figures show the slight difference between various lines.
In all four figures, because large gradients of macroscopic quantity exist around the entrance, distinct differences between the black line and other results can be found.
Whereas far away from the entrance, there is almost no difference.
Meanwhile, as shown in Figs. \ref{fig0016c} and \ref{fig0016d}, discernible differences begin to appear between the red line and results from higher-order ones (green line and blue line).
Consequently, to accurately simulate the free jet of this case, at least up to the third-order TNE effects should be included.

\begin{figure*}[htbp]
\centering
\subfigure[]{
\begin{minipage}{7cm}
\centering
	\includegraphics[width=7cm]{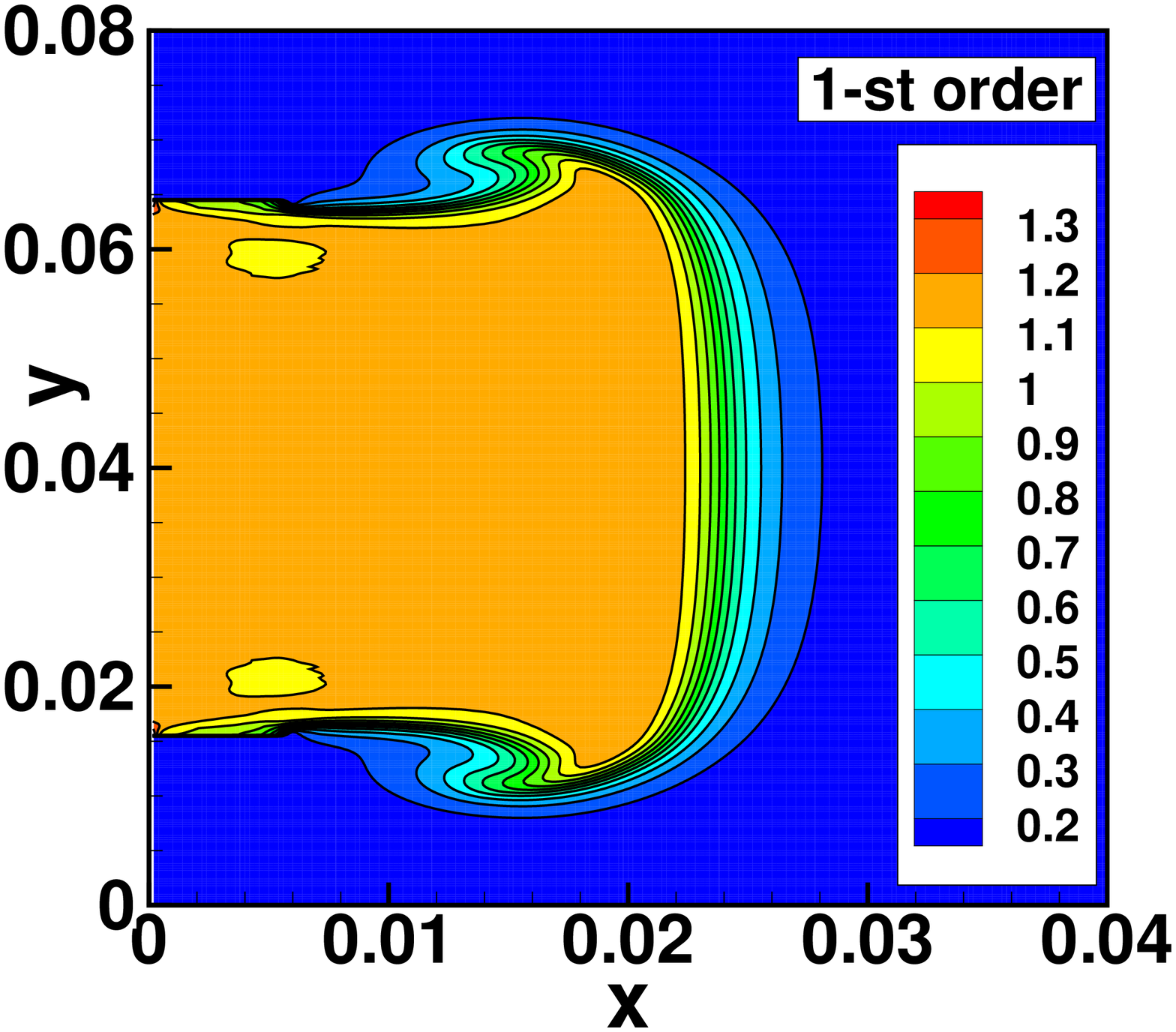}
	\label{fig0017a}
\end{minipage}
}
\subfigure[]{
\begin{minipage}{7cm}
\centering
	\includegraphics[width=7cm]{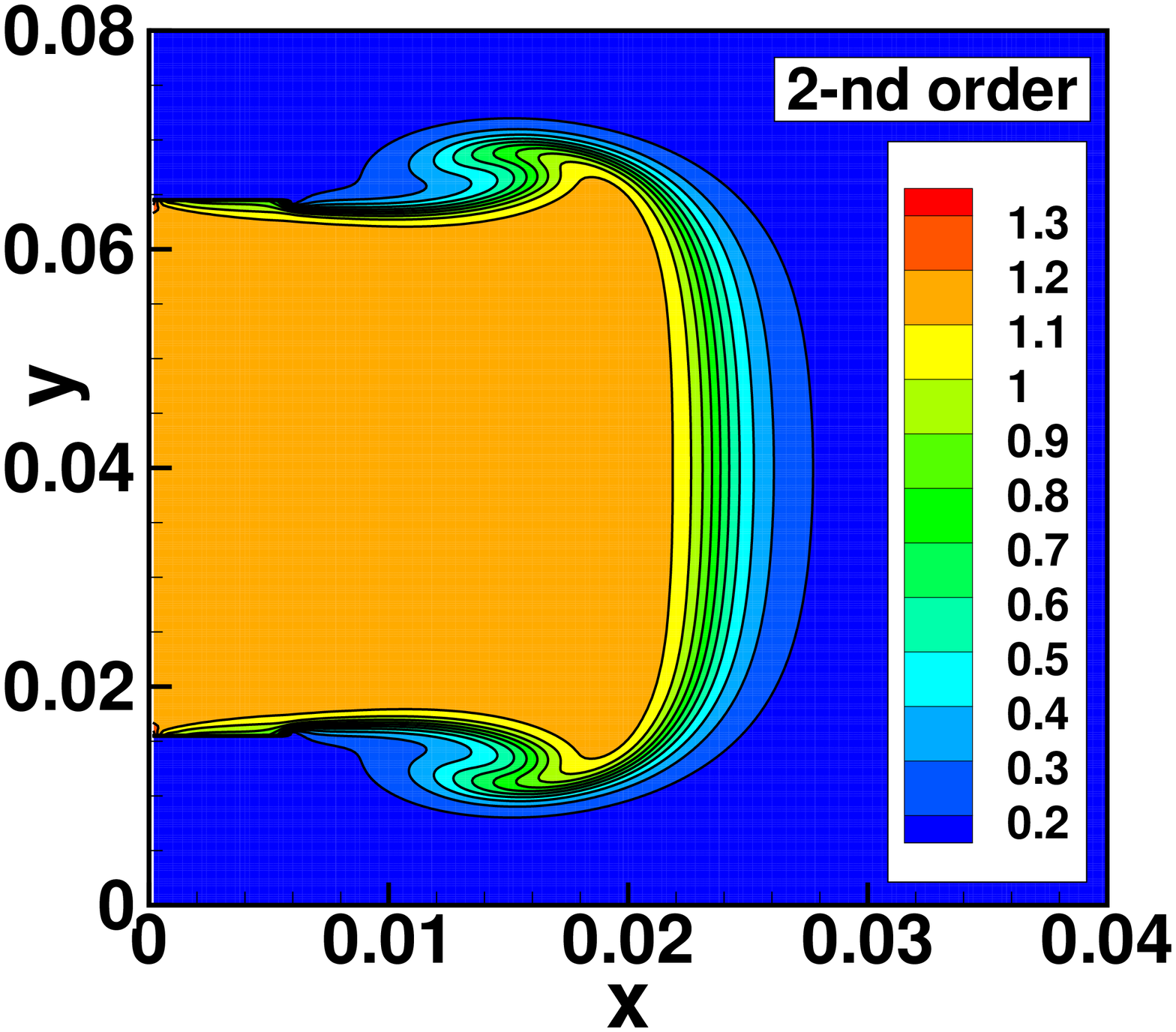}
	\label{fig0017b}
\end{minipage}
}
\subfigure[]{
\begin{minipage}{7cm}
\centering
	\includegraphics[width=7cm]{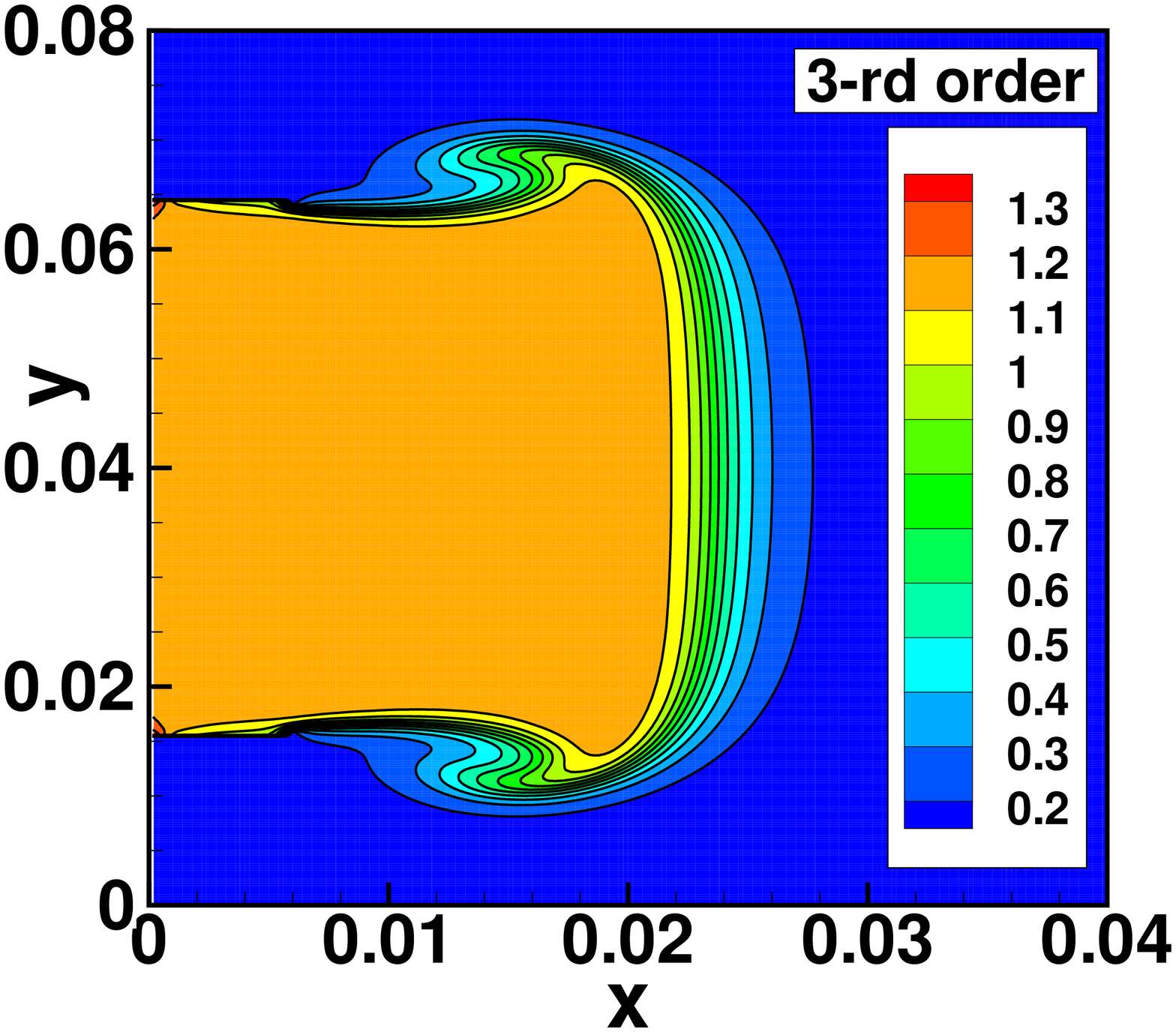}
	\label{fig0017c}
\end{minipage}
}
\subfigure[]{
\begin{minipage}{7cm}
\centering
	\includegraphics[width=7cm]{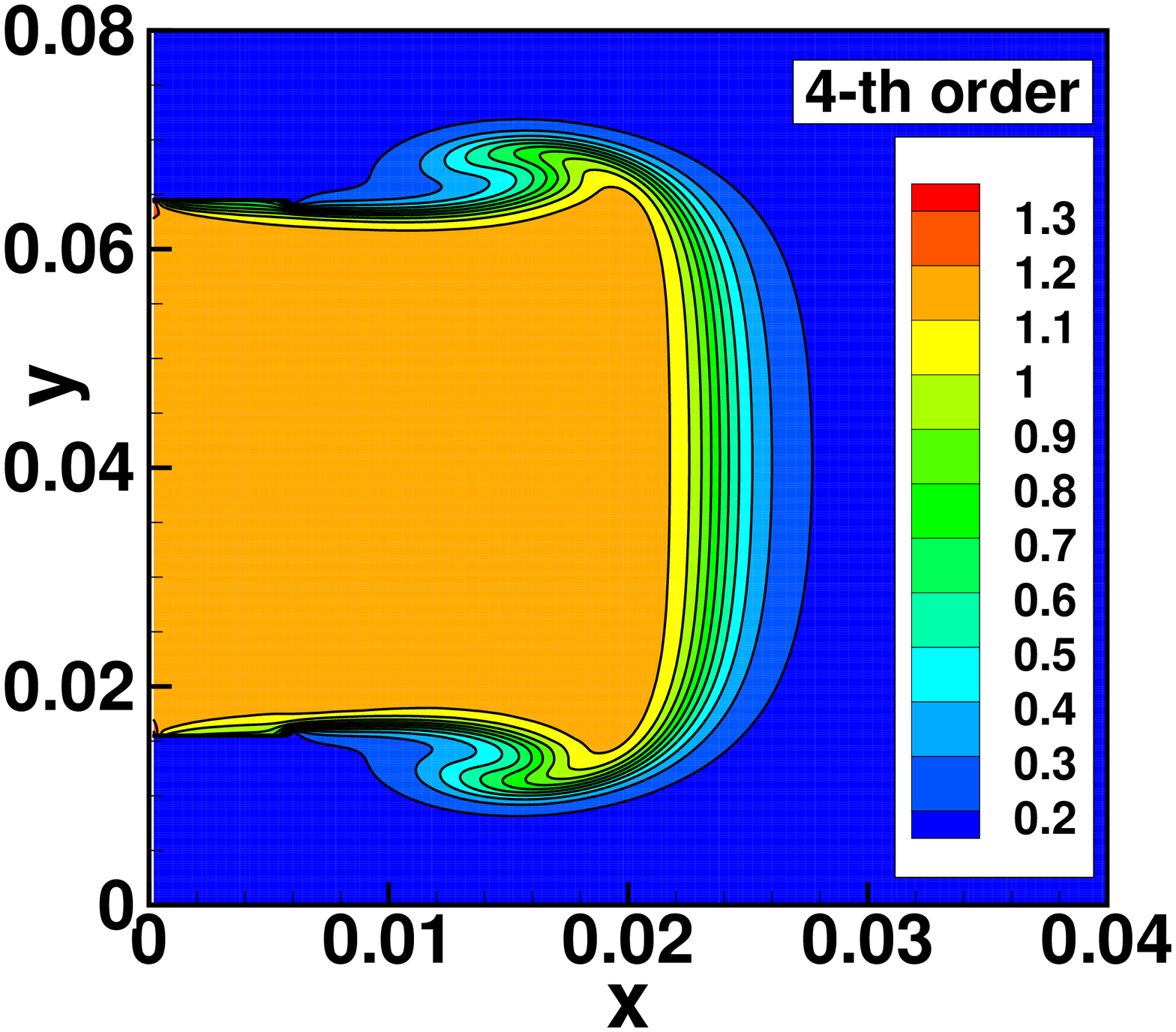}
	\label{fig0017d}
\end{minipage}
}
\caption{ Density contours at time $t$=0.04. 
(a) The contour from the 1-st order DBM, (b) from the 2-nd order DBM, (c) from the 3-rd order DBM, and (d) from the 4-th order DBM.}
\label{fig0017}
\end{figure*}

To show clearly the distinctions of simulation results between various DBMs, the density contours at time $t$=0.04 are demonstrated in Fig. \ref{fig0017}.
Figures \ref{fig0017a}-\ref{fig0017d} represent: the results from 1-st, 2-nd, 3-rd, and 4-th order DBMs, respectively.
It is observed that the isolines between 1-st order DBM and higher-order DBMs show discernible differences.

\section{Conclusions}\label{Conclusions}

DBMs that consider sufficient higher-order non-equilibrium effects have been developed to investigate the high-speed compressible flow in various depths of non-equilibrium.
In the process of constructing a DBM, the CE analysis is only used to quickly fix the kinetic moments which should keep values unchanged instead of deriving complicated high-order hydrodynamic equations.
As model examples, DBMs considering, up to from the first to the sixth order, TNE effects are examined.
Numerical tests cover a wide range, including the Riemann problem and the Couette flow (which corresponds to the large structure of macroscopic quantities), shock wave structure (which corresponds to the fine structure of density profile), the impact of two flows with various viscous effects and colliding velocities (or with various heat conduction and colliding pressures, which correspond to the fine structures of TNE quantities), and two-dimensional high-speed free jet (which correspond to macroscopic quantities).
The last two tests contain various degrees of velocity gradients which triggers various degrees of TNE effects.
It is demonstrated that the non-equilibrium depth cannot be fully described by a single parameter. Consequently, we propose to use a vector $\mathbf{S}_{TNE}$ to describe the TNE from various aspects under investigation.
With increasing TNE, more higher-order non-conserved moments should be included in the DBM to describe TNE behavior.
DBM with high order TNE may bring substantial contribution in studying the kinetic physics in ICF, aerospace field, microscale flow, etc.

\begin{acknowledgments}
The authors thank Chuandong Lin, Feng Chen, Ge Zhang, Jiahui Song, Yiming Shan, Cheng Chen, Jie Chen, Hanwei Li, and Yingqi Jia on helpful discussions on DBM.
This work was supported by the National Natural Science Foundation of China (under Grant Nos. 12172061 and 11875001), the Strategic Priority Research Program of Chinese Academy of Sciences (under Grant No. XDA25051000), the Opening Project of State Key
Laboratory of Explosion Science and Technology (Beijing Institute of Technology) (under Grant No. KFJJ21-16M), and Science Foundation of Hebei Province (Grant Nos. A2021409001, A202105005, and 226Z7601G), and fund of  Laboratory of Computational Physics.
\end{acknowledgments}

\section*{Data availability}
The data that support the findings of this study are available from the corresponding author upon reasonable request.

\appendix
\label{sec:Appendixes}
\section{Appendixes A: constitutive relationships of viscous stress and heat flux in fluid equations} \label{sec:AppendixesA}
From the CE multiscale analysis, by retaining various orders of Kn number, the Boltzmann equation can reduce to the corresponding macroscopic fluid equations which can be used for flows in the corresponding flow regimes.
In the following, we used the analytic formulas from previous literature\cite{2018Gan-pre,Zhang2017Discrete}.
The completed single-fluid macroscopic fluid equations are as follows:
\begin{equation}
\frac{\partial \rho}{\partial t}+\frac{\partial (\rho u_{\alpha})}{\partial r_{\alpha}}=0
\end{equation}
\begin{eqnarray}
\begin{aligned}
\frac{\partial}{\partial t}(\rho u_{\alpha})
&+\frac{\partial (p\delta_{\alpha\beta}+\rho
u_{\alpha}u_{\beta})}{\partial r_{\beta}}+\frac{\partial \Delta^{*}_{2,\alpha\beta}}{\partial r_{\beta}}=0
\end{aligned}
\end{eqnarray}
\begin{eqnarray}
\begin{aligned}
\frac{\partial}{\partial t}\rho E_{T}&+\frac{\partial}{\partial r_{\alpha}}(\rho E_{T}+p)u_{\alpha}
+\frac{\partial}{\partial r_{\beta}}[u_{\alpha}\Delta^{*}_{2,\alpha\beta}\\
&+\Delta_{3,1,\beta}^{*}]=0
\tt{.}
\end{aligned}
\end{eqnarray}
where $p=nkT$, $E_T=\frac{1}{2}[DT/m+u_{\alpha}^2]$ are the pressure and energy per unit mass in the case of a fixed extra degrees of freedom, i.e., $I=0$. The complete analytical solution of viscous stress and heat flux are as follows:
\begin{equation}
\Delta^{*}_{2,\alpha\beta}=\Delta^{(1)*}_{2,\alpha\beta}+\Delta^{*(2)}_{2,\alpha\beta}+\cdots+\Delta^{*(n)}_{2,\alpha\beta}
\tt{,}
\end{equation}
\begin{equation}
\Delta_{3,1,\beta}^{*}=\Delta_{3,1,\beta}^{*(1)}+\Delta_{3,1,\beta}^{*(2)}+\cdots+\Delta_{3,1,\beta}^{*(n)}
\end{equation}
By retaining different orders of Kn number, various orders of macroscopic fluid models can be obtained.
For convenience, we call $\Delta_{2,\alpha \beta}^{*(n)}$ ($\Delta_{3,1,\beta}^{*(n)}$) the $\text{n-}th$ order term of viscous stress (heat flux).
For example, when retaining up to order $O(Kn^0)$, the Euler equations which do not consider viscosity and heat conduction are obtained, i.e., $\Delta^{*}_{2,\alpha\beta}=0$ and $\Delta_{3,1,\beta}^{*}=0$.
When retaining up to order $O(Kn^1)$, the NS equation are derived, in which $\Delta^{*}_{2,\alpha\beta}=\Delta^{*(1)}_{2,\alpha\beta}$ and $\Delta_{3,1,\beta}^{*}=\Delta_{3,1,\beta}^{*(1)}$.
The analytical expressions of first-order term of viscous stress and heat flux are in the following (in the case of $D=2$ and $I=0$):
\begin{equation}
\Delta^{*(1)}_{2,\alpha\beta}=-\mu(\frac{\partial u_{\alpha}}{\partial r_{\beta}}+\frac{\partial u_{\beta}}{\partial r_{\alpha}}-\frac{2}{D}\frac{\partial u_{\gamma}}{\partial r_{\gamma}}\delta_{\alpha\beta})
\tt{,} \label{Eq.analysis-1}
\end{equation}
\begin{equation}
\Delta_{3,1,\beta}^{*(1)}=-\kappa \frac{\partial (T/m)}{\partial r_{\beta}}\label{Eq.analysis2-1}
\end{equation}
where $\mu$ is the viscosity coefficient and $\kappa$ represents the heat conductivity coefficient.
When retaining up to order $O(Kn^2)$, the Burnett equation are derived, in which $\Delta^{*}_{2,\alpha\beta}=\Delta^{*(1)}_{2,\alpha\beta}+\Delta^{*(2)}_{2,\alpha\beta}$ and $\Delta_{3,1,\beta}^{*}=\Delta_{3,1,\beta}^{*(1)}+\Delta_{3,1,\beta}^{*(2)}$.
Expressions of $\Delta^{*(2)}_{2,\alpha\beta}$ and $\Delta_{3,1,\beta}^{*(2)}$ are as follows:
\begin{eqnarray}
\begin{aligned}
\Delta_{2,xx}^{*(2)}
&=\frac{\tau^{2}}{(1-b)^2}\{(1-b)\rho [(\frac{\partial T}{\partial x})^2-(\frac{\partial T}{\partial y})^2 ]- \rho T b (\frac{\partial ^2 T}{\partial x^2}-\frac{\partial ^2 T}{\partial y^2})\\
&-\rho T [[(\frac{\partial  u_x}{\partial x})^2-(\frac{\partial u_x}{\partial y})^2]+[(\frac{\partial u_y}{\partial x})^2-\frac{\partial u_y}{\partial y})^2]] \\
&+\frac{T^2}{\rho}[(\frac{\partial \rho}{\partial x})^2-(\frac{\partial \rho}{\partial y})^2]-bT[\frac{\partial T}{\partial x}\frac{\partial \rho}{\partial x}-\frac{\partial T}{\partial y}\frac{\partial \rho}{\partial y}]\\
&-T^2 (\frac{\partial ^2 \rho}{\partial x^2}-\frac{\partial ^2 \rho}{\partial y^2})
\}
\tt{.} \label{Eq.analysis-2}
\end{aligned}
\end{eqnarray}

The $xy$ and $yy$ components of viscous stress are as follows:
\begin{eqnarray}
\begin{aligned}
\Delta_{2,yy}^{*(2)}&=\frac{\tau^{2}}{(1-b)^2}\{(1-b)\rho [(\frac{\partial T}{\partial y})^2-(\frac{\partial T}{\partial x})^2 ]- \rho T b (\frac{\partial ^2 T}{\partial y^2}-\frac{\partial ^2 T}{\partial x^2})\\
&-\rho T [[(\frac{\partial  u_x}{\partial y})^2-(\frac{\partial u_x}{\partial x})^2]+[(\frac{\partial u_y}{\partial y})^2-\frac{\partial u_y}{\partial x})^2]] \\
&+\frac{T^2}{\rho}[(\frac{\partial \rho}{\partial y})^2-(\frac{\partial \rho}{\partial x})^2]+bT[\frac{\partial T}{\partial x}\frac{\partial \rho}{\partial x}-\frac{\partial T}{\partial y}\frac{\partial \rho}{\partial y}]\\
&-T^2 (\frac{\partial ^2 \rho}{\partial y^2}-\frac{\partial ^2 \rho}{\partial x^2})
\}
\tt{.}
\end{aligned}
\end{eqnarray}
\begin{eqnarray}
\begin{aligned}
\Delta_{2,xy}^{*(2)}&=2\frac{\tau^{2}}{(1-b)^2}[(1-b)\rho \frac{\partial T}{\partial x}\frac{\partial T}{\partial y}- \rho T b \frac{\partial ^2 T}{\partial x \partial y}\\
&-\rho T (\frac{\partial u_x}{\partial x}\frac{\partial u_x}{\partial y}+\frac{\partial u_y}{\partial x}\frac{\partial u_y}{\partial y}) \\
&+\frac{T^2}{\rho}\frac{\partial \rho}{\partial x}\frac{\partial \rho}{\partial y}-bT\frac{\partial T}{\partial x}\frac{\partial \rho}{\partial y}-T^2\frac{\partial^2 \rho}{\partial x \partial y}
]
\tt{.}
\end{aligned}
\end{eqnarray}

The constitutive relationship of the second-order term of heat flux  are as follows:
\begin{eqnarray}
\begin{aligned}
\Delta_{3,1,x}^{*(2)}&=\frac{\tau^2}{1-b}\rho T [ (2+b)(\frac{\partial T}{\partial x}\frac{\partial u_x}{\partial x}+\frac{\partial T}{\partial y}\frac{\partial u_y}{\partial x})\\
&+(6-3b)(\frac{\partial u_x}{\partial x}\frac{\partial T}{\partial x}
+\frac{\partial u_x}{\partial y}\frac{\partial T}{\partial y})\\
&-(6-3b)(\frac{\partial u_x}{\partial x}\frac{\partial T}{\partial x}+\frac{\partial u_y}{\partial y}\frac{\partial T}{\partial x})\\
&-2(1-b)T(\frac{\partial^2 u_x}{\partial x^2}+\frac{\partial^2 u_y}{\partial x \partial y})+T(\frac{\partial ^2 u_x}{\partial x^2}+\frac{\partial ^2 u_x}{\partial y^2})
]
\tt{.} \label{Eq.analysis2-2}
\end{aligned}
\end{eqnarray}
\begin{eqnarray}
\begin{aligned}
\Delta_{3,1,y}^{*(2)}&=\frac{\tau^2}{1-b}\rho T [ (2+b)(\frac{\partial T}{\partial x}\frac{\partial u_x}{\partial y}+\frac{\partial T}{\partial y}\frac{\partial u_y}{\partial y})\\
&+(6-3b)(\frac{\partial u_y}{\partial x}\frac{\partial T}{\partial x}+\frac{\partial u_y}{\partial y}\frac{\partial T}{\partial y})\\
&-(6-3b)(\frac{\partial u_x}{\partial x}\frac{\partial T}{\partial y}+\frac{\partial u_y}{\partial y}\frac{\partial T}{\partial y})\\
&-2(1-b)T(\frac{\partial^2 u_x}{\partial y \partial x}+\frac{\partial^2 u_y}{\partial y^2 })+T(\frac{\partial ^2 u_y}{\partial y \partial x}+\frac{\partial ^2 u_y}{\partial y^2})
]
\tt{.}
\end{aligned}
\end{eqnarray}

Obviously, when retaining more higher orders of Kn number, it is too difficult to derive the analytical expressions of viscous stress and heat flux through CE analysis because of its complexity.

\section{ Appendixes B: experssions of the kinetic moments} \label{sec:AppendixesB}
When considering only the zeroth-order TNE effects, five kinetic moments($\mathbf{M}_0$, $\mathbf{M}_1$, $\mathbf{M}_{2,0}$, $\mathbf{M}_2$, $\mathbf{M}_{3,1}$) are needed.
When considering up to the first-order TNE effects, seven kinetic moments($\mathbf{M}_0$, $\mathbf{M}_1$, $\mathbf{M}_{2,0}$, $\mathbf{M}_2$, $\mathbf{M}_{3,1}$, $\mathbf{M}_3$, $\mathbf{M}_{4,2}$) are needed.
When considering up to the second-order TNE effects, at least the zeroth-order to (5,3)th order kinetic moments (i.e., $\mathbf{M}_0$, $\mathbf{M}_1$, $\mathbf{M}_{2,0}$, $\mathbf{M}_2$, $\mathbf{M}_{3,1}$, $\mathbf{M}_3$, $\mathbf{M}_{4,2}$, $\mathbf{M}_4$, $\mathbf{M}_{5,3}$) are necessary, according to CE multiscale expansion, where ``5,3'' means that the fifth-order tensor is contracted to a third-order tensor.
Similarly, when developing a DBM in which the third-order (fourth-, fifth-, and sixth-order) TNE effects are considered, two more moments, i.e., $\mathbf{M}_5$ ($\mathbf{M}_6$, $\mathbf{M}_7$ and $\mathbf{M}_8$) and $\mathbf{M}_{6,4}$ ($\mathbf{M}_{7,5}$, $\mathbf{M}_{8,6}$, and $\mathbf{M}_{9,7}$), should be retained, respectively.
The kinetic moments are written as follows:

\begin{equation}
M^{ES}_{0}=\sum_{i}f^{ES}_{i}=n \tt{,}
\end{equation}
\begin{equation}
M^{ES}_{1,x}=\sum_{i}f^{ES}_{i}v_{ix}=nu_{x} \tt{,}
\end{equation}
\begin{equation}
M^{ES}_{1,y}=\sum_{i}f^{ES}_{i}v_{iy}=nu_{y} \tt{,}
\end{equation}
\begin{eqnarray}
\begin{aligned}
M^{ES}_{2,0}&=\sum_{i}f^{ES}_{i}(v_{ix}^2+v_{iy}^2+\eta^{2}_{i})=\frac{n}{m}[\lambda_{xx}+\lambda_{yy}\\
&+m(u_{x}^2+u_{y}^2)]+nI\frac{T}{m}   \tt{,}
\end{aligned}
\end{eqnarray}
\begin{equation}
M^{ES}_{2,xx}=\sum_{i}f^{ES}_{i}v_{ix}v_{ix}=n[\frac{\lambda_{xx}}{m}+u_{x}u_{x}] \tt{,}
\end{equation}
\begin{equation}
M^{ES}_{2,xy}=\sum_{i}f^{ES}_{i}v_{ix}v_{ix}=\frac{n(\lambda_{xy}+mu_x u_y)}{m} \tt{,}
\end{equation}
\begin{equation}
M^{ES}_{2,yy}=\sum_{i}f^{ES}_{i}v_{iy}v_{iy}=n[\frac{\lambda_{yy}}{m}+u_{y}u_{y}] \tt{,}
\end{equation}
\begin{eqnarray}
\begin{aligned}
M^{ES}_{3,1,x}&=\sum_{i}f^{ES}_{i}( v_{ix}^2+v_{iy}^2+\eta^{2}_{i} ) v_{ix}\\
&=\frac{n}{m}[u_x \left(3\lambda_{xx}+\lambda_{xx}+IT+m \left(u_x^2+u_y^2\right)\right)+2\lambda_{xy}u_y]
\tt{,}
\end{aligned}
\end{eqnarray}
\begin{eqnarray}
\begin{aligned}
M^{ES}_{3,1,y}&=\sum_{i}f^{ES}_{i}( v_{ix}^2+v_{iy}^2+\eta^{2}_{i} ) v_{iy}\\
&=\frac{n}{m} \left[u_y \left(\lambda_{xx}+3\lambda_{yy}+IT+m\left(u_x^2+u_y^2\right)\right)+2 \lambda_{xy} u_x \right]
\tt{,}
\end{aligned}
\end{eqnarray}
\begin{eqnarray}
\begin{aligned}
M_{3,xxx}^{ES}&=\sum_{i}f^{ES}_{i}v_{ix}v_{ix}v_{ix}=n u_x \left(\frac{3\lambda_{xx}}{m}+u_x^2\right)
\tt{,}
\end{aligned}
\end{eqnarray}
\begin{eqnarray}
\begin{aligned}
M_{3,xxy}^{ES}&=\sum_{i}f^{ES}_{i}v_{ix}v_{ix}v_{iy}\\
&=\frac{n}{m}\left[u_y \left(\lambda_{xx}+m u_x^2\right)+2 \lambda_{xy} u_x \right]
\tt{,}
\end{aligned}
\end{eqnarray}
\begin{eqnarray}
\begin{aligned}
M_{3,xyy}^{ES}&=\sum_{i}f^{ES}_{i}v_{ix}v_{iy}v_{iy}\\
&=\frac{n}{m} [u_y (2 \lambda_{xy}+m u_x u_y)+\lambda_{yy} u_x]
\tt{,}
\end{aligned}
\end{eqnarray}
\begin{eqnarray}
\begin{aligned}
M_{3,yyy}^{ES}&=\sum_{i}f^{ES}_{i}v_{iy}v_{iy}v_{iy}=n u_y \left(\frac{3\lambda_{yy}}{m}+u_y^2\right)
\tt{,}
\end{aligned}
\end{eqnarray}
\begin{equation}
\begin{aligned}
M^{ES}_{4,2,xx}&=\sum_{i}f^{ES}_{i}( v_{ix}^2+v_{iy}^2+\eta^{2}_{i} )v_{ix}v_{ix}\\
&=\frac{n}{m^2} [3 \lambda_{xx}^2+\lambda_{xx} (\lambda_{yy}+IT+m (6 u_x^2+u_y^2))+2 \lambda_{xy}^2\\
&+m u_x (4 \lambda_{xy} u_y+u_x (\lambda_{yy}+IT+m (u_x^2+u_y^2)))]
\tt{,}
\end{aligned}
\end{equation}
\begin{equation}
\begin{aligned}
M^{ES}_{4,2,xy}&=\sum_{i}f^{ES}_{i}( v_{ix}^2+v_{iy}^2+\eta^{2}_{i} )v_{ix}v_{iy}\\
&=\frac{n}{m^2} [3 \lambda_{xx} (\lambda_{xy}+m u_x u_y)+\lambda_{xy} (3 \lambda_{yy}+I T+3 m (u_x^2+u_y^2))\\
&+m u_x u_y (3 \lambda_{yy}+I T+m (u_x^2+u_y^2))]
\tt{,}
\end{aligned}
\end{equation}
\begin{equation}
\begin{aligned}
M^{ES}_{4,2,yy}&=\sum_{i}f^{ES}_{i}( v_{ix}^2+v_{iy}^2+\eta^{2}_{i} )v_{iy}v_{iy}\\
&=\frac{n}{m^2} (m u_y^2 (\lambda_{xx}+6 \lambda_{yy}+I T
+m (u_x^2+u_y^2))\\
&+\lambda_{yy} (\lambda_{xx}+3 \lambda_{yy}+I T+m u_x^2)+2 \lambda_{xy}^2+4 \lambda_{xy} m u_x u_y)
\tt{,}
\end{aligned}
\end{equation}
\begin{equation}
\begin{aligned}
M^{ES}_{4,xxxx}&=\sum_{i}f^{ES}_{i}v_{ix}v_{ix}v_{ix}v_{ix}=\frac{n (3 \lambda_{xx}^2+6 \lambda_{xx} m u_x^2+m u_x^4)}{m^2}
\tt{,}
\end{aligned}
\end{equation}
\begin{equation}
\begin{aligned}
M^{ES}_{4,xxxy}&=\sum_{i}f^{ES}_{i}v_{ix}v_{ix}v_{ix}v_{iy}\\
&=\frac{n}{m^2} [3 \lambda_{xx} (\lambda_{xy} + m u_x u_y) + m u_x^2 (3 \lambda_{xy} + m u_x u_y))]
\tt{,}
\end{aligned}
\end{equation}
\begin{equation}
\begin{aligned}
M^{ES}_{4,xxyy}&=\sum_{i}f^{ES}_{i}v_{ix}v_{ix}v_{iy}v_{iy}\\
&=\frac{n}{m^2} [(\lambda_{xx}+m u_x^2) (\lambda_{yy}+m u_y^2)+2 \lambda_{xy}^2+4 \lambda_{xy}m u_x u_y]
\tt{,}
\end{aligned}
\end{equation}
\begin{equation}
\begin{aligned}
M^{ES}_{4,xyyy}&=\sum_{i}f^{ES}_{i}v_{ix}v_{iy}v_{iy}v_{iy}\\
&=\frac{n}{m^2} \left[3 \lambda_{xy} \left(\lambda_{yy}+m u_y^2\right)+m u_x u_y \left(3 \lambda_{yy}+m u_y^2\right)\right]
\tt{,}
\end{aligned}
\end{equation}
\begin{equation}
\begin{aligned}
M^{ES}_{4,yyyy}&=\sum_{i}f^{ES}_{i}v_{iy}v_{iy}v_{iy}v_{iy}=\frac{n}{m^2} \left(3 \lambda_{yy}^2+6 \lambda_{yy} m u_y^2+m^2 u_y^4\right)
\tt{,}
\end{aligned}
\end{equation}
\begin{equation}
\begin{aligned}
M^{ES}_{5,3,xxx}&=\sum_{i}f^{ES}_{i}( v_{ix}^2+v_{iy}^2+\eta^{2}_{i} )v_{ix}v_{ix}v_{ix}\\
&=\frac{n}{m^2} [6 \lambda_{xy} u_y \left(\lambda_{xx}+m u_x^2\right)+m u_x^3 (10 \lambda_{xx}+\lambda_{yy}+IT)\\
&+3 \lambda_{xx} u_x (5 \lambda_{xx}+\lambda_{yy}+IT)+m u_x u_y^2 \left(3 \lambda_{xx}+m u_x^2\right)\\
&+6 \lambda_{xy}^2 u_x+m^2 u_x^5]
\tt{,}
\end{aligned}
\end{equation}
\begin{equation}
\begin{aligned}
M^{ES}_{5,3,xxy}&=\sum_{i}f^{ES}_{i}( v_{ix}^2+v_{iy}^2+\eta^{2}_{i} )v_{ix}v_{ix}v_{iy}\\
&=\frac{n}{m^2}  [u_y (m u_x^2 (6 \lambda_{xx}+3 \lambda_{yy}+I T)+3 \lambda_{xx} (\lambda_{xx}+\lambda_{yy})\\
&+\lambda_{xx} I T+6 \lambda_{xy}^2+m^2 u_x^4)+2 \lambda_{xy} u_x (6 \lambda_{xx}+3 \lambda_{yy}+I T\\
&+2 m u_x^2)+m u_y^3 (\lambda_{xx}+m u_x^2)+6 \lambda_{xy} m u_x u_y^2]
\tt{,}
\end{aligned}
\end{equation}
\begin{equation}
\begin{aligned}
M^{ES}_{5,3,xyy}&=\sum_{i}f^{ES}_{i}( v_{ix}^2+v_{iy}^2+\eta^{2}_{i} )v_{ix}v_{iy}v_{iy}\\
&=\frac{n}{m^2} [2 \lambda_{xy} u_y (3 \lambda_{xx}+6 \lambda_{yy}+I T+3 m u_x^2+2 m u_y^2)\\
&+u_x (m u_y^2 (3 \lambda_{xx}+6 \lambda_{yy}+I T+m u_x^2)\\
&+\lambda_{yy} (3 \lambda_{xx}+3 \lambda_{yy}+I T+m u_x^2)+m u_y^4)+6 \lambda_{xy}^2 u_x]
\tt{,}
\end{aligned}
\end{equation}
\begin{equation}
\begin{aligned}
M^{ES}_{5,3,yyy}&=\sum_{i}f^{ES}_{i}( v_{ix}^2+v_{iy}^2+\eta^{2}_{i} )v_{iy}v_{iy}v_{iy}\\
&=\frac{n}{m^2} [m  u_y^3 (\lambda_{xx}+10 \lambda_{yy}+I T+m  u_x^2)+3 \lambda_{yy} u_y (\lambda_{xx}\\
&+5 \lambda_{yy}+I  T+m  u_x^2)+6 \lambda_{xy}^2 u_y+6 \lambda_{xy} u_x (\lambda_{yy}+m  u_y^2)\\
&+m u_y^5]
\tt{,}
\end{aligned}
\end{equation}
\begin{equation}
\begin{aligned}
M^{ES}_{5,xxxxx}&=\sum_{i}f^{ES}_{i}v_{ix}v_{ix}v_{ix}v_{ix}v_{ix}\\
&=\frac{n}{m^2} (15 \lambda_{xx}^2 u_x+10 \lambda_{xx} m u_x^3+m^2 u_x^5)
\tt{,}
\end{aligned}
\end{equation}
\begin{equation}
\begin{aligned}
M^{ES}_{5,xxxxy}&=\sum_{i}f^{ES}_{i}v_{ix}v_{ix}v_{ix}v_{ix}v_{iy}\\
&=\frac{n}{m^2} [3 \lambda_{xx}^2 u_y+6 \lambda_{xx} u_x (2 \lambda_{xy}+m u_x u_y)\\
&+m u_x^3 (4 \lambda_{xy}+m u_x u_y)]
\tt{,}
\end{aligned}
\end{equation}
\begin{equation}
\begin{aligned}
M^{ES}_{5,xxxyy}&=\sum_{i}f^{ES}_{i}v_{ix}v_{ix}v_{ix}v_{iy}v_{iy}=\frac{n}{m^2} [6 \lambda_{xy} u_y (\lambda_{xx}\\
&+m u_x^2)+u_x (3 \lambda_{xx}+m u_x^2) (\lambda_{yy}+m u_y^2)+6 \lambda_{xy}^2 u_x]
\tt{,}
\end{aligned}
\end{equation}
\begin{equation}
\begin{aligned}
M^{ES}_{5,xxyyy}&=\sum_{i}f^{ES}_{i}v_{ix}v_{ix}v_{iy}v_{iy}v_{iy}\\
&=\frac{n}{m^2} [u_y (\lambda_{xx}+m u_x^2) (3 \lambda_{yy}+m u_y^2)+6 \lambda_{xy}^2 u_y\\
&+6 \lambda_{xy} u_x (\lambda_{yy}+m u_y^2)]
\tt{,}
\end{aligned}
\end{equation}
\begin{equation}
\begin{aligned}
M^{ES}_{5,xyyyy}&=\sum_{i}f^{ES}_{i}v_{ix}v_{iy}v_{iy}v_{iy}v_{iy}\\
&=\frac{n}{m^2} [6 \lambda_{yy} u_y (2 \lambda_{xy}+m u_x u_y)\\
&+m u_y^3 (4 \lambda_{xy}+m u_x u_y)+3 \lambda_{yy}^2 u_x]
\tt{,}
\end{aligned}
\end{equation}
\begin{equation}
\begin{aligned}
M^{ES}_{5,yyyyy}&=\sum_{i}f^{ES}_{i}v_{iy}v_{iy}v_{iy}v_{iy}v_{iy}\\
&=\frac{n}{m^2} u_y (15 \lambda_{yy}^2+10 \lambda_{yy} m u_y^2+m^2 u_y^4)
\tt{,}
\end{aligned}
\end{equation}
\begin{equation}
\begin{aligned}
M^{ES}_{6,4,xxxx}&=\sum_{i}f^{ES}_{i}( v_{ix}^2+v_{iy}^2+\eta^{2}_{i} )v_{ix}v_{ix}v_{ix}v_{ix}\\
&=\frac{n}{m^{3}} [15 \lambda_{xx}^3+3 \lambda_{xx}^2 (\lambda_{yy}+I T+m (15 u_x^2+u_y^2))\\
&+3 \lambda_{xx} (4 \lambda_{xy}^2+8 \lambda_{xy} m u_x u_y+m u_x^2 (2 \lambda_{yy}+2 I T \\
&+5 m u_x^2+2 m u_y^2))+m u_x^2 (12 \lambda_{xy}^2+8 \lambda_{xy} m u_x u_y\\
&+m u_x^2 (\lambda_{yy}+I T+m (u_x^2+u_y^2)))]
\tt{,}
\end{aligned}
\end{equation}
\begin{equation}
\begin{aligned}
M^{ES}_{6,4,xxxy}&=\sum_{i}f^{ES}_{i}( v_{ix}^2+v_{iy}^2+\eta^{2}_{i} )v_{ix}v_{ix}v_{ix}v_{iy}\\
&=\frac{n}{m^{3}} [m u_x u_y (m u_x^2 (10 \lambda_{xx}+3 \lambda_{yy}+I T)\\
&+3 \lambda_{xx} (5 \lambda_{xx}+3 \lambda_{yy}+I T)+18 \lambda_{xy}^2+m^2 u_x^4)\\
&+\lambda_{xy} (3 m u_x^2 (10 \lambda_{xx}+3 \lambda_{yy}+I T)+3 \lambda_{xx} (5 \lambda_{xx}\\
&+3 \lambda_{yy}+I T)+6 \lambda_{xy}^2+5 m^2 u_x^4)\\
&+9 \lambda_{xy} m u_y^2 (\lambda_{xx}+m u_x^2)+m^2 u_x u_y^3 (3 \lambda_{xx}+m u_x^2)]
\tt{,}
\end{aligned}
\end{equation}
\begin{equation}
\begin{aligned}
M^{ES}_{6,4,xxyy}&=\sum_{i}f^{ES}_{i}( v_{ix}^2+v_{iy}^2+\eta^{2}_{i} )v_{ix}v_{ix}v_{iy}v_{iy}\\
&=\frac{n}{m^{3}} [3 \lambda_{xx}^2 (\lambda_{yy}+m u_y^2)+\lambda_{xx} (12 \lambda_{xy}^2+24 \lambda_{xy} m u_x u_y\\
&+m u_y^2 (6 \lambda_{yy}+I T+6 m u_x^2+m u_y^2)+\lambda_{yy} (3 \lambda_{yy}\\
&+I T+6 m u_x^2))+2 \lambda_{xy}^2 (6 \lambda_{yy}+I T+6 m (u_x^2+u_y^2))\\
&+4 \lambda_{xy} m u_x u_y (6 \lambda_{yy}+I T+2 m (u_x^2+u_y^2))\\
&+m u_x^2 (m u_y^2 (6 \lambda_{yy}+I T+m u_x^2)+\lambda_{yy} (3 \lambda_{yy}\\
&+I T+m u_x^2)+m^2 u_y^4)]
\tt{,}
\end{aligned}
\end{equation}
\begin{equation}
\begin{aligned}
M^{ES}_{6,4,xyyy}&=\sum_{i}f^{ES}_{i}( v_{ix}^2+v_{iy}^2+\eta^{2}_{i} )v_{ix}v_{iy}v_{iy}v_{iy}\\
&=\frac{n}{m^{3}} [\lambda_{xy} (3 m u_y^2 (3 \lambda_{xx}+10 \lambda_{yy}+I T+3 m u_x^2)\\
&+3 \lambda_{yy} (3 \lambda_{xx}+5 \lambda_{yy}+I T+3 m u_x^2)+5 m^2 u_y^4)\\
&+m u_x u_y (m u_y^2 (3 \lambda_{xx}+10 \lambda_{yy}+I T+m u_x^2)\\
&+3 \lambda_{yy} (3 \lambda_{xx}+5 \lambda_{yy}+I T+m u_x^2)+m^2 u_y^4)\\
&+6 \lambda_{xy}^3+18 \lambda_{xy}^2 m u_x u_y]
\tt{,}
\end{aligned}
\end{equation}
\begin{equation}
\begin{aligned}
M^{ES}_{6,4,yyyy}&=\sum_{i}f^{ES}_{i}( v_{ix}^2+v_{iy}^2+\eta^{2}_{i} )v_{iy}v_{iy}v_{iy}v_{iy}\\
&=\frac{n}{m^{3}} [3 \lambda_{yy}^2 (\lambda_{xx}+5 \lambda_{yy}+I T\\
&+m u_x^2)+m^2 u_y^4 (\lambda_{xx}+15 \lambda_{yy}+I T+m u_x^2)\\
&+3 \lambda_{yy} m u_y^2 (2 \lambda_{xx}+15 \lambda_{yy}+2 I T+2 m u_x^2)\\
&+12 \lambda_{xy}^2 (\lambda_{yy}+m u_y^2)+8 \lambda_{xy} m u_x u_y (3 \lambda_{yy}+m u_y^2)\\
&+m^{3} u_y^6]
\tt{,}
\end{aligned}
\end{equation}
\begin{equation}
\begin{aligned}
M^{ES}_{6,xxxxxx}&=\sum_{i}f^{ES}_{i}v_{ix}v_{ix}v_{ix}v_{ix}v_{ix}v_{ix}\\
&=\frac{n}{m^{3}} (15 \lambda_{xx}^3+45 \lambda_{xx}^2 m u_x^2+15 \lambda_{xx} m^2 u_x^4+m^{3} u_x^6)
\tt{,}
\end{aligned}
\end{equation}
\begin{equation}
\begin{aligned}
M^{ES}_{6,xxxxxy}&=\sum_{i}f^{ES}_{i}v_{ix}v_{ix}v_{ix}v_{ix}v_{ix}v_{iy}\\
&=\frac{n}{m^{3}} [15 \lambda_{xx}^2 (\lambda_{xy}+m u_x u_y)+10 \lambda_{xx} m u_x^2 (3 \lambda_{xy}\\
&+m u_x u_y)+m^2 u_x^4 (5 \lambda_{xy}+m u_x u_y)]
\tt{,}
\end{aligned}
\end{equation}
\begin{equation}
\begin{aligned}
M^{ES}_{6,xxxxyy}&=\sum_{i}f^{ES}_{i}v_{ix}v_{ix}v_{ix}v_{ix}v_{iy}v_{iy}\\
&=\frac{n}{m^{3}} [3 \lambda_{xx}^2 (\lambda_{yy}+m u_y^2)+6 \lambda_{xx} (2 \lambda_{xy}^2+4 \lambda_{xy} m u_x u_y\\
&+m u_x^2 (\lambda_{yy}+m u_y^2))+m u_x^2 ((2 \lambda_{xy}+m u_x u_y) (6 \lambda_{xy}\\
&+m u_x u_y)+\lambda_{yy} m u_x^2)]
\tt{,}
\end{aligned}
\end{equation}
\begin{equation}
\begin{aligned}
M^{ES}_{6,xxxyyy}&=\sum_{i}f^{ES}_{i}v_{ix}v_{ix}v_{ix}v_{iy}v_{iy}v_{iy}\\
&=\frac{n}{m^{3}} [9 \lambda_{xy} (\lambda_{xx}+m u_x^2) (\lambda_{yy}+m u_y^2)\\
&+m u_x u_y (3 \lambda_{xx}+m u_x^2) (3 \lambda_{yy}+m u_y^2)\\
&+6 \lambda_{xy}^3+18 \lambda_{xy}^2 m u_x u_y]
\tt{,}
\end{aligned}
\end{equation}
\begin{equation}
\begin{aligned}
M^{ES}_{6,xxyyyy}&=\sum_{i}f^{ES}_{i}v_{ix}v_{ix}v_{iy}v_{iy}v_{iy}v_{iy}\\
&=\frac{n}{m^{3}} [(\lambda_{xx}+m u_x^2) (3 \lambda_{yy}^2+6 \lambda_{yy} m u_y^2+m^2 u_y^4)\\
&+12 \lambda_{xy}^2 (\lambda_{yy}+m u_y^2)+8 \lambda_{xy} m u_x u_y (3 \lambda_{yy}+m u_y^2)]
\tt{,}
\end{aligned}
\end{equation}
\begin{equation}
\begin{aligned}
M^{ES}_{6,xyyyyy}&=\sum_{i}f^{ES}_{i}v_{ix}v_{iy}v_{iy}v_{iy}v_{iy}v_{iy}\\
&=\frac{n}{m^{3}} [5 \lambda_{xy} (3 \lambda_{yy}^2+6 \lambda_{yy} m u_y^2+m^2 u_y^4)\\
&+m u_x u_y (15 \lambda_{yy}^2+10 \lambda_{yy} m u_y^2+m^2 u_y^4)]
\tt{,}
\end{aligned}
\end{equation}
\begin{equation}
\begin{aligned}
M^{ES}_{6,yyyyyy}&=\sum_{i}f^{ES}_{i}v_{iy}v_{iy}v_{iy}v_{iy}v_{iy}v_{iy}\\
&=\frac{n}{m^{3}} (15 \lambda_{yy}^3+45 \lambda_{yy}^2 m u_y^2+15 \lambda_{yy} m^2 u_y^4\\
&+m^{3} u_y^6)
\tt{,}
\end{aligned}
\end{equation}
\begin{equation}
\begin{aligned}
M^{ES}_{7,5,xxxxx}&=\sum_{i}f^{ES}_{i}( v_{ix}^2+v_{iy}^2+\eta^{2}_{i} )v_{ix}v_{ix}v_{ix}v_{ix}v_{ix}\\
&=\frac{n}{m^{3}} [105 \lambda_{xx}^3 u_x+15 \lambda_{xx}^2 (2 \lambda_{xy} u_y\\
&+u_x (\lambda_{yy}+I T+7 m u_x^2+m u_y^2))+\lambda_{xx} u_x (60 \lambda_{xy}^2\\
&+60 \lambda_{xy} m u_x u_y+m u_x^2 (10 \lambda_{yy}+10 I T+21 m u_x^2\\
&+10 m u_y^2))+m u_x^3 (20 \lambda_{xy}^2+10 \lambda_{xy} m u_x u_y\\
&+m u_x^2 (\lambda_{yy}+I T+m (u_x^2+u_y^2)))]
\tt{,}
\end{aligned}
\end{equation}
\begin{equation}
\begin{aligned}
M^{ES}_{7,5,xxxxy}&=\sum_{i}f^{ES}_{i}( v_{ix}^2+v_{iy}^2+\eta^{2}_{i} )v_{ix}v_{ix}v_{ix}v_{ix}v_{iy}\\
&=\frac{n}{m^{3}} [15 \lambda_{xx}^3 u_y+3 \lambda_{xx}^2 (30 \lambda_{xy} u_x+u_y (3 \lambda_{yy}+I T\\
&+15 m u_x^2+m u_y^2))+3 \lambda_{xx} (12 \lambda_{xy}^2 u_y+4 \lambda_{xy} u_x (3 \lambda_{yy}\\
&+I T+5 m u_x^2+3 m u_y^2)+m u_x^2 u_y (6 \lambda_{yy}+2 I T\\
&+5 m u_x^2+2 m u_y^2))+u_x (24 \lambda_{xy}^3+36 \lambda_{xy}^2 m u_x u_y\\
&+2 \lambda_{xy} m u_x^2 (6 \lambda_{yy}+2 I T+3 m (u_x^2+2 u_y^2))\\
&+m^2 u_x^3 u_y (3 \lambda_{yy}+I T+m (u_x^2+u_y^2)))]
\tt{,}
\end{aligned}
\end{equation}
\begin{equation}
\begin{aligned}
M^{ES}_{7,5,xxxyy}&=\sum_{i}f^{ES}_{i}( v_{ix}^2+v_{iy}^2+\eta^{2}_{i} )v_{ix}v_{ix}v_{ix}v_{iy}v_{iy}\\
&=\frac{n }{m^{3}} [3 u_x (5 \lambda_{xx}^2 \lambda_{yy}+20 \lambda_{xx} \lambda_{xy}^2+\lambda_{xx} \lambda_{yy} (3 \lambda_{yy}+I  T)\\
&+2 \lambda_{xy}^2 (6 \lambda_{yy}+I  T))+m  u_x u_y^2 (m  u_x^2 (10 \lambda_{xx}\\
&+6 \lambda_{yy}+I  T)+3 \lambda_{xx} (5 \lambda_{xx}+6 \lambda_{yy}+I  T)+36 \lambda_{xy}^2\\
&+m^2 u_x^4)+2 \lambda_{xy} u_y (3 m  u_x^2 (10 \lambda_{xx}+6 \lambda_{yy}+I  T)\\
&+3 \lambda_{xx} (5 \lambda_{xx}+6 \lambda_{yy}+I  T)+12 \lambda_{xy}^2+5 m^2 u_x^4)\\
&+m  u_x^3 (\lambda_{yy} (10 \lambda_{xx}+3 \lambda_{yy}+I  T)+20 \lambda_{xy}^2)\\
&+12 \lambda_{xy} m  u_y^3 (\lambda_{xx}+m  u_x^2)\\
&+m^2 u_x u_y^4 (3 \lambda_{xx}+m  u_x^2)+\lambda_{yy} m^2 u_x^5]
\tt{,}
\end{aligned}
\end{equation}
\begin{equation}
\begin{aligned}
M^{ES}_{7,5,xxyyy}&=\sum_{i}f^{ES}_{i}( v_{ix}^2+v_{iy}^2+\eta^{2}_{i} )v_{ix}v_{ix}v_{iy}v_{iy}v_{iy}\\
&=\frac{n }{m^{3}} [2 \lambda_{xy}^2 u_y (18 \lambda_{xx}+30 \lambda_{yy}+3 I  T+18 m  u_x^2\\
&+10 m  u_y^2)+2 \lambda_{xy} u_x (3 m  u_y^2 (6 \lambda_{xx}+10 \lambda_{yy}+I  T+2 m  u_x^2)\\
&+3 \lambda_{yy} (6 \lambda_{xx}+5 \lambda_{yy}+I  T+2 m  u_x^2)+5 m^2 u_y^4)\\
&+m  u_y^3 (m  u_x^2 (6 \lambda_{xx}+10 \lambda_{yy}+I  T)+\lambda_{xx} (3 \lambda_{xx}+10 \lambda_{yy}\\
&+I  T)+m^2 u_x^4)+3 \lambda_{yy} u_y (m  u_x^2 (6 \lambda_{xx}+5 \lambda_{yy}\\
&+I  T)+\lambda_{xx} (3 \lambda_{xx}+5 \lambda_{yy}+I  T)+m^2 u_x^4)\\
&+m^2 u_y^5 (\lambda_{xx}+m  u_x^2)+24 \lambda_{xy}^3 u_x]
\tt{,}
\end{aligned}
\end{equation}
\begin{equation}
\begin{aligned}
M^{ES}_{7,5,xyyyy}&=\sum_{i}f^{ES}_{i}( v_{ix}^2+v_{iy}^2+\eta^{2}_{i} )v_{ix}v_{iy}v_{iy}v_{iy}v_{iy}\\
&=\frac{n }{m^{3}} [4 \lambda_{xy} m  u_y^3 (3 \lambda_{xx}+15 \lambda_{yy}+I  T+3 m  u_x^2)\\
&+6 \lambda_{xy} \lambda_{yy} u_y (6 \lambda_{xx}+15 \lambda_{yy}+2 I  T+6 m  u_x^2)\\
&+u_x (3 \lambda_{yy}^2 (3 \lambda_{xx}+5 \lambda_{yy}+I  T+m  u_x^2)+m^2 u_y^4 (3 \lambda_{xx}\\
&+15 \lambda_{yy}+I  T+m  u_x^2)+3 \lambda_{yy} m  u_y^2 (6 \lambda_{xx}+15 \lambda_{yy}\\
&+2 I  T+2 m  u_x^2)+m^{3} u_y^6)+24 \lambda_{xy}^3 u_y\\
&+36 \lambda_{xy}^2 u_x (\lambda_{yy}+m  u_y^2)+6 \lambda_{xy} m^2 u_y^5]
\tt{,}
\end{aligned}
\end{equation}
\begin{equation}
\begin{aligned}
M^{ES}_{7,5,yyyyy}&=\sum_{i}f^{ES}_{i}( v_{ix}^2+v_{iy}^2+\eta^{2}_{i} )v_{iy}v_{iy}v_{iy}v_{iy}v_{iy}\\
&=\frac{n }{m^{3}} [15 \lambda_{yy}^2 u_y (\lambda_{xx}+7 \lambda_{yy}+I  T+m  u_x^2)\\
&+m^2 u_y^5 (\lambda_{xx}+21 \lambda_{yy}+I  T+m  u_x^2)\\
&+5 \lambda_{yy} m  u_y^3 (2 \lambda_{xx}+21 \lambda_{yy}+2 I  T+2 m  u_x^2)\\
&+20 \lambda_{xy}^2 (3 \lambda_{yy} u_y+m  u_y^3)+10 \lambda_{xy} u_x (3 \lambda_{yy}^2\\
&+6 \lambda_{yy} m  u_y^2+m^2 u_y^4)+m^{3} u_y^7]
\tt{,}
\end{aligned}
\end{equation}
\begin{equation}
\begin{aligned}
M^{ES}_{7,xxxxxxx}&=\sum_{i}f^{ES}_{i}v_{ix}v_{ix}v_{ix}v_{ix}v_{ix}v_{ix}v_{ix}\\
&=\frac{n }{m^{3}} [105 \lambda_{xx}^3 u_x+105 \lambda_{xx}^2 m  u_x^3+21 \lambda_{xx} m^2 u_x^5+m^{3} u_x^7]
\tt{,}
\end{aligned}
\end{equation}
\begin{equation}
\begin{aligned}
M^{ES}_{7,xxxxxxy}&=\sum_{i}f^{ES}_{i}v_{ix}v_{ix}v_{ix}v_{ix}v_{ix}v_{ix}v_{iy}\\
&=\frac{n }{m^{3}} [15 \lambda_{xx}^3 u_y+45 \lambda_{xx}^2 u_x (2 \lambda_{xy}+m  u_x u_y)\\
&+15 \lambda_{xx} m  u_x^3 (4 \lambda_{xy}+m  u_x u_y)+m^2 u_x^5 (6 \lambda_{xy}+m  u_x u_y)]
\tt{,}
\end{aligned}
\end{equation}
\begin{equation}
\begin{aligned}
M^{ES}_{7,xxxxxyy}&=\sum_{i}f^{ES}_{i}v_{ix}v_{ix}v_{ix}v_{ix}v_{ix}v_{iy}v_{iy}\\
&=\frac{n}{m^{3}} [15 \lambda_{xx}^2 (u_y (2 \lambda_{xy}+m  u_x u_y)+\lambda_{yy} u_x)\\
&+10 \lambda_{xx} u_x (6 \lambda_{xy}^2+6 \lambda_{xy} m  u_x u_y+m  u_x^2 (\lambda_{yy}+m  u_y^2))\\
&+m  u_x^3 (20 \lambda_{xy}^2+10 \lambda_{xy} m  u_x u_y+m  u_x^2 (\lambda_{yy}+m  u_y^2))]
\tt{,}
\end{aligned}
\end{equation}
\begin{equation}
\begin{aligned}
M^{ES}_{7,xxxxyyy}&=\sum_{i}f^{ES}_{i}v_{ix}v_{ix}v_{ix}v_{ix}v_{iy}v_{iy}v_{iy}\\
&=\frac{n}{m^{3}} [u_y (3 \lambda_{xx}^2+6 \lambda_{xx} m  u_x^2+m^2 u_x^4) (3 \lambda_{yy}+m  u_y^2)\\
&+36 \lambda_{xy}^2 u_y (\lambda_{xx}+m  u_x^2)\\
&+12 \lambda_{xy} u_x (3 \lambda_{xx}+m  u_x^2) (\lambda_{yy}+m  u_y^2)+24 \lambda_{xy}^3 u_x]
\tt{,}
\end{aligned}
\end{equation}
\begin{equation}
\begin{aligned}
M^{ES}_{7,xxxyyyy}&=\sum_{i}f^{ES}_{i}v_{ix}v_{ix}v_{ix}v_{iy}v_{iy}v_{iy}v_{iy}\\
&=\frac{n}{m^{3}} [12 \lambda_{xy} u_y (\lambda_{xx}+m  u_x^2) (3 \lambda_{yy}+m  u_y^2)\\
&+u_x (3 \lambda_{xx}+m  u_x^2) (3 \lambda_{yy}^2+6 \lambda_{yy} m  u_y^2+m^2 u_y^4)\\
&+24 \lambda_{xy}^3 u_y+36 \lambda_{xy}^2 u_x (\lambda_{yy}+m  u_y^2)]
\tt{,}
\end{aligned}
\end{equation}
\begin{equation}
\begin{aligned}
M^{ES}_{7,xxyyyyy}&=\sum_{i}f^{ES}_{i}v_{ix}v_{ix}v_{iy}v_{iy}v_{iy}v_{iy}v_{iy}\\
&=\frac{n}{m^{3}} [u_y (\lambda_{xx}+m  u_x^2) (15 \lambda_{yy}^2+10 \lambda_{yy} m  u_y^2\\
&+m^2 u_y^4)+20 \lambda_{xy}^2 (3 \lambda_{yy} u_y+m  u_y^3)\\
&+10 \lambda_{xy} u_x (3 \lambda_{yy}^2+6 \lambda_{yy} m  u_y^2+m^2 u_y^4)]
\tt{,}
\end{aligned}
\end{equation}
\begin{equation}
\begin{aligned}
M^{ES}_{7,xyyyyyy}&=\sum_{i}f^{ES}_{i}v_{ix}v_{iy}v_{iy}v_{iy}v_{iy}v_{iy}v_{iy}\\
&=\frac{n}{m^{3}} (45 \lambda_{yy}^2 u_y (2 \lambda_{xy}+m  u_x u_y)\\
&+15 \lambda_{yy} m  u_y^3 (4 \lambda_{xy}+m  u_x u_y)\\
&+m^2 u_y^5 (6 \lambda_{xy}+m  u_x u_y)+15 \lambda_{yy}^3 u_x)
\tt{,}
\end{aligned}
\end{equation}
\begin{equation}
\begin{aligned}
M^{ES}_{7,yyyyyyy}&=\sum_{i}f^{ES}_{i}v_{iy}v_{iy}v_{iy}v_{iy}v_{iy}v_{iy}v_{iy}\\
&=\frac {n}{m^{3}}u_y [105 \lambda_{yy}^3+105 \lambda_{yy}^2 m  u_y^2+21 \lambda_{yy} m^2 u_y^4\\
&+m^{3} u_y^6]
\tt{,}
\end{aligned}
\end{equation}
\begin{equation}
\begin{aligned}
M^{ES}_{8,6,xxxxxx}&=\sum_{i}f^{ES}_{i}( v_{ix}^2+v_{iy}^2+\eta^{2}_{i} )v_{ix}v_{ix}v_{ix}v_{ix}v_{ix}v_{ix}\\
&=\frac{n}{m^{4}} [105 \lambda_{xx}^4+15 \lambda_{xx}^3 (\lambda_{yy}+28 m u_x^2+m u_y^2\\
&+I T)+15 \lambda_{xx}^2 (6 \lambda_{xy}^2+12 \lambda_{xy} m u_x u_y\\
&+m u_x^2 (3 \lambda_{yy}+14 m u_x^2+3 m u_y^2+3 I T))\\
&+\lambda_{xx} m u_x^2 (15 (2 \lambda_{xy}+m u_x u_y) (6 \lambda_{xy}+m u_x u_y)\\
&+m u_x^2 (15 (\lambda_{yy}+I T)+28 m u_x^2))\\
&+m^2 u_x^4 (30 \lambda_{xy}^2+12 \lambda_{xy} m u_x u_y\\
&+m u_x^2 (\lambda_{yy}+m (u_x^2+u_y^2)+I T))]
\tt{,}
\end{aligned}
\end{equation}
\begin{equation}
\begin{aligned}
M^{ES}_{8,6,xxxxxy}&=\sum_{i}f^{ES}_{i}( v_{ix}^2+v_{iy}^2+\eta^{2}_{i} )v_{ix}v_{ix}v_{ix}v_{ix}v_{ix}v_{iy}\\
&=\frac{n}{m^{4}} [105 \lambda_{xx}^3 (\lambda_{xy}+m u_x u_y)+15 \lambda_{xx}^2 (\lambda_{xy} (3 \lambda_{yy}\\
&+3 m (7 u_x^2+u_y^2)+I T)+m u_x u_y (3 \lambda_{yy}+7 m u_x^2\\
&+m u_y^2+I T))+\lambda_{xx} (60 \lambda_{xy}^3+180 \lambda_{xy}^2 m u_x u_y\\
&+15 \lambda_{xy} m u_x^2 (6 \lambda_{yy}+7 m u_x^2+6 m u_y^2+2 I T)\\
&+m^2 u_x^3 u_y (30 \lambda_{yy}+21 m u_x^2+10 m u_y^2+10 I T))\\
&+m u_x^2 (60 \lambda_{xy}^3+60 \lambda_{xy}^2 m u_x u_y+\lambda_{xy} m u_x^2 (15 \lambda_{yy}\\
&+7 m u_x^2+15 m u_y^2+5 I T)+m^2 u_x^3 u_y (3 \lambda_{yy}\\
&+m(u_x^2+u_y^2)+IT))]
\tt{,}
\end{aligned}
\end{equation}
\begin{equation}
\begin{aligned}
M^{ES}_{8,6,xxxxyy}&=\sum_{i}f^{ES}_{i}( v_{ix}^2+v_{iy}^2+\eta^{2}_{i} )v_{ix}v_{ix}v_{ix}v_{ix}v_{iy}v_{iy}\\
&=\frac{n}{m^{4}} [15 \lambda_{xx}^3 (\lambda_{yy}+m u_y^2)+3 \lambda_{xx}^2 (30 \lambda_{xy}^2\\
&+60 \lambda_{xy} m u_x u_y+m u_y^2 (6 \lambda_{yy}+15 m u_x^2+m u_y^2+I T)\\
&+\lambda_{yy} (3 \lambda_{yy}+15 m u_x^2+I T))+3 \lambda_{xx} (4 \lambda_{xy}^2 (6 \lambda_{yy}\\
&+15 m u_x^2+6 m u_y^2+I T)+8 \lambda_{xy} m u_x u_y (6 \lambda_{yy}\\
&+5 m u_x^2+2 m u_y^2+I T)+m u_x^2 (m u_y^2 (12 \lambda_{yy}+5 m u_x^2\\
&+2 I T)+\lambda_{yy} (6 \lambda_{yy}+5 m u_x^2+2 I T)+2 m^2 u_y^4))\\
&+24 \lambda_{xy}^4+96 \lambda_{xy}^3 m u_x u_y+6 \lambda_{xy}^2 m u_x^2 (12 \lambda_{yy}\\
&+5 m u_x^2+12 m u_y^2+2 I T)+4 \lambda_{xy} m^2 u_x^3 u_y (12 \lambda_{yy}\\
&+3 m u_x^2+4 m u_y^2+2 I T)+m^2 u_x^4 (m u_y^2 (6 \lambda_{yy}\\
&+m u_x^2+I T)+\lambda_{yy} (3 \lambda_{yy}+m u_x^2+I T)+m^2 u_y^4)]
\tt{,}
\end{aligned}
\end{equation}
\begin{equation}
\begin{aligned}
M^{ES}_{8,6,xxxyyy}&=\sum_{i}f^{ES}_{i}( v_{ix}^2+v_{iy}^2+\eta^{2}_{i} )v_{ix}v_{ix}v_{ix}v_{iy}v_{iy}v_{iy}\\
&=\frac{n}{m^{4}} [3 \lambda_{xy} (15 \lambda_{xx}^2 \lambda_{yy}+20 \lambda_{xx} \lambda_{xy}^2+3 \lambda_{xx} \lambda_{yy} (5 \lambda_{yy}\\
&+10 m u_x^2+I T)+2 \lambda_{xy}^2 (10 \lambda_{yy}+10 m u_x^2+I T)\\
&+\lambda_{yy} m u_x^2 (15 \lambda_{yy}+5 m u_x^2+3 I T))\\
&+m^2 u_x u_y^3 (m u_x^2 (10 (\lambda_{xx}+\lambda_{yy})+I T)\\
&+3 \lambda_{xx} (5 \lambda_{xx}+10 \lambda_{yy}+I T)+60 \lambda_{xy}^2+m^2 u_x^4)\\
&+3 \lambda_{xy} m u_y^2 (3 m u_x^2 (10 (\lambda_{xx}+\lambda_{yy})+I T)\\
&+3 \lambda_{xx} (5 \lambda_{xx}+10 \lambda_{yy}+I T)+20 \lambda_{xy}^2+5 m^2 u_x^4)\\
&+3 m u_x u_y (m u_x^2 (\lambda_{yy} (10 \lambda_{xx}+5 \lambda_{yy}+I T)+20 \lambda_{xy}^2)\\
&+3 I T (\lambda_{xx} \lambda_{yy}+2 \lambda_{xy}^2)+15 (\lambda_{xx}+\lambda_{yy}) (\lambda_{xx} \lambda_{yy}+4 \lambda_{xy}^2)\\
&+\lambda_{yy} m^2 u_x^4)+15 \lambda_{xy} m^2 u_y^4 (\lambda_{xx}+m u_x^2)\\
&+m^{3} u_x u_y^5 (3 \lambda_{xx}+m u_x^2)]
\tt{,}
\end{aligned}
\end{equation}
\begin{equation}
\begin{aligned}
M^{ES}_{8,6,xxyyyy}&=\sum_{i}f^{ES}_{i}( v_{ix}^2+v_{iy}^2+\eta^{2}_{i} )v_{ix}v_{ix}v_{iy}v_{iy}v_{iy}v_{iy}\\
&=\frac{n}{m^{4}} [6 \lambda_{xy}^2 (2 m u_y^2 (6 \lambda_{xx}+15 \lambda_{yy}+6 m u_x^2+I T)\\
&+\lambda_{yy} (12 \lambda_{xx}+15 \lambda_{yy}+12 m u_x^2+2 I T)+5 m^2 u_y^4)\\
&+4 \lambda_{xy} m u_x u_y (2 m u_y^2 (6 \lambda_{xx}+15 \lambda_{yy}+2 m u_x^2+I T)\\
&+3 \lambda_{yy} (12 \lambda_{xx}+15 \lambda_{yy}+4 m u_x^2+2 I T)+3 m^2 u_y^4)\\
&+3 \lambda_{yy}^2 (m u_x^2 (6 \lambda_{xx}+5 \lambda_{yy}+I T)+\lambda_{xx} (3 \lambda_{xx}+5 \lambda_{yy}+I T)\\
&+m^2 u_x^4)+m^2 u_y^4 (m u_x^2 (6 \lambda_{xx}+15 \lambda_{yy}+I T)\\
&+\lambda_{xx} (3 \lambda_{xx}+15 \lambda_{yy}+I T)+m^2 u_x^4)\\
&+3 \lambda_{yy} m u_y^2 (m u_x^2 (12 \lambda_{xx}+15 \lambda_{yy}+2 I T)\\
&+\lambda_{xx} (6 \lambda_{xx}+15 \lambda_{yy}+2 I T)+2 m^2 u_x^4)\\
&+m^{3} u_y^6 (\lambda_{xx}+m u_x^2)+24 \lambda_{xy}^4+96 \lambda_{xy}^3 m u_x u_y]
\tt{,}
\end{aligned}
\end{equation}
\begin{equation}
\begin{aligned}
M^{ES}_{8,6,xyyyyy}&=\sum_{i}f^{ES}_{i}( v_{ix}^2+v_{iy}^2+\eta^{2}_{i} )v_{ix}v_{iy}v_{iy}v_{iy}v_{iy}v_{iy}\\
&=\frac{n}{m^{4}} [\lambda_{xy} (15 \lambda_{yy}^2 (3 \lambda_{xx}+7 \lambda_{yy}+3 m u_x^2+I T)\\
&+5 m^2 u_y^4 (3 \lambda_{xx}+21 \lambda_{yy}+3 m u_x^2+I T)\\
&+15 \lambda_{yy} m u_y^2 (6 \lambda_{xx}+21 \lambda_{yy}+6 m u_x^2+2 I T)\\
&+7 m^{3} u_y^6)+m u_x u_y (15 \lambda_{yy}^2 (3 \lambda_{xx}+7 \lambda_{yy}+m u_x^2+I T)\\
&+m^2 u_y^4 (3 \lambda_{xx}+21 \lambda_{yy}+m u_x^2+I T)\\
&+5 \lambda_{yy} m u_y^2 (6 \lambda_{xx}+21 \lambda_{yy}+2 m u_x^2+2 I T)\\
&+m^{3} u_y^6)+60 \lambda_{xy}^3 (\lambda_{yy}+m u_y^2)\\
&+60 \lambda_{xy}^2 m u_x u_y (3 \lambda_{yy}+m u_y^2)]
\tt{,}
\end{aligned}
\end{equation}
\begin{equation}
\begin{aligned}
M^{ES}_{8,6,yyyyyy}&=\sum_{i}f^{ES}_{i}( v_{ix}^2+v_{iy}^2+\eta^{2}_{i} )v_{iy}v_{iy}v_{iy}v_{iy}v_{iy}v_{iy}\\
&=\frac{n}{m^{4}} (15 \lambda_{yy}^3 (\lambda_{xx}+7 \lambda_{yy}+m u_x^2+I T)\\
&+15 \lambda_{yy}^2 m u_y^2 (3 \lambda_{xx}+28 \lambda_{yy}+3 m u_x^2+3 I T)\\
&+m^{3} u_y^6 (\lambda_{xx}+28 \lambda_{yy}+m u_x^2+I T)\\
&+15 \lambda_{yy} m^2 u_y^4 (\lambda_{xx}+14 \lambda_{yy}+m u_x^2+I T)\\
&+30 \lambda_{xy}^2 (3 \lambda_{yy}^2+6 \lambda_{yy} m u_y^2+m^2 u_y^4)\\
&+12 \lambda_{xy} m u_x u_y (15 \lambda_{yy}^2+10 \lambda_{yy} m u_y^2+m^2 u_y^4)\\
&+m^{4} u_y^8)
\tt{,}
\end{aligned}
\end{equation}
\begin{equation}
\begin{aligned}
M^{ES}_{8,xxxxxxxx}&=\sum_{i}f^{ES}_{i}v_{ix}v_{ix}v_{ix}v_{ix}v_{ix}v_{ix}v_{ix}v_{ix}\\
&=\frac{n}{m^{4}} (105 \lambda_{xx}^4+420 \lambda_{xx}^3 m u_x^2+210 \lambda_{xx}^2 m^2 u_x^4\\
&+28 \lambda_{xx} m^{3} u_x^6+m^{4} u_x^8)
\tt{,}
\end{aligned}
\end{equation}
\begin{equation}
\begin{aligned}
M^{ES}_{8,xxxxxxxy}&=\sum_{i}f^{ES}_{i}v_{ix}v_{ix}v_{ix}v_{ix}v_{ix}v_{ix}v_{ix}v_{iy}\\
&=\frac{n}{m^{4}} [105 \lambda_{xx}^3 (\lambda_{xy}+m u_x u_y)\\
&+105 \lambda_{xx}^2 m u_x^2 (3 \lambda_{xy}+m u_x u_y)\\
&+21 \lambda_{xx} m^2 u_x^4 (5 \lambda_{xy}+m u_x u_y)+m^{3} u_x^6 (7 \lambda_{xy}+m u_x u_y)]
\tt{,}
\end{aligned}
\end{equation}
\begin{equation}
\begin{aligned}
M^{ES}_{8,xxxxxxyy}&=\sum_{i}f^{ES}_{i}v_{ix}v_{ix}v_{ix}v_{ix}v_{ix}v_{ix}v_{iy}v_{iy}\\
&=\frac{n}{m^{4}} [15 \lambda_{xx}^3 (\lambda_{yy}+m u_y^2)+45 \lambda_{xx}^2 (2 \lambda_{xy}^2+4 \lambda_{xy} m u_x u_y\\
&+m u_x^2 (\lambda_{yy}+m u_y^2))+15 \lambda_{xx} m u_x^2 ((2 \lambda_{xy}+m u_x u_y) (6 \lambda_{xy}\\
&+m u_x u_y)+\lambda_{yy} m u_x^2)+m^2 u_x^4 (30 \lambda_{xy}^2\\
&+12 \lambda_{xy} m u_x u_y+m u_x^2 (\lambda_{yy}+m u_y^2))]
\tt{,}
\end{aligned}
\end{equation}
\begin{equation}
\begin{aligned}
M^{ES}_{8,xxxxxyyy}&=\sum_{i}f^{ES}_{i}v_{ix}v_{ix}v_{ix}v_{ix}v_{ix}v_{iy}v_{iy}v_{iy}\\
&=\frac{n}{m^{4}} [15 \lambda_{xx}^2 (3 \lambda_{xy} (\lambda_{yy}+m u_y^2)+m u_x u_y (3 \lambda_{yy}+m u_y^2))\\
&+10 \lambda_{xx} (6 \lambda_{xy}^3+18 \lambda_{xy}^2 m u_x u_y+9 \lambda_{xy} m u_x^2 (\lambda_{yy}+m u_y^2)\\
&+m^2 u_x^3 u_y (3 \lambda_{yy}+m u_y^2))\\
&+m u_x^2 (60 \lambda_{xy}^3+60 \lambda_{xy}^2 m u_x u_y\\
&+15 \lambda_{xy} m u_x^2 (\lambda_{yy}+m u_y^2)+m^2 u_x^3 u_y (3 \lambda_{yy}+m u_y^2))]
\tt{,}
\end{aligned}
\end{equation}
\begin{equation}
\begin{aligned}
M^{ES}_{8,xxxxyyyy}&=\sum_{i}f^{ES}_{i}v_{ix}v_{ix}v_{ix}v_{ix}v_{iy}v_{iy}v_{iy}v_{iy}\\
&=\frac{n}{m^{4}} [(3 \lambda_{xx}^2+6 \lambda_{xx} m u_x^2+m^2 u_x^4) (3 \lambda_{yy}^2+6 \lambda_{yy} m u_y^2\\
&+m^2 u_y^4)+72 \lambda_{xy}^2 (\lambda_{xx}+m u_x^2) (\lambda_{yy}+m u_y^2)\\
&+16 \lambda_{xy} m u_x u_y (3 \lambda_{xx}+m u_x^2) (3 \lambda_{yy}+m u_y^2)\\
&+24 \lambda_{xy}^4+96 \lambda_{xy}^3 m u_x u_y]
\tt{,}
\end{aligned}
\end{equation}
\begin{equation}
\begin{aligned}
M^{ES}_{8,xxxyyyyy}&=\sum_{i}f^{ES}_{i}v_{ix}v_{ix}v_{ix}v_{iy}v_{iy}v_{iy}v_{iy}v_{iy}\\
&=\frac{n}{m^{4}} [15 \lambda_{xy} (\lambda_{xx}+m u_x^2) (3 \lambda_{yy}^2+6 \lambda_{yy} m u_y^2\\
&+m^2 u_y^4)+m u_x u_y (3 \lambda_{xx}+m u_x^2) (15 \lambda_{yy}^2+10 \lambda_{yy} m u_y^2\\
&+m^2 u_y^4)+60 \lambda_{xy}^3 (\lambda_{yy}+m u_y^2)\\
&+60 \lambda_{xy}^2 m u_x u_y (3 \lambda_{yy}+m u_y^2)]
\tt{,}
\end{aligned}
\end{equation}
\begin{equation}
\begin{aligned}
M^{ES}_{8,xxyyyyyy}&=\sum_{i}f^{ES}_{i}v_{ix}v_{ix}v_{iy}v_{iy}v_{iy}v_{iy}v_{iy}v_{iy}\\
&=\frac{n}{m^{4}} [(\lambda_{xx}+m u_x^2) (15 \lambda_{yy}^3+45 \lambda_{yy}^2 m u_y^2\\
&+15 \lambda_{yy} m^2 u_y^4+m^{3} u_y^6)+30 \lambda_{xy}^2 (3 \lambda_{yy}^2\\
&+6 \lambda_{yy} m u_y^2+m^2 u_y^4)+12 \lambda_{xy} m u_x u_y (15 \lambda_{yy}^2\\
&+10 \lambda_{yy} m u_y^2+m^2 u_y^4)]
\tt{,}
\end{aligned}
\end{equation}
\begin{equation}
\begin{aligned}
M^{ES}_{8,xyyyyyyy}&=\sum_{i}f^{ES}_{i}v_{ix}v_{iy}v_{iy}v_{iy}v_{iy}v_{iy}v_{iy}v_{iy}\\
&=\frac{n}{m^{4}} [7 \lambda_{xy} (15 \lambda_{yy}^3+45 \lambda_{yy}^2 m u_y^2\\
&+15 \lambda_{yy} m^2 u_y^4+m^{3} u_y^6)+m u_x u_y (105 \lambda_{yy}^3\\
&+105 \lambda_{yy}^2 m u_y^2+21 \lambda_{yy} m^2 u_y^4+m^{3} u_y^6)]
\tt{,}
\end{aligned}
\end{equation}
\begin{equation}
\begin{aligned}
M^{ES}_{8,yyyyyyyy}&=\sum_{i}f^{ES}_{i}v_{iy}v_{iy}v_{iy}v_{iy}v_{iy}v_{iy}v_{iy}v_{iy}\\
&=\frac{n}{m^{4}} (105 \lambda_{yy}^4+420 \lambda_{yy}^3 m u_y^2+210 \lambda_{yy}^2 m^2 u_y^4\\
&+28 \lambda_{yy} m^{3} u_y^6+m^{4} u_y^8)
\tt{,}
\end{aligned}
\end{equation}
\begin{equation}
\begin{aligned}
M^{ES}_{9,7,xxxxxxx}&=\sum_{i}f^{ES}_{i}( v_{ix}^2+v_{iy}^2+\eta^{2}_{i} )v_{ix}v_{ix}v_{ix}v_{ix}v_{ix}v_{ix}v_{ix}\\
&=\frac{n }{m^{4}}[945 \lambda_{xx}^4 u_x+105 \lambda_{xx}^3 (2 \lambda_{xy} u_y+u_x (\lambda_{yy}\\
&+12 m u_x^2+m u_y^2+I T))+21 \lambda_{xx}^2 u_x (30 \lambda_{xy}^2\\
&+30 \lambda_{xy} m u_x u_y+m u_x^2 (5 \lambda_{yy}+18 m u_x^2+5 m u_y^2\\
&+5 I T))+3 \lambda_{xx} m u_x^3 (140 \lambda_{xy}^2+70 \lambda_{xy} m u_x u_y\\
&+m u_x^2 (7 \lambda_{yy}+12 m u_x^2+7 m u_y^2+7 I T))\\
&+m^2 u_x^5 (42 \lambda_{xy}^2+14 \lambda_{xy} m u_x u_y\\
&+m u_x^2 (\lambda_{yy}+m (u_x^2+u_y^2)+I T))]
\tt{,}
\end{aligned}
\end{equation}
\begin{equation}
\begin{aligned}
M^{ES}_{9,7,xxxxxxy}&=\sum_{i}f^{ES}_{i}( v_{ix}^2+v_{iy}^2+\eta^{2}_{i} )v_{ix}v_{ix}v_{ix}v_{ix}v_{ix}v_{ix}v_{iy}\\
&=\frac{n}{m^{4}} [u_y (15 \lambda_{xx}^2 (\lambda_{xx} (7 \lambda_{xx}+3 \lambda_{yy}+I T)+18 \lambda_{xy}^2)\\
&+15 m^2 u_x^4 (\lambda_{xx} (14 \lambda_{xx}+3 \lambda_{yy}+I T)+6 \lambda_{xy}^2)\\
&+15 \lambda_{xx} m u_x^2 (\lambda_{xx} (28 \lambda_{xx}+9 \lambda_{yy}+3 I T)+36 \lambda_{xy}^2)\\
&+m^{3} u_x^6 (28 \lambda_{xx}+3 \lambda_{yy}+I T)+m^{4} u_x^8)\\
&+18 \lambda_{xy} m u_x u_y^2 (15 \lambda_{xx}^2+10 \lambda_{xx} m u_x^2+m^2 u_x^4)\\
&+m u_y^3 (15 \lambda_{xx}^3+45 \lambda_{xx}^2 m u_x^2+15 \lambda_{xx} m^2 u_x^4+m^{3} u_x^6)\\
&+2 \lambda_{xy} u_x (30 m u_x^2 (\lambda_{xx} (14 \lambda_{xx}+3 \lambda_{yy}+I T)+2 \lambda_{xy}^2)\\
&+15 \lambda_{xx} (\lambda_{xx} (28 \lambda_{xx}+9 \lambda_{yy}+3 I T)+12 \lambda_{xy}^2)\\
&+3 m^2 u_x^4 (28 \lambda_{xx}+3 \lambda_{yy}+I T)+4 m^{3} u_x^6)]
\tt{,}
\end{aligned}
\end{equation}
\begin{equation}
\begin{aligned}
M^{ES}_{9,7,xxxxxyy}&=\sum_{i}f^{ES}_{i}( v_{ix}^2+v_{iy}^2+\eta^{2}_{i} )v_{ix}v_{ix}v_{ix}v_{ix}v_{ix}v_{iy}v_{iy}\\
&=\frac{n}{m^{4}} [105 \lambda_{xx}^3 (u_y (2 \lambda_{xy}+m u_x u_y)+\lambda_{yy} u_x)\\
&+15 \lambda_{xx}^2 (42 \lambda_{xy}^2 u_x+2 \lambda_{xy} u_y (6 \lambda_{yy}+21 m u_x^2+2 m u_y^2\\
&+I T)+u_x (m u_y^2 (6 \lambda_{yy}+7 m u_x^2+I T)\\
&+\lambda_{yy} (3 \lambda_{yy}+7 m u_x^2+I T)+m^2 u_y^4))\\
&+\lambda_{xx} (240 \lambda_{xy}^3 u_y+60 \lambda_{xy}^2 u_x (6 \lambda_{yy}+7 m u_x^2+6 m u_y^2\\
&+I T)+30 \lambda_{xy} m u_x^2 u_y (12 \lambda_{yy}+7 m u_x^2\\
&+4 m u_y^2+2 I T)+m u_x^3 (m u_y^2 (60 \lambda_{yy}\\
&+21 m u_x^2+10 I T)+\lambda_{yy} (30 \lambda_{yy}+21 m u_x^2\\
&+10 I T)+10 m^2 u_y^4))+u_x (120 \lambda_{xy}^4\\
&+240 \lambda_{xy}^3 m u_x u_y+2 \lambda_{xy}^2 m u_x^2 (60 \lambda_{yy}+21 m u_x^2\\
&+60 m u_y^2+10 I T)+2 \lambda_{xy} m^2 u_x^3 u_y (30 \lambda_{yy}\\
&+7 m u_x^2+10 m u_y^2+5 I T)+m^2 u_x^4 (m u_y^2 (6 \lambda_{yy}\\
&+m u_x^2+I T)+\lambda_{yy} (3 \lambda_{yy}+m u_x^2+I T)+m^2 u_y^4))]
\tt{,}
\end{aligned}
\end{equation}
\begin{equation}
\begin{aligned}
M^{ES}_{9,7,xxxxyyy}&=\sum_{i}f^{ES}_{i}( v_{ix}^2+v_{iy}^2+\eta^{2}_{i} )v_{ix}v_{ix}v_{ix}v_{ix}v_{iy}v_{iy}v_{iy}\\
&=\frac{n}{m^{4}} [6 \lambda_{xy} m u_x u_y^2 (45 \lambda_{xx}^2+2 m u_x^2 (15 \lambda_{xx}+10 \lambda_{yy}\\
&+I T)+60 \lambda_{xx} \lambda_{yy}+6 \lambda_{xx} I T+40 \lambda_{xy}^2+3 m^2 u_x^4)\\
&+m u_y^3 (3 \lambda_{xx} (5 (\lambda_{xx}^2+2 \lambda_{xx} \lambda_{yy}+8 \lambda_{xy}^2)+\lambda_{xx} I T)\\
&+3 m u_x^2 (\lambda_{xx} (15 \lambda_{xx}+20 \lambda_{yy}+2 I T)+40 \lambda_{xy}^2)\\
&+m^2 u_x^4 (15 \lambda_{xx}+10 \lambda_{yy}+I T)+m^{3} u_x^6)\\
&+6 \lambda_{xy} u_x (45 \lambda_{xx}^2 \lambda_{yy}+60 \lambda_{xx} \lambda_{xy}^2+6 \lambda_{xx} \lambda_{yy} (5 \lambda_{yy}\\
&+5 m u_x^2+I T)+4 \lambda_{xy}^2 (10 \lambda_{yy}+5 m u_x^2\\
&+I T)+\lambda_{yy} m u_x^2 (10 \lambda_{yy}+3 m u_x^2\\
&+2 I T))+m u_y^5 (3 \lambda_{xx}^2+6 \lambda_{xx} m u_x^2+m^2 u_x^4)\\
&+3 u_y (15 \lambda_{xx}^3 \lambda_{yy}+3 \lambda_{xx}^2 (30 \lambda_{xy}^2+\lambda_{yy} (5 \lambda_{yy}\\
&+15 m u_x^2+I T))+3 \lambda_{xx} (4 \lambda_{xy}^2 (10 \lambda_{yy}\\
&+15 m u_x^2+I T)+\lambda_{yy} m u_x^2 (10 \lambda_{yy}+5 m u_x^2\\
&+2 I T))+40 \lambda_{xy}^4+6 \lambda_{xy}^2 m u_x^2 (20 \lambda_{yy}+5 m u_x^2\\
&+2 I T)+\lambda_{yy} m^2 u_x^4 (5 \lambda_{yy}+m u_x^2+I T))\\
&+20 \lambda_{xy} m^2 u_x u_y^4 (3 \lambda_{xx}+m u_x^2)]
\tt{,}
\end{aligned}
\end{equation}
\begin{equation}
\begin{aligned}
M^{ES}_{9,7,xxxyyyy}&=\sum_{i}f^{ES}_{i}( v_{ix}^2+v_{iy}^2+\eta^{2}_{i} )v_{ix}v_{ix}v_{ix}v_{iy}v_{iy}v_{iy}v_{iy}\\
&=\frac{n}{m^{4}} [24 \lambda_{xy}^3 u_y (10 \lambda_{xx}+15 \lambda_{yy}+5 m (2 u_x^2+u_y^2)\\
&+I T)+6 \lambda_{xy}^2 u_x (2 m u_y^2 (30 \lambda_{xx}+45 \lambda_{yy}+10 m u_x^2\\
&+3 I T)+\lambda_{yy} (60 \lambda_{xx}+45 \lambda_{yy}+20 m u_x^2+6 I T)\\
&+15 m^2 u_y^4)+2 \lambda_{xy} u_y (2 m u_y^2 (3 m u_x^2 (10 \lambda_{xx}\\
&+15 \lambda_{yy}+I T)+3 \lambda_{xx} (5 \lambda_{xx}+15 \lambda_{yy}+I T)\\
&+5 m^2 u_x^4)+9 \lambda_{yy} m u_x^2 (20 \lambda_{xx}+15 \lambda_{yy}+2 I T)\\
&+9 \lambda_{xx} \lambda_{yy} (10 \lambda_{xx}+15 \lambda_{yy}+2 I T)+9 m^2 u_y^4 (\lambda_{xx}\\
&+m u_x^2)+30 \lambda_{yy} m^2 u_x^4)+u_x (3 \lambda_{yy}^2 (m u_x^2 (10 \lambda_{xx}\\
&+5 \lambda_{yy}+I T)+3 \lambda_{xx} (5 (\lambda_{xx}+\lambda_{yy})+I T)\\
&+m^2 u_x^4)+m^2 u_y^4 (m u_x^2 (10 \lambda_{xx}+15 \lambda_{yy}\\
&+I T)+3 \lambda_{xx} (5 \lambda_{xx}+15 \lambda_{yy}+I T)+m^2 u_x^4)\\
&+3 \lambda_{yy} m u_y^2 (m u_x^2 (20 \lambda_{xx}+15 \lambda_{yy}+2 I T)\\
&+3 \lambda_{xx} (10 \lambda_{xx}+15 \lambda_{yy}+2 I T)+2 m^2 u_x^4)\\&+m^{3} u_y^6 (3 \lambda_{xx}+m u_x^2))+120 \lambda_{xy}^4 u_x]
\tt{,}
\end{aligned}
\end{equation}
\begin{equation}
\begin{aligned}
M^{ES}_{9,7,xxyyyyy}&=\sum_{i}f^{ES}_{i}( v_{ix}^2+v_{iy}^2+\eta^{2}_{i} )v_{ix}v_{ix}v_{iy}v_{iy}v_{iy}v_{iy}v_{iy}\\
&=\frac{n}{m^{4}} [2 \lambda_{xy}^2 u_y (10 m u_y^2 (6 \lambda_{xx}+21 \lambda_{yy}+6 m u_x^2\\
&+I T)+15 \lambda_{yy} (12 \lambda_{xx}+21 \lambda_{yy}+12 m u_x^2\\
&+2 I T)+21 m^2 u_y^4)+2 \lambda_{xy} u_x (15 \lambda_{yy}^2 (6 \lambda_{xx}\\
&+7 \lambda_{yy}+2 m u_x^2+I T)+5 m^2 u_y^4 (6 \lambda_{xx}\\
&+21 \lambda_{yy}+2 m u_x^2+I T)+15 \lambda_{yy} m u_y^2 (12 \lambda_{xx}\\
&+21 \lambda_{yy}+4 m u_x^2+2 I T)+7 m^{3} u_y^6)\\
&+15 \lambda_{yy}^2 u_y (m u_x^2 (6 \lambda_{xx}+7 \lambda_{yy}+I T)\\
&+\lambda_{xx} (3 \lambda_{xx}+7 \lambda_{yy}+I T)+m^2 u_x^4)\\&+m^2 u_y^5 (m u_x^2 (6 \lambda_{xx}+21 \lambda_{yy}+I T)\\
&+\lambda_{xx} (3 \lambda_{xx}+21 \lambda_{yy}+I T)\\
&+m^2 u_x^4)+5 \lambda_{yy} m u_y^3 (m u_x^2 (12 \lambda_{xx}+21 \lambda_{yy}\\
&+2 I T)+\lambda_{xx} (6 \lambda_{xx}+21 \lambda_{yy}+2 I T)+2 m^2 u_x^4)\\
&+m^{3} u_y^7 (\lambda_{xx}+m u_x^2)+120 \lambda_{xy}^4 u_y\\
&+240 \lambda_{xy}^3 u_x (\lambda_{yy}+m u_y^2)]
\tt{,}
\end{aligned}
\end{equation}
\begin{equation}
\begin{aligned}
M^{ES}_{9,7,xyyyyyy}&=\sum_{i}f^{ES}_{i}( v_{ix}^2+v_{iy}^2+\eta^{2}_{i} )v_{ix}v_{iy}v_{iy}v_{iy}v_{iy}v_{iy}v_{iy}\\
&=\frac{n}{m^{4}} [15 \lambda_{yy}^2 u_x (\lambda_{yy} (3 \lambda_{xx}+7 \lambda_{yy}+m u_x^2+I T)\\
&+18 \lambda_{xy}^2)+15 m^2 u_x u_y^4 (\lambda_{yy} (3 \lambda_{xx}+14 \lambda_{yy}+m u_x^2\\
&+I T)+6 \lambda_{xy}^2)+60 \lambda_{xy} m u_y^3 (\lambda_{yy} (3 \lambda_{xx}+14 \lambda_{yy}\\
&+3 m u_x^2+I T)+2 \lambda_{xy}^2)+15 \lambda_{yy} m u_x u_y^2 (\lambda_{yy} (9 \lambda_{xx}\\
&+28 \lambda_{yy}+3 m u_x^2+3 I T)+36 \lambda_{xy}^2)\\
&+30 \lambda_{xy} \lambda_{yy} u_y (\lambda_{yy} (9 \lambda_{xx}+28 \lambda_{yy}+9 m u_x^2\\
&+3 I T)+12 \lambda_{xy}^2)+6 \lambda_{xy} m^2 u_y^5 (3 \lambda_{xx}\\
&+28 \lambda_{yy}+3 m u_x^2+I T)+m^{3} u_x u_y^6 (3 \lambda_{xx}\\
&+28 \lambda_{yy}+m u_x^2+I T)+8 \lambda_{xy} m^{3} u_y^7+m^{4} u_x u_y^8]
\tt{,}
\end{aligned}
\end{equation}
\begin{equation}
\begin{aligned}
M^{ES}_{9,7,yyyyyyy}&=\sum_{i}f^{ES}_{i}( v_{ix}^2+v_{iy}^2+\eta^{2}_{i} )v_{iy}v_{iy}v_{iy}v_{iy}v_{iy}v_{iy}v_{iy}\\
&=\frac{n}{m^{4}} [105 \lambda_{yy}^2 u_y (\lambda_{yy} (\lambda_{xx}+9 \lambda_{yy}+m u_x^2+I T)\\
&+6 \lambda_{xy}^2)+21 m^2 u_y^5 (\lambda_{yy} (\lambda_{xx}+18 \lambda_{yy}+m u_x^2+I T)\\
&+2 \lambda_{xy}^2)+105 \lambda_{yy} m u_y^3 (\lambda_{yy} (\lambda_{xx}+12 \lambda_{yy}+m u_x^2\\
&+I T)+4 \lambda_{xy}^2)+m^{3} u_y^7 (\lambda_{xx}+36 \lambda_{yy}+m u_x^2\\
&+I T)+210 \lambda_{xy} \lambda_{yy}^3 u_x+630 \lambda_{xy} \lambda_{yy}^2 m u_x u_y^2\\
&+210 \lambda_{xy} \lambda_{yy} m^2 u_x u_y^4+14 \lambda_{xy} m^{3} u_x u_y^6+m^{4} u_y^9]
\tt{.}
\end{aligned}
\end{equation}

\section{ Appendixes C: process of dimensionless}\label{sec:AppendixesC}

The initial configuration of a one-dimensional normal shock wave propagating with Ma=1.45 in a flow field that filled with Ar gas is as follows, as shown in Bird's DSMC code \cite{1994Molecular}:
\[
\left\{
\begin{array}{l}
(\rho,u_x,T)^{1}_{x}=(1.094753 \times 10^{-5} kg/m^3,270.5789 m/s,394 K) \tt{,} \\
(\rho,u_x,T)^{0}_{x}=(6.64 \times 10^{-6} kg/m^3,446.11 m/s,273 K) \tt{.}
\end{array}
\right.
\]
where the index ``0'' (``1'') indicates wavefront (wave rear).
For simulating, physical quantities should be nondimensionalized.
In this simulation, we choose reference density $\rho_{\infty}$, reference temperature $T_{\infty}$, and reference length scale $L_{\infty}$ as reference variables.
The values of reference variables are $\rho_{\infty}=6.64 \times 10^{-6} kg/m^3$, $T_{\infty}=273K$, and $L_{\infty}=\lambda_0=1.315 \times 10^{-2}m$.
The speed of sound is $c_s=\sqrt{\gamma R T_{\infty}}=307.5807m/s$ and the viscosity coefficient is $\mu=2.117\times 10^{-5}N s  m^{-2}$, where $R=208.05J/(kg\cdot K)$ and $\gamma=1.6667$ for Ar.

Through the following equations, the real physical quantities can be nondimensionalized:
\begin{equation}
\hat{\rho}=\frac{\rho}{\rho_{\infty}},\hat{T}=\frac{T}{T_{\infty}},\hat{x}=\frac{x}{L_{\infty}},\hat{t}=\frac{t}{L_{\infty}/\sqrt{RT_{\infty}}},\hat{P}=\frac{P}{\rho_{\infty} R T_{\infty}}.
\end{equation}
\begin{equation}
\hat{u}=\frac{u}{u_{\infty}},\hat{\mu}=\frac{\mu}{\rho_{\infty}L_{\infty}\sqrt{RT_{\infty}}}.
\end{equation}
where $u_{\infty}=\sqrt{ R T_{\infty}}$.
The dimensionless macroscopic quantities are:
\[
\left\{
\begin{array}{l}
(\rho,u_x,T)^{1}_{x}=(1.64871,0.736742,1.44324) \tt{,} \\
(\rho,u_x,p)^{0}_{x}=(1.0,0.0,1.0) \tt{.}
\end{array}
\right.
\]

In the variable hard sphere model, the relationship between viscosity and temperature is
\begin{equation}
\mu=\mu_{ref}(T/T_{ref})^{\omega}
\end{equation}
where $\omega=0.81$ is the viscosity index.
The viscous equation based on ideal gas is $\mu = \tau \rho T$.
Consequently,  the relaxation time at different temperatures can be obtained from
\begin{equation}
\tau = \tau_{ref}\rho_{ref}/\rho(T/T_{ref})^{\omega-1}
\end{equation}
where subscript ``ref'' means the referenced variables.

\section*{Data Availability}
The data that support the findings of this study are available from the corresponding author upon reasonable request.

\section*{References}
\bibliography{2-fluid-pof}

\begin{thebibliography}{110}%
\makeatletter
\providecommand \@ifxundefined [1]{%
 \@ifx{#1\undefined}
}%
\providecommand \@ifnum [1]{%
 \ifnum #1\expandafter \@firstoftwo
 \else \expandafter \@secondoftwo
 \fi
}%
\providecommand \@ifx [1]{%
 \ifx #1\expandafter \@firstoftwo
 \else \expandafter \@secondoftwo
 \fi
}%
\providecommand \natexlab [1]{#1}%
\providecommand \enquote  [1]{``#1''}%
\providecommand \bibnamefont  [1]{#1}%
\providecommand \bibfnamefont [1]{#1}%
\providecommand \citenamefont [1]{#1}%
\providecommand \href@noop [0]{\@secondoftwo}%
\providecommand \href [0]{\begingroup \@sanitize@url \@href}%
\providecommand \@href[1]{\@@startlink{#1}\@@href}%
\providecommand \@@href[1]{\endgroup#1\@@endlink}%
\providecommand \@sanitize@url [0]{\catcode `\\12\catcode `\$12\catcode
  `\&12\catcode `\#12\catcode `\^12\catcode `\_12\catcode `\%12\relax}%
\providecommand \@@startlink[1]{}%
\providecommand \@@endlink[0]{}%
\providecommand \url  [0]{\begingroup\@sanitize@url \@url }%
\providecommand \@url [1]{\endgroup\@href {#1}{\urlprefix }}%
\providecommand \urlprefix  [0]{URL }%
\providecommand \Eprint [0]{\href }%
\providecommand \doibase [0]{http://dx.doi.org/}%
\providecommand \selectlanguage [0]{\@gobble}%
\providecommand \bibinfo  [0]{\@secondoftwo}%
\providecommand \bibfield  [0]{\@secondoftwo}%
\providecommand \translation [1]{[#1]}%
\providecommand \BibitemOpen [0]{}%
\providecommand \bibitemStop [0]{}%
\providecommand \bibitemNoStop [0]{.\EOS\space}%
\providecommand \EOS [0]{\spacefactor3000\relax}%
\providecommand \BibitemShut  [1]{\csname bibitem#1\endcsname}%
\let\auto@bib@innerbib\@empty
\bibitem [{Note1()}]{Note1}%
  \BibitemOpen
  \bibinfo {note} {Generally, the non-equilibrium described by hydrodynamic
  equations is called hydrodynamic non-equilibrium (HNE), and the
  non-equilibrium described by kinetic theory due to deviation from
  thermodynamic equilibrium is called thermodynamic non-equilibrium (TNE).
  Clearly, the HNE is only one part of TNE.}\BibitemShut {Stop}%
\bibitem [{\citenamefont {Succi}(2001)}]{2002Boltzmann}%
  \BibitemOpen
  \bibfield  {author} {\bibinfo {author} {\bibfnamefont {S.}~\bibnamefont
  {Succi}},\ }in\ \href {\doibase {}} {\emph {\bibinfo {booktitle} {The
  {Lattice} {Boltzmann} {Equation} for {fluid} {Dynamics} and {Beyond}}}}\
  (\bibinfo  {publisher} {Oxford University Press, New York},\ \bibinfo {year}
  {2001})\ Chap.~\bibinfo {chapter} {12}, pp.\ \bibinfo {pages}
  {179--213}\BibitemShut {NoStop}%
\bibitem [{\citenamefont {Xu}, \citenamefont {Zhang},\ and\ \citenamefont
  {Zhang}(2018)}]{Xu2018-Chap2}%
  \BibitemOpen
  \bibfield  {author} {\bibinfo {author} {\bibfnamefont {A.~G.}\ \bibnamefont
  {Xu}}, \bibinfo {author} {\bibfnamefont {G.~C.}\ \bibnamefont {Zhang}}, \
  and\ \bibinfo {author} {\bibfnamefont {Y.~D.}\ \bibnamefont {Zhang}},\
  }\bibfield  {title} {\enquote {\bibinfo {title} {{Discrete} {Boltzmann}
  {Modeling} of {Compressible} {Flows}},}\ }in\ \href {\doibase
  10.5772/intechopen.70748} {\emph {\bibinfo {booktitle} {Kinetic Theory}}},\
  \bibinfo {editor} {edited by\ \bibinfo {editor} {\bibfnamefont
  {G.}~\bibnamefont {Kyzas}}\ and\ \bibinfo {editor} {\bibfnamefont
  {A.}~\bibnamefont {Mitropoulos}}}\ (\bibinfo  {publisher} {InTech},\ \bibinfo
  {address} {Rijeka},\ \bibinfo {year} {2018})\ Chap.~\bibinfo {chapter}
  {02}\BibitemShut {NoStop}%
\bibitem [{\citenamefont {Mewes}(1960)}]{1960Rarefied}%
  \BibitemOpen
  \bibfield  {author} {\bibinfo {author} {\bibfnamefont {D.}~\bibnamefont
  {Mewes}},\ }\href {\doibase {}} {\emph {\bibinfo {title} {Rarefied {Gas}
  {Dynamics}}}}\ (\bibinfo  {publisher} {Rarefied Gas Dynamics},\ \bibinfo
  {year} {1960})\BibitemShut {NoStop}%
\bibitem [{\citenamefont {Chen}\ \emph {et~al.}(2016)\citenamefont {Chen},
  \citenamefont {Zhao}, \citenamefont {Jiang},\ and\ \citenamefont
  {Liu}}]{2016ChenPOG}%
  \BibitemOpen
  \bibfield  {author} {\bibinfo {author} {\bibfnamefont {W.~F.}\ \bibnamefont
  {Chen}}, \bibinfo {author} {\bibfnamefont {W.~W.}\ \bibnamefont {Zhao}},
  \bibinfo {author} {\bibfnamefont {Z.~Z.}\ \bibnamefont {Jiang}}, \ and\
  \bibinfo {author} {\bibfnamefont {H.~L.}\ \bibnamefont {Liu}},\ }\bibfield
  {title} {\enquote {\bibinfo {title} {A {Review} of {Moment} {Equations} for
  {Rarefied} {Gas} {Dynamics} (in {Chinese})},}\ }\href {\doibase {}}
  {\bibfield  {journal} {\bibinfo  {journal} {Phys. Gases}\ }\textbf {\bibinfo
  {volume} {1}},\ \bibinfo {pages} {9--24} (\bibinfo {year}
  {2016})}\BibitemShut {NoStop}%
\bibitem [{\citenamefont {Xu}\ \emph {et~al.}(2021{\natexlab{a}})\citenamefont
  {Xu}, \citenamefont {Chen}, \citenamefont {Song}, \citenamefont {Chen},\ and\
  \citenamefont {Chen}}]{2021XuACTA}%
  \BibitemOpen
  \bibfield  {author} {\bibinfo {author} {\bibfnamefont {A.~G.}\ \bibnamefont
  {Xu}}, \bibinfo {author} {\bibfnamefont {J.}~\bibnamefont {Chen}}, \bibinfo
  {author} {\bibfnamefont {J.~H.}\ \bibnamefont {Song}}, \bibinfo {author}
  {\bibfnamefont {D.~W.}\ \bibnamefont {Chen}}, \ and\ \bibinfo {author}
  {\bibfnamefont {Z.~H.}\ \bibnamefont {Chen}},\ }\bibfield  {title} {\enquote
  {\bibinfo {title} {Progress of discrete {Boltzmann} study on multiphase
  complex flows (in {Chinese})},}\ }\href {\doibase {}} {\bibfield  {journal}
  {\bibinfo  {journal} {Acta Aerodyn. Sin.}\ }\textbf {\bibinfo {volume}
  {39}},\ \bibinfo {pages} {138--169} (\bibinfo {year}
  {2021}{\natexlab{a}})}\BibitemShut {NoStop}%
\bibitem [{\citenamefont {Xu}\ \emph {et~al.}(2021{\natexlab{b}})\citenamefont
  {Xu}, \citenamefont {Song}, \citenamefont {Chen}, \citenamefont {Xie},\ and\
  \citenamefont {Ying}}]{2021XuCJCP}%
  \BibitemOpen
  \bibfield  {author} {\bibinfo {author} {\bibfnamefont {A.~G.}\ \bibnamefont
  {Xu}}, \bibinfo {author} {\bibfnamefont {J.~H.}\ \bibnamefont {Song}},
  \bibinfo {author} {\bibfnamefont {F.}~\bibnamefont {Chen}}, \bibinfo {author}
  {\bibfnamefont {K.}~\bibnamefont {Xie}}, \ and\ \bibinfo {author}
  {\bibfnamefont {Y.~J.}\ \bibnamefont {Ying}},\ }\bibfield  {title} {\enquote
  {\bibinfo {title} {Modeling and {Analysis} {Methods} for {Complex} {Fields}
  {Based} on {Phase} {Space} (in {Chinese})},}\ }\href {\doibase {}} {\bibfield
   {journal} {\bibinfo  {journal} {Chin. J. Comput. Phys.}\ }\textbf {\bibinfo
  {volume} {38}},\ \bibinfo {pages} {631--660} (\bibinfo {year}
  {2021}{\natexlab{b}})}\BibitemShut {NoStop}%
\bibitem [{\citenamefont {Xu}\ \emph {et~al.}(2021{\natexlab{c}})\citenamefont
  {Xu}, \citenamefont {Shan}, \citenamefont {Chen}, \citenamefont {Gan},\ and\
  \citenamefont {Lin}}]{2021XuACTAA}%
  \BibitemOpen
  \bibfield  {author} {\bibinfo {author} {\bibfnamefont {A.~G.}\ \bibnamefont
  {Xu}}, \bibinfo {author} {\bibfnamefont {Y.~M.}\ \bibnamefont {Shan}},
  \bibinfo {author} {\bibfnamefont {F.}~\bibnamefont {Chen}}, \bibinfo {author}
  {\bibfnamefont {Y.~B.}\ \bibnamefont {Gan}}, \ and\ \bibinfo {author}
  {\bibfnamefont {C.~D.}\ \bibnamefont {Lin}},\ }\bibfield  {title} {\enquote
  {\bibinfo {title} {Progress of mesoscale modeling and investigation of
  combustion multiphase flow (in {Chinese})},}\ }\href {\doibase {}} {\bibfield
   {journal} {\bibinfo  {journal} {Acta Aeronaut. Astronaut. Sin.}\ }\textbf
  {\bibinfo {volume} {42}},\ \bibinfo {pages} {625842} (\bibinfo {year}
  {2021}{\natexlab{c}})}\BibitemShut {NoStop}%
\bibitem [{\citenamefont {Ding}\ \emph {et~al.}(2017)\citenamefont {Ding},
  \citenamefont {Si}, \citenamefont {Yang}, \citenamefont {Lu}, \citenamefont
  {Zhai},\ and\ \citenamefont {Luo}}]{Ding2017PRL}%
  \BibitemOpen
  \bibfield  {author} {\bibinfo {author} {\bibfnamefont {J.~C.}\ \bibnamefont
  {Ding}}, \bibinfo {author} {\bibfnamefont {T.}~\bibnamefont {Si}}, \bibinfo
  {author} {\bibfnamefont {J.~M.}\ \bibnamefont {Yang}}, \bibinfo {author}
  {\bibfnamefont {X.~Y.}\ \bibnamefont {Lu}}, \bibinfo {author} {\bibfnamefont
  {Z.~G.}\ \bibnamefont {Zhai}}, \ and\ \bibinfo {author} {\bibfnamefont
  {X.~S.}\ \bibnamefont {Luo}},\ }\bibfield  {title} {\enquote {\bibinfo
  {title} {Measurement of a {Richtmyer-Meshkov} {Instability} at an
  {Air}-{${\mathrm{SF}}_{6}$} interface in a {Semiannular} {Shock} {Tube}},}\
  }\href {\doibase 10.1103/PhysRevLett.119.014501} {\bibfield  {journal}
  {\bibinfo  {journal} {Phys. Rev. Lett.}\ }\textbf {\bibinfo {volume} {119}},\
  \bibinfo {pages} {014501} (\bibinfo {year} {2017})}\BibitemShut {NoStop}%
\bibitem [{\citenamefont {Luo}\ \emph {et~al.}(2019)\citenamefont {Luo},
  \citenamefont {Li}, \citenamefont {Ding}, \citenamefont {Zhai},\ and\
  \citenamefont {Si}}]{Luo2019JFM2}%
  \BibitemOpen
  \bibfield  {author} {\bibinfo {author} {\bibfnamefont {X.~S.}\ \bibnamefont
  {Luo}}, \bibinfo {author} {\bibfnamefont {M.}~\bibnamefont {Li}}, \bibinfo
  {author} {\bibfnamefont {J.~C.}\ \bibnamefont {Ding}}, \bibinfo {author}
  {\bibfnamefont {Z.~G.}\ \bibnamefont {Zhai}}, \ and\ \bibinfo {author}
  {\bibfnamefont {T.}~\bibnamefont {Si}},\ }\bibfield  {title} {\enquote
  {\bibinfo {title} {Nonlinear behaviour of convergent {Richtmyer-Meshkov}
  instability},}\ }\href {\doibase 10.1017/jfm.2019.610} {\bibfield  {journal}
  {\bibinfo  {journal} {J. Fluid Mech.}\ }\textbf {\bibinfo {volume} {877}},\
  \bibinfo {pages} {130--141} (\bibinfo {year} {2019})}\BibitemShut {NoStop}%
\bibitem [{\citenamefont {Ding}\ \emph {et~al.}(2018)\citenamefont {Ding},
  \citenamefont {Zhai}, \citenamefont {Si},\ and\ \citenamefont
  {Luo}}]{Ding2018CSB}%
  \BibitemOpen
  \bibfield  {author} {\bibinfo {author} {\bibfnamefont {J.~C.}\ \bibnamefont
  {Ding}}, \bibinfo {author} {\bibfnamefont {Z.~G.}\ \bibnamefont {Zhai}},
  \bibinfo {author} {\bibfnamefont {T.}~\bibnamefont {Si}}, \ and\ \bibinfo
  {author} {\bibfnamefont {X.~S.}\ \bibnamefont {Luo}},\ }\bibfield  {title}
  {\enquote {\bibinfo {title} {Progress in experiments of converging
  {Richtmyer-Meshkov} instability (in {Chinese})},}\ }\href {\doibase
  10.1360/N972017-01211} {\bibfield  {journal} {\bibinfo  {journal} {Chin. Sci.
  Bull.}\ }\textbf {\bibinfo {volume} {63}},\ \bibinfo {pages} {618--628}
  (\bibinfo {year} {2018})}\BibitemShut {NoStop}%
\bibitem [{\citenamefont {Qiu}\ \emph {et~al.}(2020)\citenamefont {Qiu},
  \citenamefont {Bao}, \citenamefont {Zhou}, \citenamefont {Che}, \citenamefont
  {Chen},\ and\ \citenamefont {You}}]{Qiu2020POF}%
  \BibitemOpen
  \bibfield  {author} {\bibinfo {author} {\bibfnamefont {R.~F.}\ \bibnamefont
  {Qiu}}, \bibinfo {author} {\bibfnamefont {Y.}~\bibnamefont {Bao}}, \bibinfo
  {author} {\bibfnamefont {T.}~\bibnamefont {Zhou}}, \bibinfo {author}
  {\bibfnamefont {H.~H.}\ \bibnamefont {Che}}, \bibinfo {author} {\bibfnamefont
  {R.~Q.}\ \bibnamefont {Chen}}, \ and\ \bibinfo {author} {\bibfnamefont
  {Y.~C.}\ \bibnamefont {You}},\ }\bibfield  {title} {\enquote {\bibinfo
  {title} {Study of regular reflection shock waves using a mesoscopic kinetic
  approach: Curvature pattern and effects of viscosity},}\ }\href {\doibase
  https://doi.org/10.1063/5.0024801} {\bibfield  {journal} {\bibinfo  {journal}
  {Phys. Fluids}\ }\textbf {\bibinfo {volume} {32}},\ \bibinfo {pages} {106106}
  (\bibinfo {year} {2020})}\BibitemShut {NoStop}%
\bibitem [{\citenamefont {Bao}\ \emph {et~al.}(2022)\citenamefont {Bao},
  \citenamefont {Qiu}, \citenamefont {Zhou}, \citenamefont {Zhou},
  \citenamefont {Weng}, \citenamefont {Lin},\ and\ \citenamefont
  {You}}]{Bao2022POF}%
  \BibitemOpen
  \bibfield  {author} {\bibinfo {author} {\bibfnamefont {Y.}~\bibnamefont
  {Bao}}, \bibinfo {author} {\bibfnamefont {R.~F.}\ \bibnamefont {Qiu}},
  \bibinfo {author} {\bibfnamefont {K.}~\bibnamefont {Zhou}}, \bibinfo {author}
  {\bibfnamefont {T.}~\bibnamefont {Zhou}}, \bibinfo {author} {\bibfnamefont
  {Y.~X.}\ \bibnamefont {Weng}}, \bibinfo {author} {\bibfnamefont
  {K.}~\bibnamefont {Lin}}, \ and\ \bibinfo {author} {\bibfnamefont {Y.~C.}\
  \bibnamefont {You}},\ }\bibfield  {title} {\enquote {\bibinfo {title} {Study
  of shock wave/boundary layer interaction from the perspective of
  nonequilibrium effects},}\ }\href {\doibase doi: 10.1063/5.0085570}
  {\bibfield  {journal} {\bibinfo  {journal} {Phys. Fluids}\ }\textbf {\bibinfo
  {volume} {34}},\ \bibinfo {pages} {046109} (\bibinfo {year}
  {2022})}\BibitemShut {NoStop}%
\bibitem [{\citenamefont {White}(2016)}]{2016Fluid-mechanics}%
  \BibitemOpen
  \bibfield  {author} {\bibinfo {author} {\bibfnamefont {F.~M.}\ \bibnamefont
  {White}},\ }\href {\doibase {}} {\emph {\bibinfo {title} {Fluid mechanics}}}\
  (\bibinfo  {publisher} {McGraw-Hill Education},\ \bibinfo {year} {2016})\
  Chap.~\bibinfo {chapter} {08}, pp.\ \bibinfo {pages} {521--590}\BibitemShut
  {NoStop}%
\bibitem [{Note2()}]{Note2}%
  \BibitemOpen
  \bibinfo {note} {Knudsen number can be defined as the ratio of the mean free
  path of molecules $\lambda $ to the characteristic length $L$ i.e.,
  $Kn=\lambda /L$, where $\lambda =c_{s}\tau $ with the relaxation time $\tau $
  and the local speed of sound $c_{s}$. The characteristic length $L$ depends
  on macroscopic quantity gradients. That is to say $L=\psi /\mid \nabla \psi
  \mid $ where $\psi $ represents the macroscopic quantities such as density
  $\rho $, temperature $T$, velocity $\protect \mathbf {u}$, and pressure $p$.
  In non-equilibrium flows, the Kn number can also be defined as the ratio of
  relaxation time $\tau $ to the characteristic time $t_0$. Kn number is one of
  the common parameters to describe the non-equilibrium degrees of fluid
  systems from its own perspective. Generally, the larger the Kn number is, the
  deeper the TNE degree of the system is. However, due to the complexity of TNE
  behaviors of the system, the Kn number is inadequate in describing the TNE
  degrees of the system in some cases.}\BibitemShut {Stop}%
\bibitem [{\citenamefont {Manuel}\ \emph {et~al.}(2021)\citenamefont {Manuel},
  \citenamefont {Khiar}, \citenamefont {Rigon}, \citenamefont {Albertazzi},
  \citenamefont {Klein}, \citenamefont {Kroll}, \citenamefont {E.Brack},
  \citenamefont {Michel}, \citenamefont {Mabey}, \citenamefont {Pikuz},
  \citenamefont {Williams}, \citenamefont {Koenig}, \citenamefont {Casner},\
  and\ \citenamefont {Kuranz}}]{Manuel2021MRE}%
  \BibitemOpen
  \bibfield  {author} {\bibinfo {author} {\bibfnamefont {M.~J.-E.}\
  \bibnamefont {Manuel}}, \bibinfo {author} {\bibfnamefont {B.}~\bibnamefont
  {Khiar}}, \bibinfo {author} {\bibfnamefont {G.}~\bibnamefont {Rigon}},
  \bibinfo {author} {\bibfnamefont {B.}~\bibnamefont {Albertazzi}}, \bibinfo
  {author} {\bibfnamefont {S.~R.}\ \bibnamefont {Klein}}, \bibinfo {author}
  {\bibfnamefont {F.}~\bibnamefont {Kroll}}, \bibinfo {author} {\bibfnamefont
  {F.}~\bibnamefont {E.Brack}}, \bibinfo {author} {\bibfnamefont
  {T.}~\bibnamefont {Michel}}, \bibinfo {author} {\bibfnamefont
  {P.}~\bibnamefont {Mabey}}, \bibinfo {author} {\bibfnamefont
  {S.}~\bibnamefont {Pikuz}}, \bibinfo {author} {\bibfnamefont {J.~C.}\
  \bibnamefont {Williams}}, \bibinfo {author} {\bibfnamefont {M.}~\bibnamefont
  {Koenig}}, \bibinfo {author} {\bibfnamefont {A.}~\bibnamefont {Casner}}, \
  and\ \bibinfo {author} {\bibfnamefont {C.~C.}\ \bibnamefont {Kuranz}},\
  }\bibfield  {title} {\enquote {\bibinfo {title} {On the study of hydrodynamic
  instabilities in the presence of background magnetic fields in
  high-energy-density plasmas},}\ }\href {\doibase 10.1063/5.0025374}
  {\bibfield  {journal} {\bibinfo  {journal} {Matter Radiat. Extrem.}\ }\textbf
  {\bibinfo {volume} {6}},\ \bibinfo {pages} {026904} (\bibinfo {year}
  {2021})}\BibitemShut {NoStop}%
\bibitem [{\citenamefont {Yao}\ \emph {et~al.}(2020)\citenamefont {Yao},
  \citenamefont {Cai}, \citenamefont {Yan}, \citenamefont {Zhang},
  \citenamefont {Du}, \citenamefont {Tian}, \citenamefont {Zhang},
  \citenamefont {Wang},\ and\ \citenamefont {Zhu}}]{Yao2020MRE}%
  \BibitemOpen
  \bibfield  {author} {\bibinfo {author} {\bibfnamefont {P.~L.}\ \bibnamefont
  {Yao}}, \bibinfo {author} {\bibfnamefont {H.~B.}\ \bibnamefont {Cai}},
  \bibinfo {author} {\bibfnamefont {X.~X.}\ \bibnamefont {Yan}}, \bibinfo
  {author} {\bibfnamefont {W.~S.}\ \bibnamefont {Zhang}}, \bibinfo {author}
  {\bibfnamefont {B.}~\bibnamefont {Du}}, \bibinfo {author} {\bibfnamefont
  {J.~M.}\ \bibnamefont {Tian}}, \bibinfo {author} {\bibfnamefont {E.~H.}\
  \bibnamefont {Zhang}}, \bibinfo {author} {\bibfnamefont {X.~W.}\ \bibnamefont
  {Wang}}, \ and\ \bibinfo {author} {\bibfnamefont {S.~P.}\ \bibnamefont
  {Zhu}},\ }\bibfield  {title} {\enquote {\bibinfo {title} {Kinetic study of
  transverse electron-scale interface instability in relativistic shear
  flows},}\ }\href {\doibase 10.1063/5.0017962} {\bibfield  {journal} {\bibinfo
   {journal} {Matter Radiat. Extrem.}\ }\textbf {\bibinfo {volume} {5}},\
  \bibinfo {pages} {054403} (\bibinfo {year} {2020})}\BibitemShut {NoStop}%
\bibitem [{\citenamefont {Cai}\ \emph {et~al.}(2021)\citenamefont {Cai},
  \citenamefont {Yan}, \citenamefont {Yao},\ and\ \citenamefont
  {Zhu}}]{Cai2021MRE}%
  \BibitemOpen
  \bibfield  {author} {\bibinfo {author} {\bibfnamefont {H.~B.}\ \bibnamefont
  {Cai}}, \bibinfo {author} {\bibfnamefont {X.~X.}\ \bibnamefont {Yan}},
  \bibinfo {author} {\bibfnamefont {P.~L.}\ \bibnamefont {Yao}}, \ and\
  \bibinfo {author} {\bibfnamefont {S.~P.}\ \bibnamefont {Zhu}},\ }\bibfield
  {title} {\enquote {\bibinfo {title} {Hybrid fluid-particle modeling of
  shock-driven hydrodynamic instabilities in a plasma},}\ }\href {\doibase
  10.1063/5.0042973} {\bibfield  {journal} {\bibinfo  {journal} {Matter Radiat.
  Extrem.}\ }\textbf {\bibinfo {volume} {6}},\ \bibinfo {pages} {035901}
  (\bibinfo {year} {2021})}\BibitemShut {NoStop}%
\bibitem [{\citenamefont {Shan}\ \emph {et~al.}(2021)\citenamefont {Shan},
  \citenamefont {Wu}, \citenamefont {Yuan}, \citenamefont {Wang}, \citenamefont
  {Cai}, \citenamefont {Tian}, \citenamefont {Zhang}, \citenamefont {Zhang},
  \citenamefont {Deng}, \citenamefont {Zhang}, \citenamefont {Teng},
  \citenamefont {Bi}, \citenamefont {Yang}, \citenamefont {Yang}, \citenamefont
  {Zhou}, \citenamefont {Gu}, \citenamefont {Zhang},\ and\ \citenamefont
  {Zhu}}]{2021Shan-kinetic-effects}%
  \BibitemOpen
  \bibfield  {author} {\bibinfo {author} {\bibfnamefont {L.~Q.}\ \bibnamefont
  {Shan}}, \bibinfo {author} {\bibfnamefont {F.~J.}\ \bibnamefont {Wu}},
  \bibinfo {author} {\bibfnamefont {Z.~Q.}\ \bibnamefont {Yuan}}, \bibinfo
  {author} {\bibfnamefont {W.~W.}\ \bibnamefont {Wang}}, \bibinfo {author}
  {\bibfnamefont {H.~B.}\ \bibnamefont {Cai}}, \bibinfo {author} {\bibfnamefont
  {C.}~\bibnamefont {Tian}}, \bibinfo {author} {\bibfnamefont {F.}~\bibnamefont
  {Zhang}}, \bibinfo {author} {\bibfnamefont {T.~K.}\ \bibnamefont {Zhang}},
  \bibinfo {author} {\bibfnamefont {Z.~G.}\ \bibnamefont {Deng}}, \bibinfo
  {author} {\bibfnamefont {W.~S.}\ \bibnamefont {Zhang}}, \bibinfo {author}
  {\bibfnamefont {J.}~\bibnamefont {Teng}}, \bibinfo {author} {\bibfnamefont
  {B.}~\bibnamefont {Bi}}, \bibinfo {author} {\bibfnamefont {S.~Q.}\
  \bibnamefont {Yang}}, \bibinfo {author} {\bibfnamefont {D.}~\bibnamefont
  {Yang}}, \bibinfo {author} {\bibfnamefont {W.~M.}\ \bibnamefont {Zhou}},
  \bibinfo {author} {\bibfnamefont {Y.~Q.}\ \bibnamefont {Gu}}, \bibinfo
  {author} {\bibfnamefont {B.~H.}\ \bibnamefont {Zhang}}, \ and\ \bibinfo
  {author} {\bibfnamefont {S.~P.}\ \bibnamefont {Zhu}},\ }\bibfield  {title}
  {\enquote {\bibinfo {title} {Research progress of kinetic effects in laser
  inertial confinement fusion(in {Chinese})},}\ }\href {\doibase
  10.11884/HPLPB202133.200235} {\bibfield  {journal} {\bibinfo  {journal} {High
  Power Laser and Particle Beams}\ }\textbf {\bibinfo {volume} {33}},\ \bibinfo
  {pages} {012004} (\bibinfo {year} {2021})}\BibitemShut {NoStop}%
\bibitem [{\citenamefont {Cai}\ \emph {et~al.}(2020)\citenamefont {Cai},
  \citenamefont {Zhang}, \citenamefont {Du}, \citenamefont {Yan}, \citenamefont
  {Shan}, \citenamefont {Hao}, \citenamefont {Li}, \citenamefont {Zhang},
  \citenamefont {Gong}, \citenamefont {Yang}, \citenamefont {Zou},
  \citenamefont {Zhu},\ and\ \citenamefont {He}}]{2020Cai-kinetic-effects}%
  \BibitemOpen
  \bibfield  {author} {\bibinfo {author} {\bibfnamefont {H.~B.}\ \bibnamefont
  {Cai}}, \bibinfo {author} {\bibfnamefont {W.~S.}\ \bibnamefont {Zhang}},
  \bibinfo {author} {\bibfnamefont {B.}~\bibnamefont {Du}}, \bibinfo {author}
  {\bibfnamefont {X.~X.}\ \bibnamefont {Yan}}, \bibinfo {author} {\bibfnamefont
  {L.~Q.}\ \bibnamefont {Shan}}, \bibinfo {author} {\bibfnamefont
  {L.}~\bibnamefont {Hao}}, \bibinfo {author} {\bibfnamefont {Z.~C.}\
  \bibnamefont {Li}}, \bibinfo {author} {\bibfnamefont {F.}~\bibnamefont
  {Zhang}}, \bibinfo {author} {\bibfnamefont {T.}~\bibnamefont {Gong}},
  \bibinfo {author} {\bibfnamefont {D.}~\bibnamefont {Yang}}, \bibinfo {author}
  {\bibfnamefont {S.~Y.}\ \bibnamefont {Zou}}, \bibinfo {author} {\bibfnamefont
  {S.~P.}\ \bibnamefont {Zhu}}, \ and\ \bibinfo {author} {\bibfnamefont
  {X.~T.}\ \bibnamefont {He}},\ }\bibfield  {title} {\enquote {\bibinfo {title}
  {Characteristic and impact of kinetic effects at interfaces of inertial
  confinement fusion hohlraums(in {Chinese})},}\ }\href {\doibase
  10.11884/HPLPB202032.200134} {\bibfield  {journal} {\bibinfo  {journal} {High
  Power Laser and Particle Beams}\ }\textbf {\bibinfo {volume} {32}},\ \bibinfo
  {pages} {092007} (\bibinfo {year} {2020})}\BibitemShut {NoStop}%
\bibitem [{\citenamefont {Tsien}(2012)}]{Tsien2012Superaerodynamics}%
  \BibitemOpen
  \bibfield  {author} {\bibinfo {author} {\bibfnamefont {H.~S.}\ \bibnamefont
  {Tsien}},\ }\bibfield  {title} {\enquote {\bibinfo {title}
  {Superaerodynamics, {Mechanics} of {Rarefied} {Gases}},}\ }\href {\doibase
  10.1016/B978-0-12-398277-3.50020-8} {\bibfield  {journal} {\bibinfo
  {journal} {Collect. Works H. S. Tsien}\ }\textbf {\bibinfo {volume} {13}},\
  \bibinfo {pages} {406--429} (\bibinfo {year} {2012})}\BibitemShut {NoStop}%
\bibitem [{\citenamefont {Arkilic}, \citenamefont {Schmidt},\ and\
  \citenamefont {Breuer}(1997)}]{Arkilic1997Microelectromechanical}%
  \BibitemOpen
  \bibfield  {author} {\bibinfo {author} {\bibfnamefont {E.~B.}\ \bibnamefont
  {Arkilic}}, \bibinfo {author} {\bibfnamefont {M.~A.}\ \bibnamefont
  {Schmidt}}, \ and\ \bibinfo {author} {\bibfnamefont {K.~S.}\ \bibnamefont
  {Breuer}},\ }\bibfield  {title} {\enquote {\bibinfo {title} {Gaseous slip
  flow in long microchannels},}\ }\href {\doibase 10.1109/84.585795} {\bibfield
   {journal} {\bibinfo  {journal} {J. Microelectromech. S.}\ }\textbf {\bibinfo
  {volume} {6}},\ \bibinfo {pages} {167--178} (\bibinfo {year}
  {1997})}\BibitemShut {NoStop}%
\bibitem [{\citenamefont {Ho}\ and\ \citenamefont {Tai}(1998)}]{1998MEMS}%
  \BibitemOpen
  \bibfield  {author} {\bibinfo {author} {\bibfnamefont {C.~M.}\ \bibnamefont
  {Ho}}\ and\ \bibinfo {author} {\bibfnamefont {Y.~C.}\ \bibnamefont {Tai}},\
  }\bibfield  {title} {\enquote {\bibinfo {title}
  {{Micro-electro-mechanical-systems(MEMS)} and fluid flows},}\ }\href
  {\doibase 10.1146/annurev.fluid.30.1.579} {\bibfield  {journal} {\bibinfo
  {journal} {Annu. Rev. Fluid Mech.}\ }\textbf {\bibinfo {volume} {30}},\
  \bibinfo {pages} {579--612} (\bibinfo {year} {1998})}\BibitemShut {NoStop}%
\bibitem [{\citenamefont {Nie}, \citenamefont {Doolen},\ and\ \citenamefont
  {Chen}(2002)}]{Nie2002JSP}%
  \BibitemOpen
  \bibfield  {author} {\bibinfo {author} {\bibfnamefont {X.~B.}\ \bibnamefont
  {Nie}}, \bibinfo {author} {\bibfnamefont {G.~D.}\ \bibnamefont {Doolen}}, \
  and\ \bibinfo {author} {\bibfnamefont {S.~Y.}\ \bibnamefont {Chen}},\
  }\bibfield  {title} {\enquote {\bibinfo {title} {Lattice-{Boltzmann}
  {Simulations} of {Fluid} {Flows} in {MEMS}},}\ }\href {\doibase
  10.1023/A:1014523007427} {\bibfield  {journal} {\bibinfo  {journal} {J. Stat.
  Phys}\ }\textbf {\bibinfo {volume} {107}},\ \bibinfo {pages} {279--289}
  (\bibinfo {year} {2002})}\BibitemShut {NoStop}%
\bibitem [{\citenamefont {Lim}\ \emph {et~al.}(2002)\citenamefont {Lim},
  \citenamefont {Shu}, \citenamefont {Niu},\ and\ \citenamefont
  {Chew}}]{2002LimPRE}%
  \BibitemOpen
  \bibfield  {author} {\bibinfo {author} {\bibfnamefont {C.~Y.}\ \bibnamefont
  {Lim}}, \bibinfo {author} {\bibfnamefont {C.}~\bibnamefont {Shu}}, \bibinfo
  {author} {\bibfnamefont {X.~D.}\ \bibnamefont {Niu}}, \ and\ \bibinfo
  {author} {\bibfnamefont {Y.~T.}\ \bibnamefont {Chew}},\ }\bibfield  {title}
  {\enquote {\bibinfo {title} {Application of lattice {Boltzmann} method to
  simulate microchannel flows},}\ }\href {\doibase 10.1063/1.1483841}
  {\bibfield  {journal} {\bibinfo  {journal} {Phys. Fluids}\ }\textbf {\bibinfo
  {volume} {14}},\ \bibinfo {pages} {2299--2308} (\bibinfo {year}
  {2002})}\BibitemShut {NoStop}%
\bibitem [{\citenamefont {Lockerby}, \citenamefont {Reese},\ and\ \citenamefont
  {Gallis}(2005)}]{Lockerby2005POF}%
  \BibitemOpen
  \bibfield  {author} {\bibinfo {author} {\bibfnamefont {D.~A.}\ \bibnamefont
  {Lockerby}}, \bibinfo {author} {\bibfnamefont {J.~M.}\ \bibnamefont {Reese}},
  \ and\ \bibinfo {author} {\bibfnamefont {M.~A.}\ \bibnamefont {Gallis}},\
  }\bibfield  {title} {\enquote {\bibinfo {title} {The usefulness of
  higher-order constitutive relations for describing the {Knudsen} layer},}\
  }\href {\doibase 10.1063/1.1897005} {\bibfield  {journal} {\bibinfo
  {journal} {Phys. Fluids}\ }\textbf {\bibinfo {volume} {17}},\ \bibinfo
  {pages} {100609} (\bibinfo {year} {2005})}\BibitemShut {NoStop}%
\bibitem [{\citenamefont {Zhang}\ \emph
  {et~al.}(2019{\natexlab{a}})\citenamefont {Zhang}, \citenamefont {Zhang},
  \citenamefont {Xu},\ and\ \citenamefont {Li}}]{2019ZhangGER}%
  \BibitemOpen
  \bibfield  {author} {\bibinfo {author} {\bibfnamefont {G.}~\bibnamefont
  {Zhang}}, \bibinfo {author} {\bibfnamefont {Y.~D.}\ \bibnamefont {Zhang}},
  \bibinfo {author} {\bibfnamefont {A.~G.}\ \bibnamefont {Xu}}, \ and\ \bibinfo
  {author} {\bibfnamefont {Y.~J.}\ \bibnamefont {Li}},\ }\bibfield  {title}
  {\enquote {\bibinfo {title} {Microflow effects on the hydraulic aperture of
  single rough fractures},}\ }\href {\doibase 10.26804/ager.2019.01.09}
  {\bibfield  {journal} {\bibinfo  {journal} {Advances in {Geo-Energy}
  {Research}}\ }\textbf {\bibinfo {volume} {3}},\ \bibinfo {pages} {104--114}
  (\bibinfo {year} {2019}{\natexlab{a}})}\BibitemShut {NoStop}%
\bibitem [{\citenamefont {Zhang}\ \emph {et~al.}(2022)\citenamefont {Zhang},
  \citenamefont {Xu}, \citenamefont {Chen}, \citenamefont {Lin},\ and\
  \citenamefont {Wei}}]{2022Zhang-AIPAdv-Slip}%
  \BibitemOpen
  \bibfield  {author} {\bibinfo {author} {\bibfnamefont {Y.}~\bibnamefont
  {Zhang}}, \bibinfo {author} {\bibfnamefont {A.}~\bibnamefont {Xu}}, \bibinfo
  {author} {\bibfnamefont {F.}~\bibnamefont {Chen}}, \bibinfo {author}
  {\bibfnamefont {C.}~\bibnamefont {Lin}}, \ and\ \bibinfo {author}
  {\bibfnamefont {Z.}~\bibnamefont {Wei}},\ }\bibfield  {title} {\enquote
  {\bibinfo {title} {Non-equilibrium characteristics of mass and heat transfers
  in the slip flow},}\ }\href {\doibase doi: 10.1063/5.0086400} {\bibfield
  {journal} {\bibinfo  {journal} {AIP Adv.}\ }\textbf {\bibinfo {volume}
  {12}},\ \bibinfo {pages} {035347} (\bibinfo {year} {2022})}\BibitemShut
  {NoStop}%
\bibitem [{\citenamefont {Chapman}, \citenamefont {Cowling},\ and\
  \citenamefont {Burnett}(1990)}]{1990Chapman}%
  \BibitemOpen
  \bibfield  {author} {\bibinfo {author} {\bibfnamefont {S.}~\bibnamefont
  {Chapman}}, \bibinfo {author} {\bibfnamefont {T.~G.}\ \bibnamefont
  {Cowling}}, \ and\ \bibinfo {author} {\bibfnamefont {D.}~\bibnamefont
  {Burnett}},\ }\bibfield  {title} {\enquote {\bibinfo {title} {The
  mathematical theory of non-uniform gases: an account of the kinetic theory of
  viscosity, thermal conduction, and diffusion in gases},}\ \ }(\bibinfo
  {publisher} {Cambridge University Press},\ \bibinfo {year} {1990})\
  Chap.~\bibinfo {chapter} {07}, pp.\ \bibinfo {pages} {110--131}\BibitemShut
  {NoStop}%
\bibitem [{\citenamefont {Struchtrup}(2005)}]{2005Macroscopic}%
  \BibitemOpen
  \bibfield  {author} {\bibinfo {author} {\bibfnamefont {H.}~\bibnamefont
  {Struchtrup}},\ }\href {\doibase https://doi.org/10.1007/3-540-32386-4}
  {\emph {\bibinfo {title} {Macroscopic {Transport} {Equations} for {Rarefied}
  {Gas} {Flows}--Approximation Methods in Kinetic Theory}}}\ (\bibinfo
  {publisher} {Springer-Verlag Berlin Heidelberg},\ \bibinfo {year}
  {2005})\BibitemShut {NoStop}%
\bibitem [{\citenamefont {Burnett}(1936)}]{1936BurnettPLMS}%
  \BibitemOpen
  \bibfield  {author} {\bibinfo {author} {\bibfnamefont {D.}~\bibnamefont
  {Burnett}},\ }\bibfield  {title} {\enquote {\bibinfo {title} {The
  {Distribution} of {Molecular} {Velocities} and the {Mean} {Motion} in a
  {Non-Uniform} {Gas}},}\ }\href {\doibase 10.1112/plms/s2-40.1.382} {\bibfield
   {journal} {\bibinfo  {journal} {P. Lond. Math. Soc.}\ }\textbf {\bibinfo
  {volume} {s2-40}},\ \bibinfo {pages} {382--435} (\bibinfo {year}
  {1936})}\BibitemShut {NoStop}%
\bibitem [{\citenamefont {Grad}(1949)}]{Grad1949CPAM}%
  \BibitemOpen
  \bibfield  {author} {\bibinfo {author} {\bibfnamefont {H.}~\bibnamefont
  {Grad}},\ }\bibfield  {title} {\enquote {\bibinfo {title} {On the kinetic
  theory of rarefied gases},}\ }\href@noop {} {\bibfield  {journal} {\bibinfo
  {journal} {Commun. Pur. Appl. Math.}\ }\textbf {\bibinfo {volume} {2}},\
  \bibinfo {pages} {331--407} (\bibinfo {year} {1949})}\BibitemShut {NoStop}%
\bibitem [{\citenamefont {Struchtrup}\ and\ \citenamefont
  {Torrilhon}(2003)}]{2003Struchtrup}%
  \BibitemOpen
  \bibfield  {author} {\bibinfo {author} {\bibfnamefont {H.}~\bibnamefont
  {Struchtrup}}\ and\ \bibinfo {author} {\bibfnamefont {M.}~\bibnamefont
  {Torrilhon}},\ }\bibfield  {title} {\enquote {\bibinfo {title}
  {Regularization of {Grad's} 13 moment equations: {Derivation} and linear
  analysis},}\ }\href {\doibase 10.1063/1.1597472} {\bibfield  {journal}
  {\bibinfo  {journal} {Phys. Fluids}\ }\textbf {\bibinfo {volume} {15}},\
  \bibinfo {pages} {2668--2680} (\bibinfo {year} {2003})}\BibitemShut {NoStop}%
\bibitem [{\citenamefont {Agarwal}, \citenamefont {Yun},\ and\ \citenamefont
  {Balakrishnan}(2001)}]{Agarwal2001POF}%
  \BibitemOpen
  \bibfield  {author} {\bibinfo {author} {\bibfnamefont {R.~K.}\ \bibnamefont
  {Agarwal}}, \bibinfo {author} {\bibfnamefont {K.~Y.}\ \bibnamefont {Yun}}, \
  and\ \bibinfo {author} {\bibfnamefont {R.}~\bibnamefont {Balakrishnan}},\
  }\bibfield  {title} {\enquote {\bibinfo {title} {Beyond {Navier-Stokes}:
  {Burnett} equations for flows in the continuum-transition regime},}\ }\href
  {\doibase 10.1063/1.1397256} {\bibfield  {journal} {\bibinfo  {journal}
  {Phys. Fluids}\ }\textbf {\bibinfo {volume} {13}},\ \bibinfo {pages}
  {3061--3085} (\bibinfo {year} {2001})}\BibitemShut {NoStop}%
\bibitem [{\citenamefont {Sun}\ \emph {et~al.}(2020)\citenamefont {Sun},
  \citenamefont {Ding}, \citenamefont {Huang}, \citenamefont {Luo},\ and\
  \citenamefont {Cheng}}]{Sun2020POF}%
  \BibitemOpen
  \bibfield  {author} {\bibinfo {author} {\bibfnamefont {P.~Y.}\ \bibnamefont
  {Sun}}, \bibinfo {author} {\bibfnamefont {J.~C.}\ \bibnamefont {Ding}},
  \bibinfo {author} {\bibfnamefont {S.~H.}\ \bibnamefont {Huang}}, \bibinfo
  {author} {\bibfnamefont {X.~S.}\ \bibnamefont {Luo}}, \ and\ \bibinfo
  {author} {\bibfnamefont {W.}~\bibnamefont {Cheng}},\ }\bibfield  {title}
  {\enquote {\bibinfo {title} {Microscopic {Richtmyer-Meshkov} instability
  under strong shock},}\ }\href {\doibase 10.1063/1.5143327} {\bibfield
  {journal} {\bibinfo  {journal} {Phys. Fluids}\ }\textbf {\bibinfo {volume}
  {32}},\ \bibinfo {pages} {024109} (\bibinfo {year} {2020})}\BibitemShut
  {NoStop}%
\bibitem [{\citenamefont {Ding}\ \emph {et~al.}(2021)\citenamefont {Ding},
  \citenamefont {Sun}, \citenamefont {Huang},\ and\ \citenamefont
  {Luo}}]{Ding2021Single}%
  \BibitemOpen
  \bibfield  {author} {\bibinfo {author} {\bibfnamefont {J.~C.}\ \bibnamefont
  {Ding}}, \bibinfo {author} {\bibfnamefont {P.~Y.}\ \bibnamefont {Sun}},
  \bibinfo {author} {\bibfnamefont {S.~H.}\ \bibnamefont {Huang}}, \ and\
  \bibinfo {author} {\bibfnamefont {X.~S.}\ \bibnamefont {Luo}},\ }\bibfield
  {title} {\enquote {\bibinfo {title} {Single-and dual-mode {Rayleigh-Taylor}
  instability at microscopic scale},}\ }\href {\doibase 10.1063/5.0042505}
  {\bibfield  {journal} {\bibinfo  {journal} {Phys. Fluids}\ }\textbf {\bibinfo
  {volume} {33}},\ \bibinfo {pages} {042102} (\bibinfo {year}
  {2021})}\BibitemShut {NoStop}%
\bibitem [{\citenamefont {Xie}\ \emph {et~al.}(2022)\citenamefont {Xie},
  \citenamefont {Shao}, \citenamefont {Liu},\ and\ \citenamefont
  {Chen}}]{Xie2022POF}%
  \BibitemOpen
  \bibfield  {author} {\bibinfo {author} {\bibfnamefont {Y.~F.}\ \bibnamefont
  {Xie}}, \bibinfo {author} {\bibfnamefont {J.~L.}\ \bibnamefont {Shao}},
  \bibinfo {author} {\bibfnamefont {R.}~\bibnamefont {Liu}}, \ and\ \bibinfo
  {author} {\bibfnamefont {P.~W.}\ \bibnamefont {Chen}},\ }\bibfield  {title}
  {\enquote {\bibinfo {title} {Chemical reaction of {Ni/Al} interface
  associated with perturbation growth under shock compression},}\ }\href
  {\doibase 10.1063/5.0089368} {\bibfield  {journal} {\bibinfo  {journal}
  {Phys. Fluids}\ }\textbf {\bibinfo {volume} {34}},\ \bibinfo {pages} {044111}
  (\bibinfo {year} {2022})}\BibitemShut {NoStop}%
\bibitem [{\citenamefont {Bird}(1994)}]{1994Molecular}%
  \BibitemOpen
  \bibfield  {author} {\bibinfo {author} {\bibfnamefont {G.~A.}\ \bibnamefont
  {Bird}},\ }\bibfield  {title} {\enquote {\bibinfo {title} {Molecular {Gas}
  {Dynamics} and {The} {Direct} {Simulation} of {Gas} {Flow}},}\ }\href
  {\doibase {}} {\bibfield  {journal} {\bibinfo  {journal} {Clarendon Press}\ }
  (\bibinfo {year} {1994}),\ {}}\BibitemShut {NoStop}%
\bibitem [{\citenamefont {Wagner}(1992)}]{Wagner1992JSP}%
  \BibitemOpen
  \bibfield  {author} {\bibinfo {author} {\bibfnamefont {W.}~\bibnamefont
  {Wagner}},\ }\bibfield  {title} {\enquote {\bibinfo {title} {A convergence
  proof for {Bird's} direct simulation {Monte} {Carlo} method for the
  {Boltzmann} equation},}\ }\href {\doibase 10.1007/BF01055714} {\bibfield
  {journal} {\bibinfo  {journal} {J. Stat. Phys.}\ }\textbf {\bibinfo {volume}
  {66}},\ \bibinfo {pages} {1011--1044} (\bibinfo {year} {1992})}\BibitemShut
  {NoStop}%
\bibitem [{\citenamefont {Xu}\ and\ \citenamefont
  {Prendergast}(1994)}]{Xu1994JCP}%
  \BibitemOpen
  \bibfield  {author} {\bibinfo {author} {\bibfnamefont {K.}~\bibnamefont
  {Xu}}\ and\ \bibinfo {author} {\bibfnamefont {K.~H.}\ \bibnamefont
  {Prendergast}},\ }\bibfield  {title} {\enquote {\bibinfo {title} {Numerical
  {Navier-Stokes} {Solutions} from {Gas} {Kinetic} {Theory}},}\ }\href
  {\doibase https://doi.org/10.1006/jcph.1994.1145} {\bibfield  {journal}
  {\bibinfo  {journal} {J. Comput. Phys.}\ }\textbf {\bibinfo {volume} {114}},\
  \bibinfo {pages} {9--17} (\bibinfo {year} {1994})}\BibitemShut {NoStop}%
\bibitem [{\citenamefont {Liu}\ and\ \citenamefont {Xu}(2020)}]{Liu2019ACTA}%
  \BibitemOpen
  \bibfield  {author} {\bibinfo {author} {\bibfnamefont {C.}~\bibnamefont
  {Liu}}\ and\ \bibinfo {author} {\bibfnamefont {K.}~\bibnamefont {Xu}},\
  }\bibfield  {title} {\enquote {\bibinfo {title} {Direct modeling methodology
  and its applications in multiscale transport process (in {Chinese})},}\
  }\href {\doibase 10.7638/kpdlxxb-2020.0018} {\bibfield  {journal} {\bibinfo
  {journal} {Acta Aerodyn. Sin.}\ }\textbf {\bibinfo {volume} {38}},\ \bibinfo
  {pages} {197--216} (\bibinfo {year} {2020})}\BibitemShut {NoStop}%
\bibitem [{\citenamefont {Xu}\ and\ \citenamefont {Huang}(2010)}]{Xu2010JCP}%
  \BibitemOpen
  \bibfield  {author} {\bibinfo {author} {\bibfnamefont {K.}~\bibnamefont
  {Xu}}\ and\ \bibinfo {author} {\bibfnamefont {J.~C.}\ \bibnamefont {Huang}},\
  }\bibfield  {title} {\enquote {\bibinfo {title} {A unified gas-kinetic scheme
  for continuum and rarefied flows},}\ }\href {\doibase
  https://doi.org/10.1016/j.jcp.2010.06.032} {\bibfield  {journal} {\bibinfo
  {journal} {J. Comput. Phys.}\ }\textbf {\bibinfo {volume} {229}},\ \bibinfo
  {pages} {7747--7764} (\bibinfo {year} {2010})}\BibitemShut {NoStop}%
\bibitem [{\citenamefont {Guo}, \citenamefont {Xu},\ and\ \citenamefont
  {Wang}(2013)}]{Guo2013PRE}%
  \BibitemOpen
  \bibfield  {author} {\bibinfo {author} {\bibfnamefont {Z.~L.}\ \bibnamefont
  {Guo}}, \bibinfo {author} {\bibfnamefont {K.}~\bibnamefont {Xu}}, \ and\
  \bibinfo {author} {\bibfnamefont {R.~J.}\ \bibnamefont {Wang}},\ }\bibfield
  {title} {\enquote {\bibinfo {title} {Discrete unified gas kinetic scheme for
  all {Knudsen} number flows: Low-speed isothermal case},}\ }\href {\doibase
  10.1103/PhysRevE.88.033305} {\bibfield  {journal} {\bibinfo  {journal} {Phys.
  Rev. E}\ }\textbf {\bibinfo {volume} {88}},\ \bibinfo {pages} {033305}
  (\bibinfo {year} {2013})}\BibitemShut {NoStop}%
\bibitem [{\citenamefont {Zhang}, \citenamefont {Qin},\ and\ \citenamefont
  {Emerson}(2005)}]{Zhang2005PRE}%
  \BibitemOpen
  \bibfield  {author} {\bibinfo {author} {\bibfnamefont {Y.~H.}\ \bibnamefont
  {Zhang}}, \bibinfo {author} {\bibfnamefont {R.~S.}\ \bibnamefont {Qin}}, \
  and\ \bibinfo {author} {\bibfnamefont {D.~R.}\ \bibnamefont {Emerson}},\
  }\bibfield  {title} {\enquote {\bibinfo {title} {Lattice {Boltzmann}
  simulation of rarefied gas flows in microchannels},}\ }\href {\doibase
  10.1103/PhysRevE.71.047702} {\bibfield  {journal} {\bibinfo  {journal} {Phys.
  Rev. E}\ }\textbf {\bibinfo {volume} {71}},\ \bibinfo {pages} {047702}
  (\bibinfo {year} {2005})}\BibitemShut {NoStop}%
\bibitem [{\citenamefont {Fei}\ \emph {et~al.}(2019)\citenamefont {Fei},
  \citenamefont {Du}, \citenamefont {Luo}, \citenamefont {Succi}, \citenamefont
  {Lauricella}, \citenamefont {Montessori},\ and\ \citenamefont
  {Wang}}]{Fei2019POF}%
  \BibitemOpen
  \bibfield  {author} {\bibinfo {author} {\bibfnamefont {L.~L.}\ \bibnamefont
  {Fei}}, \bibinfo {author} {\bibfnamefont {J.~Y.}\ \bibnamefont {Du}},
  \bibinfo {author} {\bibfnamefont {K.~H.}\ \bibnamefont {Luo}}, \bibinfo
  {author} {\bibfnamefont {S.}~\bibnamefont {Succi}}, \bibinfo {author}
  {\bibfnamefont {M.}~\bibnamefont {Lauricella}}, \bibinfo {author}
  {\bibfnamefont {A.}~\bibnamefont {Montessori}}, \ and\ \bibinfo {author}
  {\bibfnamefont {Q.}~\bibnamefont {Wang}},\ }\bibfield  {title} {\enquote
  {\bibinfo {title} {Modeling realistic multiphase flows using a non-orthogonal
  multiple-relaxation-time lattice {Boltzmann} method},}\ }\href {\doibase
  https://doi.org/10.1063/1.5087266} {\bibfield  {journal} {\bibinfo  {journal}
  {Phys. Fluids}\ }\textbf {\bibinfo {volume} {31}},\ \bibinfo {pages} {042105}
  (\bibinfo {year} {2019})}\BibitemShut {NoStop}%
\bibitem [{\citenamefont {Wang}, \citenamefont {Fei},\ and\ \citenamefont
  {Luo}(2021)}]{Wang2021POF}%
  \BibitemOpen
  \bibfield  {author} {\bibinfo {author} {\bibfnamefont {G.}~\bibnamefont
  {Wang}}, \bibinfo {author} {\bibfnamefont {L.~L.}\ \bibnamefont {Fei}}, \
  and\ \bibinfo {author} {\bibfnamefont {K.~H.}\ \bibnamefont {Luo}},\
  }\bibfield  {title} {\enquote {\bibinfo {title} {Lattice {Boltzmann}
  simulation of a water droplet penetrating a micropillar array in a
  microchannel},}\ }\href {\doibase https://doi.org/10.1063/5.0047163}
  {\bibfield  {journal} {\bibinfo  {journal} {Phys. Fluids}\ }\textbf {\bibinfo
  {volume} {33}},\ \bibinfo {pages} {043308} (\bibinfo {year}
  {2021})}\BibitemShut {NoStop}%
\bibitem [{\citenamefont {Huang}, \citenamefont {Liang},\ and\ \citenamefont
  {Xu}(2022)}]{Huang2022POF}%
  \BibitemOpen
  \bibfield  {author} {\bibinfo {author} {\bibfnamefont {B.~Q.}\ \bibnamefont
  {Huang}}, \bibinfo {author} {\bibfnamefont {H.}~\bibnamefont {Liang}}, \ and\
  \bibinfo {author} {\bibfnamefont {J.~R.}\ \bibnamefont {Xu}},\ }\bibfield
  {title} {\enquote {\bibinfo {title} {Lattice {Boltzmann} simulation of binary
  three-dimensional droplet coalescence in a confined shear flow},}\ }\href
  {\doibase 10.1063/5.0082263} {\bibfield  {journal} {\bibinfo  {journal}
  {Phys. Fluids}\ }\textbf {\bibinfo {volume} {34}},\ \bibinfo {pages} {032101}
  (\bibinfo {year} {2022})}\BibitemShut {NoStop}%
\bibitem [{\citenamefont {Wen}\ \emph {et~al.}(2020)\citenamefont {Wen},
  \citenamefont {Zhao}, \citenamefont {Qiu}, \citenamefont {Ye},\ and\
  \citenamefont {Shan}}]{Wen2020PRE}%
  \BibitemOpen
  \bibfield  {author} {\bibinfo {author} {\bibfnamefont {B.~H.}\ \bibnamefont
  {Wen}}, \bibinfo {author} {\bibfnamefont {L.}~\bibnamefont {Zhao}}, \bibinfo
  {author} {\bibfnamefont {W.}~\bibnamefont {Qiu}}, \bibinfo {author}
  {\bibfnamefont {Y.}~\bibnamefont {Ye}}, \ and\ \bibinfo {author}
  {\bibfnamefont {X.~W.}\ \bibnamefont {Shan}},\ }\bibfield  {title} {\enquote
  {\bibinfo {title} {Chemical-potential multiphase lattice {Boltzmann} method
  with superlarge density ratios},}\ }\href {\doibase
  10.1103/PhysRevE.102.013303} {\bibfield  {journal} {\bibinfo  {journal}
  {Phys. Rev. E}\ }\textbf {\bibinfo {volume} {102}},\ \bibinfo {pages}
  {013303} (\bibinfo {year} {2020})}\BibitemShut {NoStop}%
\bibitem [{\citenamefont {Gu}\ \emph {et~al.}(2022)\citenamefont {Gu},
  \citenamefont {Shang}, \citenamefont {Li}, \citenamefont {Ai}, \citenamefont
  {Zhou},\ and\ \citenamefont {Yuan}}]{Gu2022POF}%
  \BibitemOpen
  \bibfield  {author} {\bibinfo {author} {\bibfnamefont {Z.~K.}\ \bibnamefont
  {Gu}}, \bibinfo {author} {\bibfnamefont {Y.~H.}\ \bibnamefont {Shang}},
  \bibinfo {author} {\bibfnamefont {D.}~\bibnamefont {Li}}, \bibinfo {author}
  {\bibfnamefont {F.~B.}\ \bibnamefont {Ai}}, \bibinfo {author} {\bibfnamefont
  {H.}~\bibnamefont {Zhou}}, \ and\ \bibinfo {author} {\bibfnamefont
  {P.}~\bibnamefont {Yuan}},\ }\bibfield  {title} {\enquote {\bibinfo {title}
  {Lattice {Boltzmann} simulation of droplet impacting on the superhydrophobic
  surface with a suspended octagonal prism},}\ }\href {\doibase
  10.1063/5.0073258} {\bibfield  {journal} {\bibinfo  {journal} {Phys. Fluids}\
  }\textbf {\bibinfo {volume} {34}},\ \bibinfo {pages} {012015} (\bibinfo
  {year} {2022})}\BibitemShut {NoStop}%
\bibitem [{\citenamefont {Oran}, \citenamefont {Oh},\ and\ \citenamefont
  {Cybyk}(1998)}]{Oran1998DIRECT}%
  \BibitemOpen
  \bibfield  {author} {\bibinfo {author} {\bibfnamefont {E.~S.}\ \bibnamefont
  {Oran}}, \bibinfo {author} {\bibfnamefont {C.~K.}\ \bibnamefont {Oh}}, \ and\
  \bibinfo {author} {\bibfnamefont {C.~Y.}\ \bibnamefont {Cybyk}},\ }\bibfield
  {title} {\enquote {\bibinfo {title} {Direct {Simulation} {Monte} {Carlo}:
  {Recent} {Advances} and {Applications}},}\ }\href {\doibase
  https://doi.org/10.1146/annurev.fluid.30.1.403} {\bibfield  {journal}
  {\bibinfo  {journal} {Annu. Rev. Fluid Mech.}\ }\textbf {\bibinfo {volume}
  {30}},\ \bibinfo {pages} {403--441} (\bibinfo {year} {1998})}\BibitemShut
  {NoStop}%
\bibitem [{\citenamefont {Jing}\ and\ \citenamefont
  {Ching}(2001)}]{FAN2001JCP}%
  \BibitemOpen
  \bibfield  {author} {\bibinfo {author} {\bibfnamefont {F.}~\bibnamefont
  {Jing}}\ and\ \bibinfo {author} {\bibfnamefont {S.}~\bibnamefont {Ching}},\
  }\bibfield  {title} {\enquote {\bibinfo {title} {{Statistical} {Simulation}
  of {Low-Speed} {Rarefied} {Gas} {Flows}},}\ }\href {\doibase
  https://doi.org/10.1006/jcph.2000.6681} {\bibfield  {journal} {\bibinfo
  {journal} {J. Comput. Phys.}\ }\textbf {\bibinfo {volume} {167}},\ \bibinfo
  {pages} {393--412} (\bibinfo {year} {2001})}\BibitemShut {NoStop}%
\bibitem [{Note3()}]{Note3}%
  \BibitemOpen
  \bibinfo {note} {The DBM can also be interpreted as the Discrete Boltzmann
  Model or the Discrete Boltzmann Modeling method according to the
  context.}\BibitemShut {Stop}%
\bibitem [{\citenamefont {Xu}\ \emph {et~al.}(2012)\citenamefont {Xu},
  \citenamefont {Zhang}, \citenamefont {Gan}, \citenamefont {Chen},\ and\
  \citenamefont {Yu}}]{Xu2012-FoP-review}%
  \BibitemOpen
  \bibfield  {author} {\bibinfo {author} {\bibfnamefont {A.~G.}\ \bibnamefont
  {Xu}}, \bibinfo {author} {\bibfnamefont {G.~C.}\ \bibnamefont {Zhang}},
  \bibinfo {author} {\bibfnamefont {Y.~B.}\ \bibnamefont {Gan}}, \bibinfo
  {author} {\bibfnamefont {F.}~\bibnamefont {Chen}}, \ and\ \bibinfo {author}
  {\bibfnamefont {X.~J.}\ \bibnamefont {Yu}},\ }\bibfield  {title} {\enquote
  {\bibinfo {title} {Lattice {Boltzmann} modeling and simulation of
  compressible flows},}\ }\href {\doibase DOI 10.1007/s11467-012-0269-5}
  {\bibfield  {journal} {\bibinfo  {journal} {Front. Phys.}\ }\textbf {\bibinfo
  {volume} {7}},\ \bibinfo {pages} {582--600} (\bibinfo {year}
  {2012})}\BibitemShut {NoStop}%
\bibitem [{\citenamefont {Xu}\ \emph {et~al.}(2015)\citenamefont {Xu},
  \citenamefont {Lin}, \citenamefont {Zhang},\ and\ \citenamefont
  {Li}}]{2015Xu-PRE}%
  \BibitemOpen
  \bibfield  {author} {\bibinfo {author} {\bibfnamefont {A.~G.}\ \bibnamefont
  {Xu}}, \bibinfo {author} {\bibfnamefont {C.~D.}\ \bibnamefont {Lin}},
  \bibinfo {author} {\bibfnamefont {G.~C.}\ \bibnamefont {Zhang}}, \ and\
  \bibinfo {author} {\bibfnamefont {Y.~J.}\ \bibnamefont {Li}},\ }\bibfield
  {title} {\enquote {\bibinfo {title} {Multiple-relaxation-time lattice
  {Boltzmann} kinetic model for combustion},}\ }\href {\doibase
  10.1103/PhysRevE.91.043306} {\bibfield  {journal} {\bibinfo  {journal} {Phys.
  Rev. E}\ }\textbf {\bibinfo {volume} {91}},\ \bibinfo {pages} {043306}
  (\bibinfo {year} {2015})}\BibitemShut {NoStop}%
\bibitem [{\citenamefont {Xu}\ \emph {et~al.}()\citenamefont {Xu},
  \citenamefont {Zhang}, \citenamefont {Zhang},\ and\ \citenamefont
  {Gan}}]{Xu2018-RGD31}%
  \BibitemOpen
  \bibfield  {author} {\bibinfo {author} {\bibfnamefont {A.~G.}\ \bibnamefont
  {Xu}}, \bibinfo {author} {\bibfnamefont {G.~C.}\ \bibnamefont {Zhang}},
  \bibinfo {author} {\bibfnamefont {Y.~D.}\ \bibnamefont {Zhang}}, \ and\
  \bibinfo {author} {\bibfnamefont {Y.~B.}\ \bibnamefont {Gan}},\ }\href@noop
  {} {\enquote {\bibinfo {title} {Discrete {Boltzmann} {Modeling} of
  {Nonequilibrium} {Effects} in multiphase flow},}\ }\bibinfo {note}
  {\url{https://mp.weixin.qq.com/s/WwHnZNX42f7taw_zSxZO5g} Accessed July 8,
  2022}\BibitemShut {NoStop}%
\bibitem [{\citenamefont {Gan}\ \emph {et~al.}(2018)\citenamefont {Gan},
  \citenamefont {Xu}, \citenamefont {Zhang}, \citenamefont {Zhang},\ and\
  \citenamefont {Succi}}]{2018Gan-pre}%
  \BibitemOpen
  \bibfield  {author} {\bibinfo {author} {\bibfnamefont {Y.~B.}\ \bibnamefont
  {Gan}}, \bibinfo {author} {\bibfnamefont {A.~G.}\ \bibnamefont {Xu}},
  \bibinfo {author} {\bibfnamefont {G.~C.}\ \bibnamefont {Zhang}}, \bibinfo
  {author} {\bibfnamefont {Y.~D.}\ \bibnamefont {Zhang}}, \ and\ \bibinfo
  {author} {\bibfnamefont {S.}~\bibnamefont {Succi}},\ }\bibfield  {title}
  {\enquote {\bibinfo {title} {Discrete {Boltzmann} trans-scale modeling of
  high-speed compressible flows},}\ }\href {\doibase
  10.1103/PhysRevE.97.053312} {\bibfield  {journal} {\bibinfo  {journal} {Phys.
  Rev. E}\ }\textbf {\bibinfo {volume} {97}},\ \bibinfo {pages} {053312}
  (\bibinfo {year} {2018})}\BibitemShut {NoStop}%
\bibitem [{\citenamefont {Zhang}\ \emph {et~al.}(2017)\citenamefont {Zhang},
  \citenamefont {Xu}, \citenamefont {Zhang}, \citenamefont {Chen},\ and\
  \citenamefont {Wang}}]{Zhang2017Discrete}%
  \BibitemOpen
  \bibfield  {author} {\bibinfo {author} {\bibfnamefont {Y.~D.}\ \bibnamefont
  {Zhang}}, \bibinfo {author} {\bibfnamefont {A.~G.}\ \bibnamefont {Xu}},
  \bibinfo {author} {\bibfnamefont {G.~C.}\ \bibnamefont {Zhang}}, \bibinfo
  {author} {\bibfnamefont {Z.~H.}\ \bibnamefont {Chen}}, \ and\ \bibinfo
  {author} {\bibfnamefont {P.}~\bibnamefont {Wang}},\ }\bibfield  {title}
  {\enquote {\bibinfo {title} {Discrete ellipsoidal statistical {BGK} model and
  {Burnett} equations},}\ }\href {\doibase 110.1007/s11467-018-0749-3}
  {\bibfield  {journal} {\bibinfo  {journal} {Front. Phys.}\ }\textbf {\bibinfo
  {volume} {13}},\ \bibinfo {pages} {135101} (\bibinfo {year}
  {2017})}\BibitemShut {NoStop}%
\bibitem [{\citenamefont {Guo}\ and\ \citenamefont
  {Shu}(2013)}]{Guo2013lattice}%
  \BibitemOpen
  \bibfield  {author} {\bibinfo {author} {\bibfnamefont {Z.~L.}\ \bibnamefont
  {Guo}}\ and\ \bibinfo {author} {\bibfnamefont {C.}~\bibnamefont {Shu}},\
  }\href@noop {} {\emph {\bibinfo {title} {Lattice {Boltzmann} method and its
  application in engineering}}},\ Vol.~\bibinfo {volume} {3}\ (\bibinfo
  {publisher} {World Scientific},\ \bibinfo {year} {2013})\BibitemShut
  {NoStop}%
\bibitem [{\citenamefont {Huang}, \citenamefont {Sukop},\ and\ \citenamefont
  {Lu}(2015)}]{Huang2015multiphase}%
  \BibitemOpen
  \bibfield  {author} {\bibinfo {author} {\bibfnamefont {H.~B.}\ \bibnamefont
  {Huang}}, \bibinfo {author} {\bibfnamefont {M.}~\bibnamefont {Sukop}}, \ and\
  \bibinfo {author} {\bibfnamefont {X.~Y.}\ \bibnamefont {Lu}},\ }\href@noop {}
  {\emph {\bibinfo {title} {Multiphase lattice {Boltzmann} methods: {Theory}
  and application}}}\ (\bibinfo  {publisher} {John Wiley \& Sons},\ \bibinfo
  {year} {2015})\BibitemShut {NoStop}%
\bibitem [{\citenamefont {Shi}, \citenamefont {Wu},\ and\ \citenamefont
  {Shan}(2021)}]{Shi2021JFM}%
  \BibitemOpen
  \bibfield  {author} {\bibinfo {author} {\bibfnamefont {Y.~Y.}\ \bibnamefont
  {Shi}}, \bibinfo {author} {\bibfnamefont {L.}~\bibnamefont {Wu}}, \ and\
  \bibinfo {author} {\bibfnamefont {X.~W.}\ \bibnamefont {Shan}},\ }\bibfield
  {title} {\enquote {\bibinfo {title} {Accuracy of high-order lattice
  {Boltzmann} method for non-equilibrium gas flow},}\ }\href {\doibase
  10.1017/jfm.2020.813} {\bibfield  {journal} {\bibinfo  {journal} {J. Fluid
  Mech.}\ }\textbf {\bibinfo {volume} {907}},\ \bibinfo {pages} {A25} (\bibinfo
  {year} {2021})}\BibitemShut {NoStop}%
\bibitem [{\citenamefont {Qian}, \citenamefont {D{'}Humi\`{e}res},\ and\
  \citenamefont {Lallemand}(1992)}]{Qian1992EPL}%
  \BibitemOpen
  \bibfield  {author} {\bibinfo {author} {\bibfnamefont {Y.~H.}\ \bibnamefont
  {Qian}}, \bibinfo {author} {\bibfnamefont {D.}~\bibnamefont
  {D{'}Humi\`{e}res}}, \ and\ \bibinfo {author} {\bibfnamefont
  {P.}~\bibnamefont {Lallemand}},\ }\bibfield  {title} {\enquote {\bibinfo
  {title} {Lattice {BGK} models for {Navier-Stokes} equations},}\ }\href@noop
  {} {\bibfield  {journal} {\bibinfo  {journal} {Europhysics Letters}\ }\textbf
  {\bibinfo {volume} {17}},\ \bibinfo {pages} {479--484} (\bibinfo {year}
  {1992})}\BibitemShut {NoStop}%
\bibitem [{\citenamefont {Bhadauria}, \citenamefont {Dorschner},\ and\
  \citenamefont {Karlin}(2021)}]{Bhadauria2021POF}%
  \BibitemOpen
  \bibfield  {author} {\bibinfo {author} {\bibfnamefont {A.}~\bibnamefont
  {Bhadauria}}, \bibinfo {author} {\bibfnamefont {B.}~\bibnamefont
  {Dorschner}}, \ and\ \bibinfo {author} {\bibfnamefont {I.}~\bibnamefont
  {Karlin}},\ }\bibfield  {title} {\enquote {\bibinfo {title} {Lattice
  {Boltzmann} method for fluid-structure interaction in compressible flow},}\
  }\href {\doibase 10.1063/5.0062117} {\bibfield  {journal} {\bibinfo
  {journal} {Phys. Fluids}\ }\textbf {\bibinfo {volume} {33}},\ \bibinfo
  {pages} {106111} (\bibinfo {year} {2021})}\BibitemShut {NoStop}%
\bibitem [{\citenamefont {Sofonea}\ \emph {et~al.}(2018)\citenamefont
  {Sofonea}, \citenamefont {Biciu{\c{s}}c{\u{a}}}, \citenamefont {Busuioc},
  \citenamefont {Ambru{\c{s}}}, \citenamefont {Gonnella},\ and\ \citenamefont
  {Lamura}}]{Sofonea2018PRE}%
  \BibitemOpen
  \bibfield  {author} {\bibinfo {author} {\bibfnamefont {V.}~\bibnamefont
  {Sofonea}}, \bibinfo {author} {\bibfnamefont {T.}~\bibnamefont
  {Biciu{\c{s}}c{\u{a}}}}, \bibinfo {author} {\bibfnamefont {S.}~\bibnamefont
  {Busuioc}}, \bibinfo {author} {\bibfnamefont {V.~E.}\ \bibnamefont
  {Ambru{\c{s}}}}, \bibinfo {author} {\bibfnamefont {G.}~\bibnamefont
  {Gonnella}}, \ and\ \bibinfo {author} {\bibfnamefont {A.}~\bibnamefont
  {Lamura}},\ }\bibfield  {title} {\enquote {\bibinfo {title}
  {Corner-transport-upwind lattice {Boltzmann} model for bubble cavitation},}\
  }\href {\doibase 10.1103/PhysRevE.97.023309} {\bibfield  {journal} {\bibinfo
  {journal} {Phys. Rev. E}\ }\textbf {\bibinfo {volume} {97}},\ \bibinfo
  {pages} {023309} (\bibinfo {year} {2018})}\BibitemShut {NoStop}%
\bibitem [{\citenamefont {Tian}\ \emph {et~al.}(2011)\citenamefont {Tian},
  \citenamefont {Luo}, \citenamefont {Zhu}, \citenamefont {Liao},\ and\
  \citenamefont {Lu}}]{Tian2011JCP}%
  \BibitemOpen
  \bibfield  {author} {\bibinfo {author} {\bibfnamefont {F.~B.}\ \bibnamefont
  {Tian}}, \bibinfo {author} {\bibfnamefont {H.~X.}\ \bibnamefont {Luo}},
  \bibinfo {author} {\bibfnamefont {L.~D.}\ \bibnamefont {Zhu}}, \bibinfo
  {author} {\bibfnamefont {J.~C.}\ \bibnamefont {Liao}}, \ and\ \bibinfo
  {author} {\bibfnamefont {X.~Y.}\ \bibnamefont {Lu}},\ }\bibfield  {title}
  {\enquote {\bibinfo {title} {An efficient immersed boundary-lattice
  {Boltzmann} method for the hydrodynamic interaction of elastic filaments},}\
  }\href {\doibase 10.1016/j.jcp.2011.05.028} {\bibfield  {journal} {\bibinfo
  {journal} {J. Comput. Phys.}\ }\textbf {\bibinfo {volume} {230}},\ \bibinfo
  {pages} {7266--7283} (\bibinfo {year} {2011})}\BibitemShut {NoStop}%
\bibitem [{\citenamefont {Sun}\ \emph {et~al.}(2011)\citenamefont {Sun},
  \citenamefont {Zhu}, \citenamefont {Pan}, \citenamefont {Yang},\ and\
  \citenamefont {Raabe}}]{Sun2011CMA}%
  \BibitemOpen
  \bibfield  {author} {\bibinfo {author} {\bibfnamefont {D.~K.}\ \bibnamefont
  {Sun}}, \bibinfo {author} {\bibfnamefont {M.~F.}\ \bibnamefont {Zhu}},
  \bibinfo {author} {\bibfnamefont {S.~Y.}\ \bibnamefont {Pan}}, \bibinfo
  {author} {\bibfnamefont {C.~R.}\ \bibnamefont {Yang}}, \ and\ \bibinfo
  {author} {\bibfnamefont {D.}~\bibnamefont {Raabe}},\ }\bibfield  {title}
  {\enquote {\bibinfo {title} {Lattice {Boltzmann} modeling of dendritic growth
  in forced and natural convection},}\ }\href {\doibase
  10.1016/j.camwa.2010.11.001} {\bibfield  {journal} {\bibinfo  {journal}
  {Comput. Math. Appl.}\ }\textbf {\bibinfo {volume} {61}},\ \bibinfo {pages}
  {3585--3592} (\bibinfo {year} {2011})}\BibitemShut {NoStop}%
\bibitem [{\citenamefont {Chai}\ and\ \citenamefont
  {Zhao}(2013)}]{Chai2013PRE}%
  \BibitemOpen
  \bibfield  {author} {\bibinfo {author} {\bibfnamefont {Z.~H.}\ \bibnamefont
  {Chai}}\ and\ \bibinfo {author} {\bibfnamefont {T.~S.}\ \bibnamefont
  {Zhao}},\ }\bibfield  {title} {\enquote {\bibinfo {title} {Lattice
  {Boltzmann} model for the convection-diffusion equation},}\ }\href {\doibase
  10.1103/PhysRevE.87.063309} {\bibfield  {journal} {\bibinfo  {journal} {Phys.
  Rev. E}\ }\textbf {\bibinfo {volume} {87}},\ \bibinfo {pages} {063309}
  (\bibinfo {year} {2013})}\BibitemShut {NoStop}%
\bibitem [{\citenamefont {Liang}, \citenamefont {Xia},\ and\ \citenamefont
  {Huang}(2021)}]{Liang2021POF}%
  \BibitemOpen
  \bibfield  {author} {\bibinfo {author} {\bibfnamefont {H.}~\bibnamefont
  {Liang}}, \bibinfo {author} {\bibfnamefont {Z.~H.}\ \bibnamefont {Xia}}, \
  and\ \bibinfo {author} {\bibfnamefont {H.~W.}\ \bibnamefont {Huang}},\
  }\bibfield  {title} {\enquote {\bibinfo {title} {Late-time description of
  immiscible {Rayleigh-Taylor} instability: {A} lattice {Boltzmann} study},}\
  }\href {\doibase https://doi.org/10.1063/5.0057269} {\bibfield  {journal}
  {\bibinfo  {journal} {Phys. Fluids}\ }\textbf {\bibinfo {volume} {33}},\
  \bibinfo {pages} {082103} (\bibinfo {year} {2021})}\BibitemShut {NoStop}%
\bibitem [{\citenamefont {Chen}\ \emph
  {et~al.}(2021{\natexlab{a}})\citenamefont {Chen}, \citenamefont {Zhou},
  \citenamefont {Zhu}, \citenamefont {Zhu}, \citenamefont {Yan},\ and\
  \citenamefont {Yu}}]{Yu2021POF}%
  \BibitemOpen
  \bibfield  {author} {\bibinfo {author} {\bibfnamefont {R.}~\bibnamefont
  {Chen}}, \bibinfo {author} {\bibfnamefont {S.~Y.}\ \bibnamefont {Zhou}},
  \bibinfo {author} {\bibfnamefont {L.~K.}\ \bibnamefont {Zhu}}, \bibinfo
  {author} {\bibfnamefont {L.~D.}\ \bibnamefont {Zhu}}, \bibinfo {author}
  {\bibfnamefont {W.~W.}\ \bibnamefont {Yan}}, \ and\ \bibinfo {author}
  {\bibfnamefont {H.~D.}\ \bibnamefont {Yu}},\ }\bibfield  {title} {\enquote
  {\bibinfo {title} {A new criterion of coalescence-induced microbubble
  detachment in three-dimensional microfluidic channel},}\ }\href {\doibase
  10.1063/5.0043155} {\bibfield  {journal} {\bibinfo  {journal} {Phys. Fluids}\
  }\textbf {\bibinfo {volume} {33}},\ \bibinfo {pages} {043320} (\bibinfo
  {year} {2021}{\natexlab{a}})}\BibitemShut {NoStop}%
\bibitem [{\citenamefont {Swift}, \citenamefont {Osborn},\ and\ \citenamefont
  {Yeomans}(1995)}]{Swift1995PRL}%
  \BibitemOpen
  \bibfield  {author} {\bibinfo {author} {\bibfnamefont {M.~R.}\ \bibnamefont
  {Swift}}, \bibinfo {author} {\bibfnamefont {W.~R.}\ \bibnamefont {Osborn}}, \
  and\ \bibinfo {author} {\bibfnamefont {J.~M.}\ \bibnamefont {Yeomans}},\
  }\bibfield  {title} {\enquote {\bibinfo {title} {Lattice {Boltzmann}
  {Simulation} of {Nonideal} {Fluids}},}\ }\href {\doibase
  10.1103/PhysRevLett.75.830} {\bibfield  {journal} {\bibinfo  {journal} {Phys.
  Rev. Lett.}\ }\textbf {\bibinfo {volume} {75}},\ \bibinfo {pages} {830--833}
  (\bibinfo {year} {1995})}\BibitemShut {NoStop}%
\bibitem [{\citenamefont {Osborn}\ \emph {et~al.}(1995)\citenamefont {Osborn},
  \citenamefont {Orlandini}, \citenamefont {Swift}, \citenamefont {Yeomans},\
  and\ \citenamefont {Banavar}}]{Osborn1995PRL}%
  \BibitemOpen
  \bibfield  {author} {\bibinfo {author} {\bibfnamefont {W.~R.}\ \bibnamefont
  {Osborn}}, \bibinfo {author} {\bibfnamefont {E.}~\bibnamefont {Orlandini}},
  \bibinfo {author} {\bibfnamefont {M.~R.}\ \bibnamefont {Swift}}, \bibinfo
  {author} {\bibfnamefont {J.~M.}\ \bibnamefont {Yeomans}}, \ and\ \bibinfo
  {author} {\bibfnamefont {J.~R.}\ \bibnamefont {Banavar}},\ }\bibfield
  {title} {\enquote {\bibinfo {title} {Lattice {Boltzmann} {Study} of
  {Hydrodynamic} {Spinodal} {Decomposition}},}\ }\href {\doibase
  10.1103/PhysRevLett.75.4031} {\bibfield  {journal} {\bibinfo  {journal}
  {Phys. Rev. Lett.}\ }\textbf {\bibinfo {volume} {75}},\ \bibinfo {pages}
  {4031--4034} (\bibinfo {year} {1995})}\BibitemShut {NoStop}%
\bibitem [{\citenamefont {Wagner}\ and\ \citenamefont
  {Yeomans}(1998)}]{Wagner1998PRL}%
  \BibitemOpen
  \bibfield  {author} {\bibinfo {author} {\bibfnamefont {A.~J.}\ \bibnamefont
  {Wagner}}\ and\ \bibinfo {author} {\bibfnamefont {J.~M.}\ \bibnamefont
  {Yeomans}},\ }\bibfield  {title} {\enquote {\bibinfo {title} {Breakdown of
  {Scale} {Invariance} in the {Coarsening} of {Phase-Separating} {Binary}
  {Fluids}},}\ }\href {\doibase 10.1103/PhysRevLett.80.1429} {\bibfield
  {journal} {\bibinfo  {journal} {Phys. Rev. Lett.}\ }\textbf {\bibinfo
  {volume} {80}},\ \bibinfo {pages} {1429--1432} (\bibinfo {year}
  {1998})}\BibitemShut {NoStop}%
\bibitem [{\citenamefont {Lin}\ \emph {et~al.}(2016)\citenamefont {Lin},
  \citenamefont {Xu}, \citenamefont {Zhang},\ and\ \citenamefont
  {Li}}]{2016Lin-CNF}%
  \BibitemOpen
  \bibfield  {author} {\bibinfo {author} {\bibfnamefont {C.~D.}\ \bibnamefont
  {Lin}}, \bibinfo {author} {\bibfnamefont {A.~G.}\ \bibnamefont {Xu}},
  \bibinfo {author} {\bibfnamefont {G.~C.}\ \bibnamefont {Zhang}}, \ and\
  \bibinfo {author} {\bibfnamefont {Y.~J.}\ \bibnamefont {Li}},\ }\bibfield
  {title} {\enquote {\bibinfo {title} {Double-distribution-function discrete
  {Boltzmann} model for combustion},}\ }\href {\doibase
  https://doi.org/10.1016/j.combustflame.2015.11.010} {\bibfield  {journal}
  {\bibinfo  {journal} {Combust. Flame}\ }\textbf {\bibinfo {volume} {164}},\
  \bibinfo {pages} {137--151} (\bibinfo {year} {2016})}\BibitemShut {NoStop}%
\bibitem [{\citenamefont {Lin}\ and\ \citenamefont
  {Luo}(2018)}]{2018Mesoscopic}%
  \BibitemOpen
  \bibfield  {author} {\bibinfo {author} {\bibfnamefont {C.~D.}\ \bibnamefont
  {Lin}}\ and\ \bibinfo {author} {\bibfnamefont {K.~H.}\ \bibnamefont {Luo}},\
  }\bibfield  {title} {\enquote {\bibinfo {title} {Mesoscopic simulation of
  nonequilibrium detonation with discrete {Boltzmann} method},}\ }\href
  {\doibase https://doi.org/10.1016/j.combustflame.2018.09.027} {\bibfield
  {journal} {\bibinfo  {journal} {Combust. Flame}\ }\textbf {\bibinfo {volume}
  {198}},\ \bibinfo {pages} {356--362} (\bibinfo {year} {2018})}\BibitemShut
  {NoStop}%
\bibitem [{\citenamefont {Ji}, \citenamefont {Lin},\ and\ \citenamefont
  {Luo}(2022)}]{Ji2022JCP}%
  \BibitemOpen
  \bibfield  {author} {\bibinfo {author} {\bibfnamefont {Y.}~\bibnamefont
  {Ji}}, \bibinfo {author} {\bibfnamefont {C.~D.}\ \bibnamefont {Lin}}, \ and\
  \bibinfo {author} {\bibfnamefont {K.~H.}\ \bibnamefont {Luo}},\ }\bibfield
  {title} {\enquote {\bibinfo {title} {A three-dimensional discrete {Boltzmann}
  model for steady and unsteady detonation},}\ }\href {\doibase
  10.1016/j.jcp.2022.111002} {\bibfield  {journal} {\bibinfo  {journal} {J.
  Comput. Phys.}\ }\textbf {\bibinfo {volume} {455}},\ \bibinfo {pages}
  {111002} (\bibinfo {year} {2022})}\BibitemShut {NoStop}%
\bibitem [{\citenamefont {Shan}\ \emph {et~al.}(2022)\citenamefont {Shan},
  \citenamefont {Xu}, \citenamefont {Zhang}, \citenamefont {Wang},\ and\
  \citenamefont {Chen}}]{Shan2022JMES}%
  \BibitemOpen
  \bibfield  {author} {\bibinfo {author} {\bibfnamefont {Y.~M.}\ \bibnamefont
  {Shan}}, \bibinfo {author} {\bibfnamefont {A.~G.}\ \bibnamefont {Xu}},
  \bibinfo {author} {\bibfnamefont {Y.~D.}\ \bibnamefont {Zhang}}, \bibinfo
  {author} {\bibfnamefont {L.~F.}\ \bibnamefont {Wang}}, \ and\ \bibinfo
  {author} {\bibfnamefont {F.}~\bibnamefont {Chen}},\ }\bibfield  {title}
  {\enquote {\bibinfo {title} {Discrete {Boltzmann} modeling of detonation:
  {Based} on the {Shakhov} model},}\ }\href {\doibase
  10.1177/09544062221096254} {\bibfield  {journal} {\bibinfo  {journal} {J.
  Mech. Eng. Sci.}\ ,\ \bibinfo {pages} {1--15}} (\bibinfo {year}
  {2022})}\BibitemShut {NoStop}%
\bibitem [{\citenamefont {Su}\ and\ \citenamefont {Lin}(2022)}]{Su2022CTP}%
  \BibitemOpen
  \bibfield  {author} {\bibinfo {author} {\bibfnamefont {X.~L.}\ \bibnamefont
  {Su}}\ and\ \bibinfo {author} {\bibfnamefont {C.~D.}\ \bibnamefont {Lin}},\
  }\bibfield  {title} {\enquote {\bibinfo {title} {Nonequilibrium effects of
  reactive flow based on gas kinetic theory},}\ }\href {\doibase
  https://doi.org/10.1088/1572-9494/ac53a0} {\bibfield  {journal} {\bibinfo
  {journal} {Commun. Theor. Phys.}\ }\textbf {\bibinfo {volume} {74}},\
  \bibinfo {pages} {035604} (\bibinfo {year} {2022})}\BibitemShut {NoStop}%
\bibitem [{\citenamefont {Lai}\ \emph {et~al.}(2016)\citenamefont {Lai},
  \citenamefont {Xu}, \citenamefont {Zhang}, \citenamefont {Gan}, \citenamefont
  {Ying},\ and\ \citenamefont {Succi}}]{Lai2016PRE}%
  \BibitemOpen
  \bibfield  {author} {\bibinfo {author} {\bibfnamefont {H.~L.}\ \bibnamefont
  {Lai}}, \bibinfo {author} {\bibfnamefont {A.~G.}\ \bibnamefont {Xu}},
  \bibinfo {author} {\bibfnamefont {G.~C.}\ \bibnamefont {Zhang}}, \bibinfo
  {author} {\bibfnamefont {Y.~B.}\ \bibnamefont {Gan}}, \bibinfo {author}
  {\bibfnamefont {Y.~J.}\ \bibnamefont {Ying}}, \ and\ \bibinfo {author}
  {\bibfnamefont {S.}~\bibnamefont {Succi}},\ }\bibfield  {title} {\enquote
  {\bibinfo {title} {Nonequilibrium thermohydrodynamic effects on the
  {Rayleigh-Taylor} instability in compressible flows},}\ }\href {\doibase
  10.1103/PhysRevE.94.023106} {\bibfield  {journal} {\bibinfo  {journal} {Phys.
  Rev. E}\ }\textbf {\bibinfo {volume} {94}},\ \bibinfo {pages} {023106}
  (\bibinfo {year} {2016})}\BibitemShut {NoStop}%
\bibitem [{\citenamefont {Lin}\ \emph {et~al.}(2017{\natexlab{a}})\citenamefont
  {Lin}, \citenamefont {Xu}, \citenamefont {Zhang}, \citenamefont {Luo},\ and\
  \citenamefont {Li}}]{2017Lin-DDBM-RT}%
  \BibitemOpen
  \bibfield  {author} {\bibinfo {author} {\bibfnamefont {C.~D.}\ \bibnamefont
  {Lin}}, \bibinfo {author} {\bibfnamefont {A.~G.}\ \bibnamefont {Xu}},
  \bibinfo {author} {\bibfnamefont {G.~C.}\ \bibnamefont {Zhang}}, \bibinfo
  {author} {\bibfnamefont {K.~H.}\ \bibnamefont {Luo}}, \ and\ \bibinfo
  {author} {\bibfnamefont {Y.~J.}\ \bibnamefont {Li}},\ }\bibfield  {title}
  {\enquote {\bibinfo {title} {{Discrete Boltzmann} modeling of
  {Rayleigh-Taylor} instability in two-component compressible flows},}\ }\href
  {\doibase 10.1103/PhysRevE.96.053305} {\bibfield  {journal} {\bibinfo
  {journal} {Phys. Rev. E}\ }\textbf {\bibinfo {volume} {96}},\ \bibinfo
  {pages} {053305} (\bibinfo {year} {2017}{\natexlab{a}})}\BibitemShut
  {NoStop}%
\bibitem [{\citenamefont {Chen}, \citenamefont {Xu},\ and\ \citenamefont
  {Zhang}(2018)}]{Chen2018POF}%
  \BibitemOpen
  \bibfield  {author} {\bibinfo {author} {\bibfnamefont {F.}~\bibnamefont
  {Chen}}, \bibinfo {author} {\bibfnamefont {A.~G.}\ \bibnamefont {Xu}}, \ and\
  \bibinfo {author} {\bibfnamefont {G.~C.}\ \bibnamefont {Zhang}},\ }\bibfield
  {title} {\enquote {\bibinfo {title} {Collaboration and {Competition}
  {Between} {Richtmyer-Meshkov} instability and {Rayleigh-Taylor}
  instability},}\ }\href {\doibase 10.1063/1.5049869} {\bibfield  {journal}
  {\bibinfo  {journal} {Phys. Fluids}\ }\textbf {\bibinfo {volume} {30}},\
  \bibinfo {pages} {102105} (\bibinfo {year} {2018})}\BibitemShut {NoStop}%
\bibitem [{\citenamefont {Lin}\ \emph {et~al.}(2019)\citenamefont {Lin},
  \citenamefont {Luo}, \citenamefont {Gan},\ and\ \citenamefont
  {Liu}}]{2019Kinetic}%
  \BibitemOpen
  \bibfield  {author} {\bibinfo {author} {\bibfnamefont {C.~D.}\ \bibnamefont
  {Lin}}, \bibinfo {author} {\bibfnamefont {K.~H.}\ \bibnamefont {Luo}},
  \bibinfo {author} {\bibfnamefont {Y.~B.}\ \bibnamefont {Gan}}, \ and\
  \bibinfo {author} {\bibfnamefont {Z.~P.}\ \bibnamefont {Liu}},\ }\bibfield
  {title} {\enquote {\bibinfo {title} {Kinetic {Simulation} of {Nonequilibrium}
  {Kelvin-Helmholtz} {Instability}},}\ }\href {\doibase {}} {\bibfield
  {journal} {\bibinfo  {journal} {Commun. Theor. Phys}\ }\textbf {\bibinfo
  {volume} {71}},\ \bibinfo {pages} {132--142} (\bibinfo {year}
  {2019})}\BibitemShut {NoStop}%
\bibitem [{\citenamefont {Chen}\ \emph {et~al.}(2020)\citenamefont {Chen},
  \citenamefont {Xu}, \citenamefont {Zhang},\ and\ \citenamefont
  {Zeng}}]{Chen2020POF}%
  \BibitemOpen
  \bibfield  {author} {\bibinfo {author} {\bibfnamefont {F.}~\bibnamefont
  {Chen}}, \bibinfo {author} {\bibfnamefont {A.~G.}\ \bibnamefont {Xu}},
  \bibinfo {author} {\bibfnamefont {Y.~D.}\ \bibnamefont {Zhang}}, \ and\
  \bibinfo {author} {\bibfnamefont {Q.~K.}\ \bibnamefont {Zeng}},\ }\bibfield
  {title} {\enquote {\bibinfo {title} {Morphological and non-equilibrium
  analysis of coupled {Rayleigh-Taylor-Kelvin-Helmholtz} instability},}\ }\href
  {\doibase 10.1063/5.0023364} {\bibfield  {journal} {\bibinfo  {journal}
  {Phys. Fluids}\ }\textbf {\bibinfo {volume} {32}},\ \bibinfo {pages} {104111}
  (\bibinfo {year} {2020})}\BibitemShut {NoStop}%
\bibitem [{\citenamefont {Ye}\ \emph {et~al.}(2020)\citenamefont {Ye},
  \citenamefont {Lai}, \citenamefont {Li}, \citenamefont {Gan}, \citenamefont
  {Lin}, \citenamefont {Chen},\ and\ \citenamefont {Xu}}]{Ye2020Entropy}%
  \BibitemOpen
  \bibfield  {author} {\bibinfo {author} {\bibfnamefont {H.~Y.}\ \bibnamefont
  {Ye}}, \bibinfo {author} {\bibfnamefont {H.~L.}\ \bibnamefont {Lai}},
  \bibinfo {author} {\bibfnamefont {D.~M.}\ \bibnamefont {Li}}, \bibinfo
  {author} {\bibfnamefont {Y.~B.}\ \bibnamefont {Gan}}, \bibinfo {author}
  {\bibfnamefont {C.~D.}\ \bibnamefont {Lin}}, \bibinfo {author} {\bibfnamefont
  {L.}~\bibnamefont {Chen}}, \ and\ \bibinfo {author} {\bibfnamefont {A.~G.}\
  \bibnamefont {Xu}},\ }\bibfield  {title} {\enquote {\bibinfo {title} {Knudsen
  {Number} {Effects} on {Two-Dimensional} {Rayleigh-Taylor} {Instability} in
  {Compressible} {Fluid}: {Based} on a {Discrete} {Boltzmann} {Method}},}\
  }\href {\doibase 10.3390/e22050500} {\bibfield  {journal} {\bibinfo
  {journal} {Entropy}\ }\textbf {\bibinfo {volume} {22}} (\bibinfo {year}
  {2020}),\ 10.3390/e22050500}\BibitemShut {NoStop}%
\bibitem [{\citenamefont {Lin}\ \emph {et~al.}(2021)\citenamefont {Lin},
  \citenamefont {Luo}, \citenamefont {Xu}, \citenamefont {Gan},\ and\
  \citenamefont {Lai}}]{Lin2021PRE}%
  \BibitemOpen
  \bibfield  {author} {\bibinfo {author} {\bibfnamefont {C.~D.}\ \bibnamefont
  {Lin}}, \bibinfo {author} {\bibfnamefont {K.~H.}\ \bibnamefont {Luo}},
  \bibinfo {author} {\bibfnamefont {A.~G.}\ \bibnamefont {Xu}}, \bibinfo
  {author} {\bibfnamefont {Y.~B.}\ \bibnamefont {Gan}}, \ and\ \bibinfo
  {author} {\bibfnamefont {H.~L.}\ \bibnamefont {Lai}},\ }\bibfield  {title}
  {\enquote {\bibinfo {title} {Multiple-relaxation-time discrete {Boltzmann}
  modeling of multicomponent mixture with nonequilibrium effects},}\ }\href
  {\doibase 10.1103/PhysRevE.103.013305} {\bibfield  {journal} {\bibinfo
  {journal} {Phys. Rev. E}\ }\textbf {\bibinfo {volume} {103}},\ \bibinfo
  {pages} {013305} (\bibinfo {year} {2021})}\BibitemShut {NoStop}%
\bibitem [{\citenamefont {Zhang}\ \emph {et~al.}(2021)\citenamefont {Zhang},
  \citenamefont {Xu}, \citenamefont {Zhang}, \citenamefont {Li}, \citenamefont
  {Lai},\ and\ \citenamefont {Hu}}]{Zhang2021POF}%
  \BibitemOpen
  \bibfield  {author} {\bibinfo {author} {\bibfnamefont {G.}~\bibnamefont
  {Zhang}}, \bibinfo {author} {\bibfnamefont {A.~G.}\ \bibnamefont {Xu}},
  \bibinfo {author} {\bibfnamefont {D.~J.}\ \bibnamefont {Zhang}}, \bibinfo
  {author} {\bibfnamefont {Y.~J.}\ \bibnamefont {Li}}, \bibinfo {author}
  {\bibfnamefont {H.~L.}\ \bibnamefont {Lai}}, \ and\ \bibinfo {author}
  {\bibfnamefont {X.~M.}\ \bibnamefont {Hu}},\ }\bibfield  {title} {\enquote
  {\bibinfo {title} {Delineation of the flow and mixing induced by
  {Rayleigh-Taylor} instability through tracers},}\ }\href {\doibase
  10.1063/5.0051154} {\bibfield  {journal} {\bibinfo  {journal} {Phys. Fluids}\
  }\textbf {\bibinfo {volume} {33}},\ \bibinfo {pages} {076105} (\bibinfo
  {year} {2021})}\BibitemShut {NoStop}%
\bibitem [{\citenamefont {Chen}\ \emph
  {et~al.}(2021{\natexlab{b}})\citenamefont {Chen}, \citenamefont {Xu},
  \citenamefont {Zhang}, \citenamefont {Gan}, \citenamefont {Liu},\ and\
  \citenamefont {Wang}}]{Chen2022FOP}%
  \BibitemOpen
  \bibfield  {author} {\bibinfo {author} {\bibfnamefont {F.}~\bibnamefont
  {Chen}}, \bibinfo {author} {\bibfnamefont {A.~G.}\ \bibnamefont {Xu}},
  \bibinfo {author} {\bibfnamefont {Y.~D.}\ \bibnamefont {Zhang}}, \bibinfo
  {author} {\bibfnamefont {Y.~B.}\ \bibnamefont {Gan}}, \bibinfo {author}
  {\bibfnamefont {B.~B.}\ \bibnamefont {Liu}}, \ and\ \bibinfo {author}
  {\bibfnamefont {S.}~\bibnamefont {Wang}},\ }\bibfield  {title} {\enquote
  {\bibinfo {title} {Effects of the initial perturbations on the
  {Rayleigh-Taylor-Kelvin-Helmholtz} instability system},}\ }\href {\doibase
  https://doi.org/10.1007/s11467-021-1145-y} {\bibfield  {journal} {\bibinfo
  {journal} {Front. Phys.}\ }\textbf {\bibinfo {volume} {17}},\ \bibinfo
  {pages} {33505} (\bibinfo {year} {2021}{\natexlab{b}})}\BibitemShut {NoStop}%
\bibitem [{\citenamefont {Chen}\ \emph {et~al.}(2022)\citenamefont {Chen},
  \citenamefont {Xu}, \citenamefont {Chen}, \citenamefont {Zhang},\ and\
  \citenamefont {Chen}}]{Chen2022PRE}%
  \BibitemOpen
  \bibfield  {author} {\bibinfo {author} {\bibfnamefont {J.}~\bibnamefont
  {Chen}}, \bibinfo {author} {\bibfnamefont {A.~G.}\ \bibnamefont {Xu}},
  \bibinfo {author} {\bibfnamefont {D.~W.}\ \bibnamefont {Chen}}, \bibinfo
  {author} {\bibfnamefont {Y.~D.}\ \bibnamefont {Zhang}}, \ and\ \bibinfo
  {author} {\bibfnamefont {Z.~H.}\ \bibnamefont {Chen}},\ }\bibfield  {title}
  {\enquote {\bibinfo {title} {Discrete {Boltzmann} modeling of
  {Rayleigh-Taylor} instability: effects of interfacial tension, viscosity and
  heat conductivity},}\ }\href {\doibase 10.1103/PhysRevE.00.005100} {\ \textbf
  {\bibinfo {volume} {00}},\ \bibinfo {pages} {005100} (\bibinfo {year}
  {2022})}\BibitemShut {NoStop}%
\bibitem [{\citenamefont {Gan}\ \emph {et~al.}(2011)\citenamefont {Gan},
  \citenamefont {Xu}, \citenamefont {Zhang}, \citenamefont {Li},\ and\
  \citenamefont {Li}}]{gan2011PRE}%
  \BibitemOpen
  \bibfield  {author} {\bibinfo {author} {\bibfnamefont {Y.~B.}\ \bibnamefont
  {Gan}}, \bibinfo {author} {\bibfnamefont {A.~G.}\ \bibnamefont {Xu}},
  \bibinfo {author} {\bibfnamefont {G.~C.}\ \bibnamefont {Zhang}}, \bibinfo
  {author} {\bibfnamefont {Y.~J.}\ \bibnamefont {Li}}, \ and\ \bibinfo {author}
  {\bibfnamefont {H.}~\bibnamefont {Li}},\ }\bibfield  {title} {\enquote
  {\bibinfo {title} {Phase separation in thermal systems: A lattice {Boltzmann}
  study and morphological characterization},}\ }\href {\doibase
  10.1103/PhysRevE.84.046715} {\bibfield  {journal} {\bibinfo  {journal} {Phys.
  Rev. E}\ }\textbf {\bibinfo {volume} {84}},\ \bibinfo {pages} {046715}
  (\bibinfo {year} {2011})}\BibitemShut {NoStop}%
\bibitem [{\citenamefont {Gan}\ \emph {et~al.}(2015)\citenamefont {Gan},
  \citenamefont {Xu}, \citenamefont {Zhang},\ and\ \citenamefont
  {Succi}}]{Gan2015Soft}%
  \BibitemOpen
  \bibfield  {author} {\bibinfo {author} {\bibfnamefont {Y.~B.}\ \bibnamefont
  {Gan}}, \bibinfo {author} {\bibfnamefont {A.~G.}\ \bibnamefont {Xu}},
  \bibinfo {author} {\bibfnamefont {G.~C.}\ \bibnamefont {Zhang}}, \ and\
  \bibinfo {author} {\bibfnamefont {S.}~\bibnamefont {Succi}},\ }\bibfield
  {title} {\enquote {\bibinfo {title} {Discrete {Boltzmann} modeling of
  multiphase flows: {Hydrodynamic} and thermodynamic non-equilibrium
  effects},}\ }\href@noop {} {\bibfield  {journal} {\bibinfo  {journal} {Soft
  Matter}\ }\textbf {\bibinfo {volume} {11}},\ \bibinfo {pages} {5336--5345}
  (\bibinfo {year} {2015})}\BibitemShut {NoStop}%
\bibitem [{\citenamefont {Zhang}\ \emph
  {et~al.}(2019{\natexlab{b}})\citenamefont {Zhang}, \citenamefont {Xu},
  \citenamefont {Zhang}, \citenamefont {Gan}, \citenamefont {Chen},\ and\
  \citenamefont {Succi}}]{Zhang2019Matter}%
  \BibitemOpen
  \bibfield  {author} {\bibinfo {author} {\bibfnamefont {Y.~D.}\ \bibnamefont
  {Zhang}}, \bibinfo {author} {\bibfnamefont {A.~G.}\ \bibnamefont {Xu}},
  \bibinfo {author} {\bibfnamefont {G.~C.}\ \bibnamefont {Zhang}}, \bibinfo
  {author} {\bibfnamefont {Y.~B.}\ \bibnamefont {Gan}}, \bibinfo {author}
  {\bibfnamefont {Z.~H.}\ \bibnamefont {Chen}}, \ and\ \bibinfo {author}
  {\bibfnamefont {S.}~\bibnamefont {Succi}},\ }\bibfield  {title} {\enquote
  {\bibinfo {title} {Entropy production in thermal phase separation: a
  kinetic-theory approach},}\ }\href {\doibase 10.1039/C8SM02637H} {\bibfield
  {journal} {\bibinfo  {journal} {Soft Matter}\ }\textbf {\bibinfo {volume}
  {15}},\ \bibinfo {pages} {2245--2259} (\bibinfo {year}
  {2019}{\natexlab{b}})}\BibitemShut {NoStop}%
\bibitem [{\citenamefont {Zhang}\ \emph
  {et~al.}(2020{\natexlab{a}})\citenamefont {Zhang}, \citenamefont {Xu},
  \citenamefont {Qiu}, \citenamefont {Wei},\ and\ \citenamefont
  {Wei}}]{Zhang2020FOP}%
  \BibitemOpen
  \bibfield  {author} {\bibinfo {author} {\bibfnamefont {Y.~D.}\ \bibnamefont
  {Zhang}}, \bibinfo {author} {\bibfnamefont {A.~G.}\ \bibnamefont {Xu}},
  \bibinfo {author} {\bibfnamefont {J.~J.}\ \bibnamefont {Qiu}}, \bibinfo
  {author} {\bibfnamefont {H.~T.}\ \bibnamefont {Wei}}, \ and\ \bibinfo
  {author} {\bibfnamefont {Z.~H.}\ \bibnamefont {Wei}},\ }\bibfield  {title}
  {\enquote {\bibinfo {title} {Kinetic modeling of multiphase flow based on
  simplified {Enskog} equation},}\ }\href {\doibase 10.1007/s11467-020-1014-0}
  {\bibfield  {journal} {\bibinfo  {journal} {Front. Phys.}\ }\textbf {\bibinfo
  {volume} {15}},\ \bibinfo {pages} {62503} (\bibinfo {year}
  {2020}{\natexlab{a}})}\BibitemShut {NoStop}%
\bibitem [{\citenamefont {Liu}\ \emph {et~al.}(2022)\citenamefont {Liu},
  \citenamefont {Song}, \citenamefont {Xu}, \citenamefont {Zhang},\ and\
  \citenamefont {Xie}}]{Liu2022JMES}%
  \BibitemOpen
  \bibfield  {author} {\bibinfo {author} {\bibfnamefont {Z.~P.}\ \bibnamefont
  {Liu}}, \bibinfo {author} {\bibfnamefont {J.~H.}\ \bibnamefont {Song}},
  \bibinfo {author} {\bibfnamefont {A.~G.}\ \bibnamefont {Xu}}, \bibinfo
  {author} {\bibfnamefont {Y.~D.}\ \bibnamefont {Zhang}}, \ and\ \bibinfo
  {author} {\bibfnamefont {K.}~\bibnamefont {Xie}},\ }\bibfield  {title}
  {\enquote {\bibinfo {title} {Discrete {Boltzmann} modeling of plasma shock
  wave},}\ }\href {\doibase 10.1177/09544062221075943} {\bibfield  {journal}
  {\bibinfo  {journal} {J. Mech. Eng. Sci.}\ } (\bibinfo {year} {2022}),\
  10.1177/09544062221075943}\BibitemShut {NoStop}%
\bibitem [{\citenamefont {Lin}\ \emph {et~al.}(2018)\citenamefont {Lin},
  \citenamefont {Luo}, \citenamefont {Gan},\ and\ \citenamefont
  {Lai}}]{Lin2018Binary}%
  \BibitemOpen
  \bibfield  {author} {\bibinfo {author} {\bibfnamefont {C.~D.}\ \bibnamefont
  {Lin}}, \bibinfo {author} {\bibfnamefont {K.~H.}\ \bibnamefont {Luo}},
  \bibinfo {author} {\bibfnamefont {Y.~B.}\ \bibnamefont {Gan}}, \ and\
  \bibinfo {author} {\bibfnamefont {H.~L.}\ \bibnamefont {Lai}},\ }\bibfield
  {title} {\enquote {\bibinfo {title} {Thermodynamic {Nonequilibrium}
  {Features} in {Binary} {Diffusion}},}\ }\href {\doibase
  10.1088/0253-6102/69/6/722} {\bibfield  {journal} {\bibinfo  {journal}
  {Commun. Theor. Phys.}\ }\textbf {\bibinfo {volume} {69}},\ \bibinfo {pages}
  {722--726} (\bibinfo {year} {2018})}\BibitemShut {NoStop}%
\bibitem [{\citenamefont {Lin}\ \emph {et~al.}(2017{\natexlab{b}})\citenamefont
  {Lin}, \citenamefont {Luo}, \citenamefont {Fei},\ and\ \citenamefont
  {Succi}}]{Lin2017SR}%
  \BibitemOpen
  \bibfield  {author} {\bibinfo {author} {\bibfnamefont {C.~D.}\ \bibnamefont
  {Lin}}, \bibinfo {author} {\bibfnamefont {K.~H.}\ \bibnamefont {Luo}},
  \bibinfo {author} {\bibfnamefont {L.~L.}\ \bibnamefont {Fei}}, \ and\
  \bibinfo {author} {\bibfnamefont {S.}~\bibnamefont {Succi}},\ }\bibfield
  {title} {\enquote {\bibinfo {title} {A multi-component discrete {Boltzmann}
  model for nonequilibrium reactive flows},}\ }\href {\doibase
  10.1038/s41598-017-14824-9} {\bibfield  {journal} {\bibinfo  {journal} {Sci.
  Rep.}\ }\textbf {\bibinfo {volume} {7}},\ \bibinfo {pages} {14580} (\bibinfo
  {year} {2017}{\natexlab{b}})}\BibitemShut {NoStop}%
\bibitem [{\citenamefont {Zhang}\ \emph
  {et~al.}(2019{\natexlab{c}})\citenamefont {Zhang}, \citenamefont {Xu},
  \citenamefont {Zhang}, \citenamefont {Chen},\ and\ \citenamefont
  {Wei}}]{2019Zhang-Shakhov}%
  \BibitemOpen
  \bibfield  {author} {\bibinfo {author} {\bibfnamefont {Y.~D.}\ \bibnamefont
  {Zhang}}, \bibinfo {author} {\bibfnamefont {A.~G.}\ \bibnamefont {Xu}},
  \bibinfo {author} {\bibfnamefont {G.~C.}\ \bibnamefont {Zhang}}, \bibinfo
  {author} {\bibfnamefont {Z.~H.}\ \bibnamefont {Chen}}, \ and\ \bibinfo
  {author} {\bibfnamefont {P.}~\bibnamefont {Wei}},\ }\bibfield  {title}
  {\enquote {\bibinfo {title} {Discrete {Boltzmann} method for non-equilibrium
  flows: {Based} on {Shakhov} model},}\ }\href {\doibase
  https://doi.org/10.1016/j.cpc.2018.12.018} {\bibfield  {journal} {\bibinfo
  {journal} {Comput. Phys. Commun.}\ }\textbf {\bibinfo {volume} {238}},\
  \bibinfo {pages} {50--65} (\bibinfo {year} {2019}{\natexlab{c}})}\BibitemShut
  {NoStop}%
\bibitem [{\citenamefont {Gan}\ \emph {et~al.}(2022)\citenamefont {Gan},
  \citenamefont {Xu}, \citenamefont {Lai}, \citenamefont {Li}, \citenamefont
  {Sun},\ and\ \citenamefont {Succi}}]{Gan2022JFM}%
  \BibitemOpen
  \bibfield  {author} {\bibinfo {author} {\bibfnamefont {Y.~B.}\ \bibnamefont
  {Gan}}, \bibinfo {author} {\bibfnamefont {A.~G.}\ \bibnamefont {Xu}},
  \bibinfo {author} {\bibfnamefont {H.~L.}\ \bibnamefont {Lai}}, \bibinfo
  {author} {\bibfnamefont {W.}~\bibnamefont {Li}}, \bibinfo {author}
  {\bibfnamefont {G.~L.}\ \bibnamefont {Sun}}, \ and\ \bibinfo {author}
  {\bibfnamefont {S.}~\bibnamefont {Succi}},\ }\bibfield  {title} {\enquote
  {\bibinfo {title} {Discrete {Boltzmann} multi-scale modeling of
  non-equilibrium multiphase flows},}\ }\href {\doibase
  10.48550/ARXIV.2203.12458} {\  (\bibinfo {year} {2022}),\
  10.48550/ARXIV.2203.12458}\BibitemShut {NoStop}%
\bibitem [{\citenamefont {Bhatnagar}, \citenamefont {Gross},\ and\
  \citenamefont {Krook}(1954)}]{BGK1954}%
  \BibitemOpen
  \bibfield  {author} {\bibinfo {author} {\bibfnamefont {B.~L.}\ \bibnamefont
  {Bhatnagar}}, \bibinfo {author} {\bibfnamefont {E.~P.}\ \bibnamefont
  {Gross}}, \ and\ \bibinfo {author} {\bibfnamefont {M.~K.}\ \bibnamefont
  {Krook}},\ }\bibfield  {title} {\enquote {\bibinfo {title} {A model for
  collision processes in gases. {I}. {Small} amplitude processes in charged and
  neutral one-component systems},}\ }\href {\doibase
  https://doi.org/10.1103/PhysRev.94.511} {\bibfield  {journal} {\bibinfo
  {journal} {Phys. Rev.}\ }\textbf {\bibinfo {volume} {94}},\ \bibinfo {pages}
  {511--525} (\bibinfo {year} {1954})}\BibitemShut {NoStop}%
\bibitem [{\citenamefont {Holway}(1966)}]{1966ES}%
  \BibitemOpen
  \bibfield  {author} {\bibinfo {author} {\bibfnamefont {J.~L.~H.}\
  \bibnamefont {Holway}},\ }\bibfield  {title} {\enquote {\bibinfo {title} {New
  statistical methods for kinetic theory: methods of construction},}\ }\href
  {\doibase https://doi.org/10.1063/1.1761920} {\bibfield  {journal} {\bibinfo
  {journal} {Phys. Fluids}\ }\textbf {\bibinfo {volume} {9}},\ \bibinfo {pages}
  {1658} (\bibinfo {year} {1966})}\BibitemShut {NoStop}%
\bibitem [{\citenamefont {Zhang}\ \emph
  {et~al.}(2020{\natexlab{b}})\citenamefont {Zhang}, \citenamefont {Xu},
  \citenamefont {Zhang},\ and\ \citenamefont {Li}}]{Zhang2020POF}%
  \BibitemOpen
  \bibfield  {author} {\bibinfo {author} {\bibfnamefont {D.~J.}\ \bibnamefont
  {Zhang}}, \bibinfo {author} {\bibfnamefont {A.~G.}\ \bibnamefont {Xu}},
  \bibinfo {author} {\bibfnamefont {Y.~D.}\ \bibnamefont {Zhang}}, \ and\
  \bibinfo {author} {\bibfnamefont {Y.~J.}\ \bibnamefont {Li}},\ }\bibfield
  {title} {\enquote {\bibinfo {title} {Two-fluid discrete {Boltzmann} model for
  compressible flows: based on {Ellipsoidal} {Statistical}
  {Bhatnagar-Gross-Krook}},}\ }\href {\doibase 10.1063/5.0017673} {\bibfield
  {journal} {\bibinfo  {journal} {Phys. Fluids}\ }\textbf {\bibinfo {volume}
  {32}},\ \bibinfo {pages} {126110} (\bibinfo {year}
  {2020}{\natexlab{b}})}\BibitemShut {NoStop}%
\bibitem [{Note4()}]{Note4}%
  \BibitemOpen
  \bibinfo {note} {The BGK-like model refers to the model of Boltzmann equation
  which is similar in form to the BGK model.}\BibitemShut {Stop}%
\bibitem [{\citenamefont {Xu}()}]{Xu20220304-DBM-note}%
  \BibitemOpen
  \bibfield  {author} {\bibinfo {author} {\bibfnamefont {A.~G.}\ \bibnamefont
  {Xu}},\ }\href@noop {} {\enquote {\bibinfo {title} {Questions \& {Replies} on
  {D}{B}{M} (continued)},}\ }\bibinfo {note}
  {\url{https://www.koushare.com/post/postdetail/5267} Accessed July 8,
  2022}\BibitemShut {NoStop}%
\bibitem [{\citenamefont {Li}\ \emph {et~al.}(2022)\citenamefont {Li},
  \citenamefont {Xu}, \citenamefont {Zhang},\ and\ \citenamefont
  {Shan}}]{Li2022CTP}%
  \BibitemOpen
  \bibfield  {author} {\bibinfo {author} {\bibfnamefont {H.~W.}\ \bibnamefont
  {Li}}, \bibinfo {author} {\bibfnamefont {A.~G.}\ \bibnamefont {Xu}}, \bibinfo
  {author} {\bibfnamefont {G.}~\bibnamefont {Zhang}}, \ and\ \bibinfo {author}
  {\bibfnamefont {Y.~M.}\ \bibnamefont {Shan}},\ }\bibfield  {title} {\enquote
  {\bibinfo {title} {Rayleigh-taylor instability under multi-mode perturbation:
  discrete {Boltzmann} modeling with tracers},}\ }\href {\doibase
  https://doi.org/10.48550/arXiv.2205.14316} {\bibfield  {journal} {\bibinfo
  {journal} {Commun. Theor. Phys.}\ } (\bibinfo {year} {2022}),\
  https://doi.org/10.48550/arXiv.2205.14316}\BibitemShut {NoStop}%
\bibitem [{\citenamefont {Shakhov}(1968)}]{Shakhov1968FD}%
  \BibitemOpen
  \bibfield  {author} {\bibinfo {author} {\bibfnamefont {E.~M.}\ \bibnamefont
  {Shakhov}},\ }\bibfield  {title} {\enquote {\bibinfo {title} {Generalization
  of the {Krook} kinetic relaxation equation},}\ }\href {\doibase
  10.1007/BF01029546} {\bibfield  {journal} {\bibinfo  {journal} {Fluid
  Dynam.}\ }\textbf {\bibinfo {volume} {3}},\ \bibinfo {pages} {95--96}
  (\bibinfo {year} {1968})}\BibitemShut {NoStop}%
\bibitem [{\citenamefont {Larina}\ and\ \citenamefont
  {Rykov}(2010)}]{2010LarinaCMMP}%
  \BibitemOpen
  \bibfield  {author} {\bibinfo {author} {\bibfnamefont {I.~N.}\ \bibnamefont
  {Larina}}\ and\ \bibinfo {author} {\bibfnamefont {V.~A.}\ \bibnamefont
  {Rykov}},\ }\bibfield  {title} {\enquote {\bibinfo {title} {Kinetic model of
  the {Boltzmann} equation for a diatomic gas with rotational degrees of
  freedom},}\ }\href {\doibase https://doi.org/10.1134/S0965542510120134}
  {\bibfield  {journal} {\bibinfo  {journal} {Comp. Math. Math. Phys.}\
  }\textbf {\bibinfo {volume} {50}},\ \bibinfo {pages} {2118--2130} (\bibinfo
  {year} {2010})}\BibitemShut {NoStop}%
\bibitem [{\citenamefont {Liu}(1990)}]{1990LiuPOF}%
  \BibitemOpen
  \bibfield  {author} {\bibinfo {author} {\bibfnamefont {G.~J.}\ \bibnamefont
  {Liu}},\ }\bibfield  {title} {\enquote {\bibinfo {title} {A method for
  constructing a model form for the {Boltzmann} equation},}\ }\href {\doibase
  https://doi.org/10.1063/1.857777} {\bibfield  {journal} {\bibinfo  {journal}
  {Phys. Fluids}\ }\textbf {\bibinfo {volume} {2}},\ \bibinfo {pages} {277}
  (\bibinfo {year} {1990})}\BibitemShut {NoStop}%
\bibitem [{\citenamefont {Zhang}\ and\ \citenamefont
  {Zhuang}(1991)}]{zhang1991nnd}%
  \BibitemOpen
  \bibfield  {author} {\bibinfo {author} {\bibfnamefont {H.~X.}\ \bibnamefont
  {Zhang}}\ and\ \bibinfo {author} {\bibfnamefont {F.~G.}\ \bibnamefont
  {Zhuang}},\ }\bibfield  {title} {\enquote {\bibinfo {title} {N{ND} schemes
  and their applications to numerical simulation of two-and three-dimensional
  flows},}\ }in\ \href {\doibase https://doi.org/10.1016/S0065-2156(08)70165-0}
  {\emph {\bibinfo {booktitle} {Adv. Appl. Mech.}}},\ Vol.~\bibinfo {volume}
  {29}\ (\bibinfo  {publisher} {Elsevier},\ \bibinfo {year} {1991})\ pp.\
  \bibinfo {pages} {193--256}\BibitemShut {NoStop}%
\bibitem [{\citenamefont {Zhang}\ \emph {et~al.}(2018)\citenamefont {Zhang},
  \citenamefont {Xu}, \citenamefont {Zhang},\ and\ \citenamefont
  {Chen}}]{Zhang2018CTP}%
  \BibitemOpen
  \bibfield  {author} {\bibinfo {author} {\bibfnamefont {Y.~D.}\ \bibnamefont
  {Zhang}}, \bibinfo {author} {\bibfnamefont {A.~G.}\ \bibnamefont {Xu}},
  \bibinfo {author} {\bibfnamefont {G.~C.}\ \bibnamefont {Zhang}}, \ and\
  \bibinfo {author} {\bibfnamefont {Z.~H.}\ \bibnamefont {Chen}},\ }\bibfield
  {title} {\enquote {\bibinfo {title} {Discrete {Boltzmann} {Method} with
  {Maxwell-Type} {Boundary} {Condition} for {Slip} {Flow}},}\ }\href {\doibase
  10.1088/0253-6102/69/1/77} {\bibfield  {journal} {\bibinfo  {journal}
  {Commun. Theor. Phys.}\ }\textbf {\bibinfo {volume} {69}},\ \bibinfo {pages}
  {77} (\bibinfo {year} {2018})}\BibitemShut {NoStop}%
\bibitem [{\citenamefont {Torrilhon}\ and\ \citenamefont
  {Struchtrup}(2004)}]{Torrilhon04regularized13-moment}%
  \BibitemOpen
  \bibfield  {author} {\bibinfo {author} {\bibfnamefont {M.}~\bibnamefont
  {Torrilhon}}\ and\ \bibinfo {author} {\bibfnamefont {H.}~\bibnamefont
  {Struchtrup}},\ }\bibfield  {title} {\enquote {\bibinfo {title} {Regularized
  13-moment equations: shock structure calculations and comparison to {Burnett}
  models},}\ }\href {\doibase https://doi.org/10.1017/S0022112004009917}
  {\bibfield  {journal} {\bibinfo  {journal} {J. Fluid Mech.}\ }\textbf
  {\bibinfo {volume} {513}},\ \bibinfo {pages} {171--198} (\bibinfo {year}
  {2004})}\BibitemShut {NoStop}%
\bibitem [{\citenamefont {Li}\ and\ \citenamefont {Zhang}(2007)}]{Li2007Gas}%
  \BibitemOpen
  \bibfield  {author} {\bibinfo {author} {\bibfnamefont {Z.~H.}\ \bibnamefont
  {Li}}\ and\ \bibinfo {author} {\bibfnamefont {H.~X.}\ \bibnamefont {Zhang}},\
  }\bibfield  {title} {\enquote {\bibinfo {title} {Gas-kinetic description of
  shock wave structures by solving {Boltzmann} model equation},}\ }\href
  {\doibase https://doi.org/10.1080/10618560802395117} {\bibfield  {journal}
  {\bibinfo  {journal} {Acta Aerody. Sin.}\ }\textbf {\bibinfo {volume} {25}},\
  \bibinfo {pages} {411--418} (\bibinfo {year} {2007})}\BibitemShut {NoStop}%
\bibitem [{\citenamefont {Eggers}\ and\ \citenamefont
  {Villermaux}(2008)}]{Eggers2008Physics}%
  \BibitemOpen
  \bibfield  {author} {\bibinfo {author} {\bibfnamefont {J.}~\bibnamefont
  {Eggers}}\ and\ \bibinfo {author} {\bibfnamefont {E.}~\bibnamefont
  {Villermaux}},\ }\bibfield  {title} {\enquote {\bibinfo {title} {Physics of
  {Liquid} {Jets}},}\ }\href {\doibase {}} {\bibfield  {journal} {\bibinfo
  {journal} {Reports on Progress in Physics}\ }\textbf {\bibinfo {volume}
  {71}},\ \bibinfo {pages} {036601} (\bibinfo {year} {2008})}\BibitemShut
  {NoStop}%
\bibitem [{\citenamefont {Gan}\ \emph {et~al.}(2019)\citenamefont {Gan},
  \citenamefont {Xu}, \citenamefont {Zhang}, \citenamefont {Lin}, \citenamefont
  {Lai},\ and\ \citenamefont {Liu}}]{Gan2019FOP}%
  \BibitemOpen
  \bibfield  {author} {\bibinfo {author} {\bibfnamefont {Y.~B.}\ \bibnamefont
  {Gan}}, \bibinfo {author} {\bibfnamefont {A.~G.}\ \bibnamefont {Xu}},
  \bibinfo {author} {\bibfnamefont {G.~C.}\ \bibnamefont {Zhang}}, \bibinfo
  {author} {\bibfnamefont {C.~D.}\ \bibnamefont {Lin}}, \bibinfo {author}
  {\bibfnamefont {H.~L.}\ \bibnamefont {Lai}}, \ and\ \bibinfo {author}
  {\bibfnamefont {Z.~P.}\ \bibnamefont {Liu}},\ }\bibfield  {title} {\enquote
  {\bibinfo {title} {Nonequilibrium and morphological characterizations of
  {Kelvin-Helmholtz} instability in compressible flows},}\ }\href {\doibase
  10.1007/s11467-019-0885-4} {\bibfield  {journal} {\bibinfo  {journal} {Front.
  Phys.}\ }\textbf {\bibinfo {volume} {14}},\ \bibinfo {pages} {43602}
  (\bibinfo {year} {2019})}\BibitemShut {NoStop}%
\end{thebibliography}%

\end{document}